\documentclass[10pt]{article}
\usepackage{amsfonts,amsmath,color,graphicx,stmaryrd,mathabx,amssymb,mathdots,esint,tikz,lscape}
\usepackage{pgfplots}
\usepackage{float}

\DeclareMathAlphabet{\mathpzc}{OT1}{pzc}{m}{it}

\makeatletter
\@addtoreset{equation}{section}
\makeatother
\newtheorem{theorem}{Theorem}[section]
\newtheorem{conjecture}[theorem]{Conjecture}
\newtheorem{proposition}[theorem]{Proposition}
\newtheorem{lemma}[theorem]{Lemma}
\newtheorem{corollary}[theorem]{Corollary}

\newtheorem{definition}[theorem]{Definition}

\newtheorem{remark}[theorem]{Remark}
\newtheorem{problem}{Problem}
\newtheorem{question}{Question}
\newtheorem{hypothesis}[theorem]{Hypothesis}
\newcommand{\be}{\begin{equation}}
\newcommand{\ee}{\end{equation}}
\newcommand{\ba}{\begin{aligned}}
\newcommand{\ea}{\end{aligned}}
\def\br{\begin{remark}\rm\small}
\def\er{\end{remark}}
\def\bt{\begin{theorem}}
\def\et{\end{theorem}}
\def\bcj{\begin{conjecture}}
\def\ecj{\end{conjecture}}
\def\bd{\begin{definition}}
\def\ed{\end{definition}}
\def\bp{\begin{proposition}}
\def\ep{\end{proposition}}
\def\bl{\begin{lemma}}
\def\el{\end{lemma}}
\def\bc{\begin{corollary}}
\def\ec{\end{corollary}}
\def\bh{\begin{hypothesis}}
\def\eh{\end{hypothesis}}
\def\beaq{\begin{eqnarray}}
\def\eeaq{\end{eqnarray}}

\newcommand{\sheet}[2]{\stackrel{{#2}}{#1}}

\newcommand{\beq}{\begin{equation}}
\newcommand{\eeq}{\end{equation}}
\newcommand{\bea}{\begin{eqnarray}}
\newcommand{\eea}{\end{eqnarray}}

\newcommand{\dd}{\mathrm{d}}

\newcommand{\Res}{\mathop{\,\rm Res\,}}

\definecolor{rouge}{rgb}{0.84,0.18,0.07}
\definecolor{bleu}{rgb}{0.22,0.41,0.74}
\definecolor{vertf}{rgb}{0.08,0.46,0.07}
\usepackage[pdftex]{hyperref}
\hypersetup{colorlinks,urlcolor=vertf,citecolor=rouge,linkcolor=bleu,filecolor=black}

\makeatletter
\newcommand\footnoteref[1]{\protected@xdef\@thefnmark{\ref{#1}}\@footnotemark}
\makeatother

\begin{document}

\sloppy

\begin{flushright}
IPhT T14/100 \\ CRM-3339
\end{flushright}

\pagestyle{empty}
\addtolength{\baselineskip}{0.20\baselineskip}
\begin{center}

\vspace{26pt}

{\Large \textsf{\textbf{Root systems, spectral curves, and analysis of a Chern-Simons matrix model for Seifert fibered spaces}}}

\vspace{26pt}

\textsl{Ga\"etan Borot}\footnote{Section de Math\'ematiques, Universit\'e de Gen\`eve, 2-4 rue du Li\`evre 1206 Gen\`eve, Switzerland.}\footnote{MIT, Maths Department, Massachusetts Avenue 77, Cambridge 02139, USA}\footnote{\label{note1}Max Planck Institut f\"ur Mathematik, Vivatsgasse 7, 53111 Bonn, Germany.}, \textsl{Bertrand Eynard} \!\footnote{Institut de Physique Th\'eorique, CEA Saclay, Orme des Merisiers, 91191 Gif-sur-Yvette, France.}\footnote{Centre de Recherches Math\'ematiques, 2920, Chemin de la Tour, Montr\'eal, QC, Canada.} \\ with a section by \textsl{Alexander Weisse}\footnoteref{note1}
\end{center}

\vspace{20pt}

\begin{center}
\textbf{Abstract}
\end{center}

\noindent We study a class of scalar, linear, non-local Riemann-Hilbert problems (RHP) involving finite subgroups of $\mathrm{PSL}_2(\mathbb{C})$. We associate to such problems a (maybe infinite) root system and describe the relevance of the orbits of the Weyl group in the construction of its solutions. As an application, we study in detail the large $N$ expansion of $\mathrm{SU}(N)$ and $\mathrm{SO}(N)/\mathrm{Sp}(2N)$ Chern-Simons partition function $Z_N(M)$ of $3$-manifolds $M$ that are either rational homology spheres or more generally Seifert fibered spaces. It has a matrix model-like representation, whose spectral curve can be characterized in terms of a RHP as above. When $\pi_1(M)$ is finite (i.e. for manifolds $M$ that are quotients of $\mathbb{S}_{3}$ by a finite isometry group of type ADE), the Weyl group associated to the RHP is finite and the spectral curve is algebraic and can be in principle computed. We then show that the large $N$ expansion of $Z_N(M)$ is computed by the topological recursion. This has consequences for the analyticity properties of $\mathrm{SU}/\mathrm{SO}/\mathrm{Sp}$ perturbative invariants of knots along fibers in $M$.

\tableofcontents*

%





\vspace{26pt}
\pagestyle{plain}
\setcounter{page}{1}


\subsection*{Acknowledgements}

G.B. thanks Andrea Brini, R\'eda Chhaibi, Thomas Haettel, Neil Hoffman, Thang L\^{e}, Alexei Oblomkov, George Thompson, and especially Albrecht Klemm, Stavros Garoufalidis and Marcos Mari{\~{n}}o for comments and fruitful discussions, as well as the organizers of the Quantum Topology and Hyperbolic Geometry conference in Nha Trang (May 2013), where a preliminary version of this work was presented. Marcos Mari{\~{n}}o and Alba Grassi collaborated in an earlier phase to the Section 8 concerning Kauffman invariants. This work benefited from the support of Fonds Europ\'een S16905 (UE7--CONFRA), the Swiss NSF, the Max-Planck-Gesellschaft, and the Simons foundation.

B.E. thanks T.~Kimura, the CRM Montr\'eal, the FQRNT grant from the Qu\'ebec government, P.~Su\l{}kowski and the ERC starting grant Fields-Knots, S.~Smirnov and the University of Geneva. 

\vspace{0.2cm}

\noindent \textbf{AMS Classification:} 14Hxx, 45Exx, 51Pxx, 57M27, 60B20, 81T45.

\section{Introduction}

Unless mentioned otherwise, the orbit space of Seifert fibered spaces in this article is assumed to be a sphere.

\subsection{Scope}

$\mathfrak{g}$ denotes a semi-simple Lie algebra, and $\mathfrak{h}$ its Cartan subalgebra, identified with $\mathbb{R}^N$. We shall study the probability measure over $\mathbf{t} = (t_1,\ldots,t_N) \in \mathfrak{h}$:
\beq
\label{themodel0}\dd\mu(\mathbf{t}) = \frac{1}{Z^{\mathfrak{g}}(M)} \Bigg[\prod_{\alpha > 0} \mathrm{sinh}^{2 - r}\big(\frac{\alpha\cdot \mathbf{t}}{2}\big) \prod_{m = 1}^r \mathrm{sinh}\big(\frac{\alpha\cdot\mathbf{t}}{2a_m}\big)\Bigg] \prod_{j = 1}^N e^{-NV(t_j)}\,\dd t_j,\qquad V(t) = \frac{t^2}{2u}.
\eeq
where  the product ranges over positive roots of $\mathfrak{g}$, $u$ is a positive parameter,
and the partition function $Z^{\mathfrak{g}}(M)$ is such that $\int_{\mathfrak{h}} \dd\mu(\mathbf{t}) = 1$. We are interested in the $A_N, B_N, C_N, D_N$ series of Lie algebras, in the regime where the rank $N$ goes to $\infty$.

The model \eqref{themodel0} arises in Chern-Simons theory with gauge group $\exp(\mathfrak{g})$ on a simple class of $3$-manifolds $M$ called Seifert fibered spaces \cite{MarinoCSM}: roughly speaking, these are $\mathbb{S}_{1}$-fibrations over a surface orbifold, here assumed to be topologically a sphere. \eqref{themodel0} captures the contribution of the trivial flat connection in perturbative Chern-Simons theory, or equivalent, the evaluation of the LMO invariants of $M$ in the weight system defined by $\mathfrak{g}$ \cite{LMO,AarhusI,AarhusII,AarhusII,AarhusIII}. Moreover, the correlation functions in \eqref{themodel0} compute the colored HOMFLY (in any representation of fixed size) of the knots going along the fibers of $M$, see \S~\ref{Wilson}. The regime $N \rightarrow \infty$ has aroused interest since it allows a rigorous definition of perturbative invariants ensuing from $Z_N^{\mathfrak{g}}(M)$ and the colored HOMFLY of links in $M$, which should be related via geometric transitions to topological strings invariants in $T^*M$ according to the physics literature \cite{WittenCS}, see \S~\ref{13para}.

General arguments \cite{BEO} show that the large $N$ asymptotic expansion of models of the type \eqref{themodel0} -- once it is proven to exist using the tools of \cite{BGK} -- can be computed by the topological recursion of \cite{EOFg}. It then remains to compute the initial data $(\omega_{0,1},\omega_{0,2})$ of the recursion : $\omega_{0,1}$ is related to the equilibrium measure of \eqref{themodel}, also called \emph{spectral curve}, and $\omega_{0,2}$ is related to the large $N$ limit of the $2$-point correlation function. Both are characterized by a saddle point equation which takes the form of a linear but non-local Riemann-Hilbert problem (RHP) on a cut locus to determine. Addressing its solution is an important part of our present work. We shall devise in Section~\ref{S3} a general method to construct the solution of a large class of RHP where the jumps are obtained by action of a finite subgroup of M\"obius transformations. We actually build the -- maybe infinite -- monodromy group of the solution, and relate it to the Weyl group of a root system. The solution can be studied in more details by algebro-geometric means when this Weyl group is finite. This construction applies to the computation of $(\omega_{0,1},\omega_{0,2})$ relevant for \eqref{themodel}, but has also its own interest and can be used as a tool in other problems.

\vspace{0.2cm}

Our work has two noteworthy consequences for knot theory.

\vspace{0.2cm}

Firstly, we obtain analyticity results on perturbative invariants of knots in manifolds different from the $3$-sphere. The colored HOMFLY of links in $\mathbb{S}_{3}$ are polynomials (after a suitable normalization) in $q$ and $q^{N}$. This is clear from skein theory, and can also be explained from quantum field theory perspective \cite{Wittenlecture}. This implies that the coefficients in the $\hbar \rightarrow 0$ expansion while keeping $q^{N} = e^{u_0}$ fixed, produce polynomials in $e^{u_0}$. These properties are not expected to be true for invariants of links in other $3$-manifolds. We show how to compute invariants of fiber knots in Seifert fibered spaces. Remarkably, the position of the singularities in the $u = N\hbar$-complex plane only depend on the ambient $3$-manifold. They occur at the singularities of the family of spectral curves (parametrized by $u$) associated to \eqref{themodel0}. We conjecture more generally that for any link in a rational homology sphere $M$, the singularities of the perturbative invariants only depend on the ambient manifold $M$. More precise statements of the conjecture described in \S~\ref{thelink}.

Secondly, we find that the spectral curves and the perturbative invariants for the colored Kauffman perturbative invariants (associated to the $B_N, C_N, D_N$ Lie algebra) are closely related to the more conventional $\mathrm{SU}(N)$ invariants ($A_{N - 1}$ series), and we show they can all be computed by the topological recursion. The only difference is that the topological grading is not respected for the $BCD$ cases. We think this has an interest, since very few was known so far on perturbative Kauffman invariants. 

\subsection{Outline and main results: matrix model}

We establish in Section~\ref{S2} the large $N \rightarrow \infty$ behavior and asymptotic expansion of the partition function and moments of \eqref{themodel0} for $\mathfrak{g} = A_N = \mathfrak{su}(N + 1)$ (Section~\ref{S2}). Some of the technical proofs are postponed to Appendix~\ref{S2proof}. Section~\ref{S3} is independent of the main body of the text: we introduce in \eqref{thep} a class of linear non-local RHP, to which we associate a (maybe infinite) root system. Many information on the solution -- and sometimes its full and explicit form -- can be extracted from the analysis of this root system.

We review the geometry of Seifert spaces and Chern-Simons theory at large $N$ in Section~\ref{S4}. An important geometric invariant of Seifert spaces is the orbifold Euler characteristic of their orbit space, here always assumed to be topologically a sphere :
\beq
\chi = 2 - r + \sum_{m = 1}^r \frac{1}{a_m}
\eeq
where $a_1,\ldots,a_m$ are integers prescribing the orders of extraordinary fibers. We also introduce:
\beq
a = \mathrm{lcm}(a_1,\ldots,a_m)
\eeq

Section~\ref{SSei}-\ref{S7} are devoted to the $A_N$ Chern-Simons matrix model for Seifert fibered spaces, and their extension to the $\mathrm{Sp}$ or $\mathrm{SO}$ Lie algebra is the matter of Section~\ref{S8}. Chern-Simons theory around the trivial flat connection depends on the single parameter:
\beq
u = N\hbar/\sigma
\eeq
where $\sigma \in \mathbb{Q}$ is another geometric parameter of the Seifert spaces. $u_0 = N\hbar$ is sometimes called the string coupling constant. In Section~\ref{SSei}, we apply the techniques of Section~\ref{S3} to the construction of the spectral curves for Seifert spaces. We find in Theorem~\ref{negative} that the finite quotients $M = \mathbb{Z}_{p}\backslash\mathbb{S}_3/\mathfrak{H}$ corresponding to $\chi > 0$ can be described in terms of an algebraic spectral curve $\mathcal{S} \hookrightarrow \mathbb{C}^{\times}\times\mathbb{C}^{\times}$, whereas the spectral curve -- if it exists -- is never algebraic when $\chi < 0$. We will not insist on the cases $\chi = 0$, which are resonant.
\begin{proposition}
The Chern-Simons matrix model for Seifert fibered spaces with $\chi > 0$ admits a spectral curve $\mathcal{S}$ of the form $P(x,y) = 0$ for a $u$-dependent polynomial $P$ whose Newton polygon is known. The coefficients on the boundary of the Newton polygon are known monomials in $e^{-\chi u/2a}$. Besides, the spectral curve comes with the action of a finite Weyl group on the sheets of $x\,:\,\mathcal{S} \rightarrow \widehat{\mathbb{C}}$, as tabulated below.
\beq
\begin{array}{|c|c|cc|c|c|}
\hline {\rm Geometry} & {\rule{0pt}{2ex}}{\rule[-1ex]{0pt}{0pt}} (a_1,\ldots,a_r) & \deg_{x} P & \deg_{y} P & {\rm genus} & {\rm Weyl}\,\,{\rm group} \\
\hline L(a_1,a_2) & (a_1,a_2) & a_1 a_2 & {\rule{0pt}{2ex}}{\rule[-1ex]{0pt}{0pt}} a_1 + a_2 & (a_1 - 1)(a_2 - 1) &  A_{a_1 + a_2 - 1} \\
\hline\mathbb{S}_{3}/D_{p + 2}\,\,(p\,\,\mathrm{odd}) & {\rule{0pt}{2ex}}{\rule[-1ex]{0pt}{0pt}} (2,2,p) & 4p & 2(p + 1) & 2p + 1& D_{p + 1} \\
\hline\mathbb{S}_{3}/D_{p + 2}\,\,(p\,\,\mathrm{even}) &{\rule{0pt}{2ex}}{\rule[-1ex]{0pt}{0pt}}(2,2,p) & 2p & p + 1 & 0 & A_{p} \\
\hline\mathbb{S}_{3}/E_6 & (2,3,3) & {\rule{0pt}{2ex}}{\rule[-1ex]{0pt}{0pt}} 8 & 8 & 5 & D_{4} \\
\hline\mathbb{S}_{3}/E_7 & (2,3,4) &{\rule{0pt}{2ex}}{\rule[-1ex]{0pt}{0pt}} 36 & 27 & 46 & E_{6} \\
\hline\mathbb{S}_{3}/E_8 & (2,3,5) & {\rule{0pt}{2ex}}{\rule[-1ex]{0pt}{0pt}} 540 & 240 & 1471 & E_{8} \\
\hline
 \end{array} \nonumber
 \eeq
This table assumes $\mathrm{gcd}(a_1,a_2) = 1$ for the lens spaces.
\end{proposition}
We were able in Section~\ref{S6}:
\begin{itemize}
\item[$\bullet$] to determine completely the curves in the cases $\mathbb{S}_3/D_{p + 2}$ -- corresponding to $(2,2,p - 2)$, in \S~\ref{even} and \ref{dodd}.
\item[$\bullet$] to determine the curves in the cases $\mathbb{S}_3/E_6$ (resp. $\mathbb{S}_{3}/E_7$) up to $2$ (resp. $4$) parameters fixed by complicated algebraic constraints.
\end{itemize}
Our spectral curve computations are supported by Monte-Carlo simulations of Alexander Weisse presented in Section~\ref{sec:numerics}. Bases on numerics, we also give conjectures about the spectral curve in the cases $\chi < 0$ (Conjecture~\ref{conj2}).

Although the genus of the spectral curve can be quite large, it does not prevent them to cover in a simple way curves of a lower genus. The completely solved cases mentioned above and our computations leads us to:
\begin{conjecture}
Assume $\chi > 0$. In the spectral curves describing the contribution of the trivial flat connection to the Chern-Simons partition function, if one eliminates $x$ from the equations $X = x^{a}$ and $P(x,y) = 0$, we obtain an equation $R(X,y) = 0$ describing a (in general : singular and with several components) curve of genus $0$.
\end{conjecture}
This is true for the $\mathbb{S}_{3}/D_p$ cases. Some formulas and definitions met during those computations are collected in Appendices~\ref{235more}-\ref{singes}.

\subsection{Outline and main results: knot theory}

In Section~\ref{S7}, we explain a practical consequence for knot theory: we can provide some informations on the analyticity properties of the coefficients (seen as functions of $e^{u}$) of the large $N$ expansion of the colored HOMFLY polynomials of fiber knots in $M$, and of the Chern-Simons free energy. So far, they were only known to be analytic in a vicinity of $u = 0$ \cite{GManalyticity}. We prove in Section~\ref{S7}:
\begin{proposition}
\label{popor}The perturbative colored HOMFLY invariants of fiber knots in a Seifert fibered space with geometry $\mathbb{S}_3/D_{p + 2}$ for $p$ even, defined initially as elements of $\mathbb{Q}[[\hbar]][[u_0]]$, are actually the $u_0 \rightarrow 0$ Taylor expansion of an element $\mathbb{Q}[[\hbar]](\kappa^2)[\sqrt{\beta(\kappa)}]$, where:
\beq
\label{kappae1}\frac{2\kappa^{1 + 1/p}}{1 + \kappa^2} = e^{-u_0/(4\sigma p^2)},\qquad \beta(\kappa) = \frac{(\kappa^2 + 1)((p + 1) - (p - 1)\kappa^2)}{\kappa^2}
\eeq
$u_0 > 0$ corresponds to $\kappa(u_0) \in ]0,1[$, and $\sigma \in \mathbb{Q}$ introduced in \eqref{sigmadef} is a geometric parameter of the Seifert spaces. The singularities in the $\kappa$-complex plane occur when $\kappa^4 \in \{1,(\kappa^*)^4\}$ where:
\beq
\label{kappaet}\kappa^* = \sqrt{\frac{p + 1}{p - 1}}
\eeq
\end{proposition}
A result similar to Proposition~\ref{popor}, with $u$ replaced by $2u$, can be deduced from Section~\ref{S7}-\ref{S8} for the perturbative colored Kauffman invariant of fiber knots in $\mathbb{S}_{3}/D_{p + 2}$ with $p$ even.

The situation is different from links in lens spaces, where the perturbative invariants are polynomials in $e^{u}$ \cite{BEMknots}, and thus had no singularities in the $u$-complex plane. Our work suggests the general conjectures:
\begin{conjecture}
\label{iun}If $M$ is a Seifert fibered space, the $F^{(g)}$ and the perturbative invariants of any link in $M$ exist as a function of $u_0 > 0$, and are real-analytic on the positive real line.
\end{conjecture}
\begin{conjecture} Moreover, if $\chi > 0$, there exists a finite degree extension $L$ of $\mathbb{Q}(e^{-\chi u_0/2\sigma})$ depending only on the ambient manifold and the series $X \in \{A, BCD\}$ of Lie algebra, such that, at least for fiber knots, the perturbative invariants colored in any representation are the Laurent expansion at $u_0 \rightarrow 0$ of elements of $L$.
\end{conjecture}
From this perspective, for the Seifert space $M = \mathbb{S}_3/D_{p}$ with $p$ even, one may ask for the geometric interpretation of the function $\kappa(u_0)$ in \eqref{kappae1}, and of the location of the dominant singularity $u^*_0 = u_0(\kappa^*)$ determined by \eqref{kappaet}.

 \subsection{Motivation from topological strings}
\label{13para}
The present work generalizes the analysis of the model $r = 2$ \cite{Halma,BEMknots} relevant to study lens spaces, and the invariant of fiber knots in lens spaces are equal to invariants of torus knots in $\mathbb{S}_3$. Chern-Simons theory on $M$ at large $N$ is dual to type A open topological strings on $T^*M$ \cite{WittenCS}, and through geometric transitions, this can sometimes be related to closed topological strings on another target space $X_M$. This program has been completed for $M = \mathbb{S}_3$ \cite{GopaV} and the lens spaces $\mathbb{Z}_{p_2}\backslash\mathbb{S}_3/\mathbb{Z}_{p_1}$ \cite{Halma,BEMknots} and $X_M$ is obtained by fractional framing transformations from the resolved conifold in both cases. So far, it has remained elusive for general Seifert fibered spaces. Since we establish the existence of an algebraic spectral curve for all Seifert spaces with $\chi \geq 0$, this gives hope for the construction of their mirror geometries $X_M$. This direction, and the connection with $5d$ gauge theories and quantum spectral curves, is under investigation \cite{BBK}.

A natural strategy to establish a duality to a closed string geometry would then be to prove that Gromov-Witten invariants of $X_M$ are also computed by the topological recursion, with same curve $\mathcal{S}$. So far, the topological recursion is indeed known to compute Gromov-Witten invariants when $X$ is a toric $3$-fold Calabi-Yau \cite{BKMP,EOBKMP}, but this might be generalized in the future to a more general class of manifolds.

\section{The matrix model}
\label{S2}
\subsection{Equilibrium measures}

We study the statistical mechanics of $N$ particles, of position $t_1,\ldots,t_n \in \mathbb{R}^N$, with joint probability distribution:
\beq
\label{themodel}\dd\mu(\mathbf{t}) = \frac{1}{Z_N} \Big[\prod_{1 \leq i < j \leq N} \mathrm{sinh}^{2 - r}\big(\frac{t_i - t_j}{2}\big) \prod_{m = 1}^r \mathrm{sinh}\big(\frac{t_i - t_j}{2a_m}\big)\Big]\,\prod_{j = 1}^N e^{-NV(t_j)}\,\dd t_j,\qquad V(t) = \frac{t^2}{2u}.
\eeq
At this stage, $a_1,\ldots,a_r$ are arbitrary positive parameters. The dominant contribution to the partition function when $N \rightarrow \infty$ should come from configurations $\mathbf{t}$ maximizing the probability density. It is reasonable to think that the empirical distribution:
\beq
L_N = \frac{1}{N} \sum_{i = 1}^N \delta_{t_i}
\eeq
of such configurations will be close to a minimizer of the energy functional:
\beq
\mathcal{E}[\lambda] = -\frac{1}{2}\iint \dd\lambda(t)\dd\lambda(t')\Big[(2 - r)\ln\big|\mathrm{sinh}\big(\frac{t - t'}{2}\big)\big| + \sum_{m = 1}^r \ln\big|\mathrm{sinh}\big(\frac{t - t'}{2a_m}\big)\big|\Big] + \int \dd\lambda(t)\,V_0(t)
\eeq
among probability measures $\lambda$. $\mathcal{E}_0$ is actually defined in $\mathbb{R}\cup\{+\infty\}$ because of the singularity of the logarithm. It is a lower semi-continuous functional, so has compact level sets for the weak-* topology, therefore it achieves its minimum. We call \emph{equilibrium measure} and denote $\lambda_{{\rm eq}}$ any minimizer of $\mathcal{E}$. It must satisfy the saddle point equation: there exist a constant $C^{\lambda_{{\rm eq}}}$ such that
\beq
\label{Veff}V_{{\rm eff}}^{\lambda_{{\rm eq}}}(t) := V(t) - \int \dd\lambda_{{\rm eq}}(t')\Big[(2 - r)\ln\big|\mathrm{sinh}\big(\frac{t - t'}{2}\big)\big| + \sum_{m = 1}^r \ln\big|\mathrm{sinh}\big(\frac{t - t'}{2a_m}\big)\big|\Big] \geq C^{\lambda_{{\rm eq}}}
\eeq
with equality $\lambda_{{\rm eq}}$-everywhere. $V_{{\rm eff}}^{\lambda}(t)$ is the effective potential felt by a particle at position $t = t_i$, taking into account the collective effect of all other $t_j$'s distributed according to $\lambda$. Equilibrium measures are characterized by the property that the effective potential achieves its minimum on the locus of $\mathbb{R}$ where the particles accumulate. Classical techniques\footnote{The arguments of \cite{SaffTotik} were developed for the pairwise interaction $\prod_{i < j} (t_i - t_j)^2$ which has a zero at coinciding points, but it is easy to generalize to any interaction which has the same singularity and is smooth elsewhere.} of potential theory \cite{SaffTotik} show that the random measure $L_N$'s limit points are equilibrium measures, and that:
\beq
Z_N = \exp\big(-N^2\,\min \mathcal{E}+ o(N^2)\big)
\eeq
Some qualitative properties of the equilibrium measures can be derived from their characterization: $V$ grows sufficiently fast at infinity to ensure that equilibrium measures have compact support ; since $V$ and the pairwise interactions are analytic (away from the singularity at $x = y$), it can be shown (see \cite{BGK}, generalizing \cite{Deift} where $-\ln|t - t'|$ was considered)
that equilibrium measures are supported on a finite number of segments, have a density which is analytic away from the edges, and is $1/2$-H\"older at the edges. When the density vanishes exactly like a squareroot at the edges, it is said \emph{off-critical}.

The question of uniqueness of $\mu_{{\rm eq}}$ has no easy answer. But, since the functional is quadratic, if the quadratic form over the space $\mathcal{M}^0$ of differences of probability measures:
\beq
\mathcal{Q}[\nu] = - \iint \dd\nu(t)\dd\nu(t')\Big[[(2 - r)\ln\big|\mathrm{sinh}\big(\frac{t - t'}{2}\big)\big| + \sum_{m = 1}^r \ln\big|\mathrm{sinh}\big(\frac{t - t'}{2a_m}\big)\big|\Big]
\eeq
is positive definite, then $\mathcal{E}$ is strictly concave and this guarantees uniqueness of its minimizer. Notice again that, a priori, $\mathcal{Q}$ takes values in $\mathbb{R}\cup\{+\infty\}$. Indeed, the singular part of $\mathcal{Q}$ is:
\beq
-2\iint \dd\nu(t)\dd\nu(t')\ln|t - t'| = \int_{0}^{\infty} \frac{|\mathcal{F}[\nu](k)|^2}{k} \geq 0
\eeq
and the right-hand side can be infinite if the measure $\nu$ is not regular enough. Here $\mathcal{F}[\nu] = \int \dd\nu(x)\,e^{{\rm i}k x}$ is the Fourier transform of the finite mesure $\nu$. As a matter of fact, it is enough to consider $\mathcal{E}$ and $\mathcal{Q}$ as functionals over measures with compact support, because $V$ grows fast at infinity compared to the pairwise interactions.

\begin{lemma}
\label{lemm1}For any $\nu \in \mathcal{M}^0$:
\beq
\mathcal{Q}[\nu] = \int_{0}^{\infty}\Big[(2 - r)\gamma(k) + \sum_{m = 1}^r \gamma(a_mk)\Big]|\mathcal{F}[\nu](k)|^2 \frac{\dd k}{k},\qquad \gamma(k) = \mathrm{cotanh}(\pi k).
\eeq
\end{lemma}
$\mathcal{Q}$ is definite positive whenever the function $(2 - r)\gamma(k) + \sum_{m = 1}^r \gamma(a_ik)$ is almost everywhere positive. In particular, it must be positive at $k \rightarrow \infty$, which gives the necessary condition:
\beq
\label{chiorb}\chi = 2 - r + \sum_{m = 1}^r \frac{1}{a_m} \geq 0
\eeq
In the model for Seifert fibered spaces, the $a_m$ are assigned integer values. We recognize in \eqref{chiorb} the orbifold Euler characteristics of the orbit space, and the list of uples leading to $\chi \geq 0$ is finite. For $r = 1$ and $r = 2$, $\mathcal{Q}$ is obviously positive definite, and for the remaining $r = 3$ cases having $\chi \geq 0$, positivity can be checked by a direct computation.
\begin{corollary}
\label{cor1}If $a_1,\ldots,a_r$ are integers, $\mathcal{Q}$ is definite positive iff $2 - r + \sum_{m = 1}^r a_m^{-1} \geq 0$. In those cases, there exists a unique equilibrium measure.
\end{corollary}
The proof of Lemma~\ref{lemm1} and Corollary~\ref{cor1} are presented in Appendix~\ref{S23} and \ref{sproof1}. We can also establish -- see Appendix~\ref{sproof2} -- some qualitative properties of equilibrium measures:
\begin{theorem}
\label{1cut} For any $a_1,\ldots,a_r > 1$, and any $u > 0$, any equilibrium measure $\check{\lambda}_{{\rm eq}}$ is supported on a single segment, and vanishes exactly as a squareroot at the edges (generic edge).
\end{theorem}
\begin{corollary}
\label{cocococo} If $\chi \geq 0$, since the equilibrium measure is unique, it must be invariant under $t \mapsto -t$. In particular, its support is a segment centered at the origin.
\end{corollary}

In Section~\ref{sec:numerics}, Weisse proposes a Monte-Carlo simulation to obtain the dominant configurations of probability densities like \eqref{themodel}. It is observed that, for any value of $a_m$ integers and $u > 0$, independently of the sign of $\chi$, the empirical measure seem to have a unique limit point. It supports the
\begin{conjecture}
\label{conj2}For any $a_1,\ldots,a_r > 1$ and $u > 0$, the equilibrium measure is unique, and its density away from the edges of the support is a real-analytic function of $u > 0$.
\end{conjecture}

\subsection{Change of variable and saddle point equation}
\label{change}
We introduce the exponential variables:
\beq
\label{expvar}\boxed{s_i = e^{t_i/a},\qquad a = \mathrm{lcm}(a_1,\ldots,a_r),\qquad \check{a}_i = a/a_i}
\eeq
The measure $\mu$ \eqref{themodel} on $\mathbf{t}\in \mathbb{R}^N$ transforms into a measure $\check{\mu}$ on $(\mathbb{R}_+^{\times})^N$:
\beq
\label{themodel2}\dd\check{\mu}(\mathbf{s}) = \frac{1}{\check{Z}_N}\prod_{1 \leq i < j \leq N} (s_i - s_j)^2 \prod_{1 \leq i,j \leq N} \check{R}(s_i,s_j)^{1/2} \prod_{j = 1}^N e^{-N\check{V}(s_i)}\dd s_i
\eeq
where:
\beq
\check{V}(s) = \frac{a^2(\ln s)^2}{2u} + \frac{a\chi}{2}\,\ln s,\quad\check{R}(s,s') = \prod_{\ell = 1}^{a - 1} (s - \zeta_{a}^{\ell}\,s')^{2 - r} \prod_{m = 1}^{r} \prod_{\ell_m = 1}^{\check{a}_m - 1} (s - \zeta_{\check{a}_m}^{\ell_m}\,s'),
\eeq
and $\zeta_{a}$ denotes a primitive $a^{{\rm th}}$-root of unity. 

$\check{\lambda}_{{\rm eq}}$ denote an equilibrium measure for $\check{\mu}$. It is the image via \eqref{expvar} of an equilibrium measure $\lambda_{{\rm eq}}$ for $\mu$. It is characterized by the saddle point equation:
\beq
\label{saddleb}\check{V}_{{\rm eff}}(x):= \check{V}(x) - \int \dd \check{\lambda}_{{\rm eq}}(s)\,\big[2\ln|x - s| + \ln \check{R}(x,s)\big] \geq C
\eeq
for some constant $C$, with equality on the support $\Gamma \subseteq \mathbb{R}_+^{\times}$ of $\lambda_{{\rm eq}}$. We shall rewrite the characterization of an equilibrium measure in term of its Stieltjes transform:
\beq
\label{Stiel}W(x) = \int \frac{x\,\dd\check{\lambda}_{{\rm eq}}(s)}{x - s} = \lim_{N \rightarrow \infty} \check{\mu}\Big[\frac{1}{N}\,\mathrm{Tr}\,\frac{x}{x - \mathbf{S}}\Big],\qquad \mathbf{S} = \mathrm{diag}(s_1,\ldots,s_N)
\eeq
$W$ is a bounded, holomorphic function on $\mathbb{C}\setminus\Gamma$, such that:
\beq
\label{growth}\big(W(x - {\rm i}0) - W(x + {\rm i}0)\big)\dd x = 2{\rm i}\pi x\,\dd\check{\lambda}_{{\rm eq}}(x),\qquad \lim_{x \rightarrow \infty} W(x) = 1,\qquad \lim_{x \rightarrow 0} W(x) = 0.
\eeq
The saddle point equation \eqref{saddleb} implies a functional equation:
\beq
\label{RH1}\forall x \in \mathring{\Gamma},\qquad W(x + {\rm i}0) + W(x - {\rm i}0) + \oint_{\Gamma} x\partial_{x}\ln \check{R}(x,s)\,\frac{W(s)\dd s}{2{\rm i}\pi s}  = x\check{V}'(x)
\eeq
with:
\beq
x\check{V}'(x) = \frac{a^2\ln x}{u} + \frac{a\chi}{2}.
\eeq
The contour of integration in \eqref{RH1} can be moved to pick up residues at rotations of $x$ of order $a$:
\beq
\label{218}\forall x \in \Gamma,\qquad \boxed{W(x + {\rm i}0) + W(x - {\rm i}0) + (2 - r) \sum_{\ell = 1}^{a - 1} W(\zeta_{a}^{\ell}x) + \sum_{m = 1}^r  \sum_{\ell_m = 1}^{\check{a}_m - 1} W(\zeta_{\check{a}_m}^{\ell_m}x) = x\check{V}'(x).}
\eeq

Since $\mu$ is invariant under $(t_1,\ldots,t_N) \rightarrow (-t_1,\ldots,-t_N)$, in the case where the equilibrium measure is unique, it must have the same symmetry. This translates into the \emph{palindrome symmetry}:
\beq
\label{symm}W(x) + W(1/x) = 1
\eeq

When $\mathcal{Q}$ is definite positive and $\Gamma$ is fixed, \eqref{218} characterizes $W$ among the holomorphic functions in $\mathbb{C}\setminus\Gamma$ satisfying \eqref{growth} and whose discontinuity on $\Gamma$ defines an integrable measure (see e.g. the argument of \cite[Lemma 3.10]{BEO}). When $\mathcal{Q}$ is not definite positive, we do not know how to address the question of uniqueness.

\subsection{Asymptotic analysis}

We rely on the results of \cite{BGK} to study the asymptotic expansion when $N \rightarrow \infty$ of the model \eqref{themodel2}.
We would like to compute the partition function $\check{Z}_N$ and the $n$-point correlators:
\beq
W_n(x_1,\ldots,x_n) = \check{\mu}\Big[\prod_{i = 1}^n \mathrm{Tr}\,\frac{x_i}{x_i - \mathbf{S}}\Big]_{{\rm conn}}
\eeq
where ${\rm conn}$ stands for "connected expectation value".

\begin{definition}
We say that $\check{\lambda}_{{\rm eq}}$ is off-critical if its density is nowhere vanishing on its support $\Gamma$, and it vanishes exactly like a squareroot at the edges of $\Gamma$.
\end{definition}

\begin{theorem}\cite{BGK}
Assume $\mathcal{E}$ is strictly concave and $\check{\lambda}_{{\rm eq}}$ is off-critical. Then, we have an asymptotic expansion of the form:
\bea
\check{Z}_{N} & = & N^{}\exp\Big(\sum_{g \geq 0} N^{2g - 2}\,F^{(g)}\Big) \nonumber \\
W_n(x_1,\ldots,x_n) & = & \sum_{g \geq 0} N^{2 - 2g - n}\,W_n^{(g)}(x_1,\ldots,x_n)
\eea
The coefficients of expansion are real analytic functions of $u \geq 0$, and $W_n^{(g)}(x_1,\ldots,x_n)$ are holomorphic functions in $(\mathbb{C}\setminus\Gamma)^n$.
\end{theorem}

In particular, this confirms for Seifert spaces, by a different method, the analyticity of Chern-Simons free energies proved in general for rational homology spheres in \cite{GManalyticity} (see \S~\ref{CSavatar}). From \eqref{themodel2}, we have the basic relation:
\beq
\label{basicerq} u^2\partial_{u} F^{(g)} = \oint_{\Gamma} \frac{\dd x}{x}\,\frac{a^2(\ln x)^2}{2}\,W_1^{(g)}(x)
\eeq

\subsection{Two point function}
\label{2pot}

\begin{definition}
We call $2$-point function:
\bea
W_2^{(0)}(x_1,x_2) & = & \lim_{N \rightarrow \infty} \check{\mu}\Big[\mathrm{Tr}\,\frac{x_1}{x_1 - \mathbf{S}}\,\mathrm{Tr}\,\frac{x_2}{x - \mathbf{S}}\Big]_{{\rm conn}} \nonumber \\
& = & \lim_{N \rightarrow \infty} \Bigg(\check{\mu}\Big[\mathrm{Tr}\,\frac{x_1}{x_1 - \mathbf{S}}\,\mathrm{Tr}\,\frac{x_2}{x_2 - \mathbf{S}}\Big] - \check{\mu}\Big[\mathrm{Tr}\,\frac{x_1}{x_1 - \mathbf{S}}\Big]\check{\mu}\Big[\mathrm{Tr}\,\frac{x_2}{x_2 - \mathbf{S}}\Big]\Bigg)
\eea
\end{definition}
It can be obtained formally from $W(x) = W_1^{(0)}(x)$ by an infinitesimal variation of the potential $\check{V}$:
\beq
W_2^{(0)}(x_1,x_2)  = \frac{\partial}{\partial \varepsilon}\check{\mu}^{\check{V}_{\varepsilon,x_2}}\Big[\sum_{i_1 = 1}^N \frac{x_1}{x_1 - s_{i_1}}\Big]\Big|_{\varepsilon = 0},\qquad \check{V}_{\varepsilon,x_2}(x) = \check{V}(x) - \frac{1}{N}\,\frac{x_2}{x_2 - x}
\eeq
It also satisfies also a saddle point equation, which can be obtained by formally applying the variation of the potential to the saddle point equation \eqref{218} satisfied by $W(x)$. The result is that $W_2^{(0)}(x_1,x_2)$ is holomorphic in $(\mathbb{C}\setminus\Gamma)^2$, and has a discontinuity when $x_1 \in \mathbb{C}\setminus\Gamma$ and $x_2 \in \Gamma$ satisfying:
\bea
\label{W2potf} & & W_2^{(0)}(x_1 + {\rm i}0,x_2) + W_2^{(0)}(x_1 - {\rm i}0,x_2) + (2 - r)\sum_{\ell = 1}^{a - 1} W_2^{(0)}(\zeta_{a}^{\ell}x_1,x_2) + \sum_{m = 1}^r \sum_{\ell_m = 1}^{\check{a}_m - 1} W_2^{(0)}(\zeta_{\check{a}_m}^{\ell_m}x_1,x_2) \nonumber \\
& = & -\frac{x_1x_2}{(x_1 - x_2)^2}
\eea

The data of $(x,W(x))$ is called the spectral curve. Together with the two-point function $W_2^{(0)}(x_1,x_2)$, it provides the initial data for the the recursive computation of $F^{(g)}$ and $W_n^{(g)}$. We give in Section~\ref{S3} general principles to solve the homogeneous linear equation, that will be put in practice to compute these data in the Seifert models (Section~\ref{S4}-\ref{SSei}).

\section{Algebraic theory of sheet transitions}
\label{S3}

\subsection{The problem}

The form taken by the saddle point equation \eqref{RH1} motivates a general study of homogeneous functional relations of the type:
\beq
\label{thep}\forall x \in \Gamma,\qquad \phi(x + {\rm i}0) + \phi(x - {\rm i}0) + \sum_{g \in S} \alpha(g)\,\phi(g(x)) = 0,
\eeq
where:
\begin{itemize}
\item[$\bullet$] $G$ is a finite subgroup of $\mathrm{PSL}_2(\mathbb{C})$, acting on the Riemann sphere by M\"obius transformations (their classification is reminded in Appendix~\ref{subgop}).
\item[$\bullet$] $S$ is a generating subset of $G$, not containing $\mathrm{id}$, and stable by inverse. $(\alpha(s))_{s \in S}$ is a sequence of numbers in a number field $\mathbb{K}$ (for instance $\mathbb{K} = \mathbb{R}$), and we assume:
\beq
\label{syme} \forall s \in S,\qquad \alpha(s^{-1}) = \alpha(s).
\eeq
\item[$\bullet$] $\Gamma$ is a collection of arcs on the Riemann sphere such that $g(\mathring{\Gamma}) \cap \Gamma = \emptyset$ whenever $g \neq 1$. $\mathring{\Gamma}$ denotes the set of interior points of $\Gamma$.
\item[$\bullet$] $U$ is an open subset of $\widehat{\mathbb{C}}$ containing $\Gamma$ and stable under the action of $G$, and $\phi$ is a holomorphic function on $U\setminus\Gamma$, that admits boundary values at any interior point of $\Gamma$. For simplicity, $U$ will be $\mathbb{C}$ or $\mathbb{C}^{\times}$ here.
\end{itemize}
This problem is usually supplemented with some growth prescription for $\phi(x)$ at the edges of $\Gamma$ and at the boundary of $U$. For instance, if $U = \mathbb{C}^{\times}$, one can ask for prescribed singular behavior at $0$ and $\infty$.
This problem can also be studied with few modifications for $\phi = $ a section of a vector bundle over $U$, in particular for $\phi =$ a holomorphic 1-form.

This problem for $\phi =$ 1-form\footnote{For Seifert spaces, $G$ consists of linear maps (rotations), and the extra factor of $x$ put in the definition of the Stieltjes transform \eqref{Stiel} is the trick allowing the formulation of the linear equation with constant coefficients $\alpha(g)$'s in terms of a function instead of a $1$-form.}
 actually appears in the study of equilibrium measures for models of the form \eqref{themodel2} with arbitrary (analytic) pairwise interaction $\check{R}$. The dependence in the potential $\check{V}$ only affects the right-hand side of such an equation, which can be set to $0$ by subtracting a particular solution, thus affecting the growth prescriptions for the solution $\phi$ we are looking for. In general, $G$ is the Galois group of the equation $\check{R}(x,y) = 0$, and may be complicated even for simple $\check{R}$. It may not be finite, and the description of the orbit of $\Gamma$ has to deal with the rich question of iterating functions in the complex plane. Here, we restrict to a subclass of model where the iteration problem is trivial, in the sense that $G$ is a finite group of (globally defined) automorphisms of the Riemann sphere. The assumption that $S$ is stable under inversion comes from the fact that pairwise interactions are symmetric $\check{R}(x,y) = \check{R}(y,x)$. One can go beyond this assumption and actually consider coupled linear systems of the form, see \cite{BEO,BBGPotts} for examples.
 
A complete, satisfactory solution of \eqref{thep} would be a description, for any fixed $\Gamma$ and $\alpha$'s, of an elementary basis of solutions which generate any solution of \eqref{thep} with prescribed meromorphic or logarithmic singularities. So far, the non-trivial case for which this program has been achieved is $G = \mathbb{Z}_2$, i.e. $G$ consists of the identity and an homographic involution $\iota$, and only if $\Gamma$ is a segment. This occurs in the $O(n)$ matrix model, and $n = -\alpha(\iota)$ here. The solution was essentially found for all values of $n$ by the second author \cite{EKOn,EKOn2} in terms of Jacobi theta functions. Apart from a few cases which reduce technically to the latter, it seems hopeless, even when $G$ is finite or even cyclic, to find a complete solution in the above sense. It would be very interesting to solve any problem in which $G \simeq \mathbb{Z}$ is a group of translation in the complex plane, and $S$ consists of a generator and its inverse.

We now undertake the general study of \eqref{thep}. The outcome will be a method to decide if the solution of \eqref{thep} can be expressed in terms of algebraic functions, and in this case, the answer can be in principle computed. It does not happen for generic $\alpha$'s, but it can actually lead actually to some explicit results for interesting models. The techniques leading to an algebraic solution of the $O(n)$ model equation when $n = -2\cos(\pi b)$ for a rational number $b$ \cite{TheseBE} can be regarded as a special case of our construction. Obviously, the methods we describe also allows the multiplicative monodromies.

The theory will be applied to the Seifert models in Section~\ref{S4}, for which the Galois group is $G = \mathbb{Z}_{a}$. 

\vspace{0.2cm}

\noindent \textbf{Remark.} If $U = \widehat{\mathbb{C}}$ and $\phi(x)$ is a function solution to \eqref{thep} with meromorphic singularities at prescribed points in $U$, and if for instance $\Gamma$ consists of finite numbers of arcs in $U$, it follows from the finiteness of the group $G$ that $\phi(x)$ has endless analytic continuation. This observation might provide another useful point of view for the computation of $\phi(x)$: if $\phi(x)$ can be complicated, its Laplace transform on certain contours might have a simpler form.

\subsection{Action of the group $G$ algebra}

Let $\mathbb{K}$ be a number field (here $\mathbb{Q}$ is enough) and let $\hat{E} = \mathbb{K}[G]$ be the group algebra of $G$. It is a vector space\footnote{We put a small hat on vectors in $\hat{E}$ to distinguish them from another vector space $E$ defined later.} with a basis $(\hat{e}_{g})_{g \in G}$ indexed by elements of $G$, and endowed with a bilinear multiplication law:
\beq
\hat{e}_{g}\cdot\hat{e}_{h} = \hat{e}_{gh}.
\eeq
$\hat{E}$ can also be identified with the algebra of $\mathbb{K}$-valued functions on $G$, with multiplication given by the left convolution:
\beq
(\hat{v}\cdot \hat{w})(g) = \sum_{h \in G} \hat{v}(gh^{-1})\hat{w}(h).
\eeq
We denote $(\ell_{g})_{g \in G}$ the dual basis, i.e. $\ell_{g}(\hat{v}) = \hat{v}(g)$ for any $v \in \mathbb{K}[G]$ and $g \in G$. If $\hat{v} \in \hat{E}$, we call $\mathrm{supp}\,\hat{v} = \{g \in G,\quad v(g) \neq 0\}$ its support. We denote $g\cdot\Gamma = g^{-1}(\Gamma)$, and if $\phi$ is a function on $\widehat{\mathbb{C}}\setminus\Gamma$, we associate to any $\hat{v} \in \hat{E}$ the following function on $U\setminus(\bigcup_{g \in \mathrm{supp}\,\hat{v}} g\cdot\Gamma)$:
\beq
(\hat{v}\cdot\phi)(x) = \sum_{g \in G} \hat{v}(g)\,\phi(g(x)).
\eeq
$\phi(x)$ can be retrieved modulo holomorphic functions in $U$ as the "part of $(\hat{v}\cdot\phi)(x)$ which has a discontinuity on $\Gamma$ only". Indeed, for any $g$ in the support of $\hat{v}$, 
\beq
\label{projections}\phi(x) -\frac{1}{\hat{v}(g)}\oint_{\Gamma} \frac{\dd s\,(\hat{v}\cdot\phi)(g^{-1}(s))}{2{\rm i}\pi\,(x - s)}
\eeq
is holomorphic for $x \in U$. This piece of information stresses that it has no discontinuity on $G\cdot\Gamma$ nor on its preimages.

\subsection{Analytic continuation and algebraic rewriting}
\label{anala}
Let us denote:
\beq
\hat{\alpha} = 2\hat{e}_{{\rm id}} + \sum_{g \in S} \alpha(g)\,\hat{e}_{g} \in \hat{E}.
\eeq
If $\phi$ satisfies the functional relation \eqref{thep}, we can rewrite:
\beq
\label{thep2}\forall x \in \mathring{\Gamma},\qquad \phi(x + {\rm i}0) = (\hat{e}_{{\rm id}} - \hat{\alpha})\cdot \phi(x - {\rm i}0).
\eeq
Therefore, we can define the analytic continuation -- denoted $\varphi$ -- of $\phi$ on two copies of $U$ equipped with a coordinate $x$, and identified along the cut $\Gamma$. In the first copy, $\varphi(x) = \phi(x)$, and in the second copy, $\varphi(x) = \big((\hat{e}_{{\rm id}} - \hat{\alpha})\cdot \phi\big)(x)$. Now, in the second copy, $\varphi(x)$ has cuts on $\bigcup_{g \in S} g\cdot \Gamma$. Actually, since $\phi(x)$ itself is continuous across $\bigcup_{g \in G} g\cdot\sigma$, we deduce from \eqref{thep2} the functional relation\footnote{This equation is initially derived for $g$ in the support of $\hat{v}$, turns out to be trivially valid for any $g \in G$. It just expresses the continuity of $\hat{v}\cdot \phi$ across $g\cdot\Gamma$ whenever $\hat{v}(g) = 0$.} for $\hat{v} \cdot \phi(x)$ for any vector $\hat{v} \in \hat{E}$:
\beq
\forall x \in g\cdot\mathring{\Gamma},\qquad (\hat{v}\cdot\phi)(x + {\rm i}0) = (\hat{v} - \ell_{g}(\hat{v})\hat{\alpha}\cdot \hat{e}_{g})\cdot\phi(x - {\rm i}0).
\eeq
Therefore, gluing copies of $U$ along the cuts encountered, we obtain a maximal (and maybe with infinitely many sheets) Riemann surface $\Sigma$ on which $\varphi$ is an analytic function.  We may have chosen an initial vector $\hat{v}_0 \in \hat{E}$ and started the same process with the function $(\hat{v}_0\cdot \phi)(x)$. We would obtain another analytic function $\varphi_{\hat{v}_0}$.

We therefore need to study the dynamics of the linear maps in $\hat{E}$:
\beq
\label{defq}\hat{T}_{g}(\hat{v}) = \hat{v} - \ell_{g}(\hat{v})\,\hat{\alpha}\cdot \hat{e}_{g}
\eeq
\eqref{defq} was defined such that:
\beq
\forall x \in g\cdot\mathring{\Gamma},\qquad (\hat{v}\cdot \phi)(x + {\rm i}0) = (\hat{T}_{g}(\hat{v})\cdot \phi)(x - {\rm i}0).
\eeq
Thanks to $\ell_{{\rm id}}(\hat{\alpha}) = 2$, we have $\ell_g(\hat T_g(\hat v))=-\ell_g(\hat v)$, and $\hat{T}_{g}$ are involutive. More precisely, they are pseudoreflections, i.e. $\mathrm{Ker}(\hat{T}_{g} + \mathrm{id})$ is generated by a single vector, namely $\hat{e}_{g}$.
\begin{definition}
We call \emph{group of sheet transitions} $\hat{\mathfrak{G}}$ the discrete subgroup of $\mathrm{GL}(\hat{E})$ generated by $(\hat{T}_g)_{g \in G}$.
\end{definition}
$\hat{\mathfrak{G}}$ is non-commutative since
\beq
[\hat{T}_{g},\hat{T}_{h}] = \hat{\alpha}(gh^{-1})\hat{\alpha}(\ell_{g}\otimes \hat{e}_{g} - \ell_{g}\otimes\hat{e}_{h}).
\eeq
However, if $g,h \in G$ such that $gh^{-1} \notin S$, we observe that $[\hat{T}_{g},\hat{T}_{h}] = 0$. $\hat{\mathfrak{G}}$ is in general infinite. 
\begin{question}
\label{Q1} Does there exists a non-zero vector $\hat{v}_0 \in \hat{E}$ with finite $\hat{\mathfrak{G}}$-orbit ? If yes, can one classify the finite orbits, and find the orbits of minimal order ?
\end{question}
Indeed, for such vectors the function $(\hat{v}_0\cdot\phi)(x)$ has a finite monodromy group around $\Gamma$. For instance, if we were looking for solutions $\phi(x)$ with meromorphic singularities in the Riemann sphere away from $\Gamma$, this implies that $\hat{v}_0\cdot\phi(x)$ is an algebraic function, i.e. $\varphi_{\hat{v}_0}$ is a meromorphic function defined on a compact Riemann surface. The order of the orbit is related to the degree of the algebraic function, and it is appealing to choose $\hat{v}_0$ so that the degree is minimal. The nice thing about algebraic functions is that they can be efficiently identified by their divergent parts at a finite number of poles. And in our problem, there are by construction lots of symmetries between the sheets due to the existence of $G$, which can help to compute the solution.

\subsection{Orbits and skeleton graphs}
\label{orbitp}
$\hat{\mathfrak{G}}$ acts transitively on the orbit of any $\hat{v}_0 \in \hat{E}$, but not freely. Let $\mathfrak{G}_{\hat{v}_0}$ denote the stabilizer of $\hat{v}_0$. It is a subgroup of $\hat{\mathfrak{G}}$, with the property:
\beq
\hat{\mathfrak{G}}_{g\cdot\hat{v}_0} = g^{-1}\hat{\mathfrak{G}}_{\hat{v}_0}g.
\eeq
We may construct the \emph{sketelon graph} $\mathcal{G}_{\hat{v}_0}$, whose vertices are elements of the orbits of $\hat{v}_0$, and edges are $\{\hat{v},\hat{v}'\}$ decorated by an element $g \in G$ whenever $\hat{v}' = \hat{T}_{g}(\hat{v})$. The labels of the edges incident to a vertex $\hat{v}$ are actually the elements of the support of $\hat{v}$. The graph $\mathcal{G}_{\hat{v}_0}$ is isomorphic to the quotient of the Cayley graph of $\hat{\mathfrak{G}}$ with generating set $(\hat{T}_{g})_{g \in G}$, by the relation:
\beq
\forall \hat{r},\hat{r}' \in \hat{\mathfrak{G}},\qquad \hat{r} \sim \hat{r}' \quad \Leftrightarrow \quad \hat{r} \in \hat{\mathfrak{G}}_{\hat{v}_0}\hat{r}'.
\eeq

Following the procedure of \S~\ref{anala}, we can analytically continue $\hat{v}_0\cdot\phi(x)$ as a function $\varphi_{\hat{v}_0}$ on a maximal Riemann surface $\Sigma_{\hat{v}_0}$. It is obtained from $\mathcal{G}_{\hat{v}_0}$ by:
\begin{itemize}
\item[$\bullet$] blowing a copy of $U$ (denoted $U_{\hat{v}}$) equipped with a coordinate $x$, at every vertex $\hat{v}$ of $\mathcal{G}_{\hat{v}_0}$.
\item[$\bullet$] for any edge $\{\hat{v},\hat{v}'\}$ decorated by $g \in G$, opening a cut along $x \in g\cdot\Gamma$ in $U_{\hat{v}}$ and $U_{\hat{v}'}$, and gluing them along the cut with opposite orientation.
\end{itemize}
The question~\ref{Q1} is then reduced to the description of the quotient $\hat{\mathfrak{G}}/\hat{\mathfrak{G}}_{\hat{v}_0}$, where in general $\hat{\mathfrak{G}}$ is infinite. In this perspective, question~\ref{Q1} seems far from obvious. We will see in the next paragraph that we can use the action of a smaller (and in some cases, finite) group $\mathfrak{G}$, which is a reflection group and actually the Weyl group of a root system, in order to understand the $\hat{\mathfrak{G}}$-orbits.

\subsection{Construction of a root system}
\label{rot}
We endow $\hat{E} = \mathbb{K}[G]$ with the scalar product $(\hat{e}_{g}|\hat{e}_{h}) = \delta_{g,h}$. The left multiplication by $\hat{\alpha}$ defines an endomorphism:
\beq
\boxed{\hat{A}(\hat{v}) = \hat{\alpha}\cdot\hat{v}}
\eeq
Since $\hat{\alpha}(g) = \hat{\alpha}(g^{-1})$, we have:
\beq
\forall \hat{v},\hat{w} \in \hat{E},\qquad \hat{A}(\hat{v})|\hat{w} = \hat{v}|\hat{A}(\hat{w}),
\eeq
i.e. $\hat{A}$ is symmetric. Therefore, we have a decomposition in orthogonal sum:
\beq
\hat{E} = E \oplus E_0,\qquad E = \mathrm{Im}\,\hat{A},\qquad E_0 = \mathrm{Ker}\,\hat{A}.
\eeq
Let us denote:
\begin{itemize}
\item[$\bullet$] $\pi_{E}$, the orthogonal projection on $E$, and $e_{g} = \pi_{E}(\hat{e}_g)$ for $g \in G$. None of them can be zero.  Since $(\hat{e}_{g})_{g \in G}$ is a basis of $\hat{E}$, their projections $(e_g)_{g \in G}$ span $E$.
\item[$\bullet$] $A \in \mathrm{GL}(E)$, the invertible endomorphism induced by $\hat{A}$ on $E$.
\item[$\bullet$] $T_{g} = A^{-1}\hat{T}_{g}A \in \mathrm{GL}(E)$. 
\end{itemize}
It can be computed as follows:
\bea
\hat{T}_{g}(\hat{\alpha}\cdot v) & = & \hat{\alpha}\cdot v - (\hat{\alpha}\cdot v)(g)\,\hat{\alpha}\cdot \hat{e}_{g} \nonumber \\
& = & \hat{\alpha}\cdot\Big(v - (\hat{\alpha}\cdot v)(g)\,\hat{e}_{g}\Big) \nonumber \\
& = & \hat{\alpha}\cdot\Big(v - (\hat{\alpha}\cdot v)(g)\,e_{g}\Big). \nonumber
\eea
Therefore:
\beq
\forall g \in G,\quad \forall v \in E,\qquad T_{g}(v) = v - (\hat{\alpha}\cdot v)(g)\,e_{g}.
\eeq
When studying the dynamics of $(T_g)_{g \in G}$, we use unhatted notations for vectors in $E$. This is to remind that, if we want to come back to the dynamics of $(\hat{T}_g)_{g \in G}$, we have:
\beq
\label{comebak}\hat{T}_{g}(\hat{v}) = A(T_{g}(v)),\quad \hat{v} = \hat{\alpha}\cdot v.
\eeq
We now introduce a symmetric bilinear form on $\hat{E}$:
\beq
\label{quadr}\langle\hat{v},\hat{w}\rangle = \frac{\hat{A}(\hat{v})|\hat{w}}{2}.
\eeq
By construction, its restriction to $E$ is non-degenerate. The projections $e_g$ have the properties:
\bea
\langle \hat{e}_{g},\hat{e}_{h} \rangle & = & \frac{1}{2}\,(\hat{\alpha}\cdot \hat{e}_{g}|\hat{e}_{h}) = \frac{1}{2}\,\ell_h(\hat{\alpha}\cdot \hat{e}_{g}) = \frac{1}{2}\,\hat{\alpha}(gh^{-1}) \nonumber \\
& = & \frac{1}{2}\,(\hat{\alpha}\cdot e_{g}| e_h) = \langle e_g,e_h \rangle
\eea
Since $\hat{\alpha}(1) = 2$, we have:
\beq
\forall g \in G,\qquad \langle\hat{e}_{g},\hat{e}_{g}\rangle = \langle e_g,e_g\rangle = 1.
\eeq
This bilinear form allows the rewriting:
\beq
\boxed{\forall g \in G,\quad \forall v \in E,\qquad T_{g}({v}) = v - 2\langle v,e_{g}\rangle\,e_{g},}
\eeq
which shows that $T_{g}$ are reflections in the quadratic space $\big(E,\langle\cdot,\cdot\rangle\big)$.

\begin{definition} We call \emph{reduced group of sheet transitions} the reflection group $\mathfrak{G} \subseteq GL(E)$ generated by $(T_{g})_{g \in G}$.
\end{definition}

The vectors $\pm e_{g}$ are $\langle$orthogonal$\rangle$ to reflection hyperplanes, and their orbit by $\mathfrak{G}$ then form of a root system $\mathfrak{R}$. $\mathfrak{G}$ coincides with the Weyl group of $\mathfrak{R}$. If we choose a subset $I \subseteq G$ so that $(e_{g})_{g \in I}$ spans $E$, $\mathbf{A} = (A(e_{g})|e_{h})_{g,h \in I}$ plays the role of a "Cartan matrix". We add quotes since it is not a priori a generalized Cartan matrix in the usual sense: off-diagonal elements might not be nonpositive. We also stress that the bilinear form $\langle\cdot,\cdot\rangle$ might not be definite -- compared to standard definitions, we waive this condition when we speak of root systems. We have three more observations:

\vspace{-0.1cm}

\begin{remark}
Since the reflections are $\langle$isometries$\rangle$ and $\langle e_{g},e_{g}\rangle = 1$, $\mathfrak{R}$ is simply-laced.
\end{remark}
\begin{remark}
 If furthermore all $\hat{\alpha}(g)$ are integers, $\mathfrak{R}$ is crystallographic. 
\end{remark}
\begin{remark}
\label{le1} $\mathfrak{R}$ is irreducible.
\end{remark}
\textbf{Proof.} Indeed, assume that $\mathfrak{R}$ can be decomposed in a disjoint union of two mutually $\langle$orthogonal$\rangle$ root systems: $\mathfrak{R}'$ containing $e_{{\rm id}}$, and a second one $\mathfrak{R}''$. Consider \mbox{$G' = \{g \in G,\quad e_{g} \in \mathfrak{R}'\}$}. We claim that $G'$ is a subgroup of $G$ generated by $S$ ; since we assumed initially that $G$ is generated by $S$, this entails $G' = G$, thus $\mathfrak{R}'' = \emptyset$.

To justify the claim, notice that $1 \in G'$. Then, $S\cup S^{-1} \subseteq G'$ since $\langle e_{{\rm id}},e_{\varsigma} \rangle = \langle e_{{\rm id}},e_{\varsigma^{-1}} \rangle = \hat{\alpha}(\varsigma) \neq 0$ when $\varsigma \in S$, which means that $e_{\varsigma}$ (and $e_{\varsigma^{-1}}$) cannot be $\langle$orthogonal$\rangle$ to $\mathfrak{R}'$, so cannot belong to $\mathfrak{R}''$, hence must belong to $\mathfrak{R}'$. Eventually, if $g \in G'$ and $\varsigma \in S$, we have $\langle e_{g},e_{\varsigma g}\rangle = \hat{\alpha}(\varsigma) \neq 0$, so $\varsigma g \in \mathfrak{R}'$ for the same reason. Since we assumed that $S$ generates $G$, we conclude that $\mathfrak{R}''$ contains no $e_{g}$ for $g \in G$, so is empty. \hfill $\Box$

\vspace{0.2cm}

\noindent We can now come back to the action of $\hat{\mathfrak{G}}$ on $\hat{E} = \mathbb{K}[G]$. In the block decomposition $\hat{E} = E \oplus E_0$, it takes the form:
\beq
\label{e34} \hat{T}_{g} = \left(\begin{array}{cc} A & 0 \\ 0 & \mathrm{id} \end{array}\right)\left(\begin{array}{cc} T_{g} & e_g\,\ell_{g} \\ 0 & \mathrm{id} \end{array}\right)\left(\begin{array}{cc} A^{-1} & 0 \\ 0 & \mathrm{id} \end{array}\right).
\eeq
Therefore, $\hat{\mathfrak{G}}$ is isomorphic to a subgroup of the semi-direct product $\mathrm{Hom}_{\mathbb{K}}(E_0,E)\rtimes \mathfrak{G}$, where the group structure of the latter is:
\beq
\forall f,f' \in \mathrm{Hom}(E_0,E),\quad \forall \Psi,\Psi' \in \mathfrak{G},\qquad (f,\Psi)\cdot(f',\Psi') = (\Psi\circ f' + f,\Psi\circ\Psi').
\eeq
We will encounter examples where $\mathfrak{G}$ is a finite Weyl group and $\hat{\mathfrak{G}}$ is the corresponding affine Weyl group. In general, it seems fairly complicated to describe completely $\hat{\mathcal{G}}$, and we shall restrict ourselves to compute $\mathfrak{G}$.

\subsection{Bonus: intertwining by the Galois group $G$}
\label{bonus}
Because we are acting in a group algebra $\hat{E} = \mathbb{K}[G]$, we have more "symmetries" than just the Weyl group $\mathfrak{G}$. 

We first start with the observation -- independent of the group structure of $G$ -- that we have a group homomorphism:
\beq
\label{action}\begin{array}{ccc} \mathfrak{S}(G) & \longrightarrow & \mathrm{GL}(\hat{E}) \\
\pi & \longmapsto &  \Big(\sum_{g \in G} \hat{v}(g)\,\hat{e}_g \mapsto \sum_{g \in G} \hat{v}(\pi(g))\,\hat{e}_{g}\Big)
\end{array}
\eeq
i.e. an linear action of the group of permutations of $G$ on $\hat{E}$. Therefore, any action of a group $\mathsf{G}$ on $G$ (i.e. a group homomorphism $\mathsf{G} \rightarrow \mathfrak{S}(G)$) induces a linear action on $\hat{E}$ by composition with \eqref{action}. Since they just permute the elements of the canonical basis, these actions are isometries with respect to the canonical scalar product.

Now, let us take advantage of the group structure on $G$. $G$ acts by right multiplication on $\hat{E}$, and this leaves $E$ and $E_0$ stable. We denote $\varepsilon_{h}$ the endomorphism of $\hat{E}$ given by \emph{right} multiplication by $h^{-1}$:
\beq
\varepsilon_h(\hat{v}) = h^{-1}\cdot\hat{v}
\eeq
so as to have a \emph{left} action of $G$ on $\hat{E}$. For any $h \in G$, the $\varepsilon_{h}$  are isometries of $\hat{E}$ as we have seen, but one can check with \eqref{quadr} that they are $\langle$isometries$\rangle$ as well. To summarize:
\begin{remark}
$E$ carries a representation of $G$ by $\langle$isometries$\rangle$, and a computation shows that this representation intertwines the generators of $\mathfrak{G}$:
\beq
T_{g} =\varepsilon_{h}T_{gh}\varepsilon_h^{-1}.
\eeq
\end{remark}
As a matter of fact, we see from the form \eqref{quadr} of $\langle\cdot,\cdot\rangle$ that for generic $\alpha$'s, this is the only possible action on $\hat{E}$ by $\langle$isometries$\rangle$.

\begin{remark} If $E$ contains at least one element $\hat{e}_{g_0}$, by right multiplication it must contain all $(\hat{e}_{g})_{g \in G}$, thus $\hat{E} = E$, i.e. $A$ is invertible. Similarly, no element $\hat{e}_{g_0}$ belongs to $E_0$: if it was the case, $E_0$ would contain all $(\hat{e}_{g})_{g \in G}$, which is not possible since $\hat{\alpha} \neq 0$.
\end{remark}
This explains that, when $\mathrm{Ker}\,A \neq 0$, $\phi(x)$ will never have a finite monodromy group, but it does not prevent linear combinations $(\hat{v}\cdot\phi)(x)$ to have a finite monodromy group for well-chosen $\hat{v}$.

If $G$ is non-commutative, $G$ also act by left multiplication on $\hat{E}$, but it is less interesting. This action is an isometry (for the canonical scalar product). $E_0$ remains stable under $h\cdot$ iff $h$ is a central element, and in this case, $E$ is also stable.

\subsection{Orbits: description and finiteness}
\label{orbfi}

We can now reap the rewards of our algebraic discussion:
\begin{corollary}
\label{cofini}There exist a non-zero vector $\hat{v} \in \hat{E}$ with finite $\hat{\mathfrak{G}}$-orbit iff $\mathfrak{R}$ is finite. Then, $\hat{v}$ has a finite orbit iff $\hat{v} \in E$.
\hfill $\Box$
\end{corollary}
Since $\mathfrak{R}$ is crystallographic -- for integer $\alpha$'s --, simply-laced and irreducible, if we assume that it is finite, it must be of $ADE$ type. In this paragraph, we assume it is the case.

If $F[I] \subseteq E$ is a subspace generated by a subset $I$ of the roots, we denote $H_{F[I]}$ its $\langle$orthogonal$\rangle$ subspace, and:
\beq
H'_{F[I]} = H_{F}\setminus\Big(\bigcup_{r \in \mathfrak{R}\setminus F[I]} H_{F[I,r]}\Big).
\eeq
It is made of the elements of $E$ $\langle$orthogonal$\rangle$ to the roots in the set $I$ and only to them. For instance, $F[\mathfrak{R}] = 0$, and $H_{0} = E$, whereas $H'_{0}$ is the complement of the union of the reflection hyperplanes: its connected components are the Weyl chambers. In general, the connected components of $H'_{F[I]}$ define cones of dimension $\mathrm{dim}\,E - \mathrm{dim}\,F[I]$, and provide a partition of $E$ indexed by subsets of simple roots. Actually, $(H'_{F'})_{F'}$ for $F'  \subseteq F$ provides a partition of $H_{F}$. We can give a complete description of the $\mathfrak{G}$-orbits of an element $v \in E$:
\begin{lemma}
The stabilizer of $v \in H'_{F}$ is the subgroup of $\mathfrak{G}$ generated by the reflections associated to the roots $r$ which belong to $F$. The connected components of $H'_{F}$ are in bijection with points in $\mathfrak{G}/\mathfrak{G}_{v}$, and there is exactly one point of the $\mathfrak{G}$-orbit of $v$ in each of them. The $\mathfrak{G}$-orbit of $v$ spans $E$. \hfill $\Box$
\end{lemma}
The type of orbits are therefore classified by the parabolic subgroups of $\mathfrak{G}$, themselves classified up to conjugacy by subsets of a set of simple roots for $\mathfrak{R}$ (see e.g. \cite[Chapter 2]{GPbook}). Three types of orbits are remarkable:
\begin{itemize}
\item[$\bullet$] If $v$ belongs to $H'_0$ (i.e. belongs to one of the Weyl chambers), the skeleton graph $\mathcal{G}_{v}$ is isomorphic to the Cayley graph of $\mathfrak{G}$ with set of generators $(T_{g})_{g \in G}$.
\item[$\bullet$] If $v$ is colinear to a root, the set of vertices of the skeleton graph is isomorphic to the set of roots $\mathfrak{R}$.
\item[$\bullet$] Orbits of small order are obtained when $v$ is a non-zero element of $H_{F[I]}$ where $I$ is the set of simple roots minus one of them. $H_{F[I]}$ is then one-dimensional. In order to obtain orbits of minimal orders, we have to choose the simple root to delete such that $\mathfrak{R}[I]$ with a Weyl group of maximal order. Then, we call $\mathfrak{R}[I]$ a maximal sub-root system, and any generator $v^*$ of $H_{F[I]}$ a \emph{maximal element}.
\end{itemize}
We insist that, once an element $v _0\in E$ giving rise to a $\mathfrak{G}$-orbit is chosen, we are actually interested in the $\hat{\mathfrak{G}}$-orbit of $\hat{v}_0 = \hat{\alpha}\cdot v_0 = A(v_0)$ in order to describe the analytic continuation $\varphi_{\hat{v}_0}$ of a solution to the functional equation \eqref{thep}. We introduce:
\beq
\boxed{\mathcal{M} = \{\hat{v} \in \hat{E},\quad \hat{v} = A(v^*),\quad v^*\,\,\mathrm{maximal}\,\,\mathrm{element}\},}
\eeq
the set of non-zero vectors in $\hat{E}$ whose orbit has minimal order. It is obviously stable under the action of $\hat{\mathfrak{G}}$, but more interestingly, as a consequence of \S~\ref{bonus}, it is also stable by right multiplication by elements of $G$.

\subsection{Enlarging the root system}

If we waive the restriction that the quadratic form be non-degenerate, we may define "root systems" larger than $\mathfrak{R}$, which will contain more information on the full group of sheet transitions $\hat{\mathfrak{G}}$. Given \eqref{e34}, if $E_0'$ is any subspace of $E_0$ (the orthogonal of $E$ for the usual scalar product in $\hat{E}$), $E' = E_0' \oplus E$ is stable under action of $(T_g)_{g \in G}$. So, the orthogonal projection of $(\hat{e}_{g})_{g \in G}$ onto $E'$, and its images under $(T_g)_{g \in G}$ still belong to $E'$, and define a "root system" $\mathfrak{R}'$ (depending on the choice of $E_0'$). We can also take $E' = \hat{E}$, and define a "root system" $\hat{\mathfrak{R}}$.  For practical purposes, the guideline is to include as much vectors as possible in $E'$, keeping in mind that we eventually would like to describe their $\hat{\mathfrak{G}}$-orbit.

\subsection{More information on the Riemann surface}

In this paragraph, we assume that $\hat{v}_0 \in \hat{E}$ is chosen to have a finite orbit, and we want to gain some information on the Riemann surface $\Sigma = \Sigma_{\hat{v}_0}$ on which $\hat{v}_0\cdot\phi(x)$ can be analytically continued to a function $\varphi_{\hat{v}_0}$.

For simplicity, we assume that $U$ is the Riemann sphere except a finite number of points away from $\Gamma$ where $\phi$ can have meromorphic singularities. So, in the construction of $\Sigma$ from the skeleton graph $\mathcal{G}_{\hat{v}_0}$, we can blow a Riemann sphere $\mathbb{C}_{\hat{v}}$ (instead of just $U$) at each vertex $\hat{v}$ of $\mathcal{G}_{\hat{v}_0}$. The outcome is a compact Riemann surface $\Sigma$, equipped with a branched covering $x\,:\,\Sigma \rightarrow \widehat{\mathbb{C}}$. Its degree $d$ is the number of vertices in $\mathcal{G}$, i.e. the order of the orbit of $\hat{v}_0$. 
\begin{lemma}
\label{Lgenus}Let $|\Gamma|$ be the number of connected components of $\Gamma$. The genus of $\Sigma_{\hat{v}_0}$ is:
\beq
\label{countg}\mathfrak{g} = 1 - d +  \frac{|\Gamma|}{2}\sum_{\hat{v} \in \mathcal{G}} |\mathrm{supp}\,\hat{v}|
\eeq
\end{lemma}
\textbf{Proof.} By construction, the branched covering $x\,:\,\Sigma \rightarrow \widehat{\mathbb{C}}$ has simple ramification points at the edges of the cuts. Each cut has two extremities, and $g\cdot\Gamma$ is a cut in $\widehat{\mathbb{C}}_{\hat{v}}$ iff $g$ is in the support of $\hat{v}$. Remember also that this cut is identified with $g\cdot\Gamma$ in $\hat{T}_g(\hat{v})$. Hence a total of $|\Gamma| \times \sum_{\hat{v} \in \mathcal{G}} |\mathrm{supp}\,\hat{v}|$. The announced result is then deduced from Riemann-Hurwitz formula:
\beq
2 - 2\mathfrak{g} = 2d - |\Gamma|\Big(\sum_{v \in \mathcal{G}} |\mathrm{supp}\,\hat{v}|\Big)
\eeq
\hfill $\Box$

\vspace{0.2cm}

\noindent $\varphi_{\hat{v}_0}$ is a meromorphic on $\Sigma$. So, there exists a polynomial equation of degree $d$ in $y = \varphi$:
\beq
\mathcal{P}(x,\varphi_{\hat{v}_0}) = 0
\eeq
Here are some general principles to grasp more information, and hopefully compute $\mathcal{P}$.
\begin{itemize}
\item[$(1)$] We usually want to solve the problem for $\phi(x)$ with prescribed singularities when $x \rightarrow 0$ or $\infty$. In other words, we know in each sheet $\hat{v} \in \mathcal{G}$ the leading term in the Puiseux expansion of $\varphi_{\hat{v}_0}(x)$ are related at leading order when $x \rightarrow 0$ or $\infty$ in $\mathbb{C}_{\hat{v}}$. This fixes the Newton polygon of $\mathcal{P}$, and the coefficients on its boundary.
\item[$(2)$] To go further, one may give names (say $c_j$'s) to the first few subleading coefficients in the Puiseux expansion of $\phi(x)$ when $\phi$ or $x$ go to $\infty$. Then, one can identify all coefficients in $\mathcal{P}$ in terms of $c_j$'s, just by writing:
\beq
\mathcal{P}(y,x) = C(x)\,\prod_{\hat{v} \in \mathcal{G}_{\hat{v}_0}} \big(y - (\hat{v}\cdot\phi)(x)\big)
\eeq
and finding what is the Puiseux expansion of the product in the right-hand side at $x \rightarrow 0$ or $\infty$.
\item[$(3)$] If the orbit of $\hat{v}_0$ has some extra symmetries, i.e. if there exists $\psi \in \mathrm{Aut}(\hat{E})$ leaving the orbit stable, this imposes some symmetries for the polynomial $\mathcal{P}$. Notice that, since the vectors of a given orbit must span $E$, and since $T_g$ are $\langle$isometries$\rangle$, such a $\psi$ must also be an $\langle$isometry$\rangle$.
For instance, it could happen (but it is not automatic) that the right multiplication by an element $h \in G$ leaves the orbit stable. 
\item[$(4)$] If the solution itself has a priori some symmetries (like \eqref{symm}), they should also reflect on $\mathcal{P}$.
\item[$(5)$] We know what is the ramification data of $x$ on $\Sigma$: the solutions of $\mathcal{P}(y,x) = \partial_{y} P(y,x) = 0$ must all be of the form $(y_b,g(x_b))$ for some $g \in G$ and $x_b$ an edge of $\Gamma$, and they must also satisfy to $\partial_{x} P(y_b,x_b) \neq 0$.
\end{itemize}

We have derived the properties that must satisfy the analytical continuation $\varphi_{\hat{v}_0}$ of $(\hat{v}_0\cdot\phi)$ when $\phi$ is a solution of \eqref{thep}. A priori, these are not sufficient conditions, and one has to check that $\varphi_{\hat{v}_0}$ satisfying the above constraints indeed provide a solution of the initial problem via \eqref{projections}. In particular, if we are given a collection of functions $(\psi_{\hat{v}})$ with the correct branching structure and asymptotic behavior, it is not at all automatic that:
\beq
\psi_{\hat{v};g}(x) = \frac{1}{\hat{v}(g)}\oint_{\Gamma} \frac{\dd s\,\psi_{\hat{v}}(g^{-1}(s))}{2{\rm i}\pi (x - s)}\psi_{\hat{v}}(x)
\eeq
does not depend on $\hat{v}$. This is nevertheless a property of the solution we are looking for. So, we only hope that imposing enough necessary constraints will lead to a finite number of possible polynomial equations, that can be browsed to meet more subtle constraints like positivity of the spectral density, or behavior at $u \rightarrow 0$, and hopefully find a unique $\mathcal{P}(x,y)$. In a practical problem -- like Seifert -- existence and unicity a priori is guaranteed, so we may conclude to the identification of the solution if the necessary conditions described above single out of unique $\mathcal{P}(x,y) = 0$.

\section{Seifert fibered spaces and Chern-Simons theory}
\label{S4}
\subsection{Geometry of Seifert fibered spaces}
\label{Seifertclass}
We review a number of facts about the geometry of Seifert fibered spaces \cite{Seifertbook}.

\subsubsection{Definition}
\label{defar}
One first defines the \emph{standard fibered torus} with slope $b/a$: It is a cylinder $[0,1]\times\mathbb{D}_2$, so that $(0,z)$ gets identified with $(1,e^{2{\rm i}\pi b/a}z)$, and the whole space is seen as a $\mathbb{S}_1$ fibration over the disk $\mathbb{D}_2$, for some order $a \geq 1$. A closed $3$-manifold is a Seifert fibered space if it admits a foliation by $\mathbb{S}_1$, so that each leaf (also called fiber) has a tubular neighborhood isomorphic to a standard fibered torus. Two famous examples of Seifert fibered spaces are:
\begin{itemize}
\item[$\bullet$] the lens space $L(a,b) = \mathbb{S}_3/\mathbb{Z}_p$. It can be realized by considering $\mathbb{S}_3 = \{(z_1,z_2) \in \mathbb{C}^2,\quad |z_1|^2 + |z_2|^2 = 1\}$ and the identifications $(z_1,z_2) \sim (\zeta_{a}z_1,\zeta_{a}^b z_2)$. We need to assume $a$ and $b$ coprime for the space to be smooth.
\item[$\bullet$] the Poincar\'e sphere, which can be described in several ways. It is the space of configurations of a icoashedron in $\mathbb{R}^3$. It is also obtained by identifying opposite faces of a dodecahedron (the icosahedron's dual). It is the only non-trivial integer homology sphere with finite fundamental group (= the binary icosahedral group).
\end{itemize}

\subsubsection{Classification}
\label{classar}
Let $M$ be a Seifert fibered space. All but a finite number number of fibers are ordinary, i.e. they have order $a = 1$. We denote $a_1,\ldots,a_r \geq 2$ the orders of exceptional fibers. Identifying in $M$ all points of the same leaf, one obtains the orbit space $O$. Ordinary fiber project to a smooth point in $O$, whereas an exceptional fiber of order $a_i$ projects to a $\mathbb{Z}_{a_i}$-orbifold point in $O$. $O$ is a $2$-dimensional orbifold. The classification depends on the topology of $O$ and the orientability of $M$. In this article, we always assume that $O$ is topologically a sphere and $M$ is orientable. The classes of oriented Seifert fibered spaces (modulo orientation and fiber preserving maps) are in one-to-one correspondence with uples of integers
\beq
 (b;a_1,b_1;\ldots;a_r,b_r),\qquad 1 \leq b_i \leq a_i - 1,\qquad \mathrm{gcd}(a_i,b_i) = 1,\qquad b \in \mathbb{Z}
\eeq
$a_i,b_i$ characterize the neighborhood of exceptional fibers, and the integer $b$ tells how exceptional fibers sit together in the global geometry. Changing orientation has the effect:
\beq
(b;a_1,b_1;\ldots;a_r,b_r) \longrightarrow (b - r;a_1,a_1 - b_1;\ldots;a_r,a_r - b_r).
\eeq
Therefore, the integer:
\beq
\label{sigmadef}\sigma = b + \sum_{m = 1}^r \frac{b_m}{a_m}
\eeq
is an invariant of orientable Seifert fibered spaces. If there are $r \geq 3$ exceptional fibers, $a_1,\ldots,a_r$ themselves are topological invariants of $M$. This is not true anymore in presence of one or two exceptional fibers, i.e. for $r \leq 2$ there exist homeomorphism which do not preserve fibers and change $(a_1,a_2)$. $r = 1,2$ give lens spaces, and since they are well-understood from the point of view of Chern-Simons theory \cite{Halma,BEMknots}, we shall assume $r \geq 3$ in the following.

\subsubsection{Orbifold Euler characteristic}
\label{orbiar}
Another important invariant, as we have seen in the matrix model, is the orbifold Euler characteristic of the orbit space:
\beq
\chi = 2 - r + \sum_{m = 1}^r \frac{1}{a_m}
\eeq
$M$ has a finite fundamental group iff $\chi > 0$. As we have already seen, for $r \geq 3$ the only possible exceptional fiber data in this case are $(2,2,p)$, $(2,3,4)$ and $(2,3,5)$. The corresponding Seifert fibered are precisely the quotients of $\mathbb{S}_3$ by the free action of a group of isometries. The outcome is that, up to isomorphism, $M = \mathbb{Z}_{n}\backslash\mathbb{S}_3/\mathfrak{H}$, with $\mathfrak{H}$ is a finite subgroup of $\mathrm{SU}(2,\mathbb{C})$, namely a cyclic group of order $n'$ (leading to lens spaces), or a binary polyhedral group:
\begin{itemize}
\item[$\bullet$] dihedral group of order $4p$, giving exceptional fibers $(2,2,p)$ -- labeled by $D_{p + 2}$ as regards the ADE classification of subgroups of $\mathrm{SO}(3,\mathbb{R})$.
\item[$\bullet$] symmetry group of the tetrahedron, order $24$, giving $(2,3,3)$ -- case $E_6$.
\item[$\bullet$] symmetry group of the octahedron, order $48$, giving $(2,3,4)$ -- case $E_7$.
\item[$\bullet$] symmetry group of the icosahedron, order $120$, giving $(2,3,5)$ -- case $E_8$.
\end{itemize}
There are only $4$ classes of Seifert fibered spaces with $\chi = 0$, namely $(2,2,2,2)$, $(3,3,3)$, $(2,4,4)$ and $(2,4,6)$. If $\sigma \neq 0$, they have Nil geometry. 

\subsubsection{Fundamental group and homology}
\label{homar}
The fundamental group of an orientable Seifert fibered spaces whose orbit space is a sphere is generated by $c_i$ going around the $i$-th exceptional fiber ($1 \leq i \leq r$), and a central element $c_0$ with relations:
\beq
c_1\cdots c_r = c_0^{b},\qquad c_i^{a_i}c_0^{b_i} = 1\,\,\,\,(1 \leq i \leq r)
\eeq
For $r \geq 3$, one can show that $\pi_1(M)$ is finite iff $\chi > 0$ (a fortiori it requires exactly $r = 3$). 

Another natural question is to ask for Seifert fibered spaces which are integer (resp. rational) homology spheres, i.e have trivial (resp. trivial up to torsion) $H_1(M)$ but are not $\mathbb{S}_3$. The answer is that $a_1,\ldots,a_r$ must be pairwise coprime. If this constraint is satisfied, then $M$ is a rational homology sphere, with:
\beq
|H_1(M)| = a\,|\sigma|,\qquad a = \prod_{m = 1}^r a_m
\eeq
and there exists a unique $b,b_1,\ldots,b_r$ such that $a \sigma = \pm 1$, i.e. such that $M$ is an integer homology sphere.
The Poincar\'e sphere is the unique case with finite fundamental group, its data is:
\beq
(b;a_1,b_1;a_2,b_2;a_3,b_3) = (-1;2,1;3,1;5,1),\qquad \chi = \sigma = \frac{1}{30}
\eeq
and for this reason it is the most interesting geometry treated in this article, but also the most cumbersome among the $\chi > 0$ cases \ldots{}  For other values of $(b;a_1,b_1;a_2,b_2;a_3,b_3)$, one obtains the Brieskorn spheres. 

\subsection{Avatars of Chern-Simons theory}
\label{CSavatar}
As a matter of fact, \eqref{themodel} first appeared in \cite{BNrational} for the L\^{e}-Murakami-Ohtsuki invariant \cite{LMO} on Seifert spaces. 

If $M$ is a closed, framed $3$-manifold obtained by surgery on a link in $\mathbb{S}_3$, the LMO invariant $Z_{{\rm LMO}}(M)$ is a graph-valued formal power series in $\hbar$ \cite{LMO}. Its relation to the Kontsevich integral -- which is a universal formal series of finite type invariant -- and Aarhus integral for rationally framed links was exposed in \cite{AarhusI,AarhusII,AarhusIII}, see also \cite{BNrational}. For any choice of compact Lie algebra $\mathfrak{g}$, it can be evaluated to a $\mathfrak{g}$-dependent, formal power series in $\hbar$. In particular, the evaluation with the Lie algebras of the series $A_N$, $B_N$, $C_N$ or $D_N$ yields up to normalization $\ln Z_{{\rm LMO}}^{\mathfrak{u}_N}$ as an element of $\hbar^{-2}\mathbb{Q}[[\hbar^2]][[u_0]]$ with:
\beq
\boxed{u_0 = N\hbar.}
\eeq
This is no more than a repackaging of LMO invariants, which is a weaker invariant than LMO \cite{Vogel}, and gives the geometric foundation of the quantities we shall compute in this article.

LMO can be considered as a mathematical definition of the perturbative expansion of the Chern-Simons partition function:
\beq
Z_{{\rm CS}}^{\mathfrak{g}}(M) = \int \mathcal{D}A\,e^{{\rm i}S_{{\rm CS}}[A]},\qquad S_{{\rm CS}}[A] = \frac{{\rm i}}{2} \int_{M} \big(A\wedge\dd A + \frac{2}{3}\,A\wedge A\wedge A\big)
\eeq
where the path integral runs over $\mathfrak{g}$-connections $A$ modulo small gauge transformations. The saddle points of the action are the flat connections, and in principle, $Z_{{\rm CS}}^{\mathfrak{g}}$ should be given by the sum over all flat connections of its perturbative expansions. When $M$ is a rational homology sphere, the LMO invariant is tailored to capture the contribution of the trivial flat connection. In particular, we have for the $A_N$ series:
\beq
Z_{{\rm LMO}}^{A} = \exp\Big(\sum_{g \geq 0} \hbar^{2g - 2}\,F_{g}^{A}\Big)
\eeq
and for any of the $X \in \{B_N,C_N,D_N\}$ series:
\beq
Z_{{\rm LMO}}^{X} = \exp\Big(\sum_{g \in \mathbb{N}/2} \hbar^{2g - 2}\,F_{g}^{X}\Big)
\eeq
where $F_{g}^{X} \in \mathbb{Q}[[u_0]]$ are the Chern-Simons free energies. It is known \cite{GManalyticity} hat $F_g^{A}$ has a finite radius of convergence independent of $g$, i.e. can be seen as the power series expansion at $u_0 \rightarrow 0$ of an analytic function.
\begin{problem}
\label{P1} Describe the singularities of $F_g^{X}$ considered as a function of $u_0$.
\end{problem}

Chern-Simons theory is a cornerstone in quantum topology, because of Witten's discovery \cite{Witten89} that the expectation value $\big\langle \mathrm{Tr}_{R} \exp\big(\oint_{K} A\big)\big\rangle$ with respect to the Chern-Simons measure (in principle computed by a path integral), is an invariant of framed knots in $M$, called "Wilson loops". Depending on the $3$-manifold, there are several ways to define rigorously those invariants, as functions of $q$, as elements of the Habiro ring, or as formal series, see e.g. the review \cite{BeliaT}.

\subsubsection{Link invariants in the $3$-sphere}

When $K$ is a knot in $M = \mathbb{S}_3$, Wilson loops turn out to be Laurent polynomials in:
\beq
q = e^{\hbar}
\eeq
and they coincide with $R$-colored HOMFLY polynomials. When $R$ is the fundamental representation of $\mathfrak{g} = \mathfrak{su}(N + 1)$, this retrieves the HOMFLY-PT polynomial \cite{HOMFLY}, and for $\mathfrak{g} = \mathfrak{su}(2)$, this is the Jones polynomial \cite{Jonesini}. When $R$ is the fundamental representation of a Lie algebra in the $BCD$ series, the Wilson line is related to the Kauffman polynomial \cite{Kauffman}. Reshetikhin and Turaev \cite{TuraRe} later provided the foundation for the rigorous TQFT construction of those invariants. 

The HOMFLY-PT of a link in $\mathbb{S}_{3}$ satisfies skein relation, which allows to reduce the computation of HOMFLY-PT of any link to the computation of HOMFLY-PT by resolution of crossings. Besides, the $R$-colored HOMFLY of a link can be realized as the HOMFLY-PT of a link obtained by taking parallels in a $R$-dependent way \cite{MMPle}. The Kauffman polynomial also satisfy a more general skein relation, but in general the colored HOMFLY polynomial of a link is not known to reduce to the HOMFLY of a related link. It is well-known that $F_g^{A}$ and the coefficients of a given power of $\hbar$ in the colored HOMFLY are entire functions of $e^{u}$ (see e.g. \cite{BEMknots}).

\subsubsection{Link invariants in rational homology spheres}
\label{thelink}
When $K$ is a knot in $M \neq \mathbb{S}_{3}$, Wilson loops can always be defined as a power series in $\hbar$ in perturbative Chern-Simons around a chosen flat connection, but cannot in general be upgraded to a function of $q$  and $q^N$ (see \cite{Witten89,GarouJones}). When $M$ is a rational homology sphere, one can formulate a skein theory for the HOMFLY-PT invariant of a link $K$ in $M$ considered as a formal series in $\hbar$ and $u_0 = N\hbar$. Skein theory then determines the colored HOMFLY for any link $L$, if one knows the value of the HOMFLY-PT for a set of \emph{basic knots} representing the conjugacy classes of $\pi_1(M)$ \cite{Kalfa}. 

Two interesting and open questions are:
\begin{problem}
What is the value of HOMFLY-PT for basic knots in a given rational homology sphere ?
\end{problem}
\begin{problem} \label{P3fds} Consider the coefficient of a given power of $\hbar$. Is it the power series expansion of an analytic function of $q^{N} = e^{u_0}$ when $u_0 \rightarrow 0$ ? What are the singularities in the complex plane of this function ? 
\end{problem}

In this article, we study the Wilson loops (in any representation $R$ of fixed size) for the knots going along the extraordinary fibers of Seifert spaces. Our method gives in principle a way to compute the coefficients of the power series in $\hbar$ as functions of $e^{u_0}$ for Seifert spaces with $\chi \geq 0$, and provide some partial answers to Problem~\ref{P1} and \ref{P3fds} : the perturbative invariants are algebraic functions of $e^{u_0}$, but in general not entire functions. For instance, in the case of $\mathbb{S}_{3}/D_{p + 2}$ with $p$ even, we could push the computation to the end and describe precisely the algebraic function field in which the perturbative invariants sit (see Theorem~\ref{t7}). For all $\mathbb{S}_{3}/D_{p}$, we can also show that the perturbative invariants have no singularity for $u_0$ on the positive (resp. negative) real axis if $\sigma > 0$ (resp. $\sigma < 0$), and Conjecture~\ref{conj2} would imply this is also true for all Seifert fibered spaces.

The fiber knots only form a subset of the basic knots: we are missing the knots going along a meridian of the exceptional fibers. At present, it is not known how to rewrite their Wilson lines as observables in the model \eqref{themodel}. We nevertheless propose the following:
\begin{conjecture}
\label{airh}For any Seifert space $M$ with $\chi > 0$, there exists a finite degree extension $L$ of $\mathbb{Q}(e^{-\chi u_0/2a})$, such that all perturbative colored HOMFLY invariants of links in $M$ belong to $L$. And, all perturbative colored Kauffman invariants of links in $M$ belong to $L_{2} \subseteq L$, obtained from $L$ by substitution $u_0 \rightarrow 2u_0$.
\end{conjecture}
In this case, $L$ would be an invariant of the ambient $3$-manifold. In Theorem~\ref{t7}, we show for $\mathbb{S}_{3}/D_{p + 2}$ that $L$ contains $\mathbb{Q}(u,\kappa^2,\beta)/\mathcal{I}$, where $\mathcal{I}$ is generated by:
\beq
\frac{4^{2p}\kappa^{2(p + 1)}}{(1 + \kappa^2)^{2p}} = e^{-u/(2p\sigma)},\qquad \beta^2 = \frac{(\kappa^2 + 1)((p + 1)\kappa^2 - (p - 1)}{\kappa^2}
\eeq

\subsection{Exact evaluations}

\begin{figure}[h!]
\begin{center}
\includegraphics[width=0.5\textwidth]{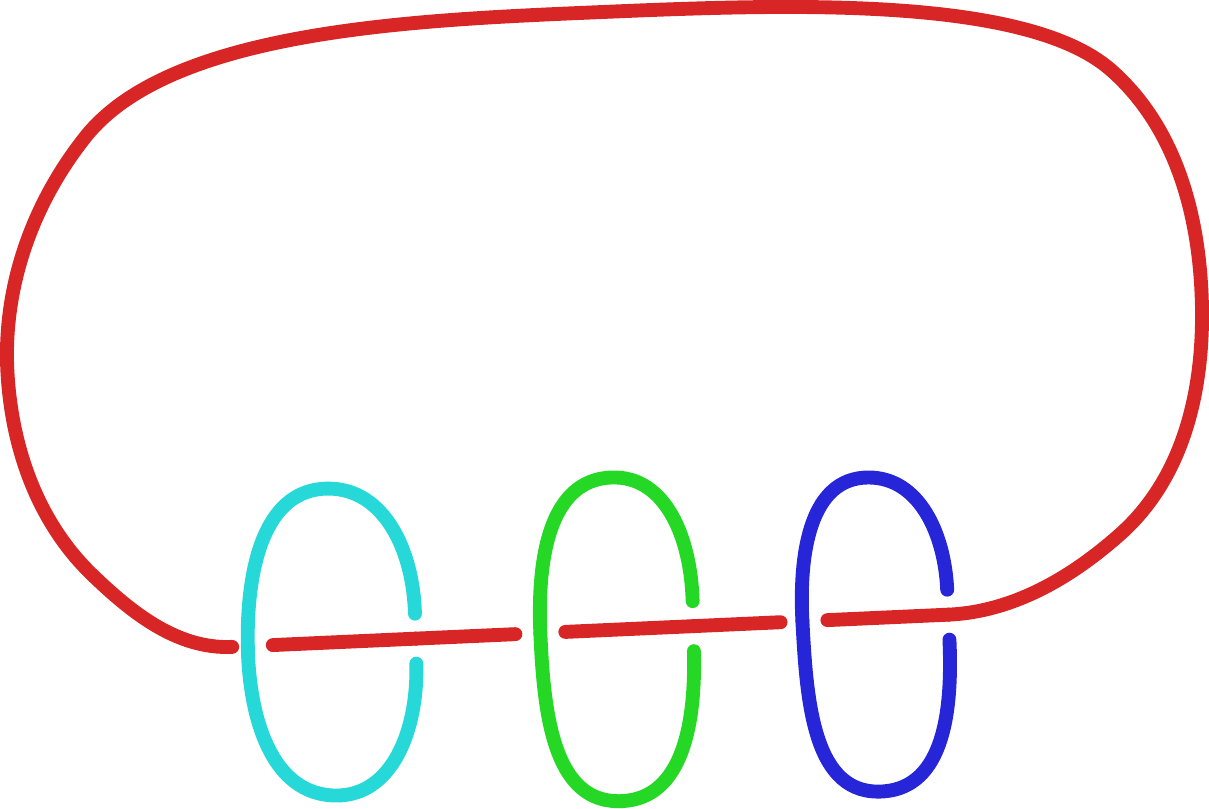}
\caption{\label{surgery} Seifert spaces which are rational homology spheres can be realized by rational surgery on this link with $(r + 1)$ components (here $r = 3$). The surgery data is $1/b$ on the horizontal component and $a_m/b_m$ on the $m$-th vertical components ($1 \leq m \leq r$). Snappy courtesy of S. Garoufalidis.}
\end{center}
\end{figure}

Exact evaluations of the Chern-Simons path integrals are rare. By "exact evaluation", we mean the reduction to a finite-dimensional sum (over dominant weights of $\mathfrak{g}$) or integral (over the real Cartan subalgebra $\mathfrak{h}$ of $\mathfrak{g}$).  Seifert fibered spaces \cite{Seifertbook} are one of the few classes of non-trivial $3$-manifolds for which it has been performed so far, and the contribution of the trivial flat connection takes the form \eqref{themodel}:

\beq
Z_{\mathrm{CS}}^{\mathfrak{g}} = C^{\mathfrak{g}}\,\int_{\mathfrak{h}} \prod_{\alpha > 0} \Big[\mathrm{sinh}^{2 - r}\big(\frac{\alpha\cdot t}{2}\big) \prod_{m = 1}^r \mathrm{sinh}\big(\frac{\alpha\cdot t}{2a_m}\big)\Big]\,\prod_{j = 1}^N e^{-\sigma t_j^2/2\hbar},\eeq
where $C^{\mathfrak{g}}$ is a known prefactor given in \cite{MarinoCSM}. Actually, \eqref{themodel} can be derived in various ways, either in the realm of LMO or of TQFT. Seifert spaces turn out to be tractable either because they can be obtained by rational surgery on a very simple link in $\mathbb{S}_3$ (see Figure~\ref{surgery}), and TQFT behaves well under surgery ; or because they carry a $U(1)$ action and localization of the path integral occurs. Here is a schematic account of the history of those exact formulae:
\begin{itemize}
\item[$\bullet$] For $\mathfrak{g} = \mathrm{su}(2)$ or $\mathfrak{so}(3)$ and $M$ a Seifert integer homology sphere, Lawrence and Rozansky \cite{Law} have used the Reshetikhin-Turaev construction to rewrite $Z_{{\rm CS}}^{\mathfrak{g}}$ as a $1$-dimensional integral \eqref{themodel}, including contributions of all flat connections.
\item[$\bullet$] Mari{\~{n}}o generalized their derivation to any simply laced-Lie algebra $\mathfrak{g}$ and Seifert rational homology spheres $M$ \cite{MarinoCSM}.
\item[$\bullet$] Bar-Natan \cite{BNrational} has computed the LMO invariant of Seifert rational homology spheres, via the Kontsevich integral.
\item[$\bullet$] Beasley and Witten \cite{BeasWitten} have developed a non-abelian localization method, allowing the computation of the contribution of isolated flat connections\footnote{The trivial flat connection in a Seifert fibered spaces is isolated iff the $a_i$ are pairwise coprime. If $\chi \geq 0$, the only cases concerned are the lens spaces $L(p,q)$, and the $(2,3,5)$ cases including the Poincar\'e sphere.} Then, correlation functions of Schur polynomials for the measure \eqref{themodel} can be interpreted in terms of Wilson loops along exceptional fibers \cite{Beas}.
\item[$\bullet$] K\"allen \cite{Kallen} derives the same results, building on earlier work of \cite{Kapustin} on a supersymmetric version of Chern-Simons theory.
\item[$\bullet$] Blau and Thompson developed a diagonalization technique, first for $U(1)$ bundles over smooth surfaces \cite{BT1}, then for $U(1)$ bundles over orbifolds \cite{BT2}, allowing the computation of the full Chern-Simons partition function. As a particular case, they retrieve the earlier results on Seifert rational homology spheres.
\end{itemize}

\subsection{Correlators and Wilson loops}

\label{Wilson}

We review the interpretation of the correlators of the model \eqref{themodel} in terms of Wilson loops. \cite{Beas} tells us that the holonomy operator $\mathcal{U}_{a_m}$ along the exceptional fiber of order $a_m$ -- on the Chern-Simons side -- gets identified with $\mathrm{diag}(e^{t_1/a_m},\ldots,e^{t_N/a_m}) = \mathbf{S}^{a/a_m}$ -- on the matrix model side, with the notations of \eqref{themodel}. Therefore, the Wilson loop in representation $R$ is equal to:
\beq
\mathcal{H}_{R;a_m} = \mu\big[\mathrm{ch}_{R}(\mathbf{S}^{a/a_m})\big],\qquad \mathbf{S} = \mathrm{diag}(e^{t_1/a},\ldots,e^{t_n/a}),
\eeq
where $\mathrm{ch}_{R}$ is the character of $R$, i.e. the Schur polynomial indexed by $R$. We prefer to work in the power-sum basis of the representation ring, and with connected observables:
\beq
\mathcal{W}_n\big(\sheet{d_1}{k_1},\ldots,\sheet{d_n}{k_n}\big) = \big\langle\mathrm{Tr}\,\,\mathcal{U}_{d_1}^{k_1}\,\cdots\,\mathrm{Tr}\,\,\mathcal{U}_{d_n}^{k_n}\big\rangle_{{\rm conn}} = \mu\big[\mathrm{Tr}\,\,\mathbf{S}^{k_1a/d_1}\,\cdots\,\mathrm{Tr}\,\,\mathbf{S}^{k_na/d_n}\big]_{{\rm conn}},\nonumber
\eeq
where:
\beq
d_j \in \{a_1,\ldots,a_r\},\qquad k_j \in \mathbb{Z}_{+}.
\eeq
The $\mathcal{H}$'s and the $\mathcal{W}$'s are related by a change of basis: to extract $\mathcal{H}_{R}$ for a representation $R$ corresponding to a Young diagram with less $n$ rows, we need to compute $\mathcal{W}_{n'}$ with $n' \leq n$.

Recalling $a = \mathrm{lcm}(a_1,\ldots,a_r)$, we define the $n$-point correlators of the matrix model as:
\beq
W_n(x_1,\ldots,x_n)  =  \mu\Big[\mathrm{Tr}\,\,\frac{x_1}{x_1 - \mathbf{S}} \cdots \mathrm{Tr}\,\,\frac{x_n}{x_n - \mathbf{S}}\Big]_{{\rm conn}} = \sum_{l_1,\ldots,l_r \geq 0} \frac{\mu\big[\mathrm{Tr}\,\,\mathbf{S}^{l_1} \cdots \mathrm{Tr}\,\,\mathbf{S}^{l_n}\big]_{{\rm conn}}}{x_1^{l_1}\cdots x_n^{l_n}}
\eeq
so that the $\mathcal{W}_n$'s can be read from the coefficients of the expansion of $W_n(x_1,\ldots,x_n)$ in Laurent series when $x_1,\ldots,x_n \rightarrow \infty$. If $a_1,\ldots,a_r$ are not coprime, the expansion of $W_n$ also records expectation values of fractional powers of the holonomy along fibers, which do not have a clear interpretation in knot theory.

In a perturbative expansion, we have a decomposition of formal power series in $u$ of the form:
\beq
\label{expf}W_n(x_1,\ldots,x_n) = \sum_{g \geq 0} N^{2 - 2g - n}\,W_n^{(g)}(x_1,\ldots,x_n),\qquad W_n^{(g)} \in \mathbb{Q}[[x_1^{-1},\ldots,x_n^{-1},u]]
\eeq
Later, we shall consider only certain linear combinations of rotations of $W_n^{(g)}$, namely:
\beq
\label{tensor}\Big(\bigotimes_{i = 1}^n \hat{v}_i\Big)\cdot W_{n}^{(g)}(x_1,\ldots,x_n) = \sum_{j_1,\ldots,j_n \in \mathbb{Z}_{a}} \hat{v}(j_1)\cdots \hat{v}(j_n)\,W_{n}^{(g)}(\zeta_{a}^{j_1}x_1,\ldots,\zeta_{a}^{j_n}x_n)
\eeq
for $\hat{v}_i$ in a certain set $\mathcal{V}$ of vectors in $\mathbb{Z}^a$. If we denote the discrete Fourier transform:
\beq
\mathcal{F}_{k}[\hat{v}] = \sum_{j \in \mathbb{Z}_{a}} \zeta_{a}^{jk}\,\hat{v}(j)
\eeq
we have the expansions when $x_i \rightarrow \infty$:
\beq
\label{buigrw2}\Big(\bigotimes_{i = 1}^n \hat{v}_i\Big)\cdot W_n^{(g)}(x_1,\ldots,x_n) = \sum_{k_1,\ldots,k_n \geq 0} \prod_{i = 1}^n \frac{\mathcal{F}_{k_i}[\hat{v}_i]}{x_i^{k_i}}\,\langle \mathrm{Tr}\,\,\mathcal{U}^{k_1}\cdots \mathrm{Tr}\,\,\mathcal{U}^{k_n}\big\rangle^{(g)}_{{\rm conn}}
\eeq
and when $x \rightarrow 0$:
\beq
\label{buigrw}\Big(\bigotimes_{i = 1}^n \hat{v}_i\Big)\cdot W_{n}^{(g)}(x_1,\ldots,x_n) = (-1)^n \sum_{k_1,\ldots,k_n \geq 0} \prod_{i = 1}^n \frac{\mathcal{F}_{k_i + 1}[\hat{v}_i]}{x_i^{k_i + 1}}\,\langle\mathrm{Tr}\,\,\mathcal{U}^{k_1}\cdots\mathrm{Tr}\,\,\mathcal{U}^{k_n}\big\rangle^{(g)}_{{\rm conn}}
\eeq
In the latter, we have used that $\mathbf{S}$ is distributed like $\mathbf{S}^{-1}$. Therefore, knowing \eqref{tensor} for $\hat{v}_i \in \mathcal{V}$ will only give access to the coefficients of expansion of $\breve{W}_n^{(g)}$ in $x_i^{-m}$ with $(m\,\,\mathrm{mod}\,\,a)$ such that there exists $\hat{v} \in \mathcal{V}$ with non-zero $\mathcal{F}_{m}[\hat{v}]$.

\subsection{Remark on formal series versus asymptotic series}

Our point of view is to consider the Chern-Simons matrix model \eqref{themodel} for $u = u_0/\sigma = N\hbar/\sigma > 0$. We thus have to assume that $0 < q = e^{\hbar} < 1$ if $\sigma > 0$, or $q > 1$ if $\sigma < 0$. The correlators $W_n(x_1,\ldots,x_n)$ of the matrix model are then defined as functions of $u$, $N$ and $q$. Our point of view is to analyze the asymptotic expansion of the correlators when $N \rightarrow \infty$ for a fixed value of $u > 0$ and $x_1,\ldots,x_n \in \mathbb{C}\setminus \mathbb{R}$. When the equilibrium measure of the matrix model has one cut $\Gamma \subseteq \mathbb{R}_+^*$ (a property guaranteed by Lemma~\ref{1cut} when $\chi \geq 0$) and is off-critical, the results of \cite{BGK} ensure that we have an \emph{asymptotic expansion} when $N \rightarrow \infty$ of the form:
\beq
\label{exouh}W_n(x_1,\ldots,x_n) = \sum_{g \geq 0} N^{2 - 2g - n}\,W_n^{(g)}(x_1,\ldots,x_n)
\eeq
where now $W_n^{(g)}(x_1,\ldots,x_n)$ is a holomorphic function of $x_1,\ldots,x_n \in \mathbb{C}\setminus\Gamma$ and of $u > 0$. Its Laurent expansion when $x_i \rightarrow \infty$ and power series expansion when $u_0 = u\sigma \rightarrow 0$ retrieves the formal series of \eqref{expf}. This approach has the extra benefit to provide $W_n^{(g)}$ as function of $u_0$, hence to allow analytic continuation in $u_0$, and thus to address Problem~\ref{P3fds} concerning the singularities in $u_0$.

Given the results of Section~\ref{S2} for $\chi \geq 0$, off-criticality boils down to checking that the density of the equilibrium measure remains positive in the interior of its support. We already know this is true for any $\chi \geq 0$ provided $u$ is small enough. We did this check for all values of $u > 0$ in the cases $(2,2,p)$ since we have an explicit expression for $W_1^{(0)}(x)$. For the remaining cases with $\chi \geq 0$, such an expression is not available because of algebraic complexity, so we were not able to check:
\begin{conjecture}
\label{c233333}For $\chi \geq 0$ (except $(2,2,p)$ and $r \leq 2$ already known), off-criticality (and thus \eqref{exouh}) holds for all values of $u > 0$.
\end{conjecture}
We checked numerically this conjecture (see Section~\ref{appAlex}), but we could not find an a priori, potential-theoretic argument ruling out zeroes of the density in all cases $\chi \geq 0$. We will assume \eqref{c233333} to continue with our reasoning. Nevertheless, all propositions and theorems stated in the text are independent of this assumption.

\subsection{Origin of the measure}

The key feature of the model \eqref{themodel} is the interaction:
\beq
\label{power}\prod_{\alpha > 0} \mathrm{sinh}^{2 - r}\big(\frac{\alpha\cdot\mathbf{t}}{2}\big) \prod_{m = 1}^r \mathrm{sinh}\big(\frac{\alpha\cdot\mathbf{t}}{2a_m}\big)
\eeq 
where the product runs over $\alpha = $ positive roots of the Lie algebra. This is a pairwise interaction between $t_j$ for the ABCD series of Lie algebras. From a geometric perspective \cite{Witten89,BT2}, \eqref{power} is essentially the Ray-Singer torsion of Seifert fibered spaces. From the LMO perspective \cite{BNrational}, \eqref{power} arises from the evaluation of the wheels in the weight system $\mathfrak{g}$:
$$
\forall \mathbf{t} \in \mathfrak{t},\qquad \Omega(x) = \mathrm{det}\Big(\frac{\mathrm{sinh}(\hbar\,\mathrm{ad}\,\mathbf{t} /2)}{\hbar\,\mathrm{ad}\,\mathbf{t}/2}\Big)^{1/2} = \prod_{\alpha > 0} \frac{\mathrm{sinh}(\hbar\,\alpha\cdot\mathbf{t}/2)}{\hbar\,\alpha\cdot\mathbf{t}/2}
$$
and the decomposition of the Lebesgue measure $\dd\mathbf{X}$ over the real Lie algebra $\mathfrak{g}_{\mathbb{R}}$ in terms of the Haar measure $\dd U$ on $\exp(\mathfrak{g})$ and the Lebesgue measure $\dd\mathbf{t}$ on $\mathfrak{t}$:
$$
\dd \mathbf{X} = C_{\mathfrak{g}}\,\dd U\,\dd \mathbf{t}\,|\mathrm{det}(\mathrm{ad}\,\mathbf{t})| = C_{\mathfrak{g}}\,\dd U\,\dd \mathbf{t}\,\prod_{\alpha > 0} (\alpha\cdot\mathbf{t})^2
$$
for some constant $C_{\mathfrak{g}}$.

\subsection{Generalizations}

We describe generalizations of \eqref{themodel}, whose study is out of scope of this article.

\subsubsection{Non-trivial flat connections}

In exact evaluations, the partition function is in general obtained as a sum of terms identified with contributions of the different flat connections. The contribution of the trivial flat connection respect the full Weyl symmetry of $\mathfrak{t}$ and correspond to \eqref{themodel} up to a known prefactor. The contribution of other reducible flat connections is the analogue of \eqref{themodel} with a potential $V$ breaking the Weyl symmetry \cite{MarinoCSM}, in a maximum of $aS$ pieces. More precisely, the $t_j$ in this case are partitioned:
\beq
\ldbrack 1,N \rdbrack = \dot{\bigcup}_{\ell = 0}^{aS - 1} I_{\ell},\qquad |I_{\ell}| = N_{\ell}
\eeq
and the term $V(t_j)$ is replaced by:
\beq
\forall j \in I_{\ell},\qquad V(t_j) = \frac{1}{u}\Big(\frac{t_j^2}{2} + \frac{2{\rm i}\pi\,t_j\ell}{\sigma}\Big)
\eeq
And, there may exist residual terms corresponding to irreducible flat connections \cite{Law,MarinoCSM}. Since the measure in \eqref{themodel} is now complex, we cannot apply stricto sensu the arguments of asymptotic analysis raised in Section~\ref{S2} and \cite{BGK}. Nevertheless, we can take the saddle point equation \eqref{218} with complex valued right-hand side as a starting point, and compute the corresponding spectral curve with the methods of Section~\ref{S3}. The only difference in the result is a rescaling of the Newton polygon, and now the coefficients inside the Newton polygon depend on the collection filling fractions $\epsilon_{l} = N_{\ell}\hbar$. This dependence is in general transcendental, since the $\epsilon_{I}$ are periods of the $1$-form $\ln\,y\,\dd \ln x$ on the spectral curve. For lens spaces, this analysis has been explicitly performed in \cite{Halma}. 

\subsubsection{Orbit space of any topology}

For Seifert fibered spaces whose orbit space $O$ is a Riemann surface of genus $h$, \eqref{power} appears to a power $1 - h$ (half the usual Euler characteristics of $O$) \cite{BT2}. For $h \geq 2$ the corresponding partition function would be ill-defined for $t_i$ integrated over $\mathbb{R}$. But, in \cite{Law,MarinoCSM}, the formula as an integral over $\mathfrak{t}$ is actually derived from a sum over dominant weights of $\mathfrak{g}$, by an Euler-MacLaurin type formula and analytical continuation in $\hbar$. In other words, the original expression is a sum over discrete $\mathbf{t}$'s where, among other details, hyperbolic functions are replaced by their trigonometric analogue, and the walls of the Weyl chamber are excluded. When $h = 0$, we can add the wall contribution since it is $0$, and arrive to an integral over $\mathfrak{t}$. When $h \geq 2$, the correct formula is the discrete sum, with pairwise interactions between the $t_i$'s behaving like $|t_i - t_j|^{2 - 2h}$ when $t_i \rightarrow t_j$. We remark that the same kind of sums appear in the partition function counting simple coverings of surfaces of genus $h$ (simple Hurwitz numbers) \cite{BEMS}. Since $t_i$'s now attract each other -- but belong to a lattice -- the large-$N$ asymptotic analysis could be very different from the repulsive case treated so far, and it is not clear how to adapt our techniques to this case. For instance, it is already not obvious that the asymptotic expansion \eqref{exouh} holds, even for a small value of $u$.

\section{Spectral curve and $2$-point function : inhomogeneous part}

\label{SSei}
\subsection{The spectral curve}
\label{S42}
Let $a_1,\ldots,a_r$ be integers. We have established in \S~\ref{change} that the spectral curve satisfies -- on top of growth constraints -- the functional relation:
\beq
\label{theeq}W(x + {\rm i}0) + W(x - {\rm i}0) + (2 - r)\sum_{l = 1}^{a - 1} W(\zeta_{a}^{l}x) + \sum_{m = 1}^r \sum_{l = 1}^{a_m - 1} W(\zeta_{a_m}^{l}x) = \frac{a^2\ln x}{u} + \frac{a\chi}{2}.
\eeq
Here $\Gamma$ is a subset of $\mathbb{R}_+^{\times}$ to determine with the solution. The first step is to get rid of the right-hand side, and the way to achieve this depends whether $\chi = 0$ or not. Then, we arrive to the problem presented in Section~\ref{S3}, with Galois group $G = \mathbb{Z}_{a}$. We denote it additively, and $(\hat{e}_0,\ldots,\hat{e}_{a - 1})$ is the canonical basis of $\hat{E} = \mathbb{Z}[G]$. The sheet transitions are ruled by the vector:
\beq
\label{alhex} \boxed{\hat{\alpha} = 2\hat{e}_0 + (2 - r)\sum_{l = 1}^{a - 1} \hat{e}_{l} + \sum_{m = 1}^r \sum_{l_m = 1}^{a_m - 1} \hat{e}_{\check{a}_ml_m},\qquad \check{a}_m = a/a_m}
\eeq

\subsubsection{$\chi \neq 0$}

It is easy find a particular solution of \eqref{theeq} which has no discontinuity on $\Gamma$, and subtracting it to $W(x)$ we find that:
\beq
\label{phiW}\phi(x) = -\frac{\chi u}{a}\big(W(x) - 1/2\big) - {\rm i}\pi + \ln x
\eeq
satisfies to the homogeneous equation bringing us back to Section~\ref{S3}:
\beq
\phi(x + {\rm i}0) + \phi(x - {\rm i}0) + (2 - r)\sum_{l = 1}^{a - 1} \phi(\zeta_{a}^{l}x) + \sum_{m = 1}^r \sum_{l = 1}^{a_m - 1} \phi(\zeta_{a_m}^{l_m}x) = 0.
\eeq
The price to pay with \eqref{phiW} is that $\phi(x)$ now has a logarithmic singularity, but we can turn into a meromorphic singularity by setting:
\beq
\boxed{Y(x) = e^{\phi(x)}.}
\eeq
The functional equation for $Y$ is now multiplicative, but it does not make much difference from the point of view of Section~\ref{S3}. If $\hat{v} \in \hat{E}$, we write:
\beq
(\hat{v}\cdot Y)(x) = e^{(\hat{v}\cdot\phi)(x)} = \prod_{l = 0}^{a - 1} \big(Y(\zeta_{a}^{l}x)\big)^{\hat{v}(l)}
\eeq
We keep the same notation but the context should make clear if the action of $\hat{E}$ should be additive (on $\phi$) or multiplicative (on $Y$). \eqref{alhex} for $\phi(x)$ translates into:
\beq
\label{analycont}\forall l \in G,\qquad(\hat{v}\cdot Y)(x - {\rm i}0) = (\hat{T}_{l}(\hat{v})\cdot Y)(x + {\rm i}0) 
\eeq
Let us introduce the parameter:
\beq
\boxed{c = e^{-\chi u/2a}}.
\eeq
The growth conditions \eqref{growth} on $\phi(x)$ imply:
\bea
\label{Puis1}\hat{v}\cdot Y(x) & \sim\,\,  (-xc)^{n_0[\hat{v}]}\,\zeta_{a}^{n_1[\hat{v}]},\qquad & x \rightarrow 0 \\
\label{Puis2}\hat{v}\cdot Y(x) & \sim\,\, (-x/c)^{n_0[\hat{v}]}\,\zeta_{a}^{n_1[\hat{v}]},\qquad & x \rightarrow \infty
\eea
where $n_0[\hat{v}] = \sum_{l = 0}^{a - 1} \hat{v}(l)$ and $n_1[\hat{v}] = \sum_{l = 0}^{a - 1} l\,\hat{v}_l$. 

\subsubsection{$\chi = 0$}
\label{chi0}
If $\chi = 0$, we can find a particular solution of \eqref{theeq} containing $\ln^3 x$. Since we prefer to avoid this type of singularities, we take another route. The function
\beq
\label{period}\phi_1(x) = x\frac{\partial}{\partial x}\Big(x\,\frac{\partial}{\partial x} W(x)\Big),\qquad W(x) = 1 + \int_{\infty}^{x} \frac{\dd x'}{x'}\Big(\int_{\infty}^{x'} \phi_2(x'')\,\frac{\dd x''}{x''}\Big)
\eeq
satisfies the homogeneous equation:
\beq
\phi_2(x + {\rm i}0) + \phi_2(x - {\rm i}0) + (2 - r)\sum_{l = 1}^{a - 1} \phi_2(\zeta_{a}^{l}x) + \sum_{m = 1}^r \sum_{l = 1}^{a_m - 1} \phi_2(\zeta_{a_m}^{l}x) = 0
\eeq
The analytic properties of $W(x)$ imply that:
\begin{itemize}
\item[$\bullet$] $\phi_2(x)$ is holomorphic in $\mathbb{C}\setminus\Gamma$.
\item[$\bullet$] $\phi_2(x) \in O(1/x)$ when $x \rightarrow \infty$, and $\phi_2(x) \in O(x)$ when $x \rightarrow 0$.
\item[$\bullet$] $\phi_2(x)$ diverges like $(x - \gamma_{\pm})^{-3/2}$ when $x \rightarrow \gamma_{\pm}$.
\item[$\bullet$] The palindrome symmetry implies $\phi_2(x) + \phi_2(1/x) = 0$.
\end{itemize}
Since $W(x)$ and $\phi_1(x)$ are continuous in $\mathbb{C}\setminus\Gamma$, the integration in \eqref{period} do not depend on the paths. And the period of $W(x)/x$ is $1$ since the equilibrium measure has total mass $1$. So we have the extra conditions:
\begin{itemize}
\item[$\bullet$] $\oint_{\Gamma} \phi_j(x)\,\frac{\dd x}{x} = 0$ for $j = 1,2$.
\item[$\bullet$] $\oint_{\Gamma} W(x)\,\frac{\dd x}{2{\rm i}\pi x} = 1$.
\end{itemize}

\subsection{The two-point function}
\label{2pin}

Let us introduce:
\beq
H(x_1;x_2) = \Big(\int^{x_2} W_2^{(0)}(x_1,x_2')\,\frac{\dd x_2'}{x_2'}\Big)\,\frac{\dd x_1}{x_1}
\eeq
which is a holomorphic function of $x_2 \in \widehat{\mathbb{C}}\setminus\Gamma$, and a holomorphic $1$-form in $x_1 \in \widehat{\mathbb{C}}\setminus\Gamma$. The functional equation \eqref{W2potf} for $W_2^{(0)}(x_1,x_2)$ implies, for $x_1 \in \mathring{\Gamma}$ and $x_2 \in \widehat{\mathbb{C}}\setminus\Gamma$:
\beq
H(x_1 + {\rm i}0;x_2) + H(x_1 - {\rm i}0;x_2) + (2 - r)\sum_{l = 1}^{a - 1} H(\zeta_{a}^{l}x_1;x_2) + \sum_{m = 1}^r \sum_{l = 1}^{a_m - 1} H(\zeta_{a_m}^{l}x_1;x_2) = \frac{\dd x_1}{x_2 - x_1}
\eeq

The strategy is to get rid of the right-hand side. We first look, for any integer $k \geq 1$, for a particular solution $h_k(x)$ of the linear equation with right-hand side $x^{k - 1}_1\dd x_1$. Then, $\sum_{k \geq 1} x_2^{-k}\,h_{k}(x_1)$ solves the linear equation with right-hand side:
\beq
\frac{\dd x_1}{x_2 - x_1} = \sum_{k \geq 1} \frac{x^{k - 1}_1\dd x_1}{x_2^k}
\eeq

One can first try to find a solution of the form $h_k(x) = c_k x^{k - 1}\dd x$. When the Fourier coefficient $\mathcal{F}_k[\hat{\alpha}]$ vanishes, such rational solutions do not exist, but one can always find a solution of the form $h_k(x) = c_k x^{k - 1}(\ln^{j_k} x)$ for a minimal value of $j_k$. Since in discrete Fourier space, the convolution by $\hat{\alpha}$ becomes a multiplication by $\mathcal{F}[\hat{\alpha}]$, there are precisely $\mathrm{dim}(\mathrm{Ker}\,\hat{A}) = a - \mathrm{dim}\,E$ values of $(k\,\,\mathrm{mod}\,\,a)$ for which a logarithm is needed (we use the notations of \S~\ref{S3}). All in all, we can construct a particular solution for $H(x;x_0)$ for the primitive of the two-point function of the form:
\beq
\label{HHH} H_{{\rm part}}(x_1;x_2) = \sum_{j \geq 0} \Big(\frac{\ln x_1}{2{\rm i}\pi}\Big)^j\,(\hat{C}^{(j)}\otimes \hat{e}_{0})\cdot \frac{\dd x_1}{x_2 - x_1}
\eeq
Here $\hat{C}^{(j)} = \sum_{i \in \mathbb{Z}_{a}} \hat{C}^{(j)}(i)\,\hat{e}_i$ represents a vector in $\hat{E}$, and with our notations:
\beq
(\hat{C}^{(j)}\otimes \hat{e}_0)\cdot \frac{\dd x_1}{x_2 - x_1} =  \sum_{i \in \mathbb{Z}_{a}} C^{(j)}(i)\,\frac{\dd(\zeta_{a}^{i}x)}{x_0 - \zeta_{a}^ix}
\eeq
i.e. $\hat{C}^{(j)}$ acts on the first variable and $\hat{e}_0$ on the second variable (its action is the identity). Then, we introduce
\bea
H_{{\rm hom}}(x_1;x_2) & = & H(x_1;x_2) - H_{{\rm part}}(x_1;x_2) \nonumber \\
\qquad \omega_2^{(0)}|_{\rm hom}(x_1,x_2) & = & W_2^{(0)}(x_1,x_2)\,\frac{\dd x_1}{x_1}\,\frac{\dd x_2}{x_2} - \dd_{x_2} H_{{\rm part}}(x_1;x_2),
\eea
which solve the homogeneous linear equation. The price to pay is that $\omega_2^{(0)}|_{{\rm hom}}$ is a $(1,1)$ form having the singularities of $-\dd_{x_2}H_{{\rm part}}(x_1;x_2)$. Therefore, for any $\hat{v}_1,\hat{v}_2 \in \hat{E}$, the singular part of:
\beq
(\hat{v}_1\otimes\hat{v}_2)\cdot\omega_2^{(0)}|_{{\rm hom}}(x_1,x_2) = \sum_{j_1,j_2 \in \mathbb{Z}_a} \hat{v}_1(j_1)\hat{v}_2(j_2)\,\omega_2^{(0)}|_{{\rm hom}}(\zeta_{a}^{j_1}x_1,\zeta_{a}^{j_2}x_2)
\eeq
is:
\beq
\sum_{j \geq 0} \sum_{i \in \mathbb{Z}_{a}} B^{(j)}(i;\hat{v}_1,\hat{v}_2)\,\frac{\dd(\zeta_{a}^ix_1)\dd x_2}{(x_2 - \zeta_{a}^ix_1)^2}\,\Big(\frac{\ln x_1}{2{\rm i}\pi}\Big)^j
\eeq
where we have:
\beq
B^{(0)}(i;\hat{v},\hat{v}_0) = \sum_{j \geq 0} \sum_{\ell,\ell' \in \mathbb{Z}_{a}} \Big(\frac{\ell}{a}\Big)^{j}\,C^{(j)}(\ell)\hat{v}_1(\ell')\hat{v}_2(\ell + \ell' - i)
\eeq
and higher $B^{(j)}(i)$ can be obtained by similar formulae.

Then, if we are given some initial vector $\hat{v}_0$, we deduce that $(\hat{v}_0\otimes\hat{v}_0)\cdot\omega_{2}^{(0)}|_{\rm hom}$ can be analytically continued as a bidifferential $(\omega_2^{(0)})_{\hat{v}_0}$ continued on the Riemann surface $\Sigma_{\hat{v}_0}\times\Sigma_{\hat{v}_0}$, and it has -- on top of the log singularities\footnote{We do not insist on the log singularities of $\hat{v}_0^{\otimes 2}\cdot\omega_{2}^{(0)}|_{\rm hom}$ : in the Seifert cases, it turns out that $\hat{B}^{(j)}(i;\hat{v}_1,\hat{v}_2) = 0$ whenever $j \geq 1$ and $\hat{v}_1,\hat{v}_2$ have finite orbits, so they are absent.}
-- double poles without residues at points so that $x^{a} = x^{a_0}$ for $x_1,x_2$ in the sheet $\mathbb{C}_{\hat{v_1}}\times \mathbb{C}_{\hat{v}_2}$ labeled by two vectors $\hat{v}_1,\hat{v}_2$ of the orbit of $\hat{v}_0$.

\begin{definition}
The coefficients of the double poles are recorded in the \emph{matrix of singularities}:
\beq
\hat{\mathbf{B}}^{(0)}_{\hat{v}_0} = \Big[\sum_{i \in \mathbb{Z}_{a}} \hat{B}^{(0)}(i;\hat{v}_1,\hat{v}_2)\,\hat{e}_i\Big]_{\hat{v_1},\hat{v}_2}
\eeq
whose rows and columns are labeled by elements $\hat{v}_1$ and $\hat{v}_2$ in the orbit of $\hat{v}_0$, and whose entries are vectors in $\hat{E}$.
\end{definition}

\subsection{Remark on the matrix of singularities}

The easiest way to produce a meromorphic form with double poles at $x^{a} = x_0^{a}$ is to consider:
\beq
(\hat{B}\otimes\hat{e}_0)\cdot\frac{\dd x_1\dd x_2}{(x_2 - x_1)^2} = \sum_{i \in \mathbb{Z}_{a}} \hat{B}(i)\,\frac{\dd(\zeta_{a}^ix_1)\dd x_2}{(x_2 - \zeta_{a}^ix_1)^2}
\eeq
Then,
\beq
\omega_2^{(0)}|_{\rm sp}(x_1,x_2) = \omega_2^{(0)}|_{{\rm hom}}(x_1,x_2) - (\hat{B}\otimes\hat{e}_0)\cdot\frac{\dd x_1\dd x_2}{(x_2 - x_1)^2} \nonumber
\eeq
is also meromorphic on $\Sigma_{\hat{v}_0}^2$ with double poles, but admits as matrix of singularities:
\beq
\hat{\mathbf{B}}^{(0)}_{{\rm sp}} = \hat{\mathbf{B}}^{(0)}_{\hat{v}_0} - \hat{B}\mathbf{J}
\eeq
where $\mathbf{J}$ is the matrix full of $1$'s. If it is possible to choose $\hat{B}$ so that $\hat{\mathbf{B}}_{{\rm sp}}$ becomes sparse, the computation of $\omega_{2}^{(0)}|_{{\rm sp}}(x_1,x_2)$ is facilitated. To summarize the relation with the two-point function:
\beq
\omega_{2}^{(0)}(x_1,x_2) = \omega_{2}^{(0)}|_{\rm sp}(x_1,x_2) + \Bigg(\Big[\hat{B} - \sum_{j \geq 0} \Big(\frac{\ln x_1}{2{\rm i}\pi}\Big)^j \hat{C}^{(j)}\Big]\otimes \hat{e}_0\Bigg)\cdot \frac{\dd x_1\dd x_2}{(x_2 - x_1)^2}
\eeq
As a matter of fact, the structure of a suitable $\hat{B}$ carries information on how to lift the action of $\mathbb{Z}_{a}$ by rotations on $\widehat{\mathbb{C}}$, to the total space $\Sigma_{\hat{v}_0}$ of the covering $x\,:\,\Sigma_{\hat{v}_0} \rightarrow \widehat{\mathbb{C}}$. This will be exemplified in the case by case study, in particular for the geometry $\mathbb{S}_3/D_{p}$.

In general, $x_1$ and $x_2$ may not play a symmetric role in $\omega_{2}^{(0)}|_{\rm sp}(x_1,x_2)$ (whereas they do in $\omega_{2}^{(0)}$ by construction). This is nevertheless the case when the entries of $\mathbf{B}_{{\rm sp}}^{(0)}$ are invariant under the inversion endomorphism $\hat{e}_{j} \rightarrow \hat{e}_{-j}$ in $\hat{E}$.

\section{Spectral curve and $2$-point function: case study}
\label{S6}
\subsection{General results}

We study in detail the sheet dynamics of a solution of the homogeneous equation in the Seifert models, characterized by:
\beq
\label{alhex2} G = \mathbb{Z}_{a},\qquad \hat{\alpha} = 2\hat{e}_0 + (2 - r)\sum_{l = 1}^{a - 1} \hat{e}_{l} + \sum_{m = 1}^r \sum_{l_m = 1}^{a_m - 1} \hat{e}_{\check{a}_ml_m},\qquad \check{a}_m = a/a_m
\eeq
Following \S~\ref{rot}, we construct a simply-laced, irreducible root system $\mathfrak{R}$ on $E$:
\beq
E = \mathrm{Im}\,A,\qquad A = \hat{\alpha}\cdot
\eeq
Since the coordinates of $\hat{\alpha}$ are integers, $\mathfrak{R}$ is crystallographic. We remind (Corollary~\ref{cofini}) that the group $\hat{\mathfrak{G}} = \langle \hat{T}_l,\quad l \in \mathbb{Z}_{a}\rangle$ has non-zero finite orbits iff $\mathfrak{R}$ is finite. We also remind the definition of the orbifold Euler characteristics of Seifert spaces $\chi = 2 - r + \sum_{m = 1}^r \frac{1}{a_m}$.

\begin{theorem}
\label{negative}If $\chi < 0$, $\mathfrak{R}$ is infinite.
\end{theorem}
\textbf{Proof.} Let $n_0[\hat{v}] = \sum_{l = 0}^{a - 1} \hat{v}(l)$ be the sum of coordinates. According to \eqref{defq}-\eqref{alhex2}, we have:
\beq
\label{evol}n_0[\hat{T}_{j}(\hat{v})] = n_0[\hat{v}] - \hat{v}(j)\Big((2 - r)a + \sum_{l = 1}^{a - 1} \frac{a}{a_m}\Big) = n_0[\hat{v}] - a\chi\,\hat{v}(j).
\eeq
Assume there exist $\hat{v} \neq 0$ such that the set $n_0[\hat{\mathfrak{G}}\cdot \hat{v}] \subseteq \mathbb{Z}$ is finite. If $\chi \neq 0$, it cannot be contained in $\{0\}$, and up to considering $-\hat{v}$ instead of $\hat{v}$, we may assume it contains a positive integer. Then, let $\hat{w} \in \hat{\mathfrak{G}}\cdot \hat{v}$ such that $n_0[\hat{w}]$ is maximal. Then, $n_0[\hat{w}] > 0$, so there exists $j \in \ldbrack 0,a - 1 \rdbrack$ such that $\hat{w}(j) > 0$. By maximality and the relation:
\beq
n_0[\hat{T}_j(\hat{w})] = n_0[\hat{w}] - a\chi\,\hat{w}(j)
\eeq
we deduce that $\chi \geq 0$. As a corollary, if $\chi < 0$, the set $n_0[\hat{\mathfrak{G}}\cdot \hat{v}]$ is never finite for $\hat{v} \neq 0$, so the orbits themselves cannot be finite. \hfill $\Box$

In the remaining of the text, we assume $\chi \geq 0$. So, the spectral curve is uniquely defined (Corollary~\ref{cor1}), has the palindrome symmetry (Corollary~\ref{cocococo}), and the cut $\Gamma$ consists of one segment (Theorem~\ref{1cut}). We will see that $\mathfrak{R}$ is finite in all cases $\chi \geq 0$. The resulting root systems are presented in Figure~\ref{finsub}. We have written a short \textsc{maple} program computing the orbits under $(\hat{T}_l)_l$ of an initial vector $\hat{v}_0$, which was very useful to guess vectors with small finite orbits -- before we were able to construct them without involving a guess, as presented below. The left action of $G$ on $\hat{E}$ discussed in \S~\ref{bonus} is generated by the shift operator:
\beq
\varepsilon_{1}(\hat{v}) = \sum_{l \in \mathbb{Z}_{a}} \hat{v}(l + 1)\,\hat{e}_l,\qquad \varepsilon_{j} = \varepsilon_{1}^j.
\eeq
which intertwines the generators as:
\beq
\forall k,j \in \mathbb{Z}_{a},\qquad T_k = \varepsilon_{j} T_{k + j}\varepsilon_{-j}.
\eeq
The shift (or its powers) will be useful to decompose further the $\mathfrak{G}$-orbits.

\begin{figure}[h!]
\begin{center}
\begin{tabular}{|c|c|c||c|c|c|c|}
\hline
{\rule{0pt}{3.2ex}}{\rule[-1.8ex]{0pt}{0pt}} Seifert manifold & $(a_1,\ldots,a_r)$ & $a$ & $\mathfrak{R}$ & $|\mathfrak{R}|$ & $|\mathfrak{G}|$ & $d$ \\
\hline\
$L(a_1,a_2)$ {\rule{0pt}{3.2ex}}{\rule[-1.8ex]{0pt}{0pt}} & $(a_1,a_2)$ & $\mathrm{lcm}(a_1,a_2)$ & $A_{a_{1,2} - 1}$ & $(a_{1,2} - 1)a_{1,2}$ & $a_{1,2}!$ & $a_{1,2}$ \\
\hline $\mathbb{S}^3/D_{p + 2}$ ($p$ odd) {\rule{0pt}{3.2ex}}{\rule[-1.8ex]{0pt}{0pt}}& $(2,2,p)$ & $2p$ & $D_{p + 1}$ & $2p(p + 1)$ & $2^{p}(p + 1)!$ & $2(p + 1)$ \\
\hline $\mathbb{S}^3/D_{p + 2}$ ($p$ even) {\rule{0pt}{3.2ex}}{\rule[-1.8ex]{0pt}{0pt}}& $(2,2,p)$ & $p$ & $A_{p}$ & $p(p + 1)$ & $(p + 1)!$&  $p + 1$ \\
\hline $\mathbb{S}^3/E_6$ {\rule{0pt}{3.2ex}}{\rule[-1.8ex]{0pt}{0pt}} & $(2,3,3)$ & $6$ & $D_{4}$ & $24$ & $192$ & $8$ \\
\hline $\mathbb{S}^3/E_7$ {\rule{0pt}{3.2ex}}{\rule[-1.8ex]{0pt}{0pt}} & $(2,3,4)$ & $12$ & $E_{6}$ & $72$ & $51840$ & $27$  \\
\hline $\mathbb{S}^3/E_8$ {\rule{0pt}{3.2ex}}{\rule[-1.8ex]{0pt}{0pt}} & $(2,3,5)$ & $30$ & $E_{8}$ &  $240$ & $696729600$ & $240$ \\
\hline
\end{tabular}  
\end{center}
\caption{\label{finsub} For Seifert manifolds with $\chi > 0$, we tabulate: (1) the manifold $\mathbb{S}_3/\mathfrak{H}$ where $\mathfrak{H}$ is a finite subgroup of $\mathrm{SU}(2,\mathbb{C})$ acting on $\mathbb{S}_3$ denoted by its ADE classification described in \S~\ref{orbiar} ; (2) the order of exceptional fibers ; (3) $a = \mathrm{lcm}(a_1,\ldots,a_r)$, which is the dimension of the group algebra \mbox{$\hat{E} = \mathbb{Z}[G]$ ;} (4) the simply-laced, crystallographic root system we have found according to Section~\ref{S3}, denoted according to its ADE classification ; (5) the number of roots ; (6) the order of the Weyl group $\mathfrak{G}$ ; (7) the minimal order of a $\hat{\mathfrak{G}}$-orbit. We included for completness the case of lens spaces treated in \cite{BEMknots,Halma}, and set $a_{1,2} = a_1 + a_2$.}
\end{figure}

\begin{figure}[h!]
\begin{center}
\begin{tabular}{|c|c||c|c|c|c|}
\hline {\rule{0pt}{3.2ex}}{\rule[-1.8ex]{0pt}{0pt}} $(a_1,\ldots,a_r)$ & $a$ & $\mathfrak{R}$ & $|\mathfrak{R}|$ & $d$ & $\hat{\mathfrak{R}}$ \\
\hline \
$(2,2,2,2)$ {\rule{0pt}{3.2ex}}{\rule[-1.8ex]{0pt}{0pt}} & $2$ & $A_{1}$ & $2$ & $2$ & $\widehat{A}_{1}$ \\
$(3,3,3)$  {\rule{0pt}{3.2ex}}{\rule[-1.8ex]{0pt}{0pt}} & $3$ & $A_{2}$ & $6$ & $3$ & $\widehat{A}_{2}$ \\
$(2,4,4)$  {\rule{0pt}{3.2ex}}{\rule[-1.8ex]{0pt}{0pt}} & $4$ & $A_{3}$ & $12$ & $4$ & $\widehat{A}_{3}$ \\
$(2,3,6)$ {\rule{0pt}{3.2ex}}{\rule[-1.8ex]{0pt}{0pt}} & $6$ & $A_{5}$ & $30$ & $6$ & $\widehat{A}_{5}$ \\
\hline
\end{tabular}
\end{center}
\caption{\label{finsub0} For Seifert manifolds with $\chi = 0$, the group of sheet transitions is an affine Weyl group.}
\end{figure}

We may wonder how the geometry of $\mathbb{Z}_{n'}\backslash\mathbb{S}_{3}/\mathfrak{H}$ is reflected in the (reduced) group of sheet transitions and in the equilibrium measure. Here are three curious observations:
\begin{itemize}
\item[$\bullet$] $a = \mathrm{dim}\,\hat{E} = |G|$ coincides with the Coxeter number of $\mathfrak{R}$. We do not know how to interprete this coincidence.
\item[$\bullet$] the ADE Dynkin diagram associated to $\mathfrak{H}$ always contains as a subdiagram the ADE Dynkin diagram associated to $\mathfrak{R}$. In particular, they are both $E_8$ for the Poincar\'e sphere, and this is the only case of equality. In the other cases, it looks like a symmetry breaking of a bigger Weyl symmetry associated to the ADE label of $\mathfrak{H}$ occuring by restriction to the contribution of the trivial flat connection. This probably can be made concrete by physics dualities between Chern-Simons theory on $\mathbb{S}_{3}/\mathfrak{H}$ and another gauge theory with gauge group associated to the ADE Dynkin diagram of $\mathfrak{H}$.
\item[$\bullet$] $(2,2,p)$ with $p$ even is the only case where $\hat{\mathcal{G}}$ is finite. It is also the only case where $a$ is also the order of an exceptional fiber, and it is tempting to speculate that this is responsible for the finiteness of $\hat{\mathfrak{G}}$, for a reason yet unknown to us.
\end{itemize}

\begin{definition}
If $\mathcal{V}$ is a $\hat{\mathfrak{G}}$-orbit, we say that it is \emph{complete} if for any $k \in \mathbb{Z}_{a}$, there exists $\hat{v} \in \mathcal{V}$ such that $\mathcal{F}_{k}[\hat{v}] \neq 0$ or $\mathcal{F}_{k + 1}[\hat{w}] \neq 0$.
\end{definition}

If $\mathcal{V}$ is a complete orbit generated by some vector $\hat{v}_0$, all moments $\mu[\mathrm{Tr}\,\,\mathbf{S}^{k}]$ can be retrieved from the Puiseux expansion of the analytical continuation $\varphi_{\hat{v}_0}(x)$ of $\hat{v}_0\cdot\phi(x)$ at a point $x \rightarrow 0$ or $\infty$ in some sheet of $\Sigma_{\hat{v}_0}$. From the case by case analysis, we find that:

\begin{lemma}
In all $\chi \geq 0$ cases, the orbits of finite minimal order are complete.
\end{lemma}

\subsection{$(2,2,p)$, $p$ even}
\label{even}
\beq
\check{R}(s,s') = \frac{\sum_{j = 0}^{p/2 - 1} s^js'^{p/2 - 1 - j}}{s^{p/2} + s'^{p/2}},\qquad a = \mathrm{dim}\,\hat{E} =  p,\qquad \chi = \frac{1}{p}.
\eeq
The dynamics are generated by:
\beq
\hat{\alpha} = 2\hat{e}_0 + \sum_{j = 1}^{p - 1} (-1)^{j}\hat{e}_{j}.
\eeq
The matrix of the left multiplication by $\hat{\alpha}$ in the canonical basis is:
\beq
\label{canrd4} \left(\begin{array}{cccccc} 2 & -1 & 1 & -1 & \ldots & -1 \\ -1 & 2 & -1 & 1 & \ddots & 1 \\ 1 & \ddots  & \ddots & \ddots & \ddots & -1 \\ \vdots & \ddots & \ddots & \ddots & \ddots & \vdots \\ \vdots & \ddots & \ddots & \ddots & \ddots & \vdots \\  -1 & \ldots & \ldots & 1 & -1 & 2 
\end{array}\right).
\eeq
This matrix has full rank, i.e. $E = \hat{E}$ and $\hat{\mathfrak{G}} \simeq \mathfrak{G}$. Let us define a new basis:
\beq
\beta_j = (-1)^{l}(e_{j - 1} + e_j),\quad (1 \leq j \leq p - 1),\qquad \beta_{p} = \hat{e}_{p - 1}.
\eeq
In this new basis, the matrix \eqref{canrd4} becomes:
\beq
\label{lamat}\left(\begin{array}{cccccc} 2 & -1 & 0 & 0 & \ldots & 0 \\ -1 & 2 & -1 & 0 & \ddots & 0 \\ 0 & \ddots  & \ddots & \ddots & \ddots & 0 \\ \vdots & \ddots & \ddots & \ddots & \ddots & \vdots \\ \vdots & \ddots & \ddots & \ddots & \ddots & \vdots \\  0 & \ldots & \ldots & 0 & -1 & 2 
\end{array}\right),
\eeq
and we recognize the Cartan matrix of the root system $A_{p}$, with $\beta$'s playing the role of simple roots. Therefore $\hat{\mathfrak{G}} = \mathfrak{G}$ is the symmetric group in $p + 1$ -- in particular $\hat{\mathfrak{G}}$ is finite.

\subsubsection{Spectral curve}

\begin{figure}[h!]
\begin{center}
\includegraphics[width=0.5\textwidth]{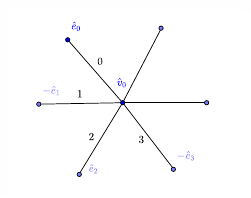}
\caption{\label{Devengraph} Case $(2,2,p)$, $p$ even. The skeleton graph generated by $\hat{e}_0$.}
\end{center}
\end{figure}

This implies that $\phi(x) = \hat{e}_0\cdot \phi(x)$ is already an algebraic function, and it is actually the only case in our list where this happens. The orbit of $\hat{e}_0$ consists of $\hat{v}_0 = \sum_{l = 0}^{p - 1} (-1)^{l + 1}\hat{e}_l$ and $\hat{v}_j' = (-1)^j\hat{e}_{j}$ for $j \in \mathbb{Z}_{p}$, its order is $p + 1$. Since it contains $\hat{e}_0$, this orbit is complete. This is actually a minimal orbit, since the maximal sub-root system of $A_{p}$ is $A_{p - 1}$, and:
\beq
d = \frac{|\mathrm{Weyl}(A_{p})|}{|\mathrm{Weyl}(A_{p - 1})|} = \frac{|\mathfrak{S}_{p + 1}|}{|\mathfrak{S}_{p}|} = p + 1.
\eeq
The skeleton graph is a star with $p$ branches connecting to the central vertex $\hat{v}_0$ (Figure~\ref{Devengraph}). Therefore, the Riemann surface $\Sigma_{\hat{v}_0}$ has genus $0$. We denote $y$ the function which is equal to $\hat{v}\cdot Y(x)$ when $x$ belongs to the sheet labeled by $\hat{v} \in \hat{\mathfrak{G}}\cdot\hat{v}_0$.  We find that:
\beq
\label{spcurveDeven}\boxed{\left\{\begin{array}{rcl} x(z) & = & z\dfrac{z^{p} - \kappa^{2}}{\kappa^{2}z^{p} - 1} \\ y(z) & = & -\dfrac{(z^{p/2} - \kappa)(\kappa z^{p/2} + 1)}{(\kappa z^{p/2} - 1)(z^{p/2} + \kappa)}\end{array}\right.}
\eeq
where 
\beq
\label{eqau2}\boxed{\frac{2\kappa^{1 + 1/p}}{1 + \kappa^{2}} = e^{-u/4p^2}:= c^2}
\eeq
is the only possibility meeting the growth requirements \eqref{Puis1}-\eqref{Puis2}, so it is the correct solution. The equilibrium measure is extracted as:
\beq
\dd\lambda_{{\rm eq}} = \frac{p^2\,\dd x}{2{\rm i}\pi\,ux}\ln\Big(\frac{Y(x + {\rm i}0)}{Y(x - {\rm i}0)}\Big)
\eeq
for the branches of $Y$ which meet on the positive real axis. The support of the equilibrium measure $\mu_{\mathrm{eq}}$ is $[1/\gamma,\gamma]$ with $\gamma = x(z_+) > 1$ such that $\dd x(z_+) = 0$. We find:
\beq
\label{thesqu} z_+ = \Big(\frac{(p + 1)\kappa^{-2} - (p - 1)\kappa^{2} \pm \sqrt{(\kappa^{-2} - \kappa^{2})((p + 1)^2\kappa^{-2} - (p - 1)^2\kappa^{2})}}{2}\Big)^{1/p}.
\eeq
By consistency, we must choose $\kappa(u)$ as the unique solution of \eqref{eqau2} such that $0 < \kappa(u) \leq 1$ when $u \geq 0$.

\subsubsection{Two-point function}

Particular solutions of the linear equation with right-hand side $x^{k - 1}\dd x$ are all rational $H_k(x) = c_k\,x^{k - 1}\dd x$ with:
\beq
c_k = \left\{\begin{array}{ccl} \frac{1}{p + 1} & & \mathrm{if}\,\,(k\,\,\mathrm{mod}\,\,p) = p/2 \\ 1 & & \mathrm{otherwise} \end{array}\right.
\eeq
This results in the residue vector:
\beq
\hat{C}^{(0)} = \hat{e}_0 + \sum_{\ell \in \mathbb{Z}_{p}} \frac{(-1)^{\ell + 1}}{p + 1}\,\hat{e}_{\ell}
\eeq
Then, we compute the matrix of singularities taking $\hat{v}_0 = \sum_{j \in \mathbb{Z}_{p}} (-1)^{\ell + 1}\hat{e}_{\ell}$ as initial vector. We found above the other vectors in the orbit are $\hat{v}_j' = (-1)^{j}\hat{e}_j$ for $j \in \mathbb{Z}_{p}$. We find the following decomposition of the matrix of singularity:
\beq
\hat{\mathbf{B}}^{(0)} = \hat{B}\mathbf{J} + \sum_{\ell \in \mathbb{Z}_{p}} (-1)^{\ell}\hat{e}_{\ell}\,\mathbf{\Pi}(\ell)
\eeq
We have defined:
\beq
\hat{B} = \frac{1}{p + 1}\Big(\sum_{\ell \in \mathbb{Z}_{p}} (-1)^{\ell + 1}\hat{e}_{\ell}\Big)
\eeq
so that the subtraction of $\hat{B}\mathbf{J}$ leaves us with a sparse matrix $\mathbf{B}^{(0)}_{{\rm sp}}$. We find that the $\mathbf{\Pi}(\ell)$ are permutations matrices of the sheets:
\beq
\mathbf{\Pi}(\ell) = \mathbf{E}_{\hat{v}_0,\hat{v}_0} + \sum_{i \in \mathbb{Z}_{p}} \mathbf{E}_{\hat{v}'_{i},\hat{v}'_{i + \ell}}
\eeq
where $\mathbf{E}_{\hat{w}_1,\hat{w}_2}$ are the elementary matrices whose only non-zero entry is a $1$ at the intersection of the row labeled $\hat{w}_1$ and column labeled $\hat{w}_2$.

We found previously that the curve $\Sigma_{\hat{v}_0}$ is rational, with uniformization variable $z$. Therefore, $\omega_{2}^{(0)}|_{\rm sp}(z_1,z_2)$ must be a rational form in $z_1$ and $z_2$, and the only possibility compatible with its matrix of singularity is:
\beq
\sum_{\ell \in \mathbb{Z}_{p}} (-1)^{\ell}\,\frac{\dd(\zeta_{p}^{\ell}z)\dd z_0}{(z_0 - z)^2}
\eeq
But we learn more from the structure of $\mathbf{B}^{(0)}_{{\rm sp}}$: indeed, if we denote $z^{(\hat{w})}(x)$ the values of $z$ in the sheet $\hat{w}$ above the point $x$, we must have
\beq
\forall \hat{w} \in \hat{\mathfrak{G}}\cdot\hat{v}_0,\qquad \forall \ell \in \mathbb{Z}_{p},\qquad z^{(\hat{w})}(\zeta_{p}^\ell x) = \zeta_{p}^j\,z^{(\mathbf{\Pi}(\ell)(\hat{w}))}(x)
\eeq
Therefore, the matrices $\mathbf{\Pi}(\ell)$ permuting the sheets tell us how the action of $\zeta_{p}^{\ell} \in \mathbb{Z}_{a}$ in the $x$-plane lifts to an action of $\mathbb{Z}_{a}$ in the $z$-plane. We thus have gained, through the computation of the two-point function, a precise description of the monodromy of the function $z(x)$ that is useful in our problem.

Putting all results together, since we observe that $\hat{B} = \hat{C}^{(0)}$, there is no contribution from the fundamental $2$-form of the $2^{{\rm nd}}$ kind of the $x$-plane, and we have:
\beq
\label{624}(\omega_2^{(0)})_{\hat{v}_0}(z_1,z_2) = \sum_{\ell \in \mathbb{Z}_{p}} \frac{(-1)^{\ell}\dd(\zeta_{p}^{\ell}z_1)\dd z_2}{(z_2 - \zeta_{p}^{\ell}z_1)^2} = \dd(z_1^{p/2})\,\dd(z_2^{p/2})\Big(\frac{1}{(z_1^{p/2} - z_2^{p/2})^2} + \frac{1}{(z_1^{p/2} + z_2^{p/2})^2}\Big)
\eeq

\subsection{$(2,2,p)$, $p$ odd}
\label{S43}
\label{dodd}
\begin{figure}[h!]
\begin{center}
\includegraphics[width=8cm]{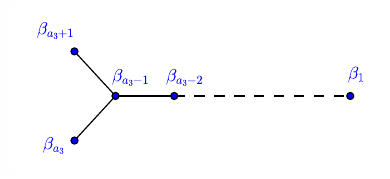}
\caption{\label{dpp1dyn} Dynkin diagram of the root system $D_{p + 1}$.}
\end{center}
\end{figure}

\beq
\check{R}(s,s') = \frac{(s + s')\big(\sum_{j = 0}^{p - 1} s^{p - 1 - j}s'^{j}\big)}{s^{p} + s'^{p}},\qquad a = \mathrm{dim}\,\hat{E} = 2p,\qquad \chi = \frac{1}{p},
\eeq
The dynamics are generated by:
\beq
\hat{\alpha} = 2\hat{e}_0 + \sum_{\substack{j = 1 \\ j \neq p}}^{2p - 1} (-1)^{j}\hat{e}_{j}.
\eeq
The subspace $E$ has rank $p + 1$, and is generated by the orthogonal projections $e_l$ on $E$ of the canonical basis $\hat{e}_l$:
\beq
e_0 = \frac{1}{2p}\Big((p + 1)\hat{e}_0 + (p - 1)\hat{e}_{p} + \sum_{\substack{j = 1 \\ j \neq p}}^{2p - 1} (-1)^{j}\hat{e}_{j}\Big), \qquad e_j = \varepsilon_{-j}(e_0)
\eeq
Out of them, we can construct of basis of $E$:
\beq
\beta_j = (-1)^{j - 1}(e_j + e_{j + 1}),\quad (1 \leq j \leq p - 1),\qquad \beta_{p} = e_0,\qquad \beta_{p + 1} = e_{p}.
\eeq
The Cartan matrix of the root system $\mathfrak{R}$ in this basis reads:
\beq
\label{themat}\mathbf{A} = \left(\begin{array}{cccccc} 2 & -1 & 0 & 0 & \ldots & 0 \\ -1 & \ddots & \ddots & \ddots & \ddots & \vdots \\ 0 & \ddots  & 2 & -1 & 0 & 0 \\ \vdots & \ddots & -1 & 2 & -1 & -1 \\ \vdots & \ddots & 0 & -1 & 2 & 0 \\  0 & \cdots & 0 & -1 & 0 & 2 
\end{array}\right).
\eeq
We recognize the Cartan matrix of the root system $D_{p + 1}$, where the $\beta$'s play the role of the simple roots, labeled as in Figure~\ref{dpp1dyn}.

\subsubsection{Minimal orbits}

A maximal sub-root system is obtained by deleting $\beta_1$ from $\mathfrak{R}$: it is of type $D_{p}$. The minimal order of an orbit is therefore:
\beq
d = \frac{|\mathrm{Weyl}(D_{p + 1})|}{|\mathrm{Weyl}(D_{p})|} = 2(p + 1)
\eeq
Vectors having such an orbit can be found of the form $\hat{v}^* = \hat{\alpha}\cdot v^*$ with $v^* \in \hat{E}$ orthogonal to the chosen maximal sub-root system:
\beq
\forall j \in \ldbrack 2,p + 1\rdbrack,\qquad \langle v^*,\beta_j\rangle = 0
\eeq
By decomposing $v^*$ on the basis of $\beta$'s, we get a linear system which can be solved, and we find:
\beq
v^* \in \mathrm{span}\Big(\sum_{\ell = 1}^{p - 1} \beta_l + \frac{\beta_{p} + \beta_{p + 1}}{2}\Big) = \mathrm{span}(e_0 + 2e_1 - e_{p}).
\eeq
We compute the image of the $\beta$'s under right multiplication by $\hat{\alpha}$, and deduce the image of $\hat{v}^*$, giving a generator for a line of vectors in $\hat{E}$ with minimal $\hat{\mathfrak{G}}$-orbit:
\beq
\hat{v}^* = \hat{e}_1 + \hat{e}_{p + 1}
\eeq 
Since any shift of the latter is also a vector with minimal orbit, and we may choose
\beq
\hat{v}_2 = \varepsilon_{1}(\hat{v}^*) = \hat{e}_0 + \hat{e}_{p}
\eeq
as initial vector.

\begin{figure}[h!]
\begin{center}
\includegraphics[width=0.45\textwidth]{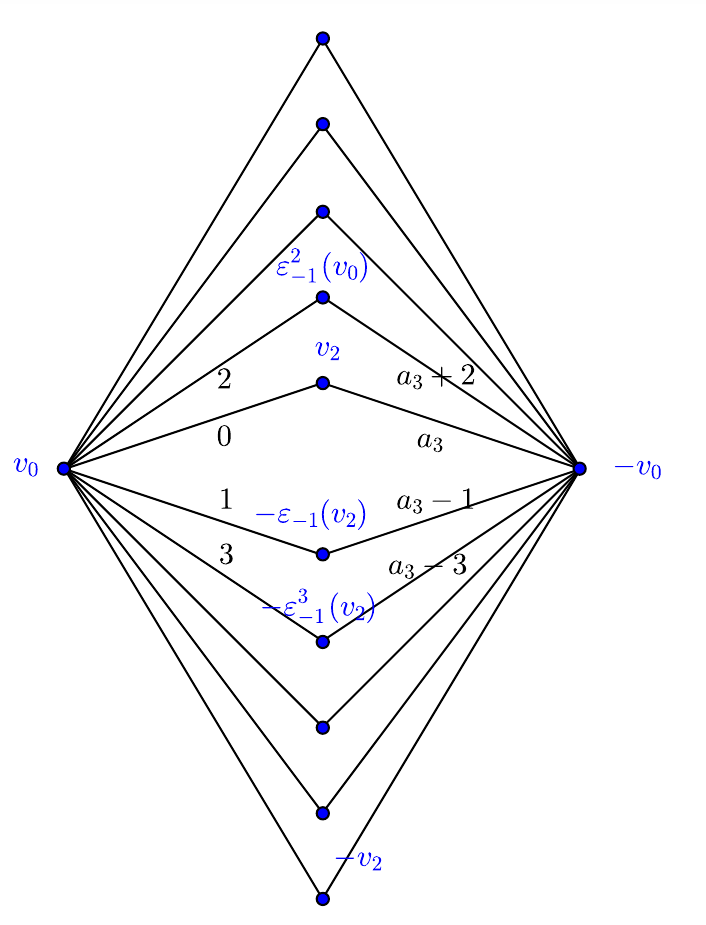}
\caption{\label{Graph-Dpodd} The skeleton graph $\mathcal{G}$ (see \S~\ref{orbitp}) generated by $\hat{v}_2$. An edge label $j$ indicate that the two sheets are connected via $\zeta_{2p}^{-j}\cdot \Gamma$. Its quotient $\mathcal{G}' = \mathcal{G}/\varepsilon_{6}$ coincide with the subgraph only made of $v_2$ and $(-\varepsilon_1)^jv_0$ for $j \in \ldbrack 0,p - 1 \rdbrack$.}
\end{center}
\end{figure}

\begin{figure}[h!]
\begin{center}
\includegraphics[width=\textwidth]{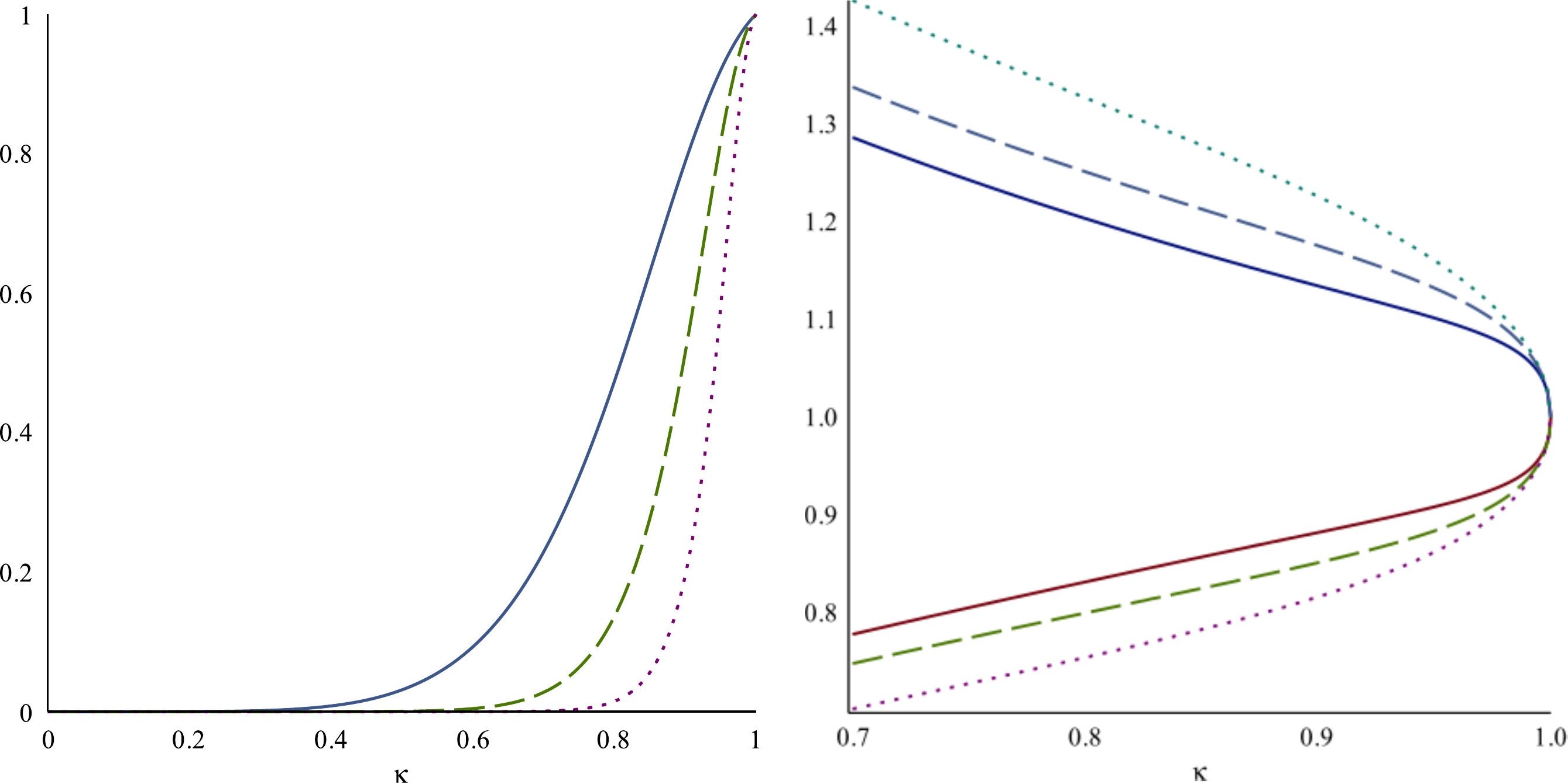}
\caption{\label{Graph-Dpoddbranch} Case $(2,2,p)$, $p$ odd = $5$ (plain line), $11$ (dashed line) and $21$ (dotted line). On the left: $c = e^{-u/4p^2}$ as a function of $\kappa$. On the right: the support $[\gamma^{-1},\gamma]$ parametrized by $\kappa$. At $u = 0$, we have $c = 1, \kappa = 1$ and the support is concentrated at $\gamma = 1$.}
\end{center}
\end{figure}

The orbit of $\hat{v}_2$ is depicted in Figure~\ref{Graph-Dpodd}. It consists of $(-\varepsilon_{-1})^{j}(\hat{v}_2)$ for $j \in \mathbb{Z}_{2p}$ and $\pm \hat{v}_0 = \pm \sum_{l = 0}^{2a_3 - 1} (-1)^{l + 1}\hat{e}_l$. The discrete Fourier transform $\mathcal{F}_k[\hat{v}_2]$ vanishes iff $k\,\,\mod\,\,2p = p$, hence the orbit is complete. We observe that all vectors in this orbit are fixed points of $\varepsilon_{1}^{p}$, i.e. that $(\hat{v}\cdot\phi)(x)$ is a actually a function of $x^2$. So, instead of $\Sigma_{\hat{v}_2}$, we consider its quotient inducing the covering $x^2\,:\,\Sigma_{\hat{v}_2}' \rightarrow \widehat{\mathbb{C}}$. Its skeleton $\mathcal{G}'_{\hat{v}_2}$ is the quotient of Figure~\ref{Graph-Dpodd} by $\varepsilon_{p}$, i.e. a star with $p$ edges connecting to the same central vertex $\hat{v}_0$.
Genus counting by Riemann-Hurwitz indicates that $\Sigma_{\hat{v}_2}'$ is rational : we must have a parametrization $z \mapsto (x^2(z),y(z))$, such that the possible values of $y$ for a fixed $x^2(z)$ are $(\hat{w}\cdot\phi)(x^2(z))$ for $\hat{w} \in \mathcal{G}'_{\hat{v}_2}$. These functions can have poles and zeroes only when $x^2 \rightarrow 0,\infty$, and their leading behavior is prescribed by \eqref{Puis1}-\eqref{Puis2}. This fixes completely the coefficients of a polynomial equation $\tilde{\mathcal{P}}(x^2(z),y(z)) = 0$. By uniqueness, it is enough to exhibit a solution with the correct leading behavior at $x \rightarrow 0$ and $\infty$. Here is one:
\beq
\label{spcurvDodd}\boxed{\left\{\begin{array}{rcl} x^2(z) & = & z^{-2}\,\dfrac{z^{2p}\kappa^{2} - 1}{z^{2p} - \kappa^{2}} \\ y(z) & = & - \dfrac{(z^{p} - \kappa)(\kappa z^{p} + 1)}{(z^{p}\kappa - 1)(z^{p} + \kappa)}\end{array}\right.}
\eeq
where $\kappa$ is the function of $u$ already encountered in \eqref{eqau2}:
\beq
\label{kappac}\frac{2\kappa^{1 + 1/p}}{1 + \kappa^{2}} = e^{-u/4p^2} := c
\eeq
The equilibrium measure can be retrieved by:
\beq
\label{tunr}\dd\lambda_{{\rm eq}} = \pm \frac{p^2\,\dd x}{{\rm i}\pi\,ux}\ln\Big(\frac{y(z(x + {\rm i}0))}{y(z(x - {\rm i}0))}\Big).
\eeq
Since all components of $\hat{w}$ in the orbit considered are $\pm 1$, \eqref{tunr} gives the density of the equilibrium measure independently of the choice of branch $z(x)$, provided the sign in prefactor is arranged to have a positive measure. The support of the equilibrium measure is of the form $[1/\gamma,\gamma]$, where $\gamma = x(z_+) > 1$  depends on $u$:
\beq
z_+ = \Bigg(\frac{(p + 1)\kappa^{-1} - (p - 1)\kappa \pm \sqrt{ (\kappa^{-1} - \kappa)((p + 1)^2\kappa^{-1} - (p - 1)^2\kappa)}}{2}\Bigg)^{1/(2p)},
\eeq
which turns out to be the squareroot of \eqref{thesqu}. By consistency, $\kappa = \kappa(u)$ should be chosen so that $0 < \kappa \leq 1$ when $u \geq 0$, hence it is the same function occuring in the $D_{p}$, $p$ even cases. See Figure~\ref{Graph-Dpoddbranch}. We can directly check that the equilibrium measure is off-critical for any $u > 0$.

\subsubsection{Two-point function}

A rational solution to the linear equation with right-hand side $x^{k - 1}\dd x$ exist whenever $(k\,\mathrm{mod}\,\,2p)$ is even or equal to $p$, and that a single logarithm is needed in the other cases. After some algebra, we arrive to the residue vectors:
\bea
\hat{C}^{(0)} & = & \frac{1}{4}\Big(\hat{e}_0 + \hat{e}_{p} + \frac{1}{p^2}\sum_{\ell \in \mathbb{Z}_{2p}} (-1)^{\ell}\hat{e}_{\ell}\Big) \\
\hat{C}^{(1)} & = & \sum_{\ell = 1}^{p - 1} (\hat{e}_{\ell + p} - \hat{e}_{\ell}) 
\eea
Taking as initial vector $\hat{v}_0 = \sum_{\ell \in \mathbb{Z}_{2p}} (-1)^{\ell + 1}\hat{e}_{\ell}$, the other vectors in the orbit are $-\hat{v}_0$ and $\pm \hat{v}'_j = (-1)^j(\hat{e}_{j} + \hat{e}_{j + p})$ for $j \in \mathbb{Z}_{2p}$, and we find that the matrix of singularity $\hat{\mathbf{B}}^{(0)}$ is already sparse, namely $\hat{B} = 0$. It can be decomposed as:
\beq
\hat{\mathbf{B}}^{(0)} = \sum_{\ell \in \mathbb{Z}_{2a_3}} (-1)^{\ell}\big(\mathbf{\Pi}_+(\ell) - \mathbf{\Pi}_-(\ell)\big)
\eeq
with the following permutation matrices:
\bea
\mathbf{\Pi}_+(\ell) & = & \mathbf{E}_{\hat{v}_0,\hat{v}_0} + \mathbf{E}_{-\hat{v}_0,-\hat{v}_0} + \sum_{i \in \mathbb{Z}_{2p}} \mathbf{E}_{\hat{v}'_i,\hat{v}'_{i + \ell}} \\
\mathbf{\Pi}_-(\ell) & = & \hat{E}_{\hat{v}_0,-\hat{v}_0} + \mathbf{E}_{-\hat{v}_0,\hat{v}_0} + \sum_{i \in \mathbb{Z}_{2p}} \mathbf{E}_{\hat{v}'_i,\hat{v}'_{i + \ell + p}}
\eea

Notice that the curve $\mathcal{\Sigma}_{\hat{v}_0}$ is not rational, but it covers a rational curve with uniformization variable $z$ by a degree $2$ covering. In a given sheet of the basis, the two choices of the squareroot correspond to a lift to a sheet of $\Sigma_{\hat{v}_2}$ labeled by $\pm \hat{v}$. Since $p$ is odd, the matrix of singularities is anti-invariant with respect to the involution $\hat{v} \rightarrow -\hat{v}$ on one line. Therefore, $\omega_{2}^{(0)}|_{\rm sp}(z_1,z_2)$ cannot be expressed as a rational form in $z_1$ and $z_2$.

\subsection{$(2,3,3)$}
\label{E6cas}
\beq
\check{R}(s,s') = \frac{s + s'}{s^2 - ss' + s'^2},\qquad a = \mathrm{dim}\,\hat{E} = 6,\qquad \chi = \frac{1}{6}
\eeq
The generator of the dynamics is:
\beq
\hat{\alpha} = 2\hat{e}_0 - \hat{e}_1 + \hat{e}_3 - \hat{e}_5.
\eeq
The subspace $E$ has dimension $4$. The orthogonal projection of $\hat{e}_0$ on $E$ is:
\beq
e_0 = \frac{1}{6}\big(4\hat{e}_0 - \hat{e}_1 + \hat{e}_2 + 2\hat{e}_3 + \hat{e}_4 - \hat{e}_5\big)
\eeq
and using the shift operator, the projections of the other vectors of the canonical basis are $e_j = \varepsilon_{-j}(\hat{e}_0)$. We observe that:
\beq
\beta_1 = e_0,\quad \beta_2 = e_1,\quad \beta_3 = e_2,\quad \beta_4 = -e_4
\eeq
is a basis of $E$. Equivalently, $E$ can be characterized:
\beq
E = \Big\{\sum_{j = 0}^3 \gamma_j\,\hat{e}_j + (\gamma_0 + \gamma_1 - \gamma_3)\hat{e}_4 + (-\gamma_0 + \gamma_2 + \gamma_3)\hat{e}_5,\qquad \gamma_j \in \mathbb{R}\Big\}.
\eeq
The Cartan matrix of the root system $\mathfrak{R}$ read in the basis of $\beta$'s is:
\beq
\label{canr}\hat{\mathbf{A}} = \left(\begin{array}{cccccc} 2 & -1 & 0 & 1 & 0 & -1 \\ -1 & 2 & -1 & 0 & 1 & 0 \\ 0 & -1 & 2 & -1 & 0 & 1 \\ 1 & 0 & -1 & 2 & -1 & 0 \\ 0 & 1 & 0 & -1 & 2 & -1 \\ -1 & 0 & 1 & 0 & -1 & 2\end{array}\right).
\eeq
We recognize the Cartan matrix of the root system $D_4$ (see Figure~\ref{dpp1dyn}), where $\beta$'s play the role of the simple roots. They can be represented in the more standard form:
\beq
\beta_1 = \delta_{1} - \delta_{2},\quad \beta_{2} = \delta_{2} - \delta_{3},\quad \beta_{3} = \delta_{3} + \delta_{4},\quad \beta_{4} = \delta_{3} - \delta_{4},
\eeq
where the $\delta$'s now form an $\langle$orthonormal$\rangle$ basis of $E$:
\bea
\delta_{1} & = & \frac{1}{4}\big(2\hat{e}_0 + \hat{e}_1 + \hat{e}_2 + 2\hat{e}_3 + \hat{e}_4 + \hat{e}_5\big),\nonumber \\
\delta_{2} & = & \frac{1}{12}\big(-2\hat{e}_0 + 5\hat{e}_1 + \hat{e}_2 + 2\hat{e}_3 + \hat{e}_4 + 5\hat{e}_5\big), \nonumber  \\
\delta_{3} & = & \frac{1}{4}\big(-\hat{e}_1 + \hat{e}_2 - \hat{e}_4 + \hat{e}_5\big), \nonumber \\
\delta_{4} & = & \frac{1}{12}\big(-2\hat{e}_0 - \hat{e}_1 - 5\hat{e}_2 + 2\hat{e}_3 - 5\hat{e}_4 - \hat{e}_5\big). \nonumber
\eea
This basis is convenient for computations.

\subsubsection{Minimal orbits}

 The $A_3$ root system generated by $I = \{\beta_1,\beta_2,\beta_4\}$ forms a maximal sub-root system of $D_4$. Since the $\delta$'s from a $\langle$orthonormal$\rangle$ basis, it is easy to see that $H_{F[I]}$ (the subspace of $E$ orthogonal to $I$) is spanned by:
\beq
v^* = \frac{1}{2}\big(\delta_{1} + \delta_{2} + \delta_{3} + \delta_{4}\big) = \frac{1}{12}\big(\hat{e}_0 + 2\hat{e}_1 + \hat{e}_2 + 5\hat{e}_3 - 2\hat{e}_4 + 5\hat{e}_5\big).
\eeq
Multiplying on the left by $\hat{\alpha}$, we find an element $\hat{v}^* \in \hat{E}$ with a minimal orbit:
\beq
\label{mine6}\hat{v}^* = \hat{\alpha}\cdot v^* = \hat{e}_3 - \hat{e}_4 + \hat{e}_5,
\eeq
and the order of its orbit is:
\beq
d = \frac{|\mathrm{Weyl}(D_4)|}{|\mathrm{Weyl}(A_3)|} = 8.
\eeq

\begin{figure}[h!]
\begin{center}
\includegraphics[width=0.6\textwidth]{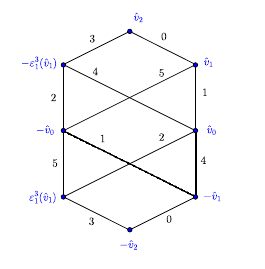}
\caption{\label{GraphE6} Case $(2,3,3)$ -- The skeleton graph of a minimal orbit.}
\end{center}
\end{figure}

We choose as initial data a shift of \eqref{mine6}, namely $\hat{v}_1 = -\hat{e}_0 + \hat{e}_1 + \hat{e}_5$. Its orbit is shown in Figure~\ref{GraphE6}, it consists of the vectors:
\beq
\label{orbe6}\hat{v}_2 = \hat{e}_0 + \hat{e}_3,\qquad \hat{v}_1,\varepsilon_{3}(\hat{v}_1),\qquad \hat{v}_0 = \hat{e}_1 - \hat{e}_2 + \hat{e}_4 - \hat{e}_{5}
\eeq
and their opposite, and one can check that the orbit is complete.

\subsubsection{Spectral curve}

There exists a polynomial $\mathcal{P}(x,y)$ of degree:
\beq
\mathrm{deg}_{y} \mathcal{P} = 8
\eeq
such that $y = \hat{w}\cdot\phi(x)$ for $\hat{w} \in \mathfrak{G}\cdot\hat{v}_1$ are the solutions of $\mathcal{P}(x,y) = 0$. Let us describe its symmetries:
\begin{itemize}
\item[$\bullet$] The orbit is stable under the $\langle$isometry$\rangle$ $\varepsilon_{3}$. Hence: $\mathcal{P}(x,y) = 0$ iff $\mathcal{P}(-x,y) = 0$. But contrarily to \S~\ref{dodd}, the orbit is not pointwise invariant under $\varepsilon_{3}$, so it will not be simpler to study the quotient by the involution $x \rightarrow -x$.
\item[$\bullet$] The orbit is stable under $\hat{v} \rightarrow -\hat{v}$, so $\mathcal{P}(x,y) = 0$ iff $\mathcal{P}(x,1/y) = 0$.
\item[$\bullet$] The palindrome symmetry adds that $\mathcal{P}(x,y) = 0$ iff $\mathcal{P}(1/x,1/y) = 0$.
\end{itemize}

\begin{figure}[h!]
\begin{center}
\includegraphics[width=0.7\textwidth]{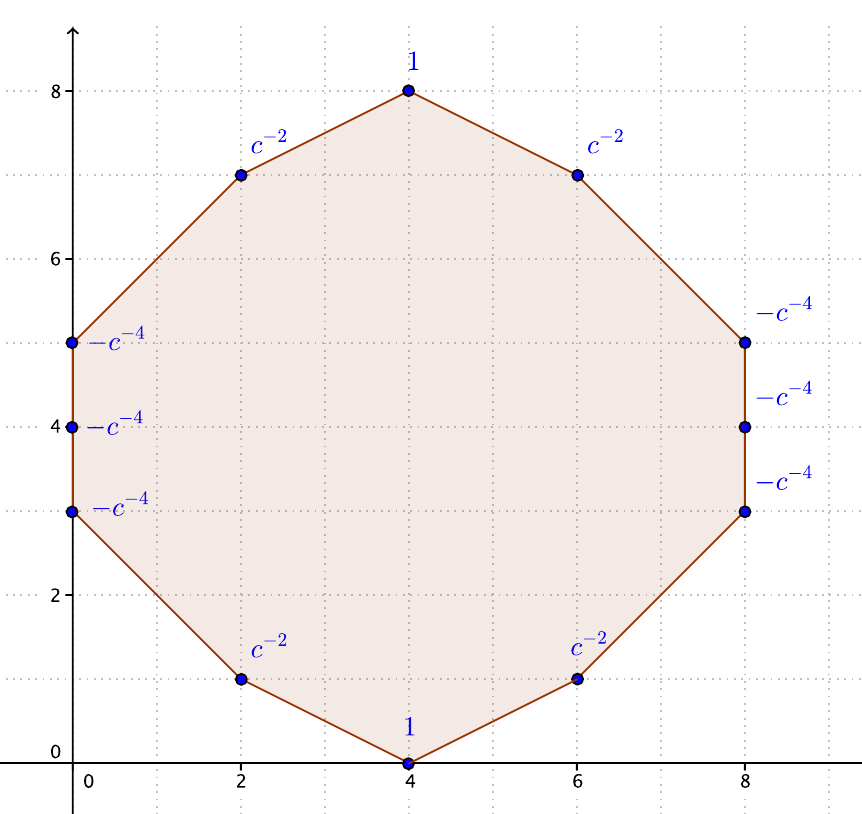}
\caption{\label{Npolye6} Case $(2,3,3)$ -- Newton polygon of $\mathcal{P}(x,Y)$. Once the coefficient $1$ at the bottom is fixed, all the coefficients in blue are determined by the slope polynomials. A rule of thumb to find the homogeneity factors of $c$: on the left (resp. right) of the picture -- the facets $x \rightarrow 0$ (resp. $x \rightarrow \infty$) -- we take a $c$ (resp. $c^{-1}$) at each step left (resp. right).}
\end{center}
\end{figure}

Since $\Gamma$ is a single segment, we count (Lemma~\ref{Lgenus}) a genus $\mathfrak{g} = 5$ for the Riemann surface obtained from the skeleton graph. Since a rational parametrization does not exist, there is no shortcut and we have to determine the coefficients of $\mathcal{P}(x,y) = 0$. It must have the Puiseux expansions at $x \rightarrow 0$ and $\infty$ in a sheet $\hat{w} \in \hat{\mathcal{G}}\cdot\hat{v}_2$:
\beq
\label{fgrowt}\hat{w}\cdot\phi(x) \mathop{\sim}_{x \rightarrow 0}\,\,c^{n_0[\hat{w}]}\,\zeta_{a}^{n_1[\hat{w}]}\,x^{n_0[\hat{w}]},\qquad (\hat{w}\cdot\phi)(x) \mathop{\sim}_{x \rightarrow \infty}\,\,c^{-n_0[\hat{w}]}\,\zeta_{a}^{n_1[\hat{w}]}\,x^{n_0[\hat{w}]}
\eeq
where:
\beq
c:= e^{-\chi u/2a} = e^{-u/72}.\nonumber
\eeq
They give rise to the slopes $(n_0[\hat{w}],-1)$ and $(n_0[\hat{w}],1)$ in the Newton polygon of $\mathcal{P}$ -- overall sign chosen so that they leave the polygon on their left --, and the leading coefficients in \eqref{fgrowt} give the roots of the slope polynomials. We observe that all leading coefficients are roots of unity, up to a homogeneity factor of $c^{n_0[\hat{w}]}$ (resp. $c^{-n_0[\hat{w}]}$) on the facets with $x \rightarrow 0$ (with $x \rightarrow \infty$). Absorbing this homogeneity factor in the definition, the slope polynomials are thus cyclotomic: 
 \beq
\begin{array}{|r|c|l|}
\hline \mathrm{sheets} & \mathrm{slope} & \mathrm{polynomial} \\
\hline
\hat{v}_2 & (2,-1) & \xi + 1 \\
\hline
\hat{v}_1,\varepsilon_{3}(\hat{v}_1) & (1,-1) & \xi^2 - 1 = (\xi - 1)(\xi + 1)  \\
\hline
\pm\hat{v}_0 & (0,-1) & \xi^2 + \xi + 1 = (\xi - \zeta_{3})(\xi - \zeta_{3}^{-1})  \\
\hline
-\hat{v_1},-\varepsilon_{3}(\hat{v}_1) & (-1,-1) & \xi^2 - 1 \\
\hline
-\hat{v}_2 & (-2,-1) & \xi + 1 \\
\hline
\end{array}
\eeq
We deduce that:
\beq
\mathrm{deg}_{x} \mathcal{P} = \sum_{\substack{\hat{w} \in \hat{\mathfrak{G}}\cdot\hat{v}_2 \\ n_0[\hat{w}] > 0}} n_0[\hat{w}] = 8
\eeq
and the slopes must be arranged in a sequence such that they form the boundary of a convex polygon. Starting with South-East steps -- belonging to facets with $x,y \rightarrow 0$ -- and following the counterclockwise order, this sequence reads:
\bea
\label{sequ1}&& \cdots \rightarrow (1,-1) \rightarrow (1,-1) \rightarrow (2,-1) \rightarrow (2,1) \rightarrow (1,1) \rightarrow (1,1) \rightarrow (0,1) \rightarrow (0,1)  \\
&& (-1,1) \rightarrow (-1,1) \rightarrow (-2,1) \rightarrow (-2,-1) \rightarrow (-1,-1) \rightarrow (-1,-1) \rightarrow (0,-1) \rightarrow (0,-1) \rightarrow \cdots \nonumber
\eea
Let us put the lexicographic order on the monomials $x^iY^j$: first compare the degree in $x$, then the degree in $y$. We infer from \eqref{sequ1} that the minimal monomial appearing in $\mathcal{P}$ is $x^4$, and we can normalize $\mathcal{P}$ so that it appears with coefficient $1$. Then, following the slope polynomials along \eqref{sequ1} determines completely the coefficients of $\mathcal{P}$ on the boundary of the Newton polygon, see Figure~\ref{Npolye6}.

With this normalization, the $\varepsilon_{3}$ symmetry of the orbit and the palindrome symmetry of our solution imply:
\beq
\label{symeq}\mathcal{P}(x,y) = \mathcal{P}(-x,y) = y^{8}x^{8}\mathcal{P}(1/x,1/y).
\eeq
If we set $\tilde{x} = x^2$ and $\tilde{P}(\tilde{x},y) = \mathcal{P}(x,y)$, it remains to find $7$ coefficients inside the Newton polygon of $\widetilde{P}$:
\bea
\widetilde{\mathcal{P}}(\tilde{x},y) =  & = & (y^8 + 1)\tilde{x}^2 + (y^7 + y)(c^2\tilde{x} + C_1\tilde{x}^2 + c^2\tilde{x}^3) + (y^6 + y^2)(C_2\tilde{x} + C_3\tilde{x}^2 + C_2\tilde{x}^3) \nonumber \\
& & + (y^3 + y^5)(-c^4 + C_4\tilde{x} + C_5\tilde{x}^2 + C_4\tilde{x}^3 - c^4\tilde{x}^4) \nonumber \\
\label{unkonc} & & y^4(-c^4 + C_6\tilde{x} + C_7\tilde{x}^2 +  C_6\tilde{x}^3 - c^4\tilde{x}^4) 
\eea
If the $C$'s were generic, the curve of equation $\mathcal{P}(x,y) = 0$ would have genus $17$. But according to \eqref{countg}, our solution must achieve genus $5$, and the $C$'s should be determined as functions of $c$.

First, we try to reduce the number of parameters. Let us give names to the first few moments of the equilibrium measure:
\beq
Y(x) \mathop{=}_{x \rightarrow 0} (-x/c)\exp\Big(-\sum_{k \geq 0} M_k\,x^{k + 1}\Big),\qquad M_k =\frac{\chi u}{a} \int s^{k}\dd\lambda_{{\rm eq}}(s)
\eeq
The $M$'s are function of $c$ and their definition implies $M_k \rightarrow 0$ when $c \rightarrow 1$. By construction we have:
\beq
\widetilde{\mathcal{P}}(x^2,y) = x^4\prod_{\hat{w} \in \mathfrak{G}\cdot\hat{v}_2} \big(y - \hat{w}\cdot Y(x))
\eeq
Expanding this polynomial when $x \rightarrow 0$, we can identify the $C$'s in terms of the $M$'s, and the form \eqref{unkonc} also imposes constraints on the $M$'s. We find that all $M$'s and $C$'s can be eliminated except $M_1$ and $M_2$.
\bea
\begin{array}{lcl} C_1 = 1 + 2M_1c^2 &\quad & \\
 C_2 = 0 &\quad & C_3 = 2 + 12M_1c^2 - 6c^3M_3 \\
C_4 = 2M_1c^4 & \quad & C_5 = c^4  - 2 - 6c^2M_1 - 4M_1^2c^4  \\
C_6 = -2c^2 - 4M_1c^4 & \quad & C_7 = 4c^4 - 4 -16M_1c^2 + 8M_1^2c^4 + 12M_2c^3
\end{array}
\eea
If $M_2$ and $M_3$ were generic, the curve of equation $\widetilde{\mathcal{P}}(\tilde{x},y) = 0$ would have genus $13$, but the solution we look for has genus $5$. It means that the curve has $8$ singular points, and they will come in two orbits under the symmetries $(\tilde{x},y) \rightarrow (\tilde{x}^{-1},y)$ and $(\tilde{x},y^{-1})$. Let us exploit the symmetries by setting:
\beq
\xi = \tilde{x} + 1/\tilde{x},\qquad \eta = y + 1/y
\eeq
and eliminate $\tilde{x}$ and $y$ to get a polynomial equation $\mathcal{Q}(\xi,\eta) = 0$:
\bea
\mathcal{Q}(\xi,\eta) & = & \eta^4 + (c^2\xi + 1 + 2M_1c^2)\eta^3 + (-2 + 12M_1c^2 - 6M_2c^3)\eta^2 \nonumber \\
& & + \big(-c^4\xi^2 + (-3c^2 + 2M_1c^4)\xi + 3c^4 - 5 -12M_1c^2 - 4M_1^2c^4\big)\eta + \nonumber \\
& & \big(-c^4\xi^2 + (-2c^2 - 4M_1c^4)\xi + 6c^4 - 6 - 40M_1c^2  + 8M_1^2c^4 + 24M_2c^3\big)
\eea
This curve should have two singular points, i.e. there exists $(\xi_1,\eta_1)$ and $(\xi_2,\eta_2)$ such that:
\beq
i = 1,2,\qquad \mathcal{P}(\xi_i,\eta_i) = \partial_{\xi} \mathcal{P}(\xi_i,\eta_i) = \partial_{\eta} \mathcal{P}_{2}(\xi_i,\eta_i) = 0
\eeq
Solving explicitly these algebraic constraints is complicated, but we do expect that they select a finite number of solutions for $M_1$ and $M_2$. Eventually, the (hopefully unique) appropriate branch should be chosen by requiring that the coefficients of $u \rightarrow 0$ expansion are rational numbers.

\subsubsection{Two-point function}

A rational solution of the linear equation with right-hand side $x^{k - 1}\dd x$ exists for $k\,\,\mathrm{mod}\,\,6 \in \{0,2,3,4\}$, otherwise we have a log. The outcome of the computation of the residue vectors of the particular solution for the primitive of the two-point function is:
\bea
\hat{C}^{(0)} & = & \frac{1}{72}\big(22\hat{e}_0 + 5\hat{e}_1 + 13\hat{e}_2 + 14\hat{e}_3 + 13\hat{e}_{4} + 5\hat{e}_5\big) \\
\nonumber \\
\hat{C}^{(1)} & = & \frac{1}{2}\big(-\hat{e}_1 - \hat{e}_2 + \hat{e}_4 + \hat{e}_5\big) \nonumber
\eea
We order the vectors of the orbit described in \eqref{orbe6} as follows:
\bea
& & \hat{w}_1 = \hat{v}_2,\quad \hat{w}_2 = \hat{v}_1,\quad \hat{w}_3 = \varepsilon_{3}(\hat{v}_1),\quad \hat{w}_4 = -\hat{v}_0 \nonumber \\
& & \hat{w}_5 = \hat{v}_0,\quad \hat{w}_6 = -\varepsilon_{3}(\hat{v}_1),\quad \hat{w}_7 = -\hat{v}_1,\quad \hat{w}_8 = -\hat{v}_2 \nonumber
\eea
in order to write down the matrix of singularities $\hat{\mathbf{B}}^{(0)}$. Substracting a matrix filled with:
\beq
\hat{B} = - \frac{\hat{e}_1 + \hat{e}_2 + \hat{e}_4 + \hat{e}_5}{2},
\eeq
we find the sparse part given in Appendix~\ref{E6sing}. It has only $0,\pm 1$ entries, and there are atmost $4$ non-zero entries in the same row. 

\begin{figure}[h!]
\begin{center}
\includegraphics[width=6cm]{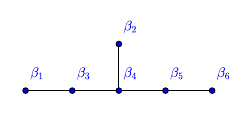}
\caption{\label{e6dyn} Dynkin diagram of the root system $E_6$}
\end{center}
\end{figure}

\subsection{$(2,3,4)$}

\beq
\check{R}(s,s') = \frac{(s^3 + s^2s' + ss'^2 + s'^3)(s^2 + ss' + s'^2)}{s^6 + s'^6},\qquad a = \mathrm{dim}\,\hat{E} = 12,\qquad \chi = \frac{1}{12}.
\eeq
The dynamics are generated by:
\beq
\hat{\alpha} = 2\hat{e}_0 - \hat{e}_1 + \hat{e}_4 - \hat{e}_5 + \hat{e}_{6} - \hat{e}_7 + \hat{e}_8 - \hat{e}_{11}.
\eeq
and the multiplication by $\hat{\alpha}$ is an operator of rank $6$. The orthogonal projections of the canonical basis on its image $E$ is:
\beq
e_0 = \frac{1}{12}\big(6\hat{e}_0 - \hat{e}_1 - \hat{e}_2 + 2\hat{e}_3 + 3\hat{e}_4 - \hat{e}_5 + 2\hat{e}_6 - \hat{e}_7 + 3\hat{e}_8 + 2\hat{e}_9 - \hat{e}_{10} - \hat{e}_{11}\big),\qquad e_j = \varepsilon_{-j}(e_0)  
\eeq
If we define a new basis of $E$:
\beq
\begin{array}{rclcrclcrcl}
\beta_1 & = & -e_1 & \qquad & e_3 & = & e_5 & \qquad & \beta_5 & = & -e_0 + e_6 \\
\beta_2 & = & e_3 & \qquad & e_4 & = & e_4 & \qquad & \beta_6 & = & - e_2 - e_7 
\end{array}
\eeq
the root system $\mathfrak{R}$ has a Cartan matrix:
\beq
\label{Aea}\mathbf{A} = \left(\begin{array}{cccccc} 2 & 0 & -1 & 0 & 0 & 0 \\ 0 & 2 & 0 & -1 & 0 & 0 \\ -1 & 0 & 2 & -1 & 0 & 0 \\ 0 & -1 & -1 & 2 & -1 & 0 \\ 0 & 0 & 0 & -1 & 2 & -1 \\ 0 & 0 & 0 & 0 & -1 & 2\end{array}\right).
\eeq
We recognize\footnote{Let us give some explanation about the introduction of the $\beta$'s. The matrix of scalar products $\langle \hat{e}_j,\hat{e}_k \rangle$ can be directly read off $\hat{\alpha}$, and it is easy to check that it is definite positive. This implies that $\mathfrak{R}$ is a finite root system, so must be of the form ADE. We already know the rank $\mathrm{dim}\,E = 4$ here, so it does not leave much choice for $\mathfrak{R}$. Then, there is a part of guess to define a basis $(\beta_j)_j$ which puts the Cartan matrix in standard form.} the Cartan matrix of the root system $E_6$, where $\beta$'s play the role of the simple roots labeled according to Figure~\ref{e6dyn}.

\subsubsection{Minimal orbits}

Deleting the root $\beta_6$ generates a maximal sub-root system, of type $D_5$. Therefore, the order of the minimal orbit is:
\beq
d = \frac{|\mathrm{Weyl}(E_6)|}{|\mathrm{Weyl}(D_5)|} = \frac{51840}{1920} = 27
\eeq
A vector with minimal orbit can be obtained as $\hat{v}^* = \hat{\alpha}\cdot v^*$ where $v^*$ is $\langle$orthogonal$\rangle$ to all $\beta$'s except $\beta_6$. Using the matrix \eqref{Aea} containing the $\langle \beta_i,\beta_j \rangle$, we can write down those equations and solve them. We find a generator:
\bea
v^* & = & \frac{1}{3}\big(2a_1 + 3a_2 + 4a_3 + 6a_4 + 5a_5 + 4a_6\big) \nonumber \\
\label{maxe7} & = & \frac{1}{36}\big(10\hat{e}_0 + 5\hat{e}_1 + \hat{e}_2 + 2\hat{e}_3 + \hat{e}_4 + 5\hat{e}_5 + 10\hat{e}_6 - 7\hat{e}_7 + \hat{e}_8 + 14\hat{e}_{9} + \hat{e}_{10} - 7\hat{e}_{11}\big).
\eea
and thus:
\beq
\hat{v}^* = \hat{e}_0 + \hat{e}_6 - \hat{e}_7 + \hat{e}_9 - \hat{e}_{11}
\eeq
The orbit of $\hat{v}^*$ consists of the following vectors and their images under the shift $\varepsilon_{2}(\hat{v}) = \sum_{j \in \mathbb{Z}_{12}} \hat{v}(j + 2)\hat{e}_j$, which make in total $27$ vectors. We indicate in the second column the size of the $\varepsilon_{2}$-orbit.
\beq
\label{orb7}\begin{array}{|r|c|l|}
\hline n_0 & \varepsilon_2 & \mathrm{vector} \\
\hline
3 & 2 & \hat{e}_2 + \hat{e}_6 + \hat{e}_{10} \\
\hline
2 & 3 & \hat{e}_1 - \hat{e}_2 + \hat{e}_3 + \hat{e}_7 - \hat{e}_8 + \hat{e}_9 \\
\hline
1 & 6 & \hat{v}^* \\
\hline
0 & 2 &  -\hat{e}_0 + \hat{e}_3 - \hat{e}_4 + \hat{e}_7 - \hat{e}_8 + \hat{e}_{11} \\
\hline\hline
0 & 1 & \sum_{j \in \mathbb{Z}_{12}} (-1)^j\hat{e}_j \\
\hline\hline
0 & 2 & -\varepsilon_{1}(\hat{v}_0') \\
\hline
-1 & 6 & -\varepsilon_{1}(\hat{v}_1) \\
\hline
-2 & 3 & -\varepsilon_{1}(\hat{v}_2) \\
\hline
-3 & 2 & -\varepsilon_{1}(\hat{v}_3) \\
\hline
\end{array}
\eeq
One can check that the orbit is complete.

There is a polynomial equation $\mathcal{P}(x,y) = 0$ of degree $27$ in $y$, whose solutions are precisely the $\hat{w}\cdot Y(x)$ for $\hat{w}$ the orbit of $\hat{v}^*$. Let us describe its symmetries:
\begin{itemize}
\item[$\bullet$] The orbit is stable under $-\varepsilon_1$, hence $\mathcal{P}(x,y) = 0$ iff $\mathcal{P}(\zeta_{12}x,1/y) = 0$.
\item[$\bullet$] The palindrome symmetry gives $\mathcal{P}(x,y) = 0$ iff $\mathcal{P}(1/x,1/y) = 0$.
\end{itemize}
Since $\Gamma$ is a segment, we find a Riemann surface of genus $46$.

\subsubsection{Spectral curve}

\begin{figure}[h!]
\begin{center}
\includegraphics[width=\textwidth]{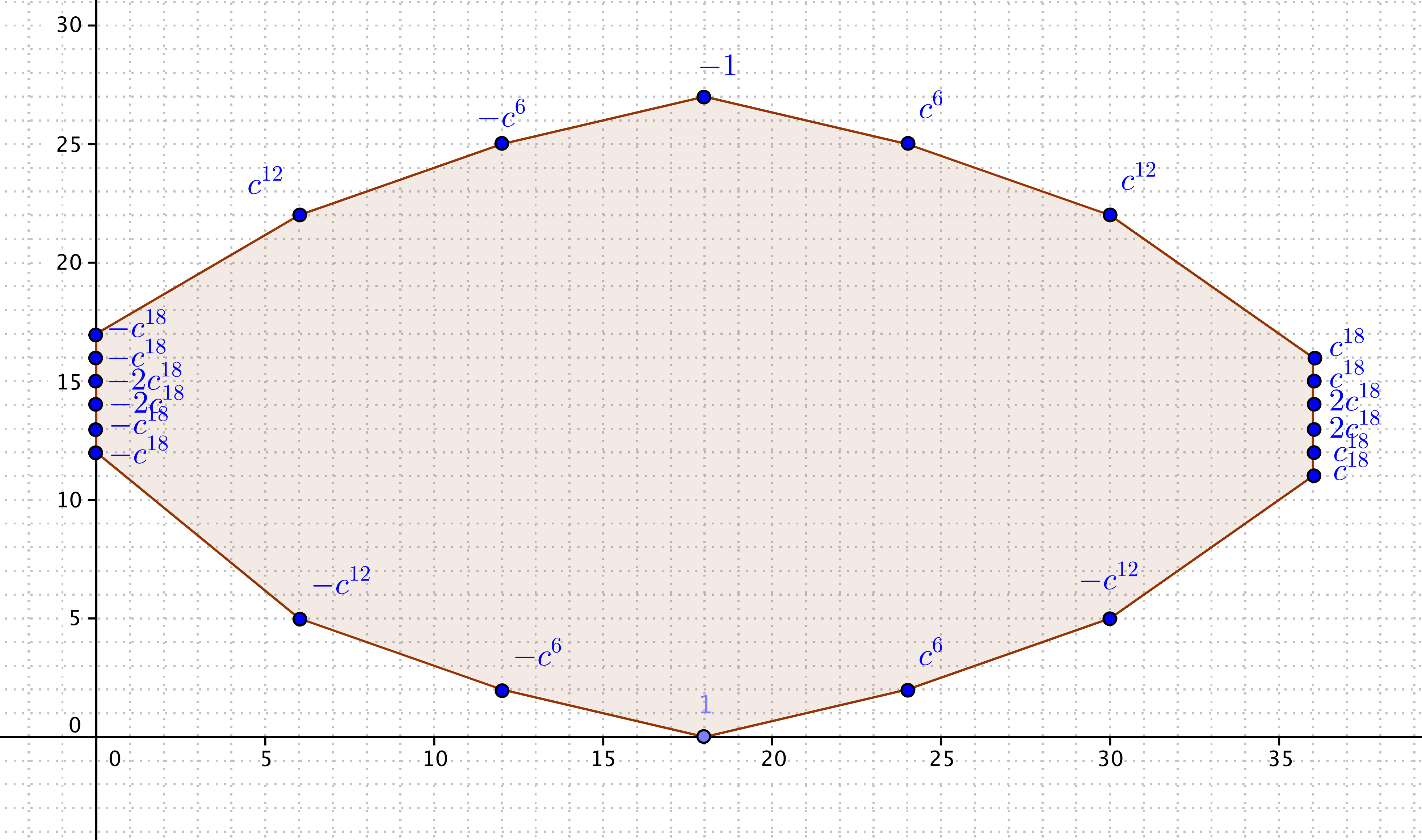}
\caption{\label{e7newtonpolygon} Case $(2,3,4)$ -- Newton polygon of $\mathcal{P}(x,y)$. Actually, only the coefficients of $x^{6j}y^k = \tilde{x}^jy^k$ are non-zero, and they satisfy the symmetries \eqref{symm7}.}
\end{center}
\end{figure}

The formulae applied to the orbit \eqref{orb7} determine the list of slopes of the Newton polygon of $\mathcal{P}$, and the list of slope polynomials:
\beq
\begin{array}{|r|l|}
\hline\mathrm{slope} & \mathrm{polynomial} \\
\hline (3,-1) & \xi^2 - 1 \\
\hline
  (2,-1) & \xi^3 + 1  \\
  \hline
(1,-1) & \xi^6 + 1  \\
\hline
 (0,-1) & (\xi + 1)(\xi^2 + 1)^2 = \xi^5 + \xi^4 + 2\xi^3 + 2\xi^2 + \xi + 1 \\
 \hline
(-1,-1) & \xi^6 - 1 \\
\hline
 (-2,-1) & \xi^3 - 1 \\
 \hline
(-3,-1) & \xi^2 + 1 \\
\hline
\end{array}
\eeq
We can then read off the coefficients on the boundary of the Newton polygon (see Figure~\ref{e7newtonpolygon}). The degree in $x$ is given by:
\beq
\mathrm{deg}_{x}\,\mathcal{P} = \sum_{\substack{\hat{w} \in \mathfrak{G}\cdot\hat{v}^*  \\ n_0[\hat{w}] > 0}} n_0[\hat{w}] = 36.
\eeq
But the $\varepsilon_{2}$-symmetry of the orbit implies:
\beq
\boxed{\mathcal{P}(x,y) = \tilde{\mathcal{P}}(\tilde{x},y),\qquad \tilde{x} = x^6}.
\eeq
and we know:
\beq
\mathrm{deg}_{\tilde{x}} \tilde{\mathcal{P}} = 6,\qquad \mathrm{deg}_{y} \tilde{\mathcal{P}} = 27.
\eeq
The residual $-\varepsilon_{1}$ symmetry and the palindrome symmetry give:
\beq
\label{symm7}\tilde{\mathcal{P}}(\tilde{x},y) = -y^{27}\tilde{\mathcal{P}}(-\tilde{x},1/y) = y^{27}\tilde{x}^{6}\tilde{\mathcal{P}}(1/\tilde{x},1/y)
\eeq
The proportionality coefficients have been found by checking the coefficients on the boundary of the Newton polygon. This gives us a polynomial involving $32$ unknown coefficients inside the Newton polygon:
\bea
\label{234Pxy} & & \widetilde{\mathcal{P}}(\tilde{x},y) \\
& = & -\tilde{x}^{3}y^{27} + C_1\tilde{x}^3y^{26}+ (c^{-6}\tilde{x}^4 + C_2\tilde{x}^3 - c^{-6}\tilde{x}^2)y^{25} + (C_3\tilde{x}^{4} + C_4\tilde{x}^3 - C_3\tilde{x}^2)y^{24} \nonumber \\
& & + (C_5\tilde{x}^4 + C_6\tilde{x}^3 - C_5\tilde{x}^2)y^{23} + (c^{-12}\tilde{x}^5 + C_7\tilde{x}^4 + C_8\tilde{x}^3 - C_7\tilde{x}^2 + c^{-12}\tilde{x})y^{22} \nonumber \\
& & + (C_9\tilde{x}^5 + C_{10}\tilde{x}^4 + C_{11}\tilde{x}^3 - C_{10}\tilde{x}^{2} + C_9\tilde{x})y^{21} + (C_{12}\tilde{x}^5 + C_{13}\tilde{x}^4 + C_{14}\tilde{x}^3 - C_{13}\tilde{x}^2 + C_{12}\tilde{x})y^{20} \nonumber \\
& & + (C_{15}\tilde{x}^5 + C_{16}\tilde{x}^4 + C_{17}\tilde{x}^3 - C_{16}\tilde{x}^2 + C_{15}\tilde{x})y^{19} + (C_{18}\tilde{x}^5 + C_{19}\tilde{x}^4 + C_{20}\tilde{x}^3 - C_{19}\tilde{x}^2 + C_{18}\tilde{x})y^{18} \nonumber \\
& & + (C_{21}\tilde{x}^5 + C_{22}\tilde{x}^4 + C_{23}\tilde{x}^3 - C_{22}\tilde{x}^2 + C_{21}\tilde{x})y^{17} \nonumber \\
& & + (c^{-18}\tilde{x}^6 + C_{24}\tilde{x}^5 + C_{25}\tilde{x}^4 + C_{26}\tilde{x}^3 - C_{25}\tilde{x}^2 + C_{24}\tilde{x} - c^{-18})y^{16} \nonumber \\
& & + (c^{-18}\tilde{x}^6 + C_{27}\tilde{x}^5 + C_{28}\tilde{x}^4 + C_{29}\tilde{x}^3 - C_{28}\tilde{x}^2 + C_{27}\tilde{x} - c^{-18})y^{15} \nonumber \\
& & + (2c^{-18}\tilde{x}^6 + C_{30}\tilde{x}^5 + C_{31}\tilde{x}^4 + C_{32}\tilde{x}^3 - C_{31}\tilde{x}^2 + C_{30}\tilde{x} - 2c^{-18})y^{14} + \cdots \nonumber 
\eea
where the $\cdots$ are completed by symmetry \eqref{symm7}. For generic $C$'s, the equation $\widetilde{\mathcal{P}}(\tilde{x},y) = 0$ defines a curve of genus $102$, but the solution we look for achieves genus $46$. By construction, we have:
\beq
\tilde{\mathcal{P}}(\tilde{x},y) = -\tilde{x}^3\prod_{\hat{w} \in \mathfrak{O}} (y - \hat{w}\cdot Y(x))
\eeq
If we write the expansion at $x \rightarrow 0$ of our solution:
\beq
Y(x) \mathop{=}_{x \rightarrow 0} (-x/c)\exp\Big(\sum_{k \geq 1} M_k\,x^k\Big),\qquad  c = e^{-u/288},\qquad M_k = -\frac{u}{144} \int s^{k}\dd\lambda_{{\rm eq}}(s).
\eeq
then plug it in $\tilde{\mathcal{P}}(\tilde{x},y) = 0$, we can identify the $C$'s in terms of the $M$'s. The constraints on the shape of the Newton polygon and the symmetry of its coefficients allows a linear elimination of all $M$'s except $4$ of them: $M_3$, $M_4$ and $M_6$ and $M_8$ ; the expression of the $C$'s is then given in Appendix~\ref{234coef}.

To determine the remaining $M_k$'s in terms of $c$, one has in principle to impose the existence of enough singular points -- modulo the symmetries $(\tilde{x},y) \rightarrow (-\tilde{x},1/y)$ and $(1/\tilde{x},1/y)$ -- to lower the genus down to $46$. This yields algebraic equations, that we were not able to write down given the length of the expressions. We expect -- although we cannot prove before writing them down -- that these equations have a finite number of solutions, and among those, one can hope -- again without proof -- that there is a unique one such that $M_{k} \in O(u)$ when $u \rightarrow 0$ and such that the coefficients of Taylor expansion in $u$ are rational numbers. The solution we seek for has all this property, and since we proceeded by necessary conditions, we know that at least one such solution exists.

\subsubsection{Two-point function}

We can find a rational solution of the linear equation with right-hand side $x^{m - 1}\dd x$ for $$m\,\,\mathrm{mod}\,\,12 \in \{0,3,4,6,8,9\}$$, and a solution with a log otherwise. We find residue vectors:
\bea
\hat{C}^{(0)} & = & \frac{1}{432}\big(82\hat{e}_0 + 34\hat{e}_6 + 23\hat{e}'_5 + 7\hat{e}'_4 + 50\hat{e}'_3 + 55\hat{e}'_2 + 23\hat{e}'_1\big)
\nonumber \\
\hat{C}^{(1)} & = & \frac{1}{3}\sum_{j = 0}^{1} (-1)^{j + 1} \big(2\hat{e}_{1 + 6j} + 2\hat{e}_{2 + 6j} + 3\hat{e}_{3 + 6j} + \hat{e}_{4 + 6j} + \hat{e}_{5 + 6j} \big) \nonumber
\eea
where $\hat{e}'_j = \hat{e}_{6 - j} + \hat{e}_{6 + j}$. On the orbit described above, we find that the plain part of the matrix of singularities is:
\beq
\hat{B} = \frac{1}{3}\sum_{\ell \in \mathbb{Z}_{12}} (-1)^{\ell}\hat{e}_{\ell}
\eeq
and the sparse part is a $27 \times 27$ $\hat{E}$-valued matrix stored in a \textsc{maple} file available on request. Its entries are $0,\pm 1$ and in the matrix $\hat{B}^{(0)}$ in front of each $\hat{e}_j$ there are exactly $11$ of them per row (or column) are non zero. Some $1$ and $-1$ could cancel to make a $0$, so we guess that the curve is a $12$-fold cover of a rational curve, which should actually be the quotient curve by the symmetries $(x,y) \rightarrow (\zeta_{6}x,y)$ and $(x,y) \rightarrow (1/x,1/y)$.

\subsection{$(2,3,5)$: Poincar\'e spheres}
\label{e8e8}
\begin{figure}[h!]
\begin{center}
\includegraphics[width=9cm]{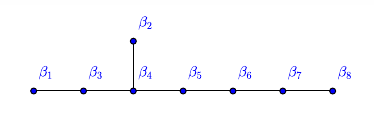}
\caption{\label{e8dyn} Dynkin diagram of the root system $E_8$}
\end{center}
\end{figure}

\beq
\check{R}(s,s') = \frac{\big(\sum_{j = 0}^9 s^{j}s'^{9 - j}\big)\big(\sum_{j = 0}^5 s^js'^{5 - j}\big)}{s^{15} + s'^{15}},\qquad a = \mathrm{dim}\,\hat{E} = 30,\qquad \chi = \frac{1}{30}
\eeq
The generator of the dynamics is:
\bea
\hat{\alpha} & = & 2\hat{e}_0 - \hat{e}_1 + \hat{e}_6 - \hat{e}_7 + \hat{e}_{10} - \hat{e}_{11} + \hat{e}_{12} - \hat{e}_{13} + \hat{e}_{15} \nonumber \\
& & - \hat{e}_{17} + \hat{e}_{18} - \hat{e}_{19} + \hat{e}_{20} - \hat{e}_{23} + \hat{e}_{24} - \hat{e}_{29}.
\eea
The subspace $E$ has dimension $8$, and it is spanned by the orthogonal projections on $E$ of canonical basis:
\bea
e_0 & = &  \frac{1}{30}\big(8\hat{e}_0 - 2\hat{e}_1 + \hat{e}_3  + 3\hat{e}_{5} + 3\hat{e}_{6} - 2\hat{e}_{7} + \hat{e}_9 + 5\hat{e}_{10} - 2\hat{e}_{11} + 3\hat{e}_{12} - 2\hat{e}_{13} + 6\hat{e}_{15} \nonumber \\
& &  - 2\hat{e}_{17} + 3\hat{e}_{18} - 2\hat{e}_{19} + 5\hat{e}_{20} + \hat{e}_{21} - 2\hat{e}_{23} + 3\hat{e}_{24} + 3\hat{e}_{25} + \hat{e}_{27} - 2\hat{e}_{29} \big),
\eea
and $e_j = \varepsilon_{j}(e_0)$ for $j \in \mathbb{Z}_{30}$. The family below defines a basis of $E$:
\beq
\begin{array}{rclcrcl} 
\beta_1 & = & e_1 & \qquad & \beta_5 & = & e_{5} \\
\beta_2 & = & e_3 & \qquad & \beta_6 & = & e_{6} \\
\beta_3 & = & e_0 - e_3 - e_4 - e_5 - e_6 + e_9 & \qquad & \beta_7 & = & -e_{11} - e_{12} \\
\beta_4 & = & -e_{15} & \qquad & \beta_8 & = & e_{2} - e_{14}.
\end{array}
\eeq
in which the Cartan matrix of $\mathfrak{R}$ reads:
\beq
\mathbf{A} = \left(\begin{array}{cccccccc} 2 & 0 & -1 & 0 & 0 & 0 & 0 & 0 \\ 0 & 2 & 0 & -1 & 0 & 0 & 0 & 0 \\
-1 & 0 & 2 & -1 & 0 & 0 & 0 & 0 \\ 0 & - 1 & -1 & 2 & -1 & 0 & 0 & 0 \\ 0 & 0 & 0 & -1 & 2 & -1 & 0 & 0 \\ 0 & 0 & 0 & 0 & -1 & 2 & -1 & 0 \\ 0 & 0 & 0 & 0 & 0 & -1 & 2 & -1 \\ 0 & 0 & 0 & 0 & 0 & 0 & -1 & 2 \end{array}\right).
\eeq
We recognize the Cartan matrix of the root system $E_8$, and the $\beta$'s play the role of the simple roots labeled as in Figure~\ref{e8dyn}. The $240$ roots generated by the $\beta$'s under action of the Weyl group $\mathfrak{G}$ are the elements $v \in E$ of the form:
\beq
v \in \mathbb{Z}[(\beta_j)_{j}] \cup (\mathbb{Z} + 1/2)[(\beta_j)_j],\qquad \langle v,v\rangle = 1,\qquad \sum_{i = 1}^8 \ell_{\beta_i}(v) = 0
\eeq
The simple roots can be represented in the standard form:
\beq
\label{betadelta}\beta_1 = -\frac{1}{2}\sum_{j = 1}^8 \delta_{1},\qquad \beta_2 = \delta_{6} + \delta_{7},\qquad \beta_j = \delta_{9 - j} - \delta_{10 - j},\quad (3 \leq j \leq 8)
\eeq
where the $\delta$ now form an $\langle$orthonormal$\rangle$ basis, expressed in terms of the canonical basis $(\hat{e}_k)_k$ in Appendix~\ref{deltaE8}.

\subsubsection{Minimal orbit}

$I = \{\beta_i\quad 1 \leq i \leq 7\}$ generates a maximal\footnote{The order of the Weyl group of $E_7$ is $2903040$. If we  delete $\beta_2$ instead of $\beta_8$, we obtain a root system of type $A_7$, whose Weyl group has order $5040$ ; if we rather delete $\beta_1$, we obtain a root system of type $D_7$, whose Weyl group has order $322560$. All other choice of roots to delete disconnect the Dynkin diagram and lead to smaller Weyl group. So, the maximal sub-root lattice of $E_8$ has type $E_7$.} sub-root system, which is of type $E_7$. We can find a vector of minimal orbit as $\hat{v}^* = \hat{\alpha}\cdot v^*$ where $v \in E$ is $\langle$orthogonal$\rangle$ to all $\beta$'s except $\beta_8$. It is easy using \eqref{betadelta} to get the decomposition of a generator on the $\langle$orthonormal$\rangle$ basis of $\delta$'s:
\beq
v^* = -\delta_{1} + \delta_{8}
\eeq
and the formulae in Appendix~\ref{deltaE8} allows its rewriting in terms of the canonical basis:
\beq
v^* = \frac{1}{6}\Big(\sum_{j \in \mathbb{Z}_{30}} \hat{e}_{j} + \sum_{j \in \mathbb{Z}_{6}} \hat{e}_{5j - 1}\Big)
\eeq
Then, the vector:
\beq
\hat{v}^* = \sum_{j \in \mathbb{Z}_6} \hat{e}_{5j - 1}
\eeq
has an orbit $\mathfrak{O}$ of minimal order, equal to:
\beq
d = \frac{|\mathrm{Weyl}(E_8)|}{|\mathrm{Weyl}(E_7)|} = 240
\eeq
We find that the orbit is stable by the shift $\varepsilon_{1}$ and the involution $\hat{v} \mapsto -\hat{v}$. This allows the decomposition of $\mathfrak{O}$ into $\varepsilon_{1}$ orbits generated by the following vectors, and their opposites. The size of the $\varepsilon_{1}$ orbits are displayed in the second column.
\beq
\label{orb8}\begin{array}{|c|c|l|}
\hline
n_0 & \varepsilon_1 & \mathrm{vector} \\
\hline
6 & 5 & \hat{v}^* \\
\hline
5 & 6 & \sum_{j \in \mathbb{Z}_{5}} \hat{e}_{6j} - \hat{e}_{6j + 1} + \hat{e}_{6j + 2} \\
\hline
4 & 15 & \hat{e}_5 - \hat{e}_{7} + \hat{e}_8 + \hat{e}_{11} - \hat{e}_{12} + \hat{e}_{14} \\
\hline
3 & 10 & -\hat{e}_2 + \hat{e}_3 + \hat{e}_5 - \hat{e}_{6} + \hat{e}_9 - \hat{e}_{12} + \hat{e}_{13} + \hat{e}_{15} - \hat{e}_{16} + \hat{e}_{19} - \hat{e}_{22} + \hat{e}_{23} + \hat{e}_{25} - \hat{e}_{26} + \hat{e}_{29} \\
3 & 10 & \hat{e}_{5} - \hat{e}_{6} + \hat{e}_7 - \hat{e}_8 + \hat{e}_{9} + \hat{e}_{15} - \hat{e}_{16} + \hat{e}_{17} - \hat{e}_{18} + \hat{e}_{19} + \hat{e}_{25} - \hat{e}_{26} + \hat{e}_{27} - \hat{e}_{28} + \hat{e}_{29}  \\
\hline
2 & 15 &  -\hat{e}_{3} + \hat{e}_4 + \hat{e}_5 - \hat{e}_6 + \hat{e}_{10} - \hat{e}_{12} + \hat{e}_{14} - \hat{e}_{18} + \hat{e}_{19} + \hat{e}_{20} - \hat{e}_{21} + \hat{e}_{25} - \hat{e}_{27} + \hat{e}_{29} \\
2 & 15 & \hat{e}_{5} - \hat{e}_6 + \hat{e}_8 - \hat{e}_9 + \hat{e}_{10} - \hat{e}_{12} + \hat{e}_{13} + \hat{e}_{20} - \hat{e}_{21} + \hat{e}_{23} - \hat{e}_{24} + \hat{e}_{25} - \hat{e}_{27} + \hat{e}_{28} \\
\hline
1 & 30 & 2\hat{e}_{0} - \hat{e}_{1} + \hat{e}_{6} - \hat{e}_{7} + \hat{e}_{10} - \hat{e}_{11} + \hat{e}_{12} - \hat{e}_{13} + \hat{e}_{15} - \hat{e}_{17} + \hat{e}_{18} - \hat{e}_{19} + \hat{e}_{20} - \hat{e}_{23} + \hat{e}_{24} - \hat{e}_{29} \\
\hline
0 & 5 & \sum_{j \in \mathbb{Z}_{6}} \hat{e}_{5j} - \hat{e}_{5j - 1} \\
0 & 5 & \sum_{j \in \mathbb{Z}_{6}} \hat{e}_{5j} - \hat{e}_{5j - 2} \\
0 & 3 & \sum_{j \in \mathbb{Z}_{10}} \hat{e}_{3j} - \hat{e}_{3j + 2} \\
\hline\hline
0 & 2* & \hat{v}_0 = \sum_{j \in \mathbb{Z}_{30}} (-1)^{j + 1}\hat{e}_{j} \\
\hline\hline 
\end{array} \nonumber
\eeq
The $2*$ in the last line means that $\varepsilon_{1}(\hat{v}_0) = -\hat{v}_0$, so we do not need to include the $\varepsilon_{1}$ orbit of $-\hat{v}_0$ when describing the other part of the orbit via $\hat{v} \rightarrow -\hat{v}$. One can check that the orbit is complete.

\subsubsection{Spectral curve}
\label{S43fin}

There exists a polynomial equation $\mathcal{P}(x,y) = 0$ of degree $240$ in $y$, whose solutions are the $\hat{w}\cdot Y(x)$ for $\hat{w} \in \mathfrak{G}\cdot\hat{v}^*$. The symmetries $\varepsilon_{1}$ and $\hat{v} \rightarrow -\hat{v}$ or the orbit, as well as the palindrome symmetry of our solution, imply:
\beq
\mathcal{P}(x,y) = 0 \quad \Longleftrightarrow \quad \mathcal{P}(\zeta_{30}x,y) = 0 \quad \Longleftrightarrow \quad \mathcal{P}(1/x,y) = 0 \quad \Longleftrightarrow \quad \mathcal{P}(1/x,1/y) = 0
\eeq
We obtain according to Lemma~\ref{Lgenus} a Riemann surface of genus $1471$. The computation of $n_1[\hat{w}]$ for all elements $\hat{w}$ of the orbit described above gives the list of slopes and slope polynomials:
\beq
\begin{array}{|r|l|}
\hline\mathrm{slope} & \mathrm{polynomial} \\
\hline (6,-1) & \xi^5 + 1 \\
\hline
  (5,-1) & \xi^6 - 1  \\
  \hline
(4,-1) & \xi^{15} - 1  \\
\hline
 (3,-1) & (\xi^{10} - 1)^2 = \xi^{20} - 2\xi^{10} + 1 \\
 \hline
(2,-1) & (\xi^{15} + 1)^2 = \xi^{30} + 2\xi^{15} + 1 \\
\hline
 (1,-1) & \xi^{30} - 1 \\
 \hline
(0,-1) & (\xi + 1)^2(\xi^2 + \xi + 1)^3(\xi^4 + \xi^3 + \xi^2 + \xi + 1)^5 \\
\hline
\end{array}
\eeq
Since all slopes are of the form $(n,\pm 1)$, the degree of the slope polynomials give the multiplicity of the corresponding slope belonging to the facet $x \rightarrow 0$. Those of the facets $x \rightarrow \infty$ form an identical set of slopes and slope polynomials thanks to the palindrome symmetry. We learn that $\mathcal{P}$ must have degree $540$ in $x$ and its minimal monomial is $x^{270}$. Choosing its coefficient to be $1$ and following the slope polynomials then determine the coefficients on the boundary of the Newton polygon of $\mathcal{P}$ (Figure~\ref{fig:e8poly}). Enforcing the symmetries yields:
\beq
\mathcal{P}(x,y) = \tilde{\mathcal{P}}(\tilde{x},y),\qquad \tilde{x} = x^{30},\qquad \mathrm{deg}_{\tilde{x}} \tilde{\mathcal{P}} = 18
\eeq
and:
\beq
\label{synu}\tilde{\mathcal{P}}(\tilde{x},y) = \tilde{x}^{18}\tilde{\mathcal{P}}(1/\tilde{x},y) = y^{240}\tilde{\mathcal{P}}(\tilde{x},1/y)
\eeq
We count that $\mathcal{P}$ depends on $801$ independent coefficients in the interior of the Newton polygon. If they were generic, the plane curve of equation $\tilde{\mathcal{P}}(\tilde{x},y) = 0$ would have genus 2949 (this is the total number of points in the interior of the Newton polygon), but our solution must achieve genus $1471$.

By construction, we have:
\beq
\tilde{\mathcal{P}}(\tilde{x},y) = \tilde{x}^{9}\prod_{\hat{w} \in \mathfrak{O}} (y - \hat{w}\cdot Y(x)),\qquad \tilde{x} = x^{30}
\eeq
If we write the expansion at $x \rightarrow 0$ of our solution:
\beq
Y(x) \mathop{=}_{x \rightarrow 0} (-x/c)\exp\Big(\sum_{k \geq 1} M_k\,x^k\Big),\qquad  c = e^{-u/1800},\qquad M_k = -\frac{u}{900} \int s^{k}\dd\lambda_{{\rm eq}}(s).
\eeq
then plug it in $\tilde{\mathcal{P}}(\tilde{x},y) = 0$ and use the information -- shape and symmetries \eqref{synu} --  on its Newton polygon, we find that $\tilde{\mathcal{P}}(\tilde{x},y)$ can entirely be expressed in terms of the $3$ moments $M_{6}$, $M_{10}$, $M_{15}$, which are however not specified.

\subsubsection{Two-point function}

We can find a rational solution of the linear equation with right-hand side $x^{m - 1}\dd x$ for $$m\,\,\mathrm{mod}\,\,30 \in \{0,6,10,12,15,18,20,24\}$$, and a solution with a log otherwise. We find residue vectors:
\bea
\hat{C}^{(0)} & = & \frac{1}{900}(58\hat{e}_{0} + 54\hat{e}_{15} + 20\hat{e}'_{14} + 24\hat{e}'_{13} + 29\hat{e}'_{12} + 24\hat{e}'_{11} + 45\hat{e}
_{10} + 33\hat{e}'_{9} + 20\hat{e}'_{8} + 24\hat{e}'_{7} + 29\hat{e}'_{6} \nonumber \\
& &  + 49\hat{e}'_{5} + 20\hat{e}'_{4} + 33\hat{e}'_{3} + 20\hat{e}'_{2} + 24\hat{e}'_{1}\big)
\eea
where $\hat{e}'_j = \hat{e}_{15 - j} + \hat{e}_{15 + j}$, while $\hat{C}^{(1)}$ is a vector having huge rational coefficients with huge denominator $2274168727208063179297410229829$ that we do not reproduce here. Nevertheless, the $240 \times 240$ matrix of singularity for the orbit described above is not so complicated. We identify its plain part as:
\beq
\hat{B} = \sum_{\ell \in \mathbb{Z}_{30}} \hat{e}_{\ell}
\eeq
and its sparse part has entries $\in \{-3,-2,-1,0,1\}$. And in each row of $\mathbf{B}^{(0)}_{\rm sp}(j)$, there are exactly $184$ non-zero entries. This is larger than the covering of degree $30\times 4 = 120$ that we get by dividing out by the symmetries.

\begin{landscape}
\addtolength{\footskip}{30pt}
\addtolength{\linewidth}{50pt}
\begin{figure}
\begin{center}
\includegraphics[width = 1.55\textwidth]{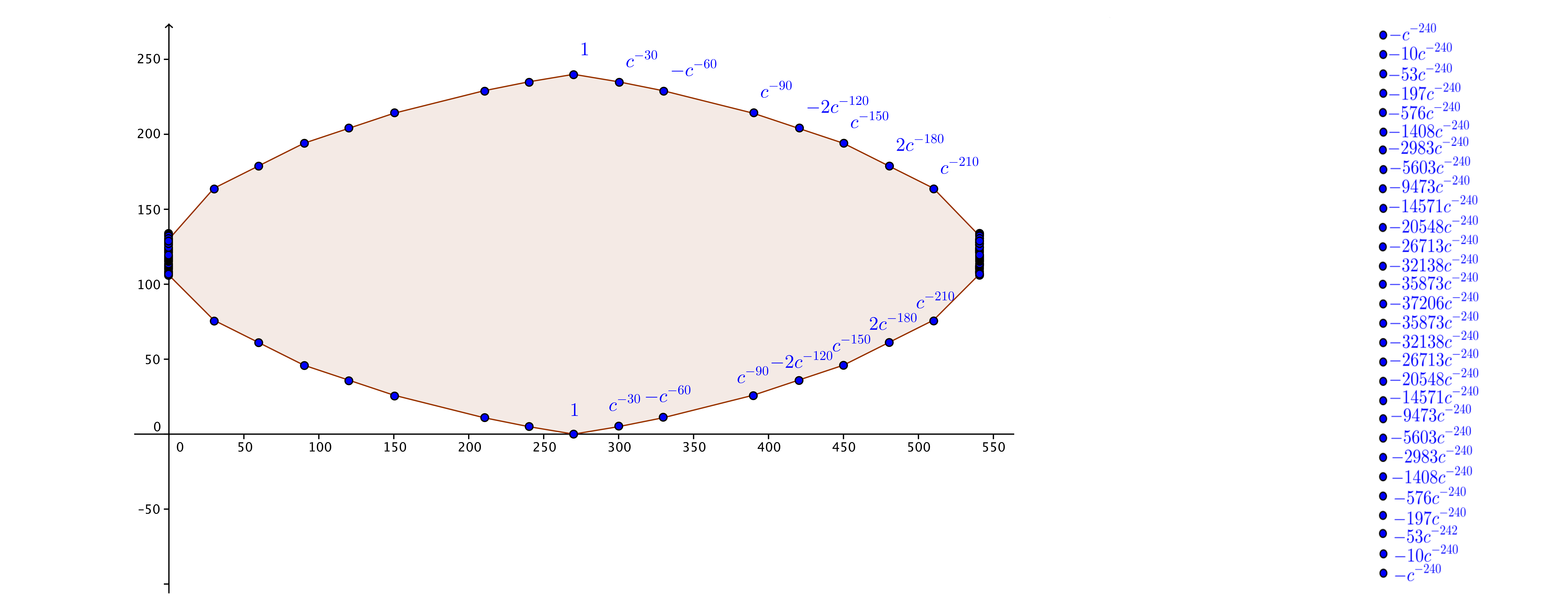}
\caption{\label{fig:e8poly} {\small The Newton polygon of $\mathcal{P}(x,y)$. The coefficients on the vertical boundaries are shown on the right. The remaining coefficients are completed by symmetry $x \rightarrow 1/x$ and $y \rightarrow 1/y$. Only monomials of the form $x^{30j}y^{k}$ can appear.}}
\end{center}
\end{figure}
\end{landscape}

\subsection{$(2,2,2,2)$}
\label{aff1}
\beq
\check{R}(s,s') = \frac{1}{(s + s')^2},\qquad a = 2,\qquad \chi = 0
\eeq
The dynamics are generated by $\hat{\alpha} = 2(\hat{e}_0 - \hat{e}_1)$. The matrix model is actually the $O(2)$ model \cite{KosOn} with a log-square potential, its saddle-point equation is:
\beq
\label{sdale22}W(x + {\rm i}0) + W(x - {\rm i}0) - 2W(-x) = \frac{a^2\ln x}{u}
\eeq
$\hat{\mathfrak{G}}$ is found to be the affine Weyl group $\widehat{A}_1$, while $\hat{\mathfrak{G}} \simeq \mathbb{Z}_2$ is the Weyl group $A_1$, and up to rescaling the finite orbit is $\pm\hat{v}_0 = \pm(\hat{e}_1 - \hat{e}_2)$ and has order $2$.


Since \eqref{sdale22} is very simple, we present a direct solution. The simplest particular solution of \eqref{sdale22} involves a log-cube, so we prefer to introduce as explained in \S~\ref{chi0}:
\beq
\phi_1(x) = x\frac{\partial}{\partial x} W(x),\qquad \phi_2(x) = x\frac{\partial}{\partial x}\phi_1(x)
\eeq
which solves, for any $x \in ]1/\gamma,\gamma[$:
\beq
\label{matchre} \phi_1(x + {\rm i}0) + \phi_1(x - {\rm i}0) - 2\phi_1(-x) = \frac{4}{u},\qquad \phi_2(x + {\rm i}0) + \phi_2(x - {\rm i}0) - 2\phi_2(-x) = 0
\eeq
Thus, $(\phi_2(x) - \phi_2(-x))^2$ has no discontinuity, and given the growth and symmetries of $phi_2(x)$, it must be of the form:
\beq
\big(\phi_2(x) - \phi_2(-x)\big)^2 = \frac{\beta_2(x^2 + x^{-2})^2 + \beta_1(x^2 + x^{-2}) + \beta_0}{(x^2 + x^{-2} - \gamma^2 - \gamma^{-2})^3}
\eeq
Since $\phi_2(x)$ has a discontinuity on $[1/\gamma,\gamma]$ only, the numerator must be a perfect square, hence an answer of the form
\beq
\phi_2(x) - \phi_2(-x) = \frac{x(\beta' (x^4 + 1) + \beta x^2)}{\big((x^2 - \gamma^2)(x^2 - \gamma^{-2})\big)^{3/2}}
\eeq
for some constants $\beta$ and $\beta'$. Notice that this function is odd/odd with respect to the symmetries $x \rightarrow -x$/$x \rightarrow x^{-1}$, because the determination\footnote{If we were choosing signs so that the squareroot is odd/odd, we find that the right-hand side of \eqref{matchre} must be $0$, which is a contradiction.}. The result can be integrated once:
\beq
\phi_1(x) = \int_{0}^{x} \frac{\phi_2(x')}{x'}\,\dd x'
\eeq
and the answer is of the form:
\beq
\boxed{\phi_1(x) - \phi_1(-x) = C_{E}\Bigg({\rm E}\big[x/\gamma;\gamma^2] - \frac{\gamma x\big(x^2 - (\gamma^2 + \gamma^{-2})/2\big)}{\sqrt{(x^2 - \gamma^2)(x^2 - \gamma^{-2})}}\Bigg) + C_{F}\,{\rm F}\big[x/\gamma;\gamma^2]}
\eeq
where $C_{E}$ and $C_{F}$ are constants to determine, and we introduce the elliptic integrals \bea
\mathrm{E}[z;k] = \int_{0}^{z} \dd t\,\sqrt{\frac{1 - k^2t^2}{1 - t^2}} & & \nonumber \\
\mathrm{E}[k] = \mathrm{E}[1;k] = \int_{0}^{1} \dd t\,\sqrt{\frac{1 - k^2t^2}{1 - t^2}} & & \mathrm{E}'[k] = \mathrm{E}[\sqrt{1 - k^2}]
\eea
and:
\bea
\mathrm{F}[z;k] = \int_{0}^{z} \frac{\dd t}{\sqrt{(1 - t^2)(1 - k^2t^2)}} & & \nonumber \\
\mathrm{K}[k] = \mathrm{F}[1;k] = \int_{0}^{1} \frac{1}{\sqrt{(1 - t^2)(1 - k^2t^2)}} & & \mathrm{K}'[k] = \mathrm{K}[\sqrt{1 - k^2}].
\eea
Since $\lim_{x \rightarrow \infty} W(x) = 1$, we must have $\lim_{x \rightarrow \infty} \phi_1^{{\rm odd}} = 0$. Since we have the asymptotic behavior when $x \rightarrow \infty$:
\bea
{\rm E}[x/\gamma;\gamma^2] & = & \gamma x + {\rm i}\big(\gamma^2{\rm E}'[\gamma^{-2}] - {\rm K}'[\gamma^2]\big) +  O(1/x) \nonumber \\
{\rm F}[x/\gamma;\gamma^2] & = & -{\rm i}{\rm K'}[\gamma^2] + O(1/x) \nonumber
\eea
we must impose a relation between $C_{E}$ and $C_{F}$:
\beq
C_{F} = C_{E}\,\frac{\gamma^2 \mathrm{E}'[\gamma^{-2}] - \mathrm{K}'[\gamma^2]}{K'[\gamma^2]}
\eeq
Next, we have to match the first equation of \eqref{matchre}. Using the functional relations:
\bea
\mathrm{E}[(x + {\rm i}0)/\gamma,\gamma^2] + \mathrm{E}[(x - {\rm i}0)/\gamma,\gamma^2] & = & 2\mathrm{E}[k]\nonumber \\
\mathrm{F}[(x + {\rm i}0)/\gamma,\gamma^2] + \mathrm{F}[(x - {\rm i}0)/\gamma,\gamma^2] & = & 2\mathrm{K}[k] 
\eea
we deduce the value:
\beq
C_{E} = \frac{2}{u}\,\frac{\mathrm{K}'}{\mathrm{E}\mathrm{K}' + \gamma^2\mathrm{E} - \mathrm{K}'}
\eeq
where all complete elliptic integrals are evaluated at $k = \gamma^2$, except for E evaluated at $\sqrt{1 - \gamma^{-4}}$. Eventually, the position of the cut $\gamma > 1$ is determined implicitly by the normalization of the spectral density:
\beq
\int_{1/\gamma}^{\gamma} \frac{\phi_1(x' - {\rm i}0) - \phi_1(x' + {\rm i}0)}{2{\rm i}\pi\,x'}\,\dd x' = 1
\eeq

\subsubsection{$2$-point function}

We find residue vectors:
\beq
\hat{C}^{(0)} = \frac{1}{8}\big(\hat{e}_0 - \hat{e}_{1}\big),\qquad  \hat{C}^{(1)} = 0,\qquad \hat{C}^{(2)} = \hat{e}_0 + \hat{e}_{1}
\eeq
Taking $\hat{v} = \hat{e}_0 - \hat{e}_1$ as initial vector, the matrix of singularity can be decomposed:
\beq
\hat{\mathbf{B}}^{(0)} = -\frac{1}{2}(\hat{e}_0 + \hat{e}_1) + \mathbf{1}_{2}\hat{e}_0 + \left(\begin{array}{cc} 0 & 1 \\ 1 & 0 \end{array}\right)\hat{e}_1
\eeq
and we find $\hat{\mathbf{B}}^{(1)} = \hat{\mathbf{B}}^{(2)} = 0$. Therefore, if we introduce a uniformization variable for the curve:
\beq
x^2(z) = \frac{\gamma^2 + \gamma^{-2}}{2} + \frac{\gamma^2 - \gamma^{-2}}{4}\Big(z + \frac{1}{z}\Big)
\eeq
the two-point function reads:
\beq
\omega_{2}^{(0)}|_{\rm hom}(z_1,z_2) = \frac{\dd z_1\,\dd z_2}{(z_1 - z_2)^2} - \frac{1}{2}\,\frac{\dd x^2(z_1)\,\dd x^2(z_2)}{(x^2(z_1) - x^2(z_2))}
\eeq

\subsection{$\mathbf{P} = (3,3,3)$}
\label{aff2}
\beq
\check{R}(s,s') = \frac{1}{s^2 + ss' + s'^2}.
\eeq
The dynamics are generated by:
\beq
\hat{\alpha} = 2\hat{e}_0 - \hat{e}_1 - \hat{e}_2.
\eeq
$E$ has dimension $2$, and it is spanned by the orthogonal projections of $(\hat{e}_0,\hat{e}_1,\hat{e}_2)$ on $E$:
\bea
e_0 & = & \frac{1}{3}\big(2\hat{e}_0 - \hat{e}_1 - \hat{e}_2\big), \\
e_1 & = & \frac{1}{3}\big(-\hat{e}_0 + 2\hat{e}_1 - \hat{e}_2\big), \\
e_2 & = & \frac{1}{3}\big(-\hat{e}_0 - \hat{e}_1 + 2\hat{e}_2\big).
\eea
So, $v \in E$ iff $v(0) + v(1) + v(2) = 0$. The Cartan matrix of the root system $\mathfrak{R} = \{\pm e_0,\pm e_1,\pm e_2\}$ in the basis $(a_1,a_2) = (e_0,e_1)$ is:
\beq
\mathbf{A} = \left(\begin{array}{cc} 2 & -1 \\ -1 & 2 \end{array}\right).
\eeq
We recognize the Cartan matrix of the root system $A_2$, where $(a_0,a_1)$ plays the role of the simple roots, and its Weyl group $\mathfrak{G}$ is the symmetric group in $3$ elements. Deleting $a_0$ (or $a_1$) gives a maximal sub-root system, of type $A_1$. A generator of $H_{\{a_0\}}$ is:
\beq
v^* = \frac{1}{2}\big(2a_1 + a_2\big) = \frac{1}{3}\big(\hat{e}_0 - \hat{e}_2\big).
\eeq
Its image is:
\beq
\hat{v}^* = A(v^*) = \hat{e}_0 - \hat{e}_2,
\eeq
The orbit of $\hat{v}^*$ is a cycle of order $d = 3$, the two other vectors begin $-\hat{e}_0 + \hat{e}_1$ and $-\hat{e}_1 + \hat{e}_2$.

We may also consider the "root system" $\hat{\mathfrak{R}}$. Its "Cartan matrix" in the canonical basis $(\hat{e}_0,\hat{e}_1,\hat{e}_2)$ is:
\beq
\hat{\mathbf{A}} = \left(\begin{array}{ccc} 2 & - 1 & - 1 \\ -1 & 2 & -1 \\ -1 & -1 & 2 \end{array}\right),
\eeq
hence $\hat{\mathfrak{R}}$ is the affine root system of type $\widehat{A}_2$.  By construction, its Weyl group is $\hat{\mathfrak{G}}$: it is therefore the symmetry group of the tiling of the plane by equilateral triangles.

\subsubsection{Spectral curve}
\label{333sP}
We set:
\beq
\phi_1(x) = x\frac{\partial}{\partial x} W(x),\qquad \phi_2(x) = x\frac{\partial}{\partial x} \phi_1(x)
\eeq
which solves, for any $x \in ]1/\gamma,\gamma[$, the linear equation:
\beq
\phi_2(x + {\rm i}0) + \phi_2(x - {\rm i}0) - \phi_2(\zeta_{3}x) - \phi_2(\zeta_{3}^{-1}x) = 0
\eeq
The discussion of \S~\ref{aff2} tells us that $\phi_2(x) - \phi_2(\zeta_{3}^{-1}x)$ can be analytically continued to an analytic function $y$ a Riemann surface with $3$ sheets. The value of $y$ in the $2$ other sheets is $\phi_2(\zeta_{3}x) - \phi_2(x)$ and $\phi_2(\zeta_{3}^{-1}x) - \phi_2(\zeta_{3}x)$, and its singularities follow from those of $\phi_2(x)$ listed in \S~\ref{chi0}. We therefore have a polynomial equation $P(x,y) = 0$, but the symmetry tells us that, if $(x,y)$ is a solution, then $(\zeta_{3}x,y)$ is a solution, and $(1/x,-y)$ is a solution. This information, and the order of singularities of $y$, prescribe the following form:
\beq
P(x,y) = (x^3 + x^{-3} - \gamma^3 - \gamma^{-3})^3\,y^3 + P_1(x^3 + x^{-3})y + P_0(x^3 + x^{-3})(x^{3} - x^{-3})
\eeq
where $P_1$ and $P_0$ are degree $2$ polynomials.

\subsubsection{$2$-point function}

We find the residue vectors:
\beq
\hat{C}^{(0)} = \frac{1}{9}\big(2\hat{e}_0 - \hat{e}_{1} - \hat{e}_{2}\big),\qquad \hat{C}^{(1)} = 0,\qquad \hat{C}^{(2)} = \frac{3}{4}\big(\hat{e}_0 + \hat{e}_{1} + \hat{e}_2\big)
\eeq
The matrix of singularities for the orbit of $\hat{v}^*$ can be decomposed as:
\beq
\hat{\mathbf{B}}^{(0)} = -\frac{1}{3}\big(\hat{e}_0 + \hat{e}_1 + \hat{e}_2\big)\mathbf{J} + \mathbf{1}_{3}\hat{e}_0 + \left(\begin{array}{ccc} 0 & 0 & 1 \\ 1 & 0 & 0 \\ 0 & 1 & 0 \end{array}\right)\hat{e}_1 + \left(\begin{array}{ccc} 0 & 1 & 0 \\ 0 & 0 & 1 \\ 1 & 0 & 0 \end{array}\right)\hat{e}_2
\eeq
and we find $\hat{\mathbf{B}}^{(1)} = \hat{\mathbf{B}}^{(2)} = 0$.

\subsection{$\mathbf{P} = (2,4,4)$}
\label{aff3}
\beq
\breve{R}(s,s') = \frac{1}{s^2 + s'^2}.
\eeq
The dynamics are generated by:
\beq
\hat{\alpha} = 2\hat{e}_0 - \hat{e}_1 - \hat{e}_3.
\eeq
$E$ has dimension $3$, and it is spanned by the orthogonal projections of $(\hat{e}_i)_{0 \leq i \leq 3}$ on $E$:
\beq
e_0 = \frac{1}{4}\big(3\hat{e}_0 - \hat{e}_1 - \hat{e}_2 - \hat{e}_3\big),\qquad e_i = S^{i}(e_0).
\eeq
So, $v \in E$ iff $\sum_{i = 0}^3 v(i) = 0$. A basis of $E$ is given by $(a_1,a_2,a_3) = (e_0,e_1,e_2)$, and the Cartan matrix of the root system in this basis is:
\beq
\mathbf{A} = \left(\begin{array}{ccc} 2 & -1 & 0 \\ -1 & 2 & -1 \\ 0 & -1 & 2 \end{array}\right).
\eeq
hence $\mathfrak{R}$ is the root system of type $A_3$. Its Weyl group $\mathfrak{G}$ is isomorphic to the symmetric group in $4$ elements. Deleting $a_1$ (or $a_3$) gives a maximal sub-root system, of type $A_2$. A generator of $H_{\{a_2,a_3\}}$ is:
\beq
v^* = \frac{1}{4}\big(3a_1 + a_2 + a_3\big) = \frac{1}{8}\big(3\hat{e}_0 + \hat{e}_1 - \hat{e}_2 - 3\hat{e}_3\big).
\eeq
Its image is:
\beq
\hat{v}^* = A(v^*) = \hat{e}_0 - \hat{e}_3 \in \mathcal{M},
\eeq
The orbit of $\hat{v}^*$ is a cycle of order $d = 4$, containing the vectors $-\hat{e}_0 + \hat{e}_1$, $-\hat{e}_1 + \hat{e}_2$ and $-\hat{e}_2 + \hat{e}_3$.

We may also consider the "root system" $\mathfrak{R}$. Its "Cartan matrix" in the canonical basis of $\hat{E}$ is:
\beq
\hat{\mathbf{A}} = \left(\begin{array}{cccc} 2 & - 1 & 0 & -1 \\ -1 & 2 & -1 & 0 \\ 0 & -1 & 2 & - 1 \\ -1 & 0 & -1 & 2 \end{array}\right),
\eeq
hence $\hat{\mathfrak{R}}$ is the affine root system of type $\widehat{A}_3$, and $\hat{\mathfrak{G}}$ is the corresponding affine Weyl group.

\subsubsection{Spectral curve}

The linear equation for $\phi_2(x)$ is:
\beq
\forall x \in ]1/\gamma,\gamma[,\qquad \phi_2(x + {\rm i}0) + \phi_2(x - {\rm i}0) - \phi_2({\rm i}x) - \phi_2(-{\rm i}x) = 0
\eeq
According to \S~\ref{aff3}, we can analytically continue $\phi_2(x) - \phi_2(-{\rm i}x)$ to a meromorphic function $y$ on a $4$-sheeted Riemann surface. Repeating the arguments of \S~\ref{333sP}, we find the polynomial equation meeting all the singularity and symmetry requirements:
\beq
(x^4 + x^{-4} - \gamma^4 - \gamma^{-4})^3\,y^4 + P_2(x^4 + x^{-4})\,y^2 + (x^4 - x^{-4})\,P_1(x^4 + x^{-4})\,y + P_0(x^4 + x^{-4}) = 0
\eeq
where:
\beq
\mathrm{deg}\,P_2 = 1,\qquad \mathrm{deg}\,P_1 = 1,\qquad \mathrm{deg}\,P_0 = 2
\eeq

\subsubsection{$2$-point function}

We find the residue vectors:
\beq
\hat{C}^{(0)} = \frac{1}{16}\big(5\hat{e}_0 - \hat{e}_{1} - 3\hat{e}_2 - \hat{e}_{3}\big),\quad \hat{C}^{(1)} = 0,\quad \hat{C}^{(2)} = \frac{2}{3}\big(\hat{e}_0 + \hat{e}_{1} + \hat{e}_2 + \hat{e}_3\big)
\eeq
The matrix of singularities for the orbit of $\hat{v}^*$ can be decomposed in a plain matrix and a sum of permutation matrices:
\bea
\hat{\mathbf{B}}^{(0)} & = & -\frac{1}{4}\Big(\sum_{\ell \in \mathbb{Z}_{4}} \hat{e}_{\ell}\Big)\mathbf{J} + \mathbf{1}_{4}\hat{e}_0  \\
& & + \left(\begin{array}{cccc} 0 & 0 & 1 & 0 \\ 1 & 0 & 0 & 0 \\ 0 & 0 & 0 & 1 \\ 0 & 1 & 0 & 0 \end{array}\right)\hat{e}_1 + 
\left(\begin{array}{cccc} 0 & 0 & 0 & 1 \\ 0 & 0 & 1 & 0 \\ 0 & 1 & 0 & 0 \\ 1 & 0 & 0 & 0 \end{array}\right)\hat{e}_2 + \left(\begin{array}{cccc} 0 & 1 & 0 & 0 \\ 0 & 0 & 0 & 1 \\ 1 & 0 & 0 & 0 \\ 0 & 0 & 1 & 0 \end{array}\right)\hat{e}_3 \nonumber
\eea
and again higher $\mathbf{B}$'s vanish.

\subsection{$\mathbf{P} = (2,3,6)$}
\label{aff4}

\beq
\breve{R}(s,s') = \frac{s + s'}{s^2 - ss' + s'^2}.
\eeq
The dynamics are generated by:
\beq
\hat{\alpha} = 2\hat{e}_0 - \hat{e}_1 - \hat{e}_5.
\eeq
$E$ has dimension $5$, and is generated by the orthogonal projections of $(\hat{e}_i)_{0 \leq i \leq 5}$ on $E$:
\beq
e_0 = \frac{1}{6}\big(5\hat{e}_0 - \hat{e}_1 - \hat{e}_2 - \hat{e}_3 - \hat{e}_4 - \hat{e}_5),\qquad e_i = S^i(e_0).
\eeq
So, $v \in E$ iff $\sum_{i = 0}^5 v(i) = 0$. $a_i = e_{i - 1}$ for $1 \leq i \leq 5$ defines a basis of $E$, and the Cartan matrix of the root system $\mathfrak{R}$ reads in this basis:
\beq
\mathbf{A} = \left(\begin{array}{ccccc} 2 & -1 & 0 & 0 & 0 \\ -1 & 2 & -1 & 0 & 0 \\ 0 & -1 & 2 & -1 & 0 \\ 0 & 0 & -1 & 2 & -1 \\ 0 & 0 & 0 & -1 & 2 \end{array}\right).
\eeq
We recognize the Cartan matrix of the $A_5$ root system. Deleting $a_1$ (or $a_5$) gives a maximal sub-root system, of type $A_4$. A generator of $H_{\{(a_2,a_3,a_4,a_5)\}}$ is:
\beq
v^* = \frac{1}{36}\sum_{i = 1}^5 (6 - i) a_i 
\eeq
Its image is:
\beq
\hat{v}^* = A(v^*) = \hat{e}_0 - \hat{e}_{5}
\eeq
The orbit of $\hat{v}^*$ is a cycle of order $d = 6$, containing the vectors $(-\hat{e}_i + \hat{e}_{i + 1})$ for $i \in \mathbb{Z}_{6}$.

We may also consider the "root system" $\mathfrak{R}$. Its "Cartan matrix" in the canonical basis of $\hat{E}$ is:
\beq
\hat{\mathbf{A}} = \left(\begin{array}{cccccc} 2 & - 1 & 0 & 0 & 0 & -1 \\ -1 & 2 & -1 & 0 & 0 & 0 \\ 0 & -1 & 2 & - 1 & 0 & 0 \\ 0 & 0 & -1 & 2 & -1 & 0 \\ 0 & 0 & 0 & -1 & 2 & -1 \\ -1 & 0 & 0 & 0 & -1 & 2 \end{array}\right),
\eeq
hence $\hat{\mathfrak{R}}$ is the affine root system of type $\widehat{A}_5$, and $\mathfrak{G}$ is the corresponding affine Weyl group.

\subsubsection{Spectral curve}

The linear equation for $\phi_2(x)$ is:
\beq
\forall x \in ]1/\gamma,\gamma[,\qquad \phi_2(x + {\rm i}0) + \phi_2(x - {\rm i}0) - \phi_2(\zeta_{6}x) - \phi_2(\zeta_{6}^{-1}x) = 0
\eeq
According to \S~\ref{aff4}, we can analytically continue $\phi_2(x) - \phi_2(\zeta_{6}^{-1}x)$ to a meromorphic function $y$ on a $6$-sheeted Riemann surface. Repeating the arguments of \S~\ref{333sP}, we find the polynomial equation meeting all the singularity and symmetry requirements:
\beq
(x^6 + x^{-6} - \gamma^6 - \gamma^{-6})^3\,y^6 + (x^{6} - x^{-6})\Bigg(\sum_{j = 0}^1 P_{2j + 1}(x^6 + x^{-6})\Bigg) + \sum_{j = 0}^{2} P_{2j}(x^6 + x^{-6}) = 0
\eeq
where:
\beq
\mathrm{deg}\,P_{2j + 1} = 2,\qquad \mathrm{deg}\,P_{2j} = 3
\eeq

\subsubsection{$2$-point function}

We find residue vectors:
\bea
\hat{C}^{(0)} & = & \frac{1}{72}(35\hat{e}_0 + 5\hat{e}_1 - 13\hat{e}_2 - 19\hat{e}_3 - 13\hat{e}_4 + 5\hat{e}_5\big),\quad \hat{C}^{(1)} = 0 \\
\hat{C}^{(2)} & = & \frac{3}{5}\Big(\sum_{\ell \in \mathbb{Z}_{6}} \hat{e}_{\ell}\Big)
\eea
We label as follows the vectors of the orbit of $-\hat{v}^* = -\hat{e}_0 + \hat{e}_{5}$:
\bea
& & \hat{v}_1 = \hat{e}_0 - \hat{e}_1,\quad \hat{v}_2 = -\hat{e}_0 + \hat{e}_5,\quad \hat{v}_3 = \hat{e}_1 - \hat{e}_2 \nonumber \\
&&  \hat{v}_4 = \hat{e}_4 - \hat{e}_5,\quad \hat{v}_5 = \hat{e}_2 - \hat{e}_3,\quad \hat{v}_6 = \hat{e}_3 - \hat{e}_4 \\
\eea
and we find a matrix of singularity:
\beq
\hat{\mathbf{B}}^{(0)} = -\frac{1}{6}\Big(\sum_{\ell \in \mathbb{Z}_{6}} \hat{e}_{\ell}\Big)\mathbf{J} + \sum_{i \in \mathbb{Z}_{6}} \hat{\mathbf{B}}^{(0)}(i)\hat{e}_i
\eeq
with:
\bea
& & \hat{\mathbf{B}}^{(0)}(0) = \mathbf{1}_{6},\quad \hat{\mathbf{B}}^{(0)}(1) = \left(\begin{array}{cccccc} 0 & 1 & 0 & 0 & 0 & 0 \\ 0 & 0 & 0 & 1 & 0 & 0 \\1 & 0 & 0 & 0 & 0 & 0 \\ 0 & 0 & 0 & 0 & 0 & 1 \\ 0 & 0 & 1 & 0 & 0 & 0 \\ 0 & 0 & 0 & 0 & 1 & 0 \end{array}\right),\quad \mathbf{B}^{(0)}(2) = \left(\begin{array}{cccccc} 0 & 0 & 0 & 1 & 0 & 0 \\ 0 & 0 & 0 & 0 & 0 & 1 \\ 0 & 1 & 0 & 0 & 0 & 0 \\ 0 & 0 & 1 & 0 & 0 & 0 \\ 1 & 0 & 0 & 0 & 0 & 0 \\ 0 & 0 & 1 & 0 & 0 & 0 \end{array}\right) \nonumber \\
& & \hat{\mathbf{B}}^{(0)}(3) = \overline{\mathbf{1}}_{6},\quad \hat{\mathbf{B}}^{(0)}(4) = \left(\begin{array}{cccccc} 0 & 0 & 0 & 0 & 1 & 0 \\ 0 & 0 & 1 & 0 & 0 & 0 \\ 0 & 0 & 0 & 0 & 0 & 1 \\ 1 & 0 & 0 & 0 & 0 & 0 \\ 0 & 0 & 0 & 1 & 0 & 0 \\ 0 & 1 & 0 & 0 & 0 & 0 \end{array}\right),\quad \hat{\mathbf{B}}^{(0)}(5) = \left(\begin{array}{cccccc} 0 & 0 & 1 & 0 & 0 & 0 \\ 1 & 0 & 0 & 0 & 0 & 0 \\ 0 & 0 & 0 & 0 & 1 & 0 \\ 0 & 1 & 0 & 0 & 0 & 0 \\ 0 & 0 & 0 & 0 & 0 & 1 \\ 0 & 0 & 0 & 1 & 0 & 0 \end{array}\right) \nonumber
\eea
where $\overline{\mathbf{1}}_{6}$ is the matrix with $1$'s on the antidiagonal. The higher $\mathbf{B}$'s vanish.

\section{Topological recursion}
\label{S7}

\subsection{General result}

It was shown in \cite{BEO} that the $1/N$ asymptotic expansion of multidimensional integrals of the form \eqref{themodel} is governed by the topological recursion introduced in \cite{EOFg}. 

We denote $\mathcal{S}$ the spectral curve : it is the Riemann surface on which $\hat{v}\cdot W(x)$ (for a certain set of $\hat{v}$) has been analytically continued on $\mathcal{C}$. We denote $\varphi$ the analytic continuation. $\mathcal{S}$ is equipped with two points $z_{\pm} \in \Sigma$, and a covering $x\,:\,\Sigma \rightarrow \widehat{\mathbb{C}}$, such that $x(z_{\pm}) = \gamma^{\pm}$ (the endpoints of the support of the equilibrium measure) are simple branchpoints. We introduce the $1$-form on $\Sigma$ defined by:
\beq
\omega_{1}^{(0)}(z) = \varphi\,\dd x
\eeq
It has simple zeroes at $z_{\pm}$.

\begin{definition}
Let $\iota$ be the deck transformation, defined in the neighborhood of $z_{\pm}$: it is the holomorphic map such that $x(\iota(z)) = x(z)$ and $\iota(z) \neq z$ for $z \neq z_{\pm}$.
\end{definition}

The spectral curve is also equipped with a bidifferential $\omega_2^{(0)}$ having (among possible other singularities) a double pole at coinciding points with coefficient $1$ and no residues. $(\mathcal{S},\omega_{1}^{(0)},\omega_{2}^{(0)})$ is the initial data needed to compute the large $N$ expansion of all correlators.

\begin{definition} We define the recursion kernel:
\beq
K(z_0,z) = \frac{1}{2}\,\frac{\int_{\iota(z)}^{z} \omega_{2}^{(0)}(\cdot,z_0)}{\omega_{1}^{(0)}(z) - \omega_{1}^{(0)}(\iota(z))}
\eeq
\end{definition}
The recursion kernel has a simple pole at $z = z_{\pm}$.

\begin{theorem}
\label{thm:toporec}For $2g - 2 + n > 0$, the differential forms $\omega_{n}^{(g)}$ are computed recursively on $2g - 2 + n$ by the formula:
\beq
\omega_{n}^{(g)}(z_0,z_I) = \Res_{z \rightarrow z_{\pm}} K(z_0,z)\Big(\omega_{n + 1}^{(g - 1)}(z,\iota(z),z_I) + \sum_{\substack{J \subseteq I \\ 0 \leq h \leq g}}^{*} \omega_{|J| + 1}^{(h)}(z,z_{J})\,\omega_{n - |J|}^{(g - h)}(\iota(z),z_{I\setminus J})\Big)
\eeq
\end{theorem}
The proof of this result follows directly from \cite[Section 1 and 3]{BEO}.

Once $\omega_1^{(0)}$ and $\omega_2^{(0)}$ are known, this formula provides an algorithm to compute the Chern-Simons free energies, and the perturbative expansion of the colored HOMFLY of fiber knots in Seifert spaces. For lens spaces, the equation of the spectral curve was a polynomial in $e^{u}$, and that implied that the perturbative invariants were polynomials in $e^{u}$. In particular, they did not have any singularity for $u$ in the complex plane.

The situation is different for the other Seifert spaces $\chi > 0$, since the interior coefficients of the equation of the spectral curves are in general algebraic function of $e^{u}$. We have seen for instance in the $\mathbb{S}_3/D_{p}$ case that the spectral curve was polynomial in $\kappa(c)$ given by \eqref{eqau2}. It would be interesting to give an interpretation of this "algebraic change of variable" as a non-trivial mirror map in the context of topological strings ($u$ should be in correspondence with some K\"ahler parameter).

We will now derive a few formulas and consequences of our results for $\mathbb{S}_3/D_{p}$. In particular, in \S~\ref{se2}-\ref{se3} we obtain hypergeometric formulas for the planar limit of the HOMFLY-PT of fiber knots generalizing those given in \cite{BEMknots} for torus knots. We also establish in \S~\ref{se2e} arithmetic properties of the all-genus perturbative invariants when $p$ is even. 

\subsection{$\mathbb{S}_3/D_{p}$, $p$ even}
\label{se2}
\subsubsection{Disk invariants}
\label{disk1}
The spectral curve is given in \eqref{spcurveDeven} The expansion when $z \rightarrow \infty$ (i.e. $x \rightarrow \infty$ in the sheet $\sum_{j} (-1)^{j + 1}\hat{e}_j$) is of the form:
\beq
\ln(-y(x)) = - \frac{u}{a_3} \sum_{m \geq 0} \langle\mathrm{Tr}\,\,\mathcal{U}^{a_3(m + 1/2)}\rangle^{(0)}\,x^{-a_3(m + 1/2)}
\eeq
We compute by Lagrange inversion:
\beq
\langle\mathrm{Tr}\,\,\mathcal{U}^{p(m + 1/2)}\rangle^{(0)} = -\frac{p}{u} \Res_{x^{-p/2} \rightarrow 0} \dd(x^{-p/2})\,x^{(m + 1)p}\ln\big(-y(x)\big) = \Res_{\zeta \rightarrow 0} \frac{\dd y(\zeta^{-2/p})}{y(\zeta^{-2/p})}\,\frac{x^{p(m + 1/2)}(\zeta^{-2/p})}{2m + 1}
\eeq
where we have defined a new coordinate $\zeta = z^{-p/2}$, which is such that $x^{p/2} \in O(\zeta^{-1})$ in the regime we are interested.
\bea
\langle\mathrm{Tr}\,\,\mathcal{U}^{p(m + 1/2)}\rangle^{(0)} = \frac{2p\kappa^{p/2}}{u(2m + 1)} \Res_{\zeta \rightarrow 0} \frac{\dd \zeta}{\zeta^{2m + 1}}\Big[\frac{(1 - \kappa^{p}\zeta^2)^{\frac{p}{2}(2m + 1)}}{(\kappa^{p} - \zeta^2)^{\frac{p}{2}(2m + 1) + 1}} - \frac{(1 - \kappa^{p}\zeta^2)^{\frac{p}{2}(2m + 1) - 1}}{(\kappa^{p} - \zeta^2)^{\frac{p}{2}(2m + 1)}}\Big] \nonumber \\
\eea
Each term can be explicitly computed and gives a Laurent polynomial in $\kappa$ that can be put in hypergeometric form:
\bea
& & \langle\mathrm{Tr}\,\,\mathcal{U}^{p(m + 1/2)}\rangle^{(0)} \nonumber \\
& = & \frac{p}{u}\,\frac{2\kappa^{p(m - p(m + 1/2))}}{2m + 1}\,(-1)^m  \\
& & \times \Bigg[{\frac{p}{2}(2m + 1) - 1 \choose m} \kappa^{p/2}\,{}_2F_{1}\big(-m,p(m + 1/2)\,;\,p(m + 1/2) - m\,;\,\kappa^{-2p}\big) \nonumber \\
& & - {\frac{p}{2}(2m + 1) \choose m}\,\kappa^{-p/2}\,{}_2F_{1}\big(-m,p(m + 1/2) + 1\,;\,p(m + 1/2) + 1 - m\,;\, \kappa^{-2p}\big)\Bigg] \nonumber 
\eea
For instance:
\beq
\langle\mathrm{Tr}\,\,\mathcal{U}^{p/2}\rangle^{(0)} = \mathrm{Planar}\,\,\mathrm{HOMFLY-PT}(\mathcal{K}_{2}^{(2,2,p)}) = \frac{2p}{u}\,\kappa^{-p^2/2}(\kappa^{p/2} - \kappa^{-p/2}) 
\eeq
for the knot $\mathcal{K}_{2}$ going along an exceptional fiber of order $2$. Similar formulas involving sum of two polynomial hypergeometric can be found for $\big\langle \mathrm{Tr}\,\,\mathcal{U}^{kp/2} \big\rangle$ for any $k \in \mathbb{N}$. Those formulae can be seen as generalizations of the large $N$ limit of Rosso-Jones formula for torus knots \cite{RossoJones}.

\subsubsection{All-genus invariants}
\label{se2e}
\begin{theorem}
\label{t7}For any $k_1,\ldots,k_n$, and genus $g$, we have:
\beq
\langle \mathrm{Tr}\,\,\mathcal{U}^{k_1}\cdots\mathrm{Tr}\,\,\mathcal{U}^{k_n}\rangle^{(g)}_{{\rm conn}} \in \mathbb{Q}(\kappa^2)[\sqrt{\beta(\kappa)}],\qquad \beta(\kappa) = \frac{(\kappa^2 + 1)((p +1)\kappa^2 - (p - 1))}{\kappa^2}
\eeq
Poles may occur at $\kappa^2 \in \{0,\pm 1,\pm (\kappa^*)^2\}$.
\end{theorem}
\begin{theorem}
The endpoints of the singularity locus of $u^2\partial_{u} F^{(g)}$ -- seen as a function of $\kappa$ -- occur at $\kappa^2 = \{0,\pm 1,\pm(\kappa^*)^2\}$. A more precise description arise follows from \eqref{731}.
\end{theorem}

\noindent \textbf{Proof.} Applying Theorem~\ref{thm:toporec} requires the compute the coefficients of expansion around the ramification points of several quantities. We only focus on the arithmetic properties of those coefficients. It is convenient to introduce the coordinate $\tilde{z} = z^{p}$:
\bea
x^{p}(z) & = & \tilde{z}\Big(\frac{\tilde{z} - \kappa^2}{\tilde{z}\kappa^2 - 1}\Big)^p \equiv \tilde{x}(\tilde{z}) \\
y(z) & = & -\frac{(\tilde{z}^{1/2} - \kappa)(\tilde{z}^{1/2}\kappa + 1)}{(\tilde{z}^{1/2} + \kappa)(\tilde{z}^{1/2}\kappa - 1)} \equiv \tilde{y}(\tilde{z})
\eea
Then, the two ramification points $\tilde{z}_{\pm}$ so that $x(\tilde{z}_{\pm}) \in \mathbb{R}_+$ are:
\beq
\tilde{z}_{\pm} = \frac{(p + 1)\kappa^{-2} - (p - 1)\kappa^{2} + \sqrt{(\kappa^2 - \kappa^{-2})\big((p + 1)\kappa^2 - (p - 1)\kappa^{-2}\big)}}{2}
\eeq
Since $\tilde{z}_{+}\tilde{z}_{-} = 1$, we may write:
\beq
\tilde{z}_{\pm} = e^{\pm \zeta}
\eeq

\noindent \textbf{Deck transformations}. The equation:
\beq
\tilde{x}(\tilde{z}_{\pm} + t) = \tilde{x}(\tilde{z}_{\pm} - t + \tau_{\pm}(t))
\eeq
defines a unique formal power series $\tau_{\pm}(t) = \sum_{k \geq 2} \tau_{\pm,k}\,t^{k}$. The coefficients $\tau_{\pm,k}$ are polynomials in $(\tilde{x}^{(m)}(\tilde{z}_{\pm}))_{3 \leq m \leq k}$ and $1/(\tilde{x}''(\tilde{z}_{\pm})$. Since $x^{a}(\tilde{z}) \in \mathbb{Q}(\kappa^2,e^{\pm\zeta})$, we deduce that $\tau_{\pm,k} \in \mathbb{Q}(\kappa^2)[e^{\pm\zeta}]$. The singularities of $\tau_{\pm,k}$ come from zeroes of $\tilde{x}''(\tilde{z}_{\pm})$. The latter only occur at $\kappa = \pm 1$ and $\kappa = \pm {\rm i}\kappa^*$ with:
\beq
\kappa^* = \sqrt{\frac{p +1}{p - 1}} 
\eeq

\vspace{0.2cm}

\noindent \textbf{Spectral density}. In the recursion kernel, we meet the quantity:
\beq
\Upsilon(t) = \frac{1}{\ln\Big(\frac{\tilde{y}(\tilde{z}_{\pm} + t)}{\tilde{y}(\tilde{z}_{\pm} - t + \tau_{\pm}(t))}\Big)} = \sum_{k \geq -1} \Upsilon_{\pm,k}\,t^{k}
\eeq
to expand in formal powers series when $t \rightarrow 1$. A priori, we have $\Upsilon_{\pm,k} \in \mathbb{Q}(\kappa,e^{\pm \zeta/2})$, but we also observe that:
\beq
\tilde{y}_{\kappa}(\tilde{z})\tilde{y}_{-\kappa}(\tilde{z}) = 1
\eeq
Thus, $\Upsilon_{\pm}(t)$ is odd in $\kappa$, i.e. $\Upsilon_{\pm,k} \in \kappa\,\mathbb{Q}(\kappa^2,e^{\pm\zeta/2})$. In particular, we have:
\beq
\Upsilon_{\pm,-1} = \frac{1}{2 (\ln \tilde{y})'(\tilde{z}_{\pm})} = \frac{\tilde{z}_{\pm}^{1/2}}{\kappa}\,\frac{(\tilde{z}_{\pm} - \kappa^2)(\tilde{z}_{\pm}\kappa^2 - 1)}{(\tilde{z}_{\pm} + 1)(\kappa^2 - 1)}
\eeq
and the other $\Upsilon_{\pm,k}$ have singularities at singularities of $\Upsilon_{\pm,-1}$. These singularities occur at $\kappa = \pm 1$, or $\tilde{z}_{+} = -1$ or $\tilde{z}_{-}$ which amounts to $\kappa \in \{\pm {\rm i},\pm {\rm i}\kappa^*,\pm \kappa^*\}$.

\vspace{0.2cm}

\noindent \textbf{Two-point function}. The two point function was given in \eqref{624}:
\beq
\omega_{2}^{(0)}(\zeta_1,\zeta_2) = \dd\tilde{z}_{1}^{1/2}\,\dd\tilde{z}_{2}^{1/2}\Bigg(\frac{1}{\big(\tilde{z}_1^{1/2}- \tilde{z}_2^{1/2}\big)^2} + \frac{1}{\big(\tilde{z}_1^{1/2} + \tilde{z}_2^{1/2}\big)^2}\Bigg)
\eeq
We need its expansion:
\begin{itemize}
\item[$\bullet$] when $\tilde{z}_1$ is a free variable, and $\tilde{z}_2 = \tilde{z}_{\pm} + t$. The coefficients of expansion are in $\mathbb{Q}(\kappa^2,e^{\pm\zeta/2},\tilde{z}_1^{1/2})$. The singularities in $\tilde{z}_1$ involve denominators which are powers of $(\tilde{z}^{1/2} - \tilde{z}_{\pm}^{1/2})$ and $(\tilde{z}^{1/2} + \tilde{z}^{1/2}_{\pm})$.
\item[$\bullet$] when $\tilde{z}_1 = \tilde{z}_{\pm} + t$ and $\tilde{z}_2 = \tilde{z}_{\pm} - t + \tau_{\pm}(t)$. The coefficients of expansion are in $\mathbb{Q}(\kappa^2,e^{\pm \zeta/2})$.
\end{itemize}
No new singularity occur in the $\kappa$-plane.

\vspace{0.2cm}

\noindent \textbf{Topological recursion.} With the formula of Thm.~\ref{thm:toporec}, the above series allows, after a sequence of $2g - 2 + n > 0$ nested residues at $\tilde{z} \rightarrow \tilde{z}_{+}$ and $\tilde{z}_{-}$, the computation of:
\beq
\frac{\omega_{n}^{(g)}(\tilde{z}_1,\ldots,\tilde{z}_n)}{\dd\tilde{z}_1\cdots\dd\tilde{z}_n} \in \mathbb{Q}(\kappa^2,e^{\pm \zeta},\tilde{z}_1,\ldots,\tilde{z}_n)
\eeq
Since we sum the residues at the two ramification points $\tilde{z}_{\pm} = e^{\pm \zeta}$ at each step, the result must be even of $\zeta$. Hence, it will be a rational function in the variables $(\mathrm{ch}(\zeta k/2))_{k \geq 0}$, which can be themselves rewritten as polynomials in:
\beq
2\mathrm{ch}(\zeta/2) = \sqrt{\tilde{z}_+ + \tilde{z}_- + 2} = \sqrt{\beta(\kappa)}\,\qquad \beta(\kappa) = \frac{(\kappa^2 + 1)((p + 1) - (p - 1)\kappa^2)}{\kappa^2}
\eeq
Subsequently:
\beq
\frac{\omega_{n}^{(g)}(\tilde{z}_1,\ldots,\tilde{z}_n)}{\dd\tilde{z}_1\cdots\dd\tilde{z}_n} \in \mathbb{Q}(\kappa^2,\sqrt{\beta(\kappa)},\tilde{z}_1^{1/2},\ldots,\tilde{z}_n^{1/2})
\eeq

\noindent \textbf{Lagrange inversion}. All perturbative invariants can be extracted from the expansion of $\omega_{n}^{(g)}$ when $\tilde{z} \rightarrow 1/\kappa^{1/2}$ as a power series involving $x$ (see \S~\ref{Wilson}), as done in \S~\ref{disk1} for the disk invariants. The $\omega_{n}^{(g)}(\tilde{z}_1,\ldots,\tilde{z}_n)$ can be decomposed with coefficients in $\mathbb{Q}(\sqrt{\beta(\kappa)},\kappa^2)$ in the basis of:
\beq
\dd\varsigma_{m,\varepsilon}(\tilde{z}_i) = \frac{\dd\tilde{z}_i^{1/2}}{(\tilde{z}_i^{1/2} + \varepsilon\tilde{z}_{\pm})^m(\tilde{z}_i^{1/2} + \varepsilon \tilde{z}_{\pm}^{1/2})^m} = \frac{\dd\tilde{z}_i^{1/2}}{(\tilde{z}_i + 1 - 2\tilde{z}_i^{1/2}\,\mathrm{ch}(\zeta/2))^{m}}
\eeq
for $m \in \mathbb{Z}_{+}$ and $\varepsilon \in \{-1,1\}$. The latter can be expanded:
\beq
\frac{\dd\varsigma_{m,\varepsilon}(\tilde{z}_i)}{\dd \tilde{z}_i} = \sum_{k \geq 0} \varsigma_{m,\varepsilon,k}\,\big(x(\tilde{z}_i^{1/p})\big)^{k}
\eeq
with:
\bea
\varsigma_{m,\varepsilon,k} & = & \Res_{z \rightarrow \kappa^{2/p}} \frac{\dd x(z)}{(x(z))^{k + 1}}\,\dd\varsigma_{m,\varepsilon}(z^{1/p}) \nonumber \\
\label{outeee}& = & \Res_{\widehat{z} \rightarrow \kappa} \frac{\dd\widehat{z}}{\widehat{z}^{2k}}\Big(\frac{\widehat{z}^2\kappa^2 - 1}{\widehat{z}^2 - \kappa^2}\Big)^{k}\Big(\frac{2}{\widehat{z}} + \frac{2p\widehat{z}}{\widehat{z}^2 - \kappa^2} - \frac{2p\widehat{z}\kappa^2}{\widehat{z}^2\kappa^2 - 1}\Big)\,\frac{1}{(\hat{z}^2 + 1 - 2\hat{z}\sqrt{\beta(\kappa)})^{m}} \nonumber \\
& \in & \mathbb{Q}(\kappa^2)[\sqrt{\beta(\kappa)}].
\eea
A singularity in $\varsigma_{m,\varepsilon,k}$ occurs when $\kappa$ is such that one of the term in \eqref{outeee}  which is not explicitly a $(\widehat{z}^2 - \kappa^2)$ has a pole at $\widehat{z} = \kappa$, that is when $\kappa = 1/\kappa$, or $\kappa = 0$, or $\kappa^2 + 1 = 2\kappa\sqrt{\beta(\kappa)}$. Therefore, the singularities of $\varsigma_{m,\varepsilon,k}$ in the $\kappa$-plane occur when $\kappa \in \{\pm 1,\pm {\rm i}\}$.

Combining all results, we find that $\big\langle \mathrm{Tr}\,\,\mathcal{U}^{k_1}\cdots\mathrm{Tr}\,\,\mathcal{U}^{k_n}\big\rangle$ belongs to $\mathbb{Q}(\kappa^2)[\sqrt{\beta(\kappa)}]$. Moreover, the denominators can only contain powers of $(\kappa^4 - 1)$ and $(\kappa^4 - \kappa_*^4)$. We remind the relation \eqref{kappac} between $\kappa$ and the Chern-Simons variable $u = N\hbar$:
\beq
\frac{2\kappa^{1 + 1/p}}{1 + \kappa^2} = e^{-u/4p^2}
\eeq
Therefore, the list of singularities of the perturbative colored HOMFLY in the $u$-complex plane occur when $(u\,\,\mathrm{mod}\,\,8{\rm i}\pi p^2\mathbb{Z})$ is equal to:
\bea
0, & & -4{\rm i}\pi p(p + 1) \nonumber \\
 -2p^2S(p), & & -2p^2S(p) - 4{\rm i}\pi p(p + 1) \nonumber \\
 -2p^2(S(p) - 2\ln p) - 2{\rm i}\pi p(p + 1), & & -2p^2(S(p) - 2\ln p) - 6{\rm i}\pi p(p + 1) \nonumber
 \eea
 where:
 \beq
S(p) = (p + 1)\ln(p + 1) + (p - 1)\ln(p - 1)
\eeq
This list is complete, except that $u = 0$ must be removed since by construction, we know that  invariants are $O(1)$ when $u \rightarrow 0$.

\vspace{0.2cm}

\noindent \textbf{Free energies.} The $F^{(g)}$ can be obtained from $W_{1}^{(g)}$ by \eqref{basicerq}. For $g \geq 1$, the topological recursion computes $W_1^{(g)}$ in terms of the two-point function. We therefore need to study the arithmetic properties of the quantities:
\beq
J_{m,\pm} = \oint \frac{\dd x}{2{\rm i}\pi x}\frac{p^2\ln^2 x}{x}\,\Bigg(\frac{1}{\big(\widehat{z} - \tilde{z}_{\pm}^{1/2}\big)^{m + 2}} + \frac{1}{\big(\widehat{z} + \tilde{z}_{\pm}^{1/2}\big)^{m + 2}}\Bigg)
\eeq
where we integrate $x = x(\tilde{z})$ around the cut and $\widehat{z} = \tilde{z}^{1/2}$. We have:
\bea
\label{clears} J_{m,\pm} & = & - \frac{m + 2}{3} \oint \frac{\dd\widehat{z}}{2{\rm i}\pi}\,\ln^3 \tilde{x}(\widehat{z}^2)\,\Bigg(\frac{1}{\big(\widehat{z} - \tilde{z}_{\pm}^{1/2}\big)^{m + 3}} + \frac{1}{\big(\widehat{z} + \tilde{z}_{\pm}^{1/2}\big)^{m + 3}}\Bigg) \nonumber \\
& = & -\frac{m + 2}{3}\Bigg(\frac{\dd^{m + 2}}{\dd \widehat{z}^{m + 2}}\Big|_{\widehat{z} = \tilde{z}_{\pm}^{1/2}} + \frac{\dd^{m +2}}{\dd \widehat{z}^{m + 2}}\Big|_{\widehat{z} = - \tilde{z}_{\pm}^{1/2}}\Bigg)\big[\ln^3 \tilde{x}(\widehat{z}^2)\big]
\eea
The derivatives can be explicitly computed, and we find:
\beq
J_{m,\pm} \in \mathbb{Q}(\kappa^2,\tilde{z}_{\pm}^{1/2})\cdot\ln^2\tilde{x}(\tilde{z}_{\pm}) + \mathbb{Q}(\kappa^2,\tilde{z}_{\pm}^{1/2})\cdot\ln\tilde{x}(\tilde{z}_{\pm}) + \mathbb{Q}(\kappa^2,\tilde{z}_{\pm}^{1/2})
\eeq
Besides, the three rational functions of $\tilde{z}_{\pm}^{1/2}$ appearing in this expression are odd if $m$ is odd (resp. even if $m$ is even). It is clear from \eqref{clears} that the singularities of $J_{m,\pm}$ as a function of $\kappa$ only occur when the singularities of $\ln \tilde{x}(\tilde{z})$ as a function of $\tilde{z}$ approach the cut $\tilde{z} \in [\tilde{z}_{-},\tilde{z}_{+}]$. The quantities $u^2\partial_{u}F^{(g)}$ are linear combinations of $J_{m,\pm}$ with coefficients in $\mathbb{Q}[\kappa^2,e^{\pm\zeta}]$, and the result should be even in $\zeta$ -- since $\tilde{z}_+ = e^{\zeta}$ and $\tilde{z}_{-} = e^{-\zeta}$ play a symmetric role. And, we remind that $\tilde{x}(\tilde{z}_{+})\tilde{x}(\tilde{z}_-) = 1$. Therefore:
\beq
\label{731}u^2\partial_{u}F^{(g)} \in \mathbb{Q}(\kappa^2)[\sqrt{\beta(\kappa)}]\cdot\ln^2\tilde{x}(\tilde{z}_{+}) + \mathbb{Q}(\kappa^2)[\sqrt{\beta(\kappa)}]\sqrt{\beta_2(\kappa)}\cdot\ln\tilde{x}(\tilde{z}_{+}) + \mathbb{Q}(\kappa^2)[\sqrt{\beta(\kappa)}]
\eeq
where:
\beq
\sqrt{\beta_2(\kappa)} = \mathrm{sinh}\,\zeta = \frac{\sqrt{(\kappa^4 - 1)((p + 1)\kappa^4 - (p - 1))}}{\kappa^2}
\eeq
The location of singularities of the free energies can easily be deduced. \hfill $\Box$

\subsection{$\mathbb{S}_3/D_{p}$, $p$ odd}
\label{se3}
\subsubsection{Disk invariants}

The spectral curve is given in \eqref{spcurvDodd}. Checking the behavior \eqref{Puis1}-\eqref{Puis2} shows that $z \rightarrow \kappa^{\pm 1/a_3}$ corresponds to $x^{\pm 1} \rightarrow \infty$ in the sheet labeled by $-(\hat{e}_0 + \hat{e}_{a_3})$. Therefore, the expansion at those points of $\ln \varphi$ in terms of the variable $x$, reads:
\bea
\ln[-\varphi\,x^2c^{2}] \mathop{=}_{x \rightarrow \infty} \sum_{m \geq 0} \frac{u}{4p^2}\,\big\langle \mathrm{Tr}\,\mathcal{U}^{2m + 1}\big\rangle^{(0)}\,x^{2m + 2} \nonumber \\
\ln[-\varphi\,x^2c^{-2}] \mathop{=}_{x \rightarrow 0} \sum_{m \geq 0} \frac{u}{4p^2}\,\big\langle\mathrm{Tr}\,\mathcal{U}^{2m} \big\rangle^{(0)}\,x^{2m} \nonumber 
\eea
By Lagrange inversion, we find:
\beq
\big\langle \mathrm{Tr}\,\mathcal{U}^{2m + 1} \big\rangle^{(0)} = \Res_{z \rightarrow \kappa^{-1/p}} \frac{a^2}{u}\,\ln[-\varphi\,x^2c^2]\,x^{-(2m + 1)}\,\dd x = \Res_{z \rightarrow \kappa^{-1/p}} \frac{\dd \varphi(z)}{\varphi(z)}\,\frac{x^{-2m}(z)}{2m} \nonumber
\eeq
After a change of variable $\zeta = z^{p}$, this residue can be explicitly computed:
\bea
\label{termsq} & & \big\langle \mathrm{Tr}\,\mathcal{U}^{2m + 1} \big\rangle^{(0)}_{(2,2,p)} = \frac{2p^2}{um}\,\frac{(1 - \kappa^2)}{\kappa^{2m(1 + 1/p)}} \nonumber \\
& & \times \Bigg[\Big(\frac{m}{2p} - \frac{1}{2}\Big)\kappa^2\,{}_{2}F_{1}\Big[\frac{3}{2} - \frac{m}{2p},1 - m; 2 ; 1 - \kappa^{4}\Big] + \Big(\frac{m}{2p} + \frac{1}{2}\Big){}_{2}F_{1}\Big[\frac{1}{2} - \frac{m}{a},1 - m; 2 ; 1 - \kappa^{4}\Big]\Bigg] \nonumber
\eea
Since $\kappa$ is a function of $u$ independent of the parity of $p$, we observe the relation:
\beq
\big\langle \mathrm{Tr}\,\mathcal{U}^{2m + 1} \big\rangle^{(0)}_{(2,2,p\,\,{\rm odd})} = \frac{1}{2}\,\big\langle \mathrm{Tr}\,\mathcal{U}^{m} \big\rangle_{(2,2,p\,\,{\rm even})}^{(0)}
\eeq
In particular, the planar limit of the HOMFLY-PT of $\mathcal{K}_{2}$ is obtained by taking $2m + 1 = p$ in \eqref{termsq}. The computation of even powers is very similar, and we get:
\bea
& & \big\langle \mathrm{Tr}\,\mathcal{U}^{2m} \big\rangle^{(0)} = \frac{a^2}{u(m/2)}\,(\kappa^{-2} - 1)\kappa^{2m(1 + 1/a_3)} \\
& &  \times \Bigg[\Big(\frac{m}{a} - \frac{1}{2}\Big){}_{2}F_1\Big[\frac{1}{2} + \frac{m}{a},1 - m;2;1 - \kappa^{-4}\Big] + \Big(\frac{m}{a} + \frac{1}{2}\Big)\kappa^{-2}\,{}_{2}F_1\Big[\frac{3}{2} + \frac{m}{a},1 - m;2;1 - \kappa^{-4}\Big]\Bigg] \nonumber 
\eea
In particular, the planar limit of the HOMFLY of the knot following the fiber of order $p$ is obtained by taking $m = 1$:
\beq
\mathrm{Planar}\,\,\mathrm{HOMFLY-PT}(\mathcal{K}_{p}) = \frac{p(1 - \kappa^{2}(u))}{2u\kappa^{2/p}(u)}\big(\kappa^{-2}(u)(2 + p) + 2 - p\big)
\eeq

\section{Generalization to $\mathrm{SO}$ and $\mathrm{Sp}$ Chern-Simons}

\label{S8}

\subsection{The model}

For more general gauge groups, Chern-Simons theory on Seifert spaces is described by the following measure on the Cartan subalgebra -- identified with $\mathbb{R}^N$:
\beq
\dd\mu(t_1,\ldots,t_N) = \frac{1}{Z}\prod_{\alpha > 0}\Big[ 2\,\mathrm{sinh}^{2 - r}\Big(\frac{t\cdot \alpha}{2}\Big) \prod_{m = 1}^r 2\,\mathrm{sinh}\Big(\frac{t\cdot\alpha}{2a_m}\Big)\Big]\,\prod_{i = 1}^N e^{-Nt_i^2/2u}\,\dd t_i
\eeq
where the product ranges over positive roots. For the $A$ series, we had:
\beq
\prod_{\alpha > 0} 2\,\mathrm{sinh}\Big(\frac{\alpha\cdot t}{2}\Big) = \prod_{1 \leq i < j \leq N} 2\,\mathrm{sinh}\Big(\frac{t_i - t_j}{2}\Big)
\eeq
Now, we would like to address the $BCD$ series:
\beq
\label{bcdmod}\prod_{\alpha > 0} 2\,\mathrm{sinh}\Big(\frac{\alpha\cdot t}{2}\Big) = \prod_{1 \leq i < j \leq N} 4\,\mathrm{sinh}\Big(\frac{t_i - t_j}{2}\Big)\mathrm{sinh}\Big(\frac{t_i + t_j}{2}\Big)\,\prod_{i = 1}^N \Pi(t_i)
\eeq
with
\beq \Pi_{B}(t) = \,\, 2\,\mathrm{sinh}(t/2),\qquad \Pi_{C}(t) = \,\, 2\,\mathrm{sinh}(t),\qquad \Pi_{D}(t) = \,\, 1.
\eeq
They correspond to the Lie algebras:
\beq
B_N = \mathfrak{so}(2N + 1),\qquad C_N = \mathfrak{sp}(2N),\qquad D_N = \mathfrak{so}(2N)
\eeq
We will explain the main points of the analysis of those models, and skip the details which are very similar to the already studied A series.

\subsubsection{First rewriting}

Let us perform the change of variable $s_i = e^{t_i/a}$ with $a = \mathrm{lcm}(a_1,\ldots,a_r)$, and put the measure \eqref{bcdmod} in the form:
\beq
\label{seconda}\prod_{1 \leq i < j \leq N} (s_i - s_j)^2 \Big(\sqrt{s_is_j} - \frac{1}{\sqrt{s_is_j}}\Big)^2 \prod_{1 \leq i,j \leq N} \check{R}^{1/2}(s_i,s_j) \prod_{i = 1}^N e^{-NV_{{\rm G}}(s_i;u)}\dd s_i
\eeq 
where the potential is:
\be\ba
\check{V}_{{\rm A}}(x;u) & = \,\, \frac{a^2(\ln x)^2}{2u} + \frac{a\chi}{2}\,\ln x \\
V_{{\rm G}}(x;u) & = 2\check{V}_{{\rm A}}(x;2u) - \frac{1}{2N}\Big((2 - r)U_{{\rm G}}(x^{a}) + \sum_{m = 1}^r U_{{\rm G}}(x^{\check{a}_m})\Big)
\ea\ee
The terms $U_{{\rm G}}$ are absent for the A series, and are equal to:
\beq
\label{UGDEF}U_{B}(x) = -\ln(x^{1/2} + x^{-1/2}),\qquad U_{C}(x) = 0,\qquad U_{D}(x) =\,\, -\ln(x - x^{-1}).
\eeq
And, the pairwise interaction is:
\beq
 \check{R}_{{\rm A}}(x,y) =  \prod_{\ell = 1}^{a - 1} (x - \zeta_{a}^{\ell}y)^{2 - r} \prod_{m = 1}^{r} \prod_{\ell_m = 1}^{\check{a}_m - 1} (x - \zeta_{\check{a}_m}^{\ell_m}y)
\eeq
 for the A series, and:
 \beq
 \check{R}(x,y) = \check{R}_{{\rm A}}(x,y)\check{R}_{{\rm A}}\Big(\sqrt{xy},\frac{1}{\sqrt{xy}}\Big)
\eeq
for the BCD series.

\subsubsection{Second rewriting}

The BCD matrix model can also be rewritten:
\be
\label{checm}\dd\check{\mu}(s_1,\ldots,s_N) = \prod_{1 \leq i \neq j \leq N} \big|R^{1/2}(s_i,s_j)\big|\,\prod_{i = 1}^N s_i^{-1}\,\Pi_{G}(s_i)\,e^{-Na^2(\ln s_i)^2/2u}\,\dd s_i
\ee
with:
\bea
\label{Rstru}R(s_i,s_j) & = & \Big[\Big(\frac{s_i}{s_j}\Big)^{a/2} - \Big(\frac{s_i}{s_j}\Big)^{-a/2}\Big]\Big[(s_is_j)^{a/2} - (s_is_j)^{-a/2}\Big] \nonumber \\
& & \times \prod_{m = 1}^r \Big[\Big(\frac{s_i}{s_j}\Big)^{\check{a}_m/2} - \Big(\frac{s_i}{s_j}\Big)^{-\check{a}_m/2}\Big]\Big[(s_is_j)^{\check{a}_m/2} - (s_is_j)^{-\check{a}_m/2}\Big]
\eea

The potential for this model is obtained by including the part of the diagonal terms $i = j$ which are regular:
\be
\label{Vtott}V_{\mathrm{tot},{\rm G}}(x) = N\check{V}(x) + \delta\check{V}_{{\rm G}}(x) = N\,\frac{a^2(\ln x)^2}{2t} + \sum_{m} \check{U}_{G}(x^{\check{a}_m}),\qquad  \check{U}_{G}(x) = U_0(x) - U_{G}(x)
\ee
with:
\beq
U_{0}(x) = \frac{1}{2}\ln(x - x^{-1})
\eeq
and $U_{G}$ were defined in \eqref{UGDEF}. In other words:
\beq
\label{Udefs}\check{U}_{B}(x) = \frac{1}{2}\ln\Big(\frac{x^{1/2} - x^{-1/2}}{x^{1/2} + x^{-1/2}}\Big),\qquad \check{U}_{C}(x) = \frac{1}{2}\ln(x - x^{-1}),\qquad \check{U}_{D}(x) = -\frac{1}{2}\ln(x - x^{-1})
\eeq

The expressions \eqref{checm}-\eqref{Rstru} are prepared in order to write down Schwinger-Dyson equation (see \S~\ref{sdeqfw}), while the more compact expression \eqref{seconda} is appropriate to study equilibrium measures.

\subsubsection{Observables and symmetries}

We observe in \eqref{bcdmod} that $t_i$ individually is distributed like $-t_i$. Therefore, the joint probability density vanishes at $t_i = \pm t_j$, and to find the dominant configuration of \eqref{bcdmod} when $N \rightarrow \infty$ of \eqref{bcdmod} via potential theory, it is enough to restrict to even configurations. One can then show uniqueness and off-criticality of the equilibrium measure at least when $\chi \geq 0$ (see Appendix~\ref{CvxSO}). We assume $\chi \geq 0$ in all what follows.

The naive definition of the $n$-point correlators is:
\bea
\overline{W}_{{\rm G}|n}(x_1,\ldots,x_n) = \mu\Big[\prod_{i = 1}^n \mathrm{Tr}\,\frac{x_i}{x_i - e^{\mathbf{T}/a}}\Big] \nonumber \\
W_{{\rm G}|n}(x_1,\ldots,x_n) = \mu\Big[\prod_{i = 1}^n \,\mathrm{Tr}\,\frac{x_i}{x_i - e^{\mathbf{T}/a}}\Big]_{{\rm conn}}
\eea
with $\mathbf{T} = \mathrm{diag}(t_1,\ldots,t_n)$. In order to emphasize later the relation between the A series and ${\rm G} \in \{{\rm B},{\rm C},{\rm D}\}$ series, it is convenient to introduce the symmetrized correlators:
\bea
\widetilde{\overline{W}}_{{\rm G}|n}(x_1,\ldots,x_n) = \sum_{\varepsilon_1,\ldots,\varepsilon_n = \pm 1} \Big[\prod_{i = 1}^n \varepsilon_i\Big]\,\overline{W}_{{\rm G}|n}(x_1^{\varepsilon_1},\ldots,x_n^{\varepsilon_n}) \nonumber \\
\widetilde{W}_{{\rm G}|n}(x_1,\ldots,x_n) = \sum_{\varepsilon_1,\ldots,\varepsilon_n = \pm 1} \Big[\prod_{i = 1}^n \varepsilon_i\Big]\,W_{{\rm G}|n}(x_1^{\varepsilon_1},\ldots,x_n^{\varepsilon_n})
\eea
By symmetry, we have:
\beq
\label{symiu1}W_{{\rm G}|1}(x) + W_{{\rm G}|1}(1/x) = N,\qquad W_{{\rm G}|n}(1/x_1,\ldots,1/x_n) = (-1)^n\,W_{{\rm G}|n}(x_1,\ldots,x_n)
\eeq
and thus for the leading order $\lim_{N \rightarrow \infty} N^{-1}\,W_{{\rm G}|1}(x) = W_{{\rm G}}(x)$:
\beq
\label{symiu2}W_{{\rm G}}(x) + W_{{\rm G}}(1/x) = 1,\qquad \check{W}_{{\rm G}}(x) = W_{{\rm G}}(x) - W_{{\rm G}}(1/x) = 2W_{{\rm G}}(x) - 1.
\eeq
Stricto sensu, the existence of an all-order asymptotic expansion of the form:
\bea
Z_{{\rm G}} & = & N^{\gamma N + \gamma'}\exp\Big(\sum_{k \geq 0} N^{2 - k}\,F^{[k]}_{{\rm G}}\Big) \nonumber \\
\label{exou}W_{{\rm G}|n}(x_1,\ldots,x_n) & = & \sum_{k \geq 0} N^{2 - n - k}\,W_{{\rm G}|n}^{[k]}(x_1,\ldots,x_n)  \\
\label{thy}\widetilde{W}_{{{\rm G}|n}}(x_1,\ldots,x_n) & = & \sum_{k \geq 0} N^{2 - n - k}\,\widetilde{W}_{{\rm G}|n}^{[k]}(x_1,\ldots,x_n) \nonumber 
\eea
does not follow from the results of \cite{BGK} since the potential has a log singularity at $t_i = 0$. It is nevertheless likely that the argument can be generalized for this model. In the remaining, we will take the asymptotic expansions \eqref{exou} as starting point\footnote{From the theory of the LMO invariants, it holds at least as equality of formal series in the variables $\hbar$ and $u_0 = N\hbar$ (see \S~\ref{CSavatar}).}, with $W_{{\rm G}|n}^{[k]}(x_1,\ldots,x_n)$ being analytic functions of $n$ variables $x_1,\ldots,x_n$ away from a cut  $[1/\gamma,\gamma]$. Compared to the expansion \eqref{expf} for the A series, since the log term of the potential contains a $1/N$ corrections (the log terms coming from $i = j$), $\check{W}_{{\rm G}|n}$ is not expected to have an expansion of parity $(-1)^n$ in $N$.

Our goal is to compute the coefficients of \eqref{exou}. We will first write linear functional equations $W_{{\rm G}|1}^{[0]} = W_{{\rm G}}$, as well as $W_{{\rm G}|2}^{[0]}$ and $W_{{\rm G}|1}^{[1]}$. If $\chi \geq 0$, the equation for $W_{{\rm G}|1}^{[0]}$ given in \S~\ref{sss1} below can be derived as a saddle point equation for the Stieltjes transform of the equilibrium measure. For $W_{{\rm G}|2}^{[0]}$ and $W_{{\rm G}|1}^{[1]}$, the equations are given in \S~\ref{Soihqt} and \S~\ref{sss2} below can be derived from the Schwinger-Dyson equations presented later in \S~\ref{SDe1}-\ref{SDe3}, after pluging the expansion \eqref{exou} and equating the discontinuities of the two sides of the Schwinger-Dyson equations.
For this purpose, we will show that the symmetrized correlators satisfy a set of loop equations, which are solved by a version of the topological recursion adapted to the grading in \eqref{exou}. This is performed by adapting the tools of \cite{BEO} to take into account the presence of a symmetry.

\subsection{Spectral curves}
\label{sss1}
We first have a look at the equilbrium measure in the model \eqref{bcdmod}. We denote $\Gamma_{{\rm G}}(u)$ the support $[1/\gamma,\gamma]$ of the equilibrium measure, emphasizing its dependence in the parameter $u$. We find, for any $x \in \Gamma_{{\rm BCD}}(u)$. The saddle point equation for $W_{{\rm G}}(x)$ is obtained as \eqref{RH1} was obtained from \eqref{themodel2}. Since it concerns only the leading order when $N \rightarrow \infty$, the only term relevant in the potential is that involving $V_{{\rm A}}$. Notice that here, the joint probability density is singular when $x_i \rightarrow x_j$ and $x_ix_j \rightarrow 1$. The second singularity is also relevant since configurations with $x_ix_j = 1$ can occur with $x_i,x_j \in\Gamma(u)$. We find, for any $x \in \Gamma_{{\rm G}}(u)$:
\bea
W_{\rm G}(x + {\rm i}0) + W_{\rm G}(x - {\rm i}0) - W_{{\rm G}}(x^{-1} + {\rm i}0) - W_{{\rm G}}(x^{-1} - {\rm i}0) & & \\
(2 - r) \sum_{\ell = 1}^{a - 1} \big(W_{{\rm G}}(\zeta_{a}^{\ell} x) - W_{{\rm G}}(\zeta_{a}^{\ell}x^{-1})\big) + \sum_{m = 1}^r \sum_{\ell_m = 1}^{\check{a}_m - 1} \big(W_{{\rm G}}(\zeta_{\check{a}_m}^{\ell_m}x) - W_{{\rm G}}(\zeta_{\check{a}_m}^{\ell_m}x^{-1})\big) & = & 2\partial_{x} V_{{\rm A}}(x;2u) \nonumber
\eea
In terms of the symmetrized correlator, we therefore find, for any $x \in \Gamma_{{\rm G}}(u)$:
\beq
\label{816}\widetilde{W}_{\rm G}(x + {\rm i}0) + \widetilde{W}_{\rm G}(x - {\rm i}0) + (2 - r) \sum_{\ell = 1}^{a - 1} \widetilde{W}_{{\rm G}}(\zeta_{a}^{\ell}x) + \sum_{m = 1}^r \sum_{\ell_m = 1}^{\check{a}_m - 1} \widetilde{W}_{{\rm G}}(\zeta_{\check{a}_m}^{\ell_m}x) = \partial_{x} V_{{\rm A}}(x;2u)
\eeq
For $\chi \geq 0$, this equation characterizes the equilibrium measures together with the condition:
\beq
\lim_{x \rightarrow \infty} x\,\widetilde{W}_{{\rm G}}(x) = 1
\eeq
This is shown by repeating the proof of Corollary~\ref{cor1} using the strict concavity proved in Appendix~\ref{CvxSO}. for the BCD series. Therefore, we deduce, if $\chi \geq 0$:
\beq
\label{P81}\boxed{\widetilde{W}_{{\rm G}}(x;u) = W_{{\rm A}}(x;2u),\qquad \Gamma_{{\rm G}}(u) = \Gamma_{{\rm A}}(2u)}
\eeq
In other words, the spectral curve are the same in the B, C, or D series. They are obtained from the spectral curve of the A series upon replacement of $u$ by $2u$, for which we can use the results of Section~\ref{SSei}-\ref{S6}.

\subsection{Two-point function}
\label{Soihqt}

Following the strategy of \S~\ref{2pot}, we can derive the saddle point equation for the leading order of the $2$-point correlator. We find, for any $x \in \Gamma_{{\rm G}}(u)$ and $x_2 \in \mathbb{C}\setminus\Gamma_{\rm G}(u)$:
\bea
W_{\rm G|2}^{[0]}(x + {\rm i}0,x_2) + W_{\rm G|2}^{[0]}(x - {\rm i}0,x_2) - W_{\rm G|2}^{[0]}(x^{-1} + {\rm i}0),x_2) - W_{\rm G|2}^{[0]}(x^{-1} - {\rm i}0,x_2) & & \nonumber \\
+ (2 - r)\sum_{\ell = 1}^{a - 1} \big(W_{\rm G|2}^{[0]}(\zeta_{a}^{\ell}x,x_2) - W_{\rm G|2}^{[0]}(\zeta_{a}^{\ell}x^{-1},x_2)\big) & & \nonumber \\
+ \sum_{m = 1}^r \sum_{\ell_m = 1}^{\check{a}_m - 1} \big(W_{\rm G|2}^{[0]}(\zeta_{\check{a}_m}^{\ell_m}x,x_2) - W_{\rm G|2}^{[0]}(\zeta_{\check{a}_m}^{\ell_m}x^{-1},x_2)\big) & = & \mathcal{B}_0(x,x_2) \nonumber
\eea
where $\mathcal{B}_0(x,x_2) = -xx_2/(x - x_2)^2$. This implies the saddle point equation for the symmetrized $2$-point function:
\bea
& & \widetilde{W}_{\rm G|2}^{[0]}(x + {\rm i}0,x_2) + \widetilde{W}_{\rm G|2}^{[0]}(x - {\rm i}0,x_2) + (2 - r)\sum_{\ell = 1}^{a - 1} \widetilde{W}_{\rm G|2}^{[0]}(\zeta_{a}^{\ell}x,x_2) + \sum_{m = 1}^r \sum_{\ell_m = 1}^{\check{a}_m - 1} \widetilde{W}_{\rm G|2}^{[0]}(\zeta_{\check{a}_m}^{\ell_m}x,x_2) \nonumber \\
\label{sad20}  & & = \mathcal{B}_{0}(x,x_1) - \mathcal{B}_0(x,x_1^{-1})
\eea
Since $\mathcal{B}_0(x_1,x_2) = \mathcal{B}_0(x_1^{-1},x_2^{-1})$, the right-hand side is a symmetric function of $x$ and $x_1$, which is transformed into its opposite when $(x,x_1) \rightarrow (1/x,x_1)$.

Let us compare \eqref{sad20} with the saddle point equation of the two-point function for the A series: for any $x \in \Gamma_{{\rm A}}(2u)$ and $x_2 \in \mathbb{C}\setminus\Gamma_{{\rm A}}(2u)$,
\bea
& & \widetilde{W}_{\rm A|2}^{[0]}(x + {\rm i}0,x_2) + \widetilde{W}_{\rm A|2}^{[0]}(x - {\rm i}0,x_2) + (2 - r)\sum_{\ell = 1}^{a - 1} \widetilde{W}_{\rm A|2}^{[0]}(\zeta_{a}^{\ell}x,x_2) + \sum_{m = 1}^r \sum_{\ell_m = 1}^{\check{a}_m - 1} \widetilde{W}_{\rm A|2}^{[0]}(\zeta_{\check{a}_m}^{\ell_m}x,x_2) \nonumber \\
\label{sad20b}  & & = \mathcal{B}_{0}(x,x_1)
\eea
Since $\Gamma_{{\rm G}}(u) = \Gamma_{{\rm A}}(2u)$ and this equation characterizes the two-point function with appropriate growth condition at $\infty$ in the cases $\chi \geq 0$, we deduce:
\beq
\label{P82}\boxed{\widetilde{W}_{\rm G|2}^{[0]}(x_1,x_2) = W_{{\rm A}|2}^{[0]}(x_1,x_2;2u) - W_{{\rm A}|2}^{[0]}(x_1^{-1},x_2;2u)}
\eeq

\noindent \textbf{Remark.} The two-point function $W_{\rm A|2}^{[0]}(x_1,x_2;2u)$ plays the role of a fundamental solution in the following sense: the function $x \mapsto \oint \frac{\dd x_2}{2{\rm i}\pi}\,f(x_2)\,W_{{\rm A}|2}^{[0]}(x_1,x_2)$ for any choice of function $f(x_2)$ holomorphic near the contour of integration which avoids the cut, is the solution of the functional equation on the cut $\Gamma_{{\rm G}}(u)$:
\beq
\label{eou}\phi(x + {\rm i}0) + \phi(x - {\rm i}0) + (2 - r) \sum_{\ell = 1}^{a - 1} \phi(\zeta_{a}^{\ell}x) + \sum_{m = 1}^r \sum_{\ell_m = 1}^{\check{a}_m - 1} \phi(\zeta_{\check{a}_m}^{\ell_m}x) = \oint \frac{\dd x_2}{2{\rm i}\pi} f(x_2)\mathcal{B}_0(x,x_2)\,\dd x_2
\eeq
for a function $\phi$ holomorphic on $\mathbb{C}\setminus\Gamma_{{\rm G}}(u)$ satisfying appropriate growth conditions at $\infty$. It is convenient to introduce:
\beq
\label{cacuhete}\mathcal{G}_{{\rm G}}(x,x_2) = \int_{\infty}^{x_2} \frac{\dd x_1}{x_1}\,\widetilde{W}_{\rm G|2}^{[0]}(x,x_1)
\eeq

\subsection{First correction to the spectral curve}
\label{sss2}
In the A series, the first correction to the spectral curve was of order $1/N^2$. In the BCD series, because the potential $V_{{\rm tot}}$ contains an extra term of order $1/N$, the first correction is $W_{\rm G|1}(x) = NW_{{\rm G}}(x) + W^{[1]}_{{\rm G}|1}(x) + o(1)$, and is obtained by formally writing down the $N$-dependent saddle point equation for $V_{\rm tot}$, and collecting the term of order $1/N$. We find, for any $x \in \Gamma_{\rm G}(u)$:
\beq
\label{whitedq}\widetilde{W}_{\rm G|1}^{[1]}(x + {\rm i}0) + \widetilde{W}_{\rm G|1}^{[1]}(x - {\rm i}0) + (2 - r) \sum_{\ell = 1}^{a - 1} \widetilde{W}_{{\rm G}|1}^{[1]}(\zeta_{a}^{\ell}x) + \sum_{m = 1}^r \sum_{\ell_m = 1}^{\check{a}_m - 1} \widetilde{W}_{{\rm G}|1}^{[1]}(\zeta_{\check{a}_m}^{\ell_m}x) = \partial_{x}(\delta\check{V}_{{\rm G}})(x)
\eeq
Given the expression of the right-hand side \eqref{whitedq}, it can be decomposed in partial fraction expansion:
\beq
\partial_{x}(\delta\check{V}_{{\rm G}}(x)) = \sum_{\ell = 0}^{a - 1} r_{\ell}\,\frac{x}{x - \zeta_{a}^{\ell}} + r_{\infty}
\eeq
Using the remark of \S~\ref{Soihqt} and \eqref{cacuhete}, we deduce that, if $\chi \geq 0$:
\beq
\label{P83}\boxed{\widetilde{W}_{{\rm G}|1}^{[1]}(x) = \sum_{\ell = 0}^{a - 1} \frac{r_{\ell}}{2}\,\mathcal{G}_{{\rm G}}(x_1,\zeta_{a}^{\ell})}
\eeq

\subsection{Schwinger-Dyson equations}
\label{sdeqfw}
\subsubsection{First equation \ldots}

\label{SDe1}

The invariance of the partition function under the infinitesimal change of variable $s_i \rightarrow s_i + \varepsilon\,\frac{1}{x - s_i}$ yields relations between correlators. Those equations can be derived rigorously by integration by parts. Here, we obtain:
\be\ba
\label{SD1}& \mu\Big[ \sum_{i} \frac{1}{(x - s_i)^2} + \frac{1}{2}\sum_{i \neq j} \Big(\frac{\partial_{s_i}(\ln R)(s_i,s_j)}{x - s_i} + \frac{\partial_{s_j}\ln R(s_i,s_j)}{x - s_j}\Big) \\
&  - \sum_{i} \Big(N\,\frac{a^2\ln s_i}{us_i} + \frac{1}{s_i} - \partial_{s_i}\ln \Pi_{G}(s_i)\Big)\frac{1}{x - s_i}\Big] = 0
\ea\ee
We first add the diagonal term $i = j$ to obtain $\sum_{i,j}$ and substract it, which has for effect to cancel the $\sum_{i} \mu\big[\frac{1}{(x - s_i)^2}\big]$: this is a consequence of vanishing of order $\beta = 2$ of the joint probability density \eqref{checm} when $s_i \rightarrow s_j$. Adding the terms $i = j$ also gives extra contributions which can be put in the form of a perturbation of the potential. Then, by symmetry between the two terms in the $\sum_{i,j}$ in \eqref{SD1}, we find:
\be
\label{SD1b} \mu\Big[\sum_{i,j} \frac{\partial_{s_i}(\ln R)(s_i,s_j)}{x - s_i} - \sum_{i} \frac{N\check{V}'(s_i) + \delta \check{V}'_{{\rm G}}(s_i)}{x - s_i}\Big] = 0
\ee
The first term is the expectation value of an analytic function in a neighborhood of the cut, so can be rewritten in terms of a contour integral of $\overline{W}_2$.
\be
\mu\Big[\sum_{i,j} \frac{\partial_{s_i}(\ln R)(s_i,s_j)}{x - s_i}\Big] = \oint \frac{\dd\xi\dd\eta}{(2{\rm i}\pi)^2}\,\frac{\partial_{\eta}(\ln R)(\xi,\eta)}{x - \eta}\,\overline{W}_2(\xi,\eta)
\ee
With the goal to compute the integral over $\xi$, we compute the partial fraction expansion of $\partial_{\eta}(\ln R)(\xi,\eta)$ as a function of $\xi$:
\be
\partial_{\eta}(\ln R)(\xi,\eta) = (2 - r)\Bigg(\sum_{\ell = 0}^{a - 1} \frac{\zeta_{a}^{\ell}}{\xi - \zeta_{a}^{\ell}\eta} + \frac{-\zeta_{a}^{\ell}\eta^{-2}}{\xi - \zeta_{a}^{\ell}\eta^{-1}}\Bigg) + \sum_{m = 1}^r \sum_{\ell_m = 0}^{\check{a}_m - 1} \Big(\frac{\zeta_{\check{a_m}}^{\ell_m}}{\xi - \zeta_{\check{a}_m}^{\ell_m}\eta} + \frac{-\zeta_{\check{a}_m}^{\ell_m}\eta^{-2}}{\xi - \zeta_{\check{a}_m}^{\ell_m}\eta^{-1}}\Big)
\ee
If we rewrite $\mu\big[\sum_{i} \cdots \big]$ in terms of $W_1$, we should keep in mind that $\delta\check{V}'(\xi)$ has poles at $\xi = 1$ which is on the cut. All in all, we find the equation:
\bea
  (2 - r) \sum_{\ell = 0}^{a - 1} \oint \frac{\dd\eta}{2{\rm i}\pi \eta^2}\,\frac{\overline{W}_{{\rm G}|2}(\eta,\zeta_{a}^{\ell}\eta) - \overline{W}_{{\rm G}|2}(\eta,\zeta_{a}^{\ell}\eta^{-1})}{x-  \eta} & &  \nonumber \\
+  \sum_{m = 1}^r \sum_{\ell_m = 0}^{\check{a}_m - 1} \oint \frac{\dd\eta}{2{\rm i}\pi \eta^2}\,\frac{\overline{W}_{{\rm G}|2}(\eta,\zeta_{\check{a}_m}^{\ell_m}\eta) - \overline{W}_{{\rm G}|2}(\eta,\zeta_{\check{a}_m}^{\ell_m}\eta^{-1})}{x - \eta} & &  \nonumber \\
\label{SDf1}   - \oint \frac{\dd\xi}{2{\rm i}\pi \xi}\,\frac{V'_{{\rm tot},{\rm G}}(\xi)}{x - \xi}\,W_1(\xi) & = & 0
\eea
The first double integral could be perform by picking up residues at $\eta = x$ and $\eta = 0$, but this expression is enough for our purposes. For comparison, the first Schwinger-Dyson equation of the model of the A series in terms of its correlators
\beq
W_{{\rm A}|n}(x_1,\ldots,x_n) = \mu_{{\rm A}}\Big[\prod_{i = 1}^n \mathrm{Tr}\,\frac{x_i}{x_i - e^{\mathbf{T}/a}}\Big]_{{\rm conn}}
\eeq
can be written:
\beq
\label{AseriesSD}(2 - r) \sum_{\ell = 0}^{a - 1} \oint \frac{\dd\eta}{2{\rm i}\pi \eta^2}\,\frac{\overline{W}_{{\rm A}|2}(\eta,\zeta_{a}^{\ell}\eta)}{x-  \eta} +  \sum_{m = 1}^r \sum_{\ell_m = 0}^{\check{a}_m - 1} \oint \frac{\dd\eta}{2{\rm i}\pi \eta^2}\,\frac{\overline{W}_{{\rm A}|2}(\eta,\zeta_{\check{a}_m}^{\ell_m}\eta)}{x - \eta}
  - \oint \frac{\dd\xi}{2{\rm i}\pi \xi}\,\frac{V'(\xi)}{x - \xi}\,W_{{\rm A}|1}(\xi) = 0
\eeq
\eqref{SDf1} only differs from \eqref{AseriesSD} by addition of terms involving $\eta^{-1}$, due to the structure of \eqref{Rstru}, and a potential $V_{{\rm tot},{\rm G}}$ instead of $V$.

\subsubsection{\ldots and its symmetrized form}
\label{SDe2}
Our next goal is to derive Schwinger-Dyson equation involving the symmetrized correlators only:
\be\ba
\widetilde{W}_{{\rm G}|1}(x) & =\,\, W_{{\rm G}|1}(x) - W_{{\rm G}|1}(x^{-1}) \\
\widetilde{\overline{W}}_{{\rm G}|2}(x_1,x_2) & = \overline{W}_{{\rm G}|2}(x_1,x_2) - \overline{W}_{{\rm G}|2}(x_1^{-1},x_2) - \overline{W}_{{\rm G}|2}(x_1,x_2^{-1}) + \overline{W}_{{\rm G}|2}(x_1^{-1},x_2^{-1})
\ea\ee
We denote $\mathcal{A}_{1}(x)$ and $\mathcal{A}_2(x)$ the terms involving $W_{{\rm G}|1}$ and $\overline{W}_{{\rm G}|2}$ in \eqref{SDf1}. We observe that:
\be
V_{{\rm tot},{\rm G}}(x) = V_{{\rm tot},{\rm G}}(x^{-1}) + \mathrm{cte}
\ee
Therefore, we have:
\be\ba
x^{-4}\,\mathcal{A}_1(x^{-1}) & =\,\, -\oint \frac{\dd\eta\,V'_{{\rm tot},{\rm G}}(\eta)}{2{\rm i}\pi}\,\frac{x^{-4}\,\eta^{-1}\,W_{{\rm G}|1}(\eta)}{x^{-1} - \eta} = - \oint \frac{\dd\eta\,V'_{{\rm tot},{\rm G}}(\eta)}{2{\rm i}\pi}\,\frac{\eta\,W_{{\rm G}|1}(\eta^{-1})}{x^{-1} - \eta^{-1}} \\
& = \,\,-\oint \frac{\dd\eta\,V'_{{\rm tot},{\rm G}}(\eta)}{2{\rm i}\pi}\,\frac{\eta^3}{x^3}\,\frac{-\eta^{-1} W_{{\rm G}|1}(\eta^{-1})}{x - \eta} \\
& = \,\,\oint \frac{\dd\eta\,V'_{{\rm tot},{\rm G}}(\eta)}{2{\rm i}\pi}\Big(\frac{1}{x^3} + \frac{1}{\eta x^2} + \frac{1}{\eta^2 x}\Big)\,\eta\,W_{{\rm G}|1}(\eta^{-1}) - \oint \frac{\dd\eta\,V'_{{\rm tot},{\rm G}}(\eta)}{2{\rm i}\pi}\,\frac{-\eta^{-1}\,W_{{\rm G}|1}(\eta^{-1})}{x - \eta}
\ea\ee
Therefore:
\be
\label{A1dq} \mathcal{A}_1(x) + x^{-4}\mathcal{A}_1(x^{-1}) = \oint \frac{\dd\eta\,V'_{{\rm tot},{\rm G}}(\eta)}{2{\rm i}\pi}\Big(\frac{1}{\eta x^3} + \frac{1}{x^2} + \frac{\eta}{x}\Big)\,W_{{\rm G}|1}(\eta) - \oint \frac{\dd\eta\,V'_{{\rm tot},{\rm G}}(\eta)}{2{\rm i}\pi}\,\frac{\widetilde{W}_{{\rm G}|1}(\eta)}{x - \eta}
\ee
Similarly, let us define $\check{a}_0 = a$ and consider a summation index $\alpha = (m,\ell_m)$ with $m \in \{0,\ldots,r\}$ and $\ell \in \mathbb{Z}_{\check{a}_m}$, and 
\beq
c_{\alpha} = \zeta_{\check{a}_m}^{\ell_m},\qquad d_{\alpha} = \left\{\begin{array}{lcl} 2 - r & \quad & m = 0 \\ 1 & \quad & \mathrm{otherwise}\end{array}\right. \nonumber
\eeq
We compute:
\be\ba
x^{-4}\,\mathcal{A}_2(x^{-1}) & = \,\,\sum_{\alpha} d_{\alpha} \oint \frac{\dd\eta}{2{\rm i}\pi}\,x^{-4}\,\frac{\eta^{-2}\,\overline{W}_{{\rm G}|2}(\eta,c_{\alpha}\eta) - \overline{W}_{{\rm G}|2}(\eta,c_{\alpha}\eta^{-1})}{x^{-1} - \eta} \\
& = \,\,\sum_{\alpha} d_{\alpha} \oint \frac{\dd\eta}{2{\rm i}\pi} x^{-4}\,\frac{-\overline{W}_{{\rm G}|2}(\eta^{-1},c_{\alpha}\eta^{-1}) + \overline{W}_{{\rm G}|2}(\eta^{-1},c_{\alpha}\eta)}{x^{-1} - \eta^{-1}} \\
& = \,\,\sum_{\alpha} d_{\alpha} \oint \frac{\dd\eta}{2{\rm i}\pi}\,\frac{\eta}{x^3}\,\frac{\overline{W}_{{\rm G}|2}(\eta^{-1},c_{\alpha}\eta^{-1}) -
\overline{W}_{{\rm G}|2}(\eta^{-1},c_{\alpha}\eta)}{x - \eta} \\
& = \,\,- \sum_{\alpha} d_{\alpha} \oint \frac{\dd\eta}{2{\rm i}\pi}\,\Big(\frac{1}{x^3} + \frac{1}{\eta x^2} + \frac{1}{\eta^2 x}\Big)\big[\overline{W}_{{\rm G}|2}(\eta^{-1},c_{\alpha}\eta^{-1}) - \overline{W}_{{\rm G}|2}(\eta^{-1},c_{\alpha}\eta)\big] \\
& \,\, + \sum_{\alpha} d_{\alpha} \oint \frac{\dd\eta}{2{\rm i}\pi \eta^2}\,\frac{\overline{W}_{{\rm G}|2}(\eta^{-1},c_{\alpha}\eta^{-1}) - \overline{W}_{{\rm G}|2}(\eta^{-1},c_{\alpha}\eta)}{x - \eta} 
\ea\ee
Since the sum over $\alpha$ contains both $c_{\alpha}$ and $c_{\alpha}^{-1}$, we find:
\be\ba
\mathcal{A}_2(x) + x^{-4}\mathcal{A}_2(x^{-1}) & =\,\, \sum_{\alpha} d_{\alpha} \oint \frac{\dd\eta}{2{\rm i}\pi \eta^2}\, \frac{\widetilde{\overline{W}}_{{\rm G}|2}(\eta,c_{\alpha}\eta)}{x - \eta} \\
\label{A2dq}& \,\, + \sum_{\alpha} d_{\alpha} \oint \frac{\dd\eta}{2{\rm i}\pi}\,\Big(\frac{1}{\eta^2 x^3} + \frac{1}{\eta x^2} + \frac{1}{x}\Big)\big[\overline{W}_{{\rm G}|2}(\eta,c_{\alpha}\eta) - \overline{W}_{{\rm G}|2}(\eta,c_{\alpha}\eta^{-1})\big]
\ea\ee
We observe that in \eqref{A1dq} (resp. \eqref{A2dq}) the second term -- call it $\mathcal{S}_1(x)$, resp. $\mathcal{S}_2(x)$ -- is the negative part of the Laurent expansion of $x^{-4}\mathcal{A}_{1}(x)$ (resp. $x^{-4}\mathcal{A}_2(x^{-4})$) at $x = 0$, whereas the first term is regular at $x = 0$. Since we have the equation:
\be
\mathcal{A}_1(x) + \mathcal{A}_2(x) = 0
\ee
this implies that $\mathcal{S}_1(x) + \mathcal{S}_2(x) = 0$. Then, summing \eqref{A1dq} and \eqref{A2dq} gives:
\be
\label{unfeq}(2 - r) \sum_{\ell = 0}^{a - 1} \oint \frac{\dd\eta}{2{\rm i}\pi \eta^2}\,\frac{\widetilde{\overline{W}}_{{\rm G}|2}(\eta,\zeta_{a}^{\ell}\eta)}{x - \eta} + \sum_{m = 1}^r \sum_{\ell_m = 0}^{\check{a}_m - 1} \oint \frac{\dd\eta}{2{\rm i}\pi \eta^2}\,\frac{\widetilde{\overline{W}}_{{\rm G}|2}(\eta,\zeta_{\check{a}_m}^{\ell_m}\eta)}{x - \eta} - \oint \frac{\dd\eta}{2{\rm i}\pi}\,\frac{V'_{{\rm tot},{\rm G}}(\eta)}{x - \eta}\,\widetilde{W}_{{\rm G}|1}(\eta) = 0 
\ee
In this form, we recognize the Schwinger-Dyson equation of the correlators of the model in the A series, in the potential $V_{{\rm tot}}$ given by \eqref{Vtott}.

\subsubsection{Higher equations}
\label{SDe3}

Let $\delta_{x}$ be the insertion operator: it computes $\partial_{\lambda = 0}$ under the perturbation $\check{V}(\xi) \rightarrow \check{V}(\xi) - \frac{\lambda}{N}\,\frac{1}{x - \xi}$. We have the properties:
\be
\delta_{x} W_{{\rm G}|n}(x_1,\ldots,x_n) = W_{{\rm G}|n + 1}(x,x_1,\ldots,x_n),\qquad \delta_{x}(N\check{V}'(y)) = -\frac{1}{(x - y)^2}
\ee
Higher Schwinger-Dyson equations involving $W_{{\rm G}|n + 1},W_{{\rm G}|n},\ldots,W_{{\rm G}|1}$ can be derived by applying the operator $\delta_{x_2}\cdots\delta_{x_n}$ to \eqref{SDf1}. One can also define an insertion operator adapted $\widetilde{\delta}_{x}$ adapted to symmetrized observables: it computed $\partial_{\lambda' = 0}$ under the perturbation $\check{V}(\xi) \rightarrow \check{V}(\xi) - \frac{\lambda'}{N}\big(\frac{x}{x - \xi} + \frac{-x^{-1}}{x^{-1} - \xi}\big)$. It has the property:
\bea
\widetilde{\delta}_{x}(N\check{V}')(y) & = & -\frac{x}{(x - y)^2} - \frac{-x^{-1}}{(x^{-1} - \xi)^2} \nonumber \\
\widetilde{\delta}_{x_0} \widetilde{W}_{{\rm G}|n}(x_1,\ldots,x_n) & = & \widetilde{W}_{{\rm G}|n + 1}(x_0,x_1,\ldots,x_n)
\eea
A word of caution: the symmetries \eqref{symiu1}-\eqref{symiu2} are only valid when the potential (and its perturbation) are invariant under $x \rightarrow x^{-1}$. Therefore, if we make any use of this relation to transform the Schwinger-Dyson equation, only the application of symmetrized insertion operators $\widetilde{\delta}_{x}$ is allowed.

\subsection{Loop equation and recursive formula}

The Schwinger-Dyson equation \eqref{unfeq} and their analogue is now similar to those treated in \cite[Section 3]{BEO}.

If one insert the expansion \eqref{thy} in the Schwinger-Dyson equations and study the discontinuity of both sides, one can prove by recursion on $(n,k)$ that $\widetilde{W}_{{\rm G}|n}^{[k]}(x_1,\ldots,x_n)$ defined for $x_i \in \mathbb{C}\setminus\Gamma_{{\rm G}}(u)$, satisfies the linear functional equation:
\beq
\label{lineqf}
\widetilde{W}_{{\rm G}|n}^{[k]}(x + {\rm i}0,x_I) + \widetilde{W}_{n}^{[k]}(x - {\rm i}0,x_I) + \sum_{\ell = 1}^{a - 1} (2 - r)\widetilde{W}_{{\rm G}|n}^{[k]}(\zeta_{a}^{\ell}x,z_I) + \sum_{m = 1}^r \sum_{\ell_m = 1}^{\check{a}_m - 1} \widetilde{W}_{{\rm G}|n}^{[k]}(\zeta_{\check{a}_m}^{\ell_m}x,x_I) = 0
\eeq
if $(n,k) \neq (1,0),(2,0),(1,1)$, and \eqref{816}-\eqref{sad20}-\eqref{whitedq} in the first three cases. One can then establish for $(n,k) \neq (1,0),(2,0),(1,1)$ that $\widetilde{W}_{{\rm G}|n}^{[k]}$ diverges polynomially when $x_i$ approaches the edges of the cut $\Gamma_{{\rm G}}(u)$, but is regular at other points, including $x_i = 1$. This contrasts with the singularity at $x = 1$ occuring for $(n,k) = (1,1)$, which was due to the log singularity in the potential.

From \eqref{lineqf} by Schwarz principle, we deduce that $W_{{\rm G}|n}^{[k]}(x_1,\ldots,x_n)$ can be analytically continued as a meromorphic function on the spectral curve $\Sigma$. The spectral curve is equipped with a covering $x\,:\,\Sigma \rightarrow \widehat{\mathbb{C}}$, and with a local involution $\iota$ which exchanges the two sheets at the vicinity of the ramification points. The initial $\mathbb{C}\setminus\Gamma_{{\rm G}}(u)$ is identified with the physical sheet in this covering. We still denote $\gamma^{\pm 1}$ the preimages ot the two endpoints of the cut in the physical sheet. We introduce the differential $n$-forms on the spectral curve:
\beq
\widetilde{\mathcal{W}}_{{\rm G}|n}^{[k]}(z_1,\ldots,z_n) = \Bigg(\widetilde{W}_{{\rm G}|n}^{[k]}(x(z_1),\ldots,x(z_n)) + \frac{\delta_{k,0}\delta_{n,2}}{\big(x(z_1) - x(z_2)\big)^2}\Bigg)\prod_{i = 1}^n \dd x(z_i)
\eeq
They are meromorphic with poles at all preimages of the endpoints of the cut $\Gamma_{{\rm G}}(u)$ in $\Sigma$. We also introduce the Cauchy kernel:
\beq
\boxed{\mathcal{H}_{{\rm G}}(z,z_0) = \int^{z} \widetilde{\mathcal{W}}_{{\rm G}|2}^{[0]}(z,z_0) = \mathcal{G}_{{\rm G}}(x(z),x(z_0)) + \frac{\dd x(z_0)}{x(z_0) - x(z)} + {\rm cte}}
\eeq
where $\mathcal{G}_{{\rm G}}$ is given by \eqref{cacuhete}.

Then, following \cite[Section 3.7]{BEO}, the Schwinger-Dyson equation -- and its higher versions -- can be written for any $n,k$, with the convention that $\widetilde{W}_{{\rm G}|n}^{[k]} = 0$ for $k < 0$:
\be
\label{Scylla}-\widetilde{\mathcal{W}}_{{\rm G}|n}^{[k - 2]}(z,\iota(z),z_I) - \sum_{\substack{J \subseteq I \\ k_1 + k_2 = k}} \widetilde{\mathcal{W}}_{{\rm G}|1 + |J|}^{[k_1]}(z,z_J)\widetilde{\mathcal{W}}_{{\rm G}||J| + 1}^{[k_2]}(\iota(z),z_{I\setminus J}) = \widetilde{P}_{{\rm G}|n}^{[k]}(z;z_I) + \delta\widetilde{P}_{{\rm G}|n}^{[k - 1]}(z;z_I)
\ee
where $\widetilde{P}_{{\rm G}|n}^{[k]}(z;z_I)$ is a holomorphic quadratic differential in $z$ in a neighborhood of the image of $\Gamma_{{\rm G}}(u)$ in $\Sigma$, and we have written apart the term which has a simple pole at $x(z) = 1$:
\be
\delta\widetilde{P}_{{\rm G}|n}^{[k - 1]}(z;z_I) = \frac{r_{0}\,(\dd x(z))^2}{x(z) - 1} \Big(\mathop{\mathrm{Res}}_{\xi \rightarrow 1} \mathcal{W}_{{\rm G}|n}^{[k - 1]}(\xi)\Big)
\ee
There are two differences with the setting of the topological recursion presented in \cite{BEO}. Firstly, we have an extra term $\delta\widetilde{P}_{{\rm G}|n}^{[k]}(z;z_I)$ which contains a pole a $x(z) = 1$. This pole thus occurs away from the branchpoints. Secondly, since the potential was perturbed by a term of order $1/N$, we have all possible powers of $N^{2 - k}$ for $k \geq 0$ in the expansion of $\widetilde{W}_{{\rm G}|n}$ -- instead of just $N^{2 - 2g - n}$ for $g \geq 0$.

We can nevertheless proceed as usual to solve the abstract loop equations. We use the notations:
\be
\Delta f(z) = f(z) - f(\iota(z)),\qquad \mathcal{S}f(z) = f(z) + f(\iota(z))
\ee
so that we can write:
\be
f(z)g(\iota(z)) + f(\iota(z))g(z) = \frac{1}{2}\Big(\mathcal{S} f(z) \mathcal{S} g(z) - \Delta f(z)\Delta g(z))
\ee
Then, we may recast \eqref{Scylla} as:
\be
\label{Chary}\frac{1}{2}\,\Delta\widetilde{\mathcal{W}}_{{\rm G}|1}^{[0]}(z)\,\Delta\widetilde{\mathcal{W}}_{{\rm G}|n}^{[k]}(z) = \mathcal{E}_{{\rm G}|n}^{[k]}(z,\iota(z);z_I) + \mathcal{Q}_{{\rm G}|n}^{[k]}(z;z_I)
\ee
with:
\be
\boxed{\mathcal{E}_{{\rm G}|n}^{[k]}(z,z';z_I) =  \widetilde{\mathcal{W}}_{{\rm G}|n + 1}^{[k - 2]}(z,z',z_I) + \sum_{\substack{J \subseteq I \\ k_1 + k_2 = k}}^{*} \widetilde{\mathcal{W}}_{{\rm G}|1 + |J|}^{[k_1]}(z,z_J)\widetilde{\mathcal{W}}_{{\rm G}|n - |J|}^{[k_2]}(z',z_{I\setminus J})}
\ee
and:
\be
\mathcal{Q}_{{\rm G}|n}^{[k]}(z;z_I) = \,\,\widetilde{\mathcal{P}}_{{\rm G}|n}^{[k]}(z;z_I)  + \delta\widetilde{\mathcal{P}}_{{\rm G}|n}^{[k]}(z;z_I) - \frac{1}{2}\Delta \widetilde{\mathcal{W}}_{{\rm G}|1}^{[0]}(z)\Delta\widetilde{\mathcal{W}}_{{\rm G}|n}^{[k]}(z;z_I)
\ee
We know that:
\be
\label{uiu}\widetilde{\mathcal{W}}_{{\rm G}|n}^{[k]}(z_0,z_I) - \mathop{\mathrm{Res}}_{z \rightarrow \mathrm{Fix}(\iota)} \frac{\Delta_{z}\mathcal{H}_{{\rm G}}(z_0,z)}{4}\,\Delta_{z}\widetilde{\mathcal{W}}_{{\rm G}|n}^{[k]}(z,z_I)
\ee
is holomorphic in a neighborhood of $\gamma^{\pm} \in \Sigma$, and satisfy the linear equation \eqref{lineqf}. Here, the poles at those on the image in $\Sigma$ of the cut. By unicity, \eqref{uiu} must vanish. Notice that $\Delta\widetilde{\mathcal{W}}_{{\rm G}|1}^{[0]}$ has a double zero at $\gamma^{\pm}$ since the equilibrium measure behaves exactly like a squareroot at the edges of the cut (this is the off-critical property). So, we find that $\mathcal{Q}_{{\rm G}|n}^{[k]}$ has the same property, and we deduce that the right-hand side of \eqref{Chary} does not contribute to the residue at $\mathrm{Fix}(\iota)$ in \eqref{uiu}, which yields:
\be
\label{reuc}\boxed{\widetilde{\mathcal{W}}_{{\rm G}|n}^{[k]}(z_0,z_I) = \mathop{\mathrm{Res}}_{z \rightarrow \gamma^{\pm}} \frac{\Delta_{z}\mathcal{H}_{{\rm G}}(z_0,z)}{2\Delta\widetilde{\mathcal{W}}_{{\rm G}|1}^{[0]}(z)}\,\mathcal{E}_{{\rm G}|n}^{[k]}(z,\iota(z);z_I)}
\ee
This is a recursion on $n + k > 2$. We observe that this recursion is universal : the only way the B,C or D series matters is through the value of $\widetilde{\mathcal{W}}_{{\rm G}|n}^{[k]}$ for $(n,k) \in \{(1,0),(2,0),(1,1)\}$ which are used at every step. These values are given in equations~\eqref{P81}-\eqref{P82}-\eqref{P83}.

This formula is a recursion on $(n,k)$, and is the generalization for the $\mathrm{SO}/\mathrm{Sp}$ Chern-Simons theory  of the topological recursion found in Theorem~\ref{thm:toporec} for the $\mathrm{SU}$ Chern-Simons theory.

\pgfplotsset{width=0.8\linewidth,height=0.35\textheight,legend style={font=\small}}

\section{Numerical estimates of equilibrium distributions}
\label{sec:numerics}
\label{appAlex}

\noindent\textsl{by Alexander Weisse}
\vspace{26pt}

\subsection{Markov chain Monte Carlo}
\label{sub:mcmc}

The probability measure
\begin{equation}
  \dd\mu_N(t_1,\ldots,t_N) = \frac{1}{Z_N} \prod_{1\le i<j\le N} R(t_i, t_j)
  \prod_{i=1}^{N} e^{-NV(t_i)} \dd t_i
\end{equation}
describes a statistical system of $N$ classical particles with
coordinates $t_i\in I\subset \mathbb{R}$, and $Z_N =\int_{I^N}
\dd\mu_N$ is its partition function. We are interested in the 
equilibrium distribution of the particle coordinates, $\mu_{\text{eq}}$,
which can be estimated from the large-$N$ limit of the
empirical measure
\begin{equation}
  L_N = \frac{1}{N}\sum_{i=1}^N \delta_{t_i} 
  \quad\xrightarrow{N\to\infty}\quad \mu_{\text{eq}}\,.
\end{equation}

The phase space of such a many-particle system is far too large to be
summed up completely or to be explored by naively selecting random
configurations. Instead, the standard simulation technique is a Monte
Carlo algorithm with \emph{importance sampling}~\cite{BH2010}. It is
based on a suitable Markov chain process which converges to the
measure $\dd\mu_N(t_1,\ldots,t_N) = p_{\text{eq}}(t_1,\ldots,t_N)\,
\dd t_1\cdots\dd t_N$.  A Markov chain is a stochastic process in
which a set of random variables $\{t_i\}$ is updated to a new
configuration $\{t_i\}'$ with a probability $W$ that depends only on
$\{t_i\}$ and $\{t_i\}'$, but not on the earlier history of the
process,
\begin{equation}
  \cdots\xrightarrow{\phantom{W}} \{t_i\} \xrightarrow{W} \{t_i\}' \xrightarrow{W} \{t_i\}''
  \xrightarrow{\phantom{W}}\cdots\,.
\end{equation}
The process converges to the distribution $\dd\mu_N$, if the
transition probability $W$ fulfills the condition of detailed balance,
\begin{equation}
  p_{\text{eq}}(\{t_i\}) W(\{t_i\}\to\{t_i\}') = p_{\text{eq}}(\{t_i\}') W(\{t_i\}'\to\{t_i\})\,.
\end{equation}
A good choice for such a transition probability is the classic
Metropolis algorithm~\cite{metro1953},
\begin{equation}
  W(\{t_i\}\to\{t_i\}') = \min\left(1, \frac{p_{\text{eq}}(\{t_i\}')}{p_{\text{eq}}(\{t_i\})}\right)\,.
\end{equation}
For large $N$ the probability ratio in the above expression is a ratio
of two very small numbers. Since this is difficult to handle
numerically, we rewrite $\dd\mu_N$ in terms of an energy function
$E(t_1,\ldots,t_N)$,
\begin{align}
  \dd\mu_N(t_1,\ldots,t_N) & = \frac{1}{Z_N} e^{-E(t_1,\ldots,t_N)} \prod_{i=1}^{N} \dd t_i\,,\\
  E(t_1,\ldots,t_N) & = -\sum_{1\le i<j\le N} \log(R(t_i,t_j)) + N \sum_{i=1}^{N} V(t_i)\,,
\end{align}
and use the energy difference of two configurations, $\Delta E =
E(\{t_i\}') - E(\{t_i\})$, to evaluate the transition probability
$W(\{t_i\}\to\{t_i\}') = \min(1, \exp(-\Delta E))$. In our simulations
we use local updates, i.e., we propose a random change of one
coordinate $x_j$. If this change lowers the energy, $\Delta E<0$, we
always accept the update. If the change increases the energy, we
accept the update only with probability $\exp(-\Delta E)$. We loop
over all coordinates and repeatedly try updates for each of them.

After a certain ``warm-up'' period the Markov process usually reaches
equilibrium with a stable coordinate distribution $L_N$, which we can
measure in a histogram. We call the successful Metropolis update of
$N$ coordinates a \emph{Monte Carlo step}. For the results presented
below we typically performed $10^4$ warm-up steps and $10^6$
measurement steps.

\subsection{Torus knots}
\label{sub:torus}

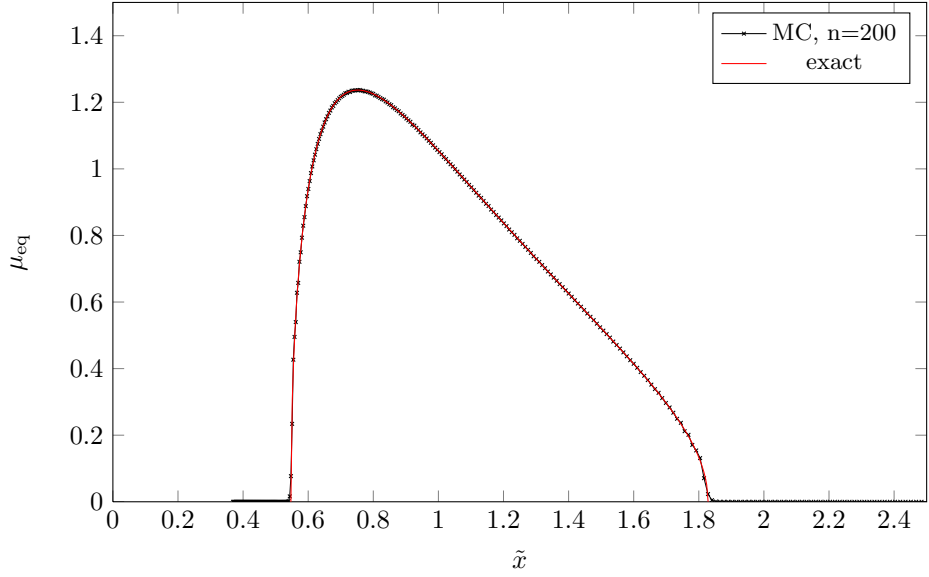
\begin{figure}
  \centering 
  \begin{tikzpicture}
    \begin{axis}[clip marker paths=true,xlabel=$\tilde{x}$,ylabel=$\mu_{\text{eq}}$,xmin=0,xmax=2.5,ymin=0,ymax=1.5]
      \addplot[color=black,mark=x,mark size=1pt] file {histo_n200_p02_q03_u00.500_lg.dat};
      \addlegendentry{MC, n=200}
      \addplot[color=red] file {exact_p02_q03_u00.500.dat};
      \addlegendentry{exact}
    \end{axis}
  \end{tikzpicture}
  \caption{Equilibrium density $\mu_{\text{eq}}$ of the statistical model for torus knots with $(p,q)=(2,3)$ and
    interaction strength $u=0.5$. The data is given in terms of the variables $\tilde{x}_i = \exp(t_i)$.
    \label{fig:torus}}
\end{figure}

The reliability of the Monte Carlo approach can be tested by first
considering the special case $r=2$ of the general
model~\eqref{themodel}, which has been studied earlier in the context
of torus knots,
\begin{equation}
  \begin{aligned}
    R(t_1,t_2) & = \sinh\left(\frac{t_1-t_2}{2 p}\right)\sinh\left(\frac{t_1-t_2}{2 q}\right)\,,\\
    V(t) & = \frac{t^2}{2 u p q}\,.
  \end{aligned}
\end{equation}
Here, $p$ and $q$ are two coprime integers labeling the knot, and $u$
is a free parameter. The equilibrium distribution~$\mu_{\text{eq}}$ of
this model is determined by the equations~\cite{BEMknots}
\begin{align}
  \tilde{x} & = e^{(1/p + 1/q)u/2} z^{-1/q} \biggl(\frac{1-e^{-u} z}{1-z}\biggr)^{1/p}\,,\\
  \mu_{\text{eq}}(\tilde{x}) &
  = \frac{p}{2\pi i u} \log\biggl(\frac{z(\tilde{x}+i0)}{z(\tilde{x}-i0)}\biggr)
  \frac{\dd x}{x}\,,
\end{align}
where the boundaries of the domain of $\tilde{x}=\exp(t)$ (the branch
points) are given by the two zeros of $\dd \tilde{x}/ \tilde{x}$.  In
Figure~\ref{fig:torus} we compare this exact result to the numerical
data obtained by simulating $N=200$ particles. The agreement is almost
perfect: only the edges of the distribution are slightly softened due
to the finite number of particles.

\subsection{Seifert fibered spaces}
\label{sub:seifert}

\subsubsection{R\^ole of strict convexity}

In \S~\ref{orbiar} it was pointed out that an important geometric invariant of the 
Seifert space is the orbifold Euler characteristic
\begin{equation}\label{euchi}
  \chi = 2 - r + \sum_{m=1}^{r} 1/a_m\,,
\end{equation}
and $\chi$ is also expected to have an influence on the properties of the
statistical model~\eqref{themodel}. The long distance behavior of the
pairwise interaction $R(t_1,t_2) \equiv R(t_1-t_2)$ is determined by $\chi$,
\begin{equation}
  \log(R(d)) \sim \chi |d| /2 
  \quad\text{for}\quad d\to\infty\,.
\end{equation}
For $\chi\ge 0$, $R(t_1,t_2)$ is always repulsive and strictly convex, and
a unique equilibrium measure $\mu_{\text{eq}}$ exists and can be
calculated. For $\chi<0$, unfortunately, the analytic methods cannot
be applied, and it was unclear whether a unique $\mu_{\text{eq}}$
exists and how it looks like.

We therefore performed numerical simulations for both situations,
$\chi\ge 0$ and $\chi<0$, in the case $r=3$. The statistical
model~\eqref{themodel} then specializes to
\begin{equation}\label{eq:modelr3}
  \begin{aligned}
    R(t_1,t_2) & = \frac{\sinh[(t_1-t_2)/(2 a_1)]\sinh[(t_1-t_2)/(2 a_2)]\sinh[(t_1-t_2)/(2 a_3)]}{\sinh[(t_1-t_2)/2]}\,,\\
    V(t) & = \frac{t^2}{2 u}\,,
  \end{aligned}
\end{equation}
where $a_i$ are three integer parameters and $u$ is real and positive.

\begin{figure}
  \centering 
  \begin{tikzpicture}
    \begin{axis}[xlabel=$t$,ylabel=$\mu_{\text{eq}}$,xmin=-4,xmax=4,ymin=0,ymax=0.25,
      ytick={0,0.1,0.2},yminorticks=true]
      \addplot[color=blue] file {histo_n200_02_03_05_u03.000_sc.dat};
      \addlegendentry{$(2,3,5)$};
      \addplot[color=red] file {histo_n200_02_03_97_u03.000_sc.dat};
      \addlegendentry{$(2,3,97)$};
      \addplot[color=black!40!green] file {histo_n200_11_11_11_u03.000_sc.dat};
      \addlegendentry{$(11,11,11)$};
    \end{axis}
  \end{tikzpicture}
  \caption{Numerical results for the equilibrium density $\mu_{\text{eq}}$ of the statistical model
    for Seifert fibered spaces. The simulations were performed with $N=200$ particles, $u=3$,
    and the parameters $(a_1,a_2,a_3)$ given in the legend.
    \label{fig:seifert}}
\end{figure}
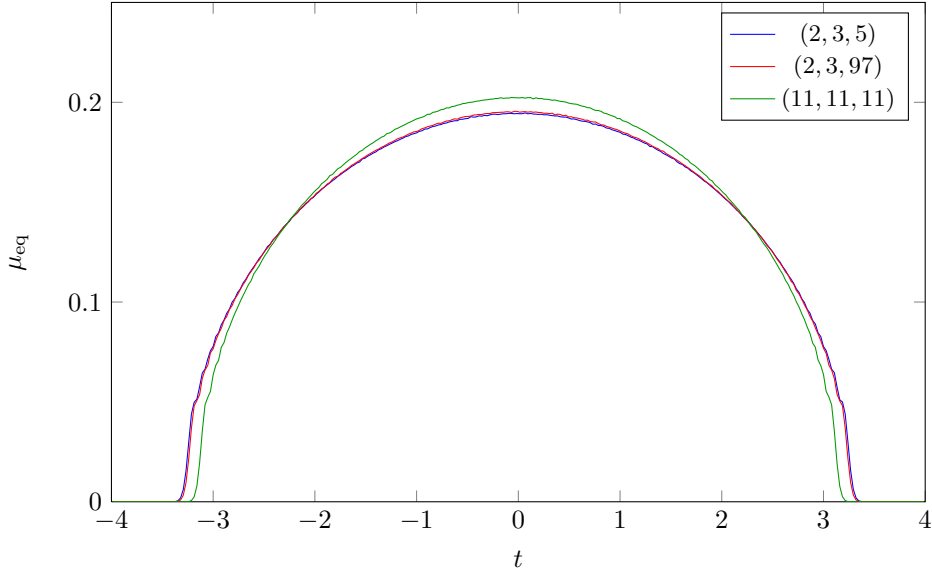

The Monte Carlo simulations with $N=200$ particles suggest that a
unique equilibrium measure $\mu_{\text{eq}}$ exists independently of
the sign of $\chi$. In Figure~\ref{fig:seifert} we show data for the
parameter sets $(a_1,a_2,a_3) = (2,3,5)$, $(2,3,97)$ and
$(11,11,11)$. The first set fulfills the condition $\chi>0$ and leads
to a unique equilibrium distribution $\mu_{\text{eq}}$. However, the
simulations converge equally well for the other parameter sets with
$\chi<0$. The shape of the resulting equilibrium distributions differs
only marginally from the first case. The small oscillations near the
edges of the distributions and the softening of the edges are caused
by the finite particle number. For $N\to\infty$ we expect smooth
distributions which at the boundaries of the domain vanish like a
half-integer power of $x$.

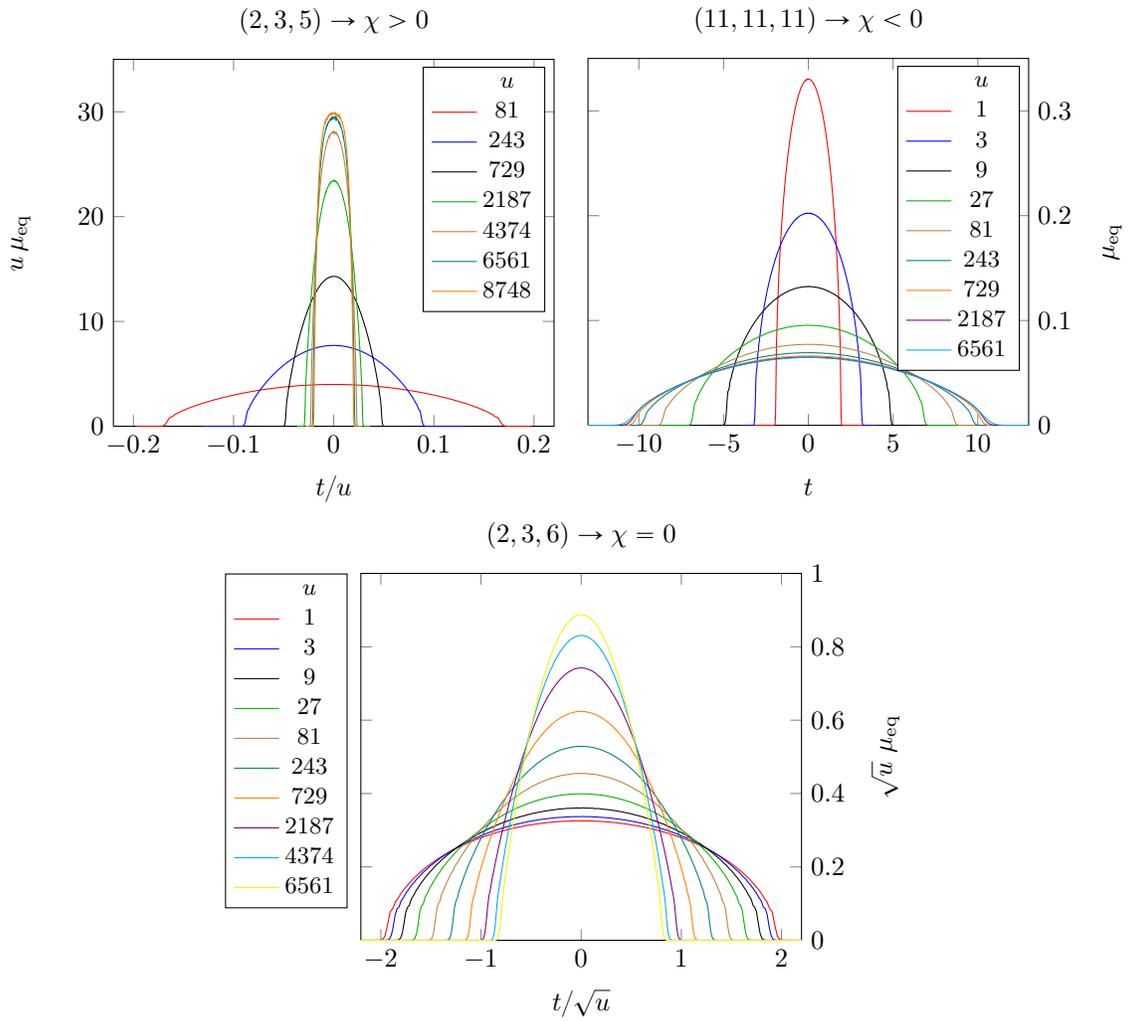
\begin{figure}
  \pgfplotsset{width=0.48\linewidth,height=0.275\textheight}
  \pgfplotscreateplotcyclelist{color list}{white,red,blue,black,green!70!black,brown,teal,orange,violet,cyan,yellow,magenta,gray}
  \centering 
  \begin{tikzpicture}
    \begin{axis}[title={$(2,3,5)\to \chi>0$},xlabel=$t/u$,ylabel=$u\,\mu_{\text{eq}}$,
      xmin=-0.22,xmax=0.22,ymin=0,ymax=35.,ylabel style={at={(0.0,0.5)}},
      cycle list name=color list]
      \addplot coordinates { (0,0) };
      \addlegendentry{$u$}
      \addplot file {histo_n200_02_03_05_u81.000_rs2.dat};
      \addlegendentry{$81$};
      \addplot file {histo_n200_02_03_05_u243.000_rs2.dat};
      \addlegendentry{$243$};
      \addplot file {histo_n200_02_03_05_u729.000_rs2.dat};
      \addlegendentry{$729$};
      \addplot file {histo_n200_02_03_05_u2187.000_rs2.dat};
      \addlegendentry{$2187$};
      \addplot file {histo_n200_02_03_05_u4374.000_rs2.dat};
      \addlegendentry{$4374$};
      \addplot file {histo_n200_02_03_05_u6561.000_rs2.dat};
      \addlegendentry{$6561$};
      \addplot file {histo_n200_02_03_05_u8748.000_rs2.dat};
      \addlegendentry{$8748$};
    \end{axis}
  \end{tikzpicture}
  \hfil
  \begin{tikzpicture}
    \begin{axis}[title={$(11,11,11)\to \chi<0$},xlabel=$t\vphantom{/\sqrt{u}}$,ylabel=$\mu_{\text{eq}}$,
      xmin=-13,xmax=13,ymin=0,ymax=0.35,
      yticklabel pos=right,ylabel style={at={(1.4,0.5)}},
      cycle list name=color list]
      \addplot coordinates { (0,0) };
      \addlegendentry{$u$}
      \addplot file {histo_n200_11_11_11_u01.000_sc.dat};
      \addlegendentry{$1$};
      \addplot file {histo_n200_11_11_11_u03.000_sc.dat};
      \addlegendentry{$3$};
      \addplot file {histo_n200_11_11_11_u09.000_sc.dat};
      \addlegendentry{$9$};
      \addplot file {histo_n200_11_11_11_u27.000_sc.dat};
      \addlegendentry{$27$};
      \addplot file {histo_n200_11_11_11_u81.000_sc.dat};
      \addlegendentry{$81$};
      \addplot file {histo_n200_11_11_11_u243.000_sc.dat};
      \addlegendentry{$243$};
      \addplot file {histo_n200_11_11_11_u729.000_sc.dat};
      \addlegendentry{$729$};
      \addplot file {histo_n200_11_11_11_u2187.000_sc.dat};
      \addlegendentry{$2187$};
      \addplot file {histo_n200_11_11_11_u6561.000_sc.dat};
      \addlegendentry{$6561$};
    \end{axis}
  \end{tikzpicture}
  \begin{tikzpicture}
    \begin{axis}[title={$(2,3,6)\to \chi=0$},xlabel=$t/\sqrt{u}$,ylabel=$\sqrt{u}\ \mu_{\text{eq}}$,
      xmin=-2.2,xmax=2.2,ymin=0,ymax=1.0,
      yticklabel pos=right,ylabel style={at={(1.4,0.5)}},
      cycle list name=color list,
      legend style={at={(-0.03,1.0)},anchor=north east}]
      \addplot coordinates { (0,0) };
      \addlegendentry{$u$}
      \addplot file {histo_n200_02_03_06_u01.000_rs.dat};
      \addlegendentry{$1$};
      \addplot file {histo_n200_02_03_06_u03.000_rs.dat};
      \addlegendentry{$3$};
      \addplot file {histo_n200_02_03_06_u09.000_rs.dat};
      \addlegendentry{$9$};
      \addplot file {histo_n200_02_03_06_u27.000_rs.dat};
      \addlegendentry{$27$};
      \addplot file {histo_n200_02_03_06_u81.000_rs.dat};
      \addlegendentry{$81$};
      \addplot file {histo_n200_02_03_06_u243.000_rs.dat};
      \addlegendentry{$243$};
      \addplot file {histo_n200_02_03_06_u729.000_rs.dat};
      \addlegendentry{$729$};
      \addplot file {histo_n200_02_03_06_u2187.000_rs.dat};
      \addlegendentry{$2187$};
      \addplot file {histo_n200_02_03_06_u4374.000_rs.dat};
      \addlegendentry{$4374$};
      \addplot file {histo_n200_02_03_06_u6561.000_rs.dat};
      \addlegendentry{$6561$};
    \end{axis}
  \end{tikzpicture}
  \caption{Scaling of $\mu_{\text{eq}}$ with $u$ for $(a_1,a_2,a_3)=(2,3,5)$ on the upper left,
    $(11,11,11)$ on the upper right, and $(2,3,6)$ on the lower central panel. 
    For all simulations we used $N=200$ particles and $u$ as given in the legend.
    \label{fig:seifert_uscal}}
\end{figure}

We also studied the scaling of $\mu_{\text{eq}}$ with increasing $u$,
i.e., with decreasing strength of the potential $V(t)$. For $\chi>0$
the interaction $-\log(R(t_1,t_2))$ is always repulsive and the expansion
of the particle distribution is limited by the potential (and thus
depends on $u$). For $\chi<0$ the interaction is repulsive at short
distance but attractive at long distance. Therefore, for large enough
$u$, it is plausible that the support of the particle distribution reaches a limit which
is independent of $u$. In Figure~\ref{fig:seifert_uscal} we illustrate
the scaling of $\mu_{\text{eq}}(t)$ for the three cases $(2,3,5)$,
$(2,3,6)$ and $(11,11,11)$, which correspond to positive, vanishing
and negative $\chi$, respectively.  For $\chi>0$ and large enough $u$
the width of the distribution scales like $u$, whereas it seems to approach a
constant for $\chi<0$. For $\chi=0$ the scaling seems to be weaker
than the first guess $\sqrt{u}$. This could be due to logarithmic corrections.

\subsubsection{The case $(2,2,\text{even})$}

\begin{figure}[H]
  \pgfplotsset{width=0.47\linewidth,height=0.25\textheight}
  \begin{tikzpicture}
    \begin{axis}[clip marker paths=true,xlabel=$t$,ylabel=$\mu_{\text{eq}}$,xmin=-3,xmax=3,ymin=0,ymax=0.4,title={$p=2$}]
      \addplot[color=black,mark=x,mark size=1pt]
      file {histo_n200_02_02_02_u01.000_rs.dat};
      \addplot[color=red]
      file {newexact_02_02_02_u1.0.dat};
    \end{axis}
  \end{tikzpicture}
  \hfil
  \begin{tikzpicture}
    \begin{axis}[clip marker paths=true,xlabel=$t$,ylabel=$\mu_{\text{eq}}$,xmin=-3,xmax=3,ymin=0,ymax=0.4,title={$p=4$}]
      \addplot[color=black,mark=x,mark size=1pt]
      file {histo_n200_02_02_04_u01.000_rs.dat};
      \addplot[color=red]
      file {newexact_02_02_04_u1.0.dat};
    \end{axis}
  \end{tikzpicture}\\[2mm]
  \begin{tikzpicture}
    \begin{axis}[clip marker paths=true,xlabel=$t$,ylabel=$\mu_{\text{eq}}$,xmin=-3,xmax=3,ymin=0,ymax=0.4,title={$p=10$}]
      \addplot[color=black,mark=x,mark size=1pt]
      file {histo_n200_02_02_10_u01.000_rs.dat};
      \addplot[color=red]
      file {newexact_02_02_10_u1.0.dat};
    \end{axis}
  \end{tikzpicture}
  \hfil
  \begin{tikzpicture}
    \begin{axis}[clip marker paths=true,xlabel=$t$,ylabel=$\mu_{\text{eq}}$,xmin=-3,xmax=3,ymin=0,ymax=0.4,title={$p=22$}]
      \addplot[color=black,mark=x,mark size=1pt]
      file {histo_n200_02_02_22_u01.000_rs.dat};
      \addplot[color=red]
      file {newexact_02_02_22_u1.0.dat};
    \end{axis}
  \end{tikzpicture}
  \caption{Simulation data obtained with $N=200$ particles compared to exact equilibrium density for
    $u=1$ and $(a_1,a_2,a_3)=(2,2,p\text{ even})$.
    \label{fig:seifert_case22even}}
\end{figure}
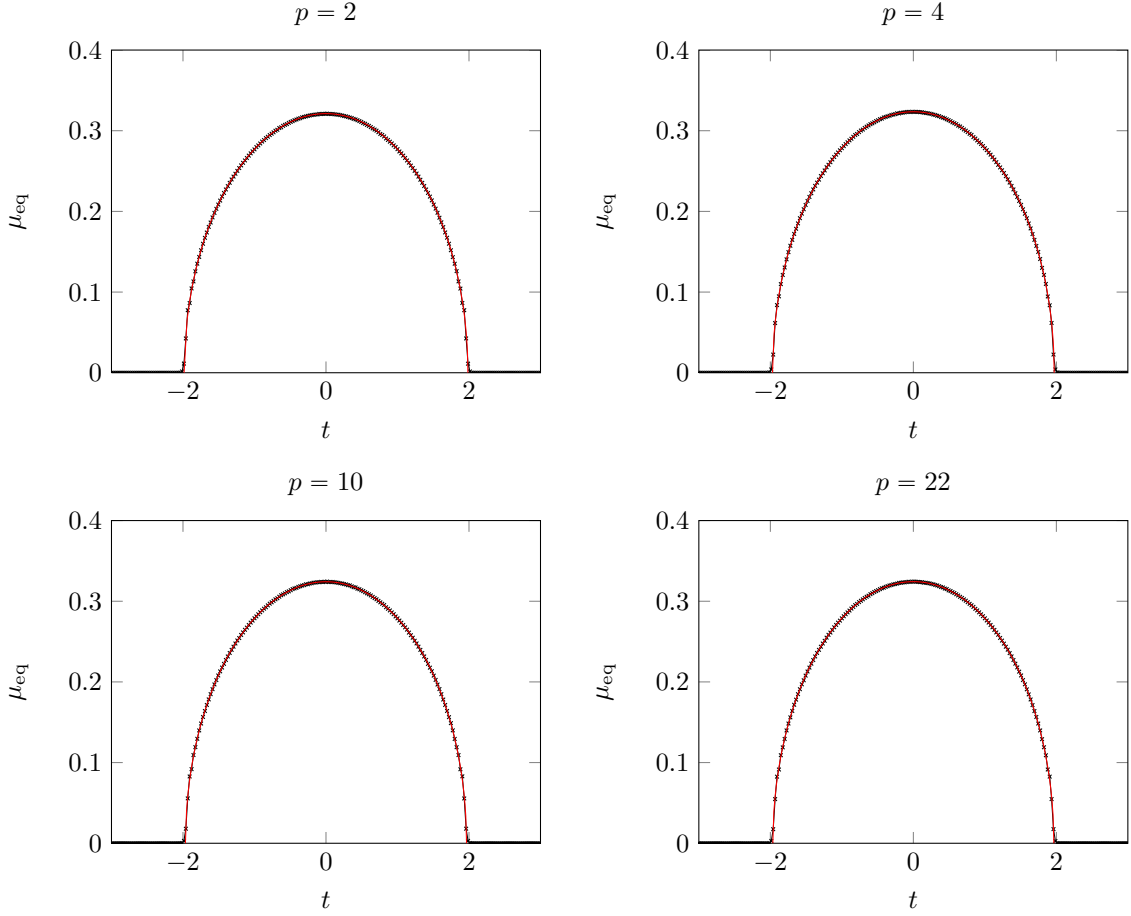

In this and the remaining sections we compare analytic results for the
model~(\ref{eq:modelr3}) with numerical data.  For the case $(2,2,p)$
with even $p$ the characteristic $\chi>0$ and the equilibrium density
$\mu_{\text{eq}}$ is given by
\begin{equation}
  \dd\mu_{\text{eq}} = \frac{p^2\,\dd \tilde x}{2 i\pi\,u \tilde x}
  \,\log\left(\frac{y(\tilde x + i0)}{y(\tilde x - i0)}\right)\,,
\end{equation}
where $\tilde x$ and $y$ are solutions to the equations (see \S~\ref{even}):
\begin{equation}
  \begin{aligned} 
    \tilde x(z) & =  z\dfrac{z^{p} - \kappa^{p}}{\kappa^{p}z^{p} - 1}\,,\\ 
    y(z) & =  -\dfrac{(z^{p/2} - \kappa^{p/2})(\kappa^{p/2}z^{p/2} + 1)}{(\kappa^{p/2}z^{p/2} - 1)(z^{p/2} + \kappa^{p/2})}\,,
  \end{aligned}
\end{equation}
with
\begin{equation}
  \frac{2\kappa^{(p + 1)/2}}{1 + \kappa^{p}} = e^{-u/(4p^2)}\,,\qquad 
  \tilde x = \exp(t/p)\,.
\end{equation}
In Figure~\ref{fig:seifert_case22even} we compare this result to
simulation data for $N=200$ particles and $u=1$. The agreement is
excellent. Note that we plot the densities as a function of $t$ rather
than $\tilde x$.

\subsubsection{The case $(2,2,\text{odd})$}

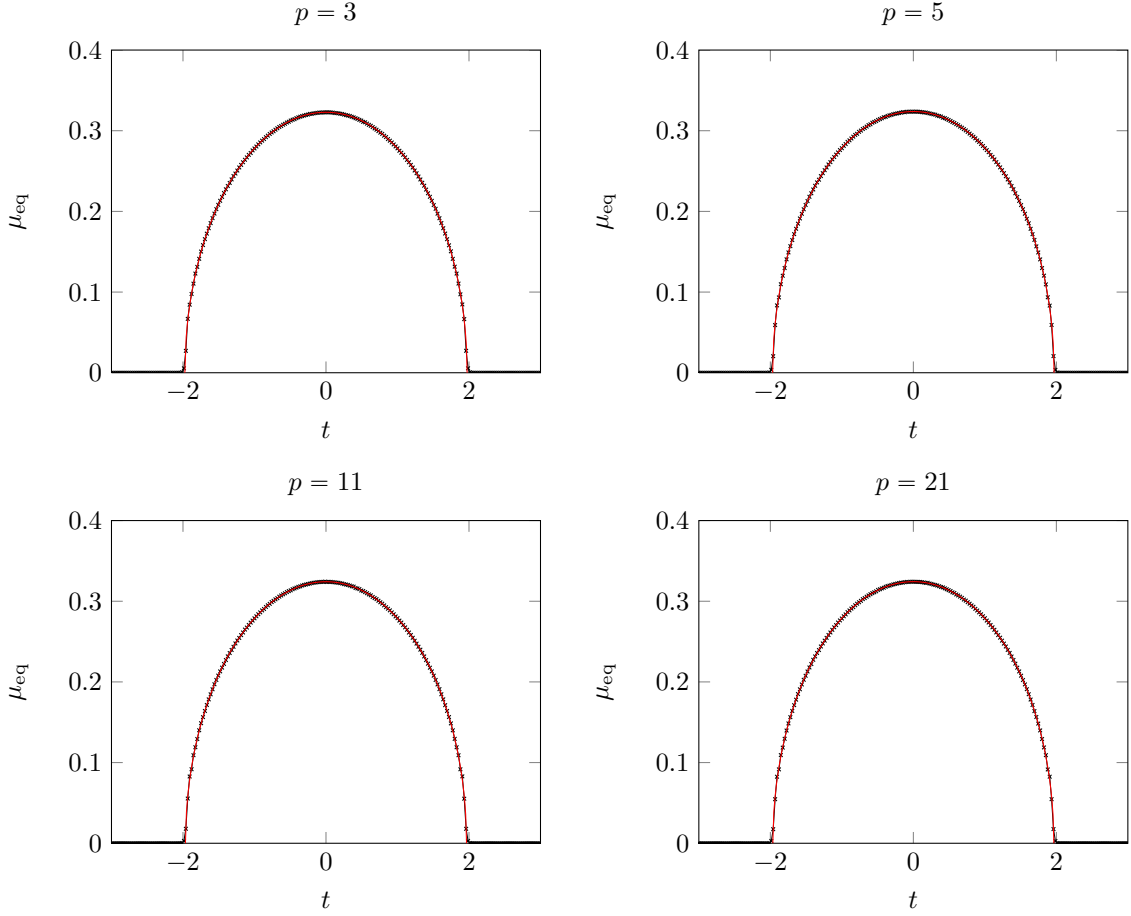
\begin{figure}[H]
  \pgfplotsset{width=0.47\linewidth,height=0.25\textheight}
  \begin{tikzpicture}
    \begin{axis}[clip marker paths=true,xlabel=$t$,ylabel=$\mu_{\text{eq}}$,xmin=-3,xmax=3,ymin=0,ymax=0.4,title={$p=3$}]
      \addplot[color=black,mark=x,mark size=1pt]
      file {histo_n200_02_02_03_u01.000_rs.dat};
      \addplot[color=red]
      file {newexact_02_02_03_u1.0.dat};
    \end{axis}
  \end{tikzpicture}
  \hfil
  \begin{tikzpicture}
    \begin{axis}[clip marker paths=true,xlabel=$t$,ylabel=$\mu_{\text{eq}}$,xmin=-3,xmax=3,ymin=0,ymax=0.4,title={$p=5$}]
      \addplot[color=black,mark=x,mark size=1pt]
      file {histo_n200_02_02_05_u01.000_rs.dat};
      \addplot[color=red]
      file {newexact_02_02_05_u1.0.dat};
    \end{axis}
  \end{tikzpicture}\\[2mm]
  \begin{tikzpicture}
    \begin{axis}[clip marker paths=true,xlabel=$t$,ylabel=$\mu_{\text{eq}}$,xmin=-3,xmax=3,ymin=0,ymax=0.4,title={$p=11$}]
      \addplot[color=black,mark=x,mark size=1pt]
      file {histo_n200_02_02_11_u01.000_rs.dat};
      \addplot[color=red]
      file {newexact_02_02_11_u1.0.dat};
    \end{axis}
  \end{tikzpicture}
  \hfil
  \begin{tikzpicture}
    \begin{axis}[clip marker paths=true,xlabel=$t$,ylabel=$\mu_{\text{eq}}$,xmin=-3,xmax=3,ymin=0,ymax=0.4,title={$p=21$}]
      \addplot[color=black,mark=x,mark size=1pt]
      file {histo_n200_02_02_21_u01.000_rs.dat};
      \addplot[color=red]
      file {newexact_02_02_21_u1.0.dat};
    \end{axis}
  \end{tikzpicture}
  \caption{Simulation data obtained with $N=200$ particles compared to exact equilibrium density for
    $u=1$ and $(a_1,a_2,a_3)=(2,2,p\text{ odd})$.
    \label{fig:seifert_case22odd}}
\end{figure}

For the case $(2,2,p)$ with odd $p$ the equilibrium density,
\begin{equation}
  \dd\mu_{\text{eq}} = \pm \frac{p^2\,\dd x}{i\pi\,ux}
  \,\log\left(\frac{y(x + i0)}{y(x - i0)}\right)\,,
\end{equation}
can be obtained analytically from
\begin{equation}
  \begin{aligned} 
    x^2(z) & = z^{-2}\,\frac{z^{2p}\kappa^{2p} - 1}{z^{2p} - \kappa^{2p}}\,,\\ 
    y(z) & = - \frac{(z^{p} - \kappa^{p})(\kappa^{p}z^{p} + 1)}{(z^{p}\kappa^{p} - 1)(z^{p} + \kappa^{p})}\,,
  \end{aligned}
\end{equation}
where
\begin{equation}
  \frac{2\kappa^{p + 1}}{1 + \kappa^{2p}} = e^{-u/(4p^2)}\,, \qquad
  \tilde x = \exp(t/(2p))\,.
\end{equation}
In Figure~\ref{fig:seifert_case22odd} we compare the analytical result to
simulation data for $N=200$ particles and $u=1$. Again, the agreement is 
excellent.

\subsubsection{The case $(2,3,3)$}

The case $(2,3,3)$ admits only a partial analytical solution. The
equilibrium density again corresponds to a branch cut,
\begin{equation}
  \dd\mu_{\text{eq}} = \pm \frac{a^2\,\dd x}{i\pi\,u x}
  \,\log\left(\frac{y(x + i0)}{y(x - i0)}\right)\,,\qquad
\end{equation}
but the equations defining the Riemann surface contain two parameters
$m_2$ and $m_3$ whose dependence on the interaction
$c:=\exp[-u/(2a^2)]$ as yet could not be determined (see \S~\ref{E6cas}).  In terms of the
variables $z=y+1/y$ and $w=\tilde{x} + 1/\tilde{x}$ with $\tilde{x} = x^2$, the branch points
fulfill:
\begin{equation}\label{eq:cond233wz}
  \begin{aligned}
    A & = c^4 z^4+c^4 z^3-2 c^4 z^2-5 c^4 z-6 c^4+2 c^2 m_{2} z^3 +12 c^2 m_{2} z^2-12 c^2 m_{2} z\\
    & -40 c^2 m_{2}+c^2 w z^3 -3 c^2 w z-2 c^2 w-6 c m_{3} z^2+24 c m_{3}-4 m_{2}^2 z \\
    & +8 m_{2}^2+2 m_{2} w z-4 m_{2} w-w^2 z-w^2+3 z+6 = 0\,,\\
    \dd A/\dd z & = 0\,.
  \end{aligned}
\end{equation}
Taking $w$ and $m_2$ as adjustable parameters, we can easily solve
these equations for $z$ and $m_3$ and fit the solution of $A(z,w)=0$
to the numerical data. For all values of $u$ we can perfectly match
the simulation results, see Figure~\ref{fig:seifert_233} below. The
corresponding parameters $m_2$ and $m_3$ are given in the legend.

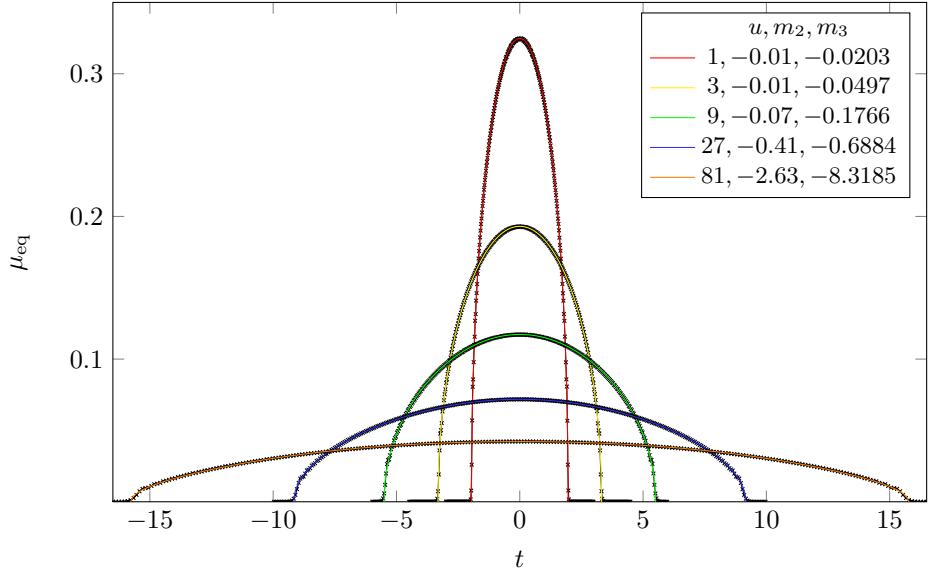
\begin{figure}[H]
  \pgfplotsset{width=0.8\linewidth,height=0.35\textheight}
  \centering 
  \begin{tikzpicture}
    \begin{axis}[clip marker paths=true,xlabel=$t$,ylabel=$\mu_{\text{eq}}$,
      xmin=-16.5,xmax=16.5,ymin=0,ymax=0.35,ytick={0.1,0.2,0.3}]
      \addplot[color=white] coordinates { (0,0) };
      \addlegendentry{$u, m_2, m_3$}
      \addplot[color=black,mark=x,mark size=1pt,forget plot]
      file {histo_n200_02_03_03_u01.000_sc.dat};
      \addplot[color=red]
      file {expexact_02_03_03_u01.000.dat};
      \addlegendentry{$1, -0.01, -0.0203$};
      \addplot[color=black,mark=x,mark size=1pt,forget plot]
      file {histo_n200_02_03_03_u03.000_sc.dat};
      \addplot[color=yellow]
      file {expexact_02_03_03_u03.000.dat};
      \addlegendentry{$3, -0.01, -0.0497$};
      \addplot[color=black,mark=x,mark size=1pt,forget plot]
      file {histo_n200_02_03_03_u09.000_sc.dat};
      \addplot[color=green]
      file {expexact_02_03_03_u09.000.dat};
      \addlegendentry{$9, -0.07, -0.1766$};
      \addplot[color=black,mark=x,mark size=1pt,forget plot]
      file {histo_n200_02_03_03_u27.000_sc.dat};
      \addplot[color=blue!80!white]
      file {expexact_02_03_03_u27.000.dat};
      \addlegendentry{$27, -0.41, -0.6884$};
      \addplot[color=black,mark=x,mark size=1pt,forget plot]
      file {histo_n200_02_03_03_u81.000_sc.dat};
      \addplot[color=orange]
      file {expexact_02_03_03_u81.000.dat};
      \addlegendentry{$81, -2.63, -8.3185$};
    \end{axis}
  \end{tikzpicture}
  \caption{For $(a_1,a_2,a_3)=(2,3,3)$ the parameters $m_2$ and $m_3$ in Eq.~\eqref{eq:cond233wz} 
    can be adjusted such that the analytical results (lines) fit the numerical data (crosses) very well.
    We simulated $N=200$ particles with interaction $u$ as given in the legend.
    \label{fig:seifert_233}}
\end{figure}

\subsubsection{The gauge groups SO and Sp}

In Section~\ref{S8} the generalization of the statistical model~\eqref{themodel} to gauge groups ${\rm SO}$ and ${\rm Sp}$ was studied, and it was shown in \S~\ref{P81} that
in the limit $N\to\infty$ the corresponding equilibrium distributions
$\mu_{\text{eq}}$ for parameter $u$ are given by the ${\rm SU}$ data with
parameter $2u$. One noticeable difference between the ${\rm SO}$ and ${\rm Sp}$
densities (BCD series) and the ${\rm SU}$ density (A series) is a
finite-size correction taking the form of a Dirac peak of order $O(1/N)$ at $t=0$ (see \eqref{P83}).

\begin{figure}[H]
  \pgfplotsset{width=0.8\linewidth,height=0.35\textheight}
  \centering 
  \begin{tikzpicture}
    \begin{axis}[title={$(2,2,2)$, case $B$},xlabel=$t$,ylabel=$\mu_{\text{eq}}$,
      xmin=-3,xmax=3,ymin=0,ymax=0.4]
      \addplot[color=white] coordinates { (0,0) };
      \addlegendentry{$N$}
      \addplot[color=black] file {histo_n050_02_02_02_u00.500_BCD0_sc.dat};
      \addlegendentry{$50$};
      \addplot[color=red] file {histo_n100_02_02_02_u00.500_BCD0_sc.dat};
      \addlegendentry{$100$};
      \addplot[color=blue] file {histo_n200_02_02_02_u00.500_BCD0_sc.dat};
      \addlegendentry{$200$};
      \addplot[color=green!70!black] file {histo_n400_02_02_02_u00.500_BCD0_sc.dat};
      \addlegendentry{$400$};
    \end{axis}
  \end{tikzpicture}
  \caption{Finite-size scaling of the equilibrium density for case B with 
    parameters $(a_1,a_2,a_3)=(2,2,2)$ and $u=0.5$. As expected, the $\delta$-singularity at $t=0$ 
    vanishes like $1/N$.
    \label{fig:bcd_scaling}}
\end{figure}
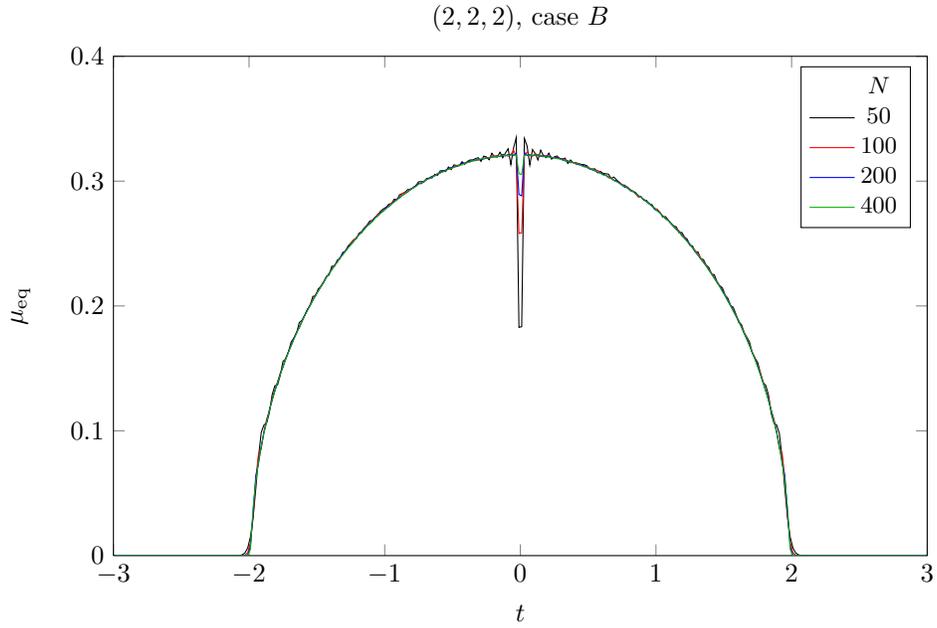

To verify these predictions, which were derived for the convex
interaction $\chi>0$, and to analyze whether they hold for $\chi\le
0$, we performed numerical simulations for various parameter sets
$(a_1, a_2, a_3)$ and increasing particle numbers $N=50,\dots,
400$. In Figure~\ref{fig:bcd_comp} we show equilibrium densities for
three parameter sets $(a_1,a_2,a_3)$ with positive, zero and negative
$\chi$. The densities for the B, C, and D cases at $u=0.5$ agree
very well with the data for the A case at $u=1$. Only in the
vicinity of $t=0$ we observe the predicted finite-size correction.
The total mass of the Dirac peak is negative for the cases B and C,
and positive for the case D, but its absolute amplitude is the same for all BCD cases. In Figure~\ref{fig:bcd_scaling} we analyze the scaling with increasing particle number $N$. Clearly, doubling $N$ reduces the $\delta$-peak by a factor $1/2$. These observations are in agreement with \eqref{Udefs}-\eqref{P83}, which predicts for $\chi > 0$ the following masses $m$ for the Dirac peak:
\beq
m_{B} = m_{C} = -\frac{a\chi}{4},\qquad m_{D} = \frac{a\chi}{4}
\eeq

\begin{figure}[H]
  \pgfplotsset{width=0.47\linewidth,height=0.25\textheight}
  \centering
  \begin{tikzpicture}
    \begin{axis}[clip marker paths=true,xlabel=$t$,ylabel=$\mu_{\text{eq}}$,xmin=-3,xmax=3,ymin=0,ymax=0.4,title={$(2,3,5)\to\chi>0$}]
      \addplot[color=black,semithick] file {histo_n200_02_03_05_u01.000_sc.dat};
      \addlegendentry{$A$}
      \addplot[color=red,thick] file {histo_n200_02_03_05_u00.500_BCD0_sc.dat};
      \addlegendentry{$B$}
      \addplot[color=blue] file {histo_n200_02_03_05_u00.500_BCD1_sc.dat};
      \addlegendentry{$C$}
      \addplot[color=green!70!black] file {histo_n200_02_03_05_u00.500_BCD2_sc.dat};
      \addlegendentry{$D$}
    \end{axis}
  \end{tikzpicture}
  \hfil
  \begin{tikzpicture}
    \begin{axis}[clip marker paths=true,xlabel=$t$,ylabel=$\mu_{\text{eq}}$,xmin=-3,xmax=3,ymin=0,ymax=0.4,title={$(11,11,11)\to\chi<0$}]
      \addplot[color=black,semithick] file {histo_n200_11_11_11_u01.000_sc.dat};
      \addlegendentry{$A$}
      \addplot[color=red,thick]   file {histo_n200_11_11_11_u00.500_BCD0_sc.dat};
      \addlegendentry{$B$}
      \addplot[color=blue] file {histo_n200_11_11_11_u00.500_BCD1_sc.dat};
      \addlegendentry{$C$}
      \addplot[color=green!70!black] file {histo_n200_11_11_11_u00.500_BCD2_sc.dat};
      \addlegendentry{$D$}
    \end{axis}
  \end{tikzpicture}\\[2mm]
  \begin{tikzpicture}
    \begin{axis}[clip marker paths=true,xlabel=$t$,ylabel=$\mu_{\text{eq}}$,xmin=-3,xmax=3,ymin=0,ymax=0.4,title={$(2,3,6)\to\chi=0$}]
      \addplot[color=black,semithick] file {histo_n200_02_03_06_u01.000_sc.dat};
      \addlegendentry{$A$}
      \addplot[color=red,thick]   file {histo_n200_02_03_06_u00.500_BCD0_sc.dat};
      \addlegendentry{$B$}
      \addplot[color=blue] file {histo_n200_02_03_06_u00.500_BCD1_sc.dat};
      \addlegendentry{$C$}
      \addplot[color=green!70!black] file {histo_n200_02_03_06_u00.500_BCD2_sc.dat};
      \addlegendentry{$D$}
    \end{axis}
  \end{tikzpicture}
  \caption{Equilibrium density for the BCD series at $u=0.5$ compared to data for the A series at $u=1.0$.
    The curves agree very well, except for a finite-size correction $\sim \delta(t)/N$, which is negative for 
    cases B and C, and positive for case D.
    We simulated $N=200$ particles with interaction parameters $(a_1,a_2,a_3)$ as given in the titles.
    \label{fig:bcd_comp}}
\end{figure}
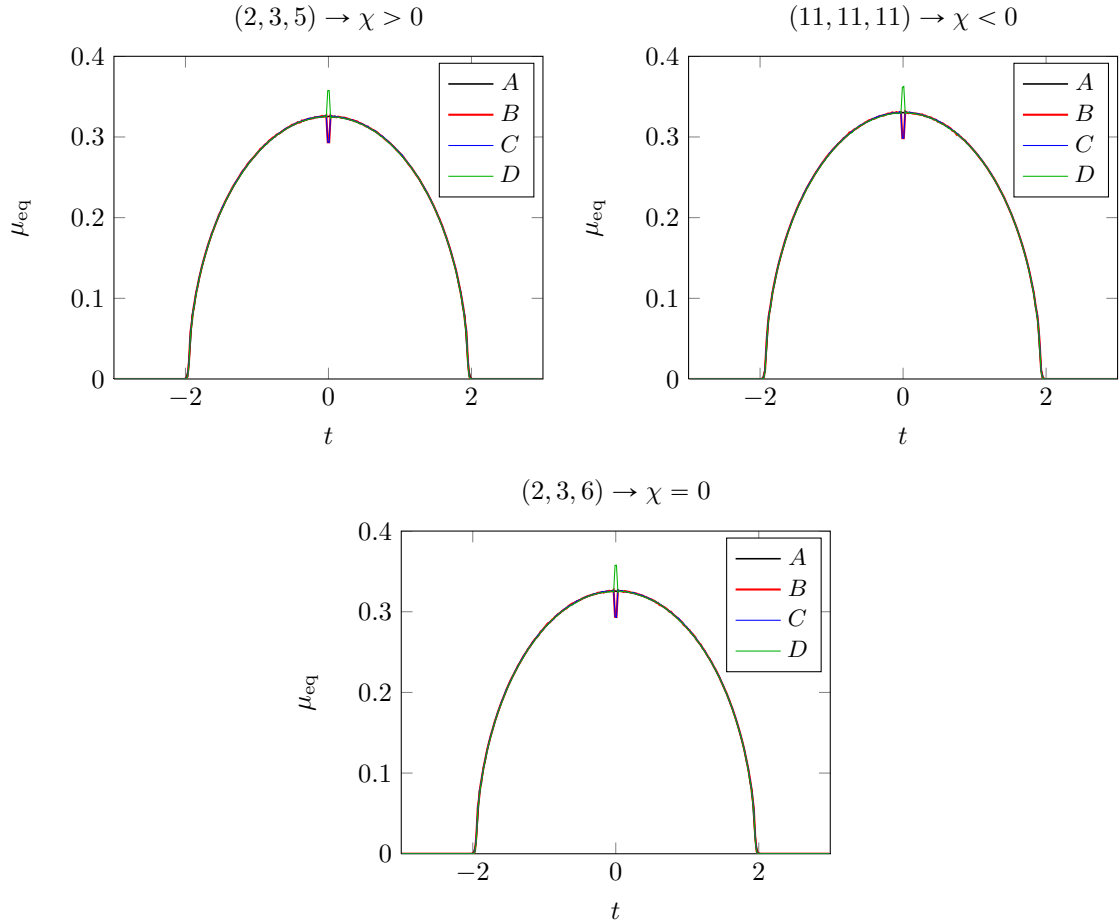

\subsection{Conclusion}
The matrix models considered in this work can reliably be simulated
with Monte Carlo methods.  The numerical data has been used to verify
analytic results and to gain further insight into cases that elude
an analytic solution.

\clearpage

\newpage

\appendix

\section{Proofs of Section~\ref{S2}}
\label{S2proof}
\subsection{Proof of Lemma~\ref{lemm1}}
\label{S23}

Let $\nu \in \mathcal{M}^0$ with compact support. We compute:
\bea
\label{rep}-\iint \dd\nu(t)\dd\nu(t')\ln\big|\mathrm{sinh}\big(\frac{t - t'}{2}\big)\big| & = & \iint\dd\nu(t)\dd\nu(t') \Big(\frac{|t - t'|}{2} + \ln(1 - e^{-|t - t'|}) - \ln 2\Big) \nonumber \\
& = & \iint \dd\nu(t)\dd\nu(t')f(t - t')
\eea
with $f(t) = -\Theta(t)\big[t + 2\ln(1 - e^{-t})\big]$ and $\Theta$ is the Heaviside step function. We compute the Fourier representation of $f$, in the distributional sense:
\bea
f(t) & = & - \int_{0}^t \dd t'\Big(\int_{0}^{t'} \dd t''\,\delta(t'')\Big) - 2 \int_{\mathbb{R}} \frac{\dd k}{2\pi}\,e^{-{\rm i}k t} \int_{0}^{\infty} \dd t'\,e^{{\rm i}kt'}\ln(1 - e^{-t'}) \nonumber \\
 & = & -\int_{0}^t \dd t'\Big(\int_{0}^{t'} \dd t''\int_{\mathbb{R}} \frac{\dd k}{2\pi}\,e^{-{\rm i}k t''}\Big) + \int_{\mathbb{R}}\frac{\dd k\,e^{-{\rm i}kt}}{{\rm i}\pi k}\Big( \int_{0}^{\infty} \frac{\dd t'\,e^{{\rm i}k t'}}{e^{t'} - 1}\Big) \nonumber \\
\label{reps} & = & - \int_{\mathbb{R}} \frac{\dd k}{2\pi}\Big(\frac{e^{-{\rm i}k t} - 1}{({\rm i}k)^2} + \frac{t}{{\rm i}k}\Big) + \int_{\mathbb{R}} \frac{\dd k\,e^{-{\rm i}k t}}{{\rm i}\pi k}\,\Big(\int_{0}^{\infty} \frac{\dd t'\,e^{{\rm i}k t'}}{e^{t'} - 1}\Big) 
\eea
The Fourier transform of the measure $\nu$ is denoted:
\beq
\mathcal{F}[\nu](k) = \int \dd\nu(t)\,e^{-{\rm i}kt}
\eeq
It is a $\mathcal{C}^{\infty}$ function of $k$ because $\nu$ is the difference of two probability measures with compact support, and since $\nu(\mathbb{R}) = 0$, we have $\hat{\nu}(k) \in O(k)$ when $k \rightarrow 0$. $\nu$ being real, we also have $\big(\mathcal{F}[\nu](k)\big)^* = \mathcal{F}[\nu](-k)$. Using the representation \eqref{reps} in \eqref{rep}, we find:
\bea
-\iint \dd\nu(t)\dd\nu(t')\,f(t - t') & = & \int_{\mathbb{R}} \dd k\,|\mathcal{F}[\nu](k)|^2\Big[\frac{1}{2\pi k^2} + \frac{1}{{\rm i}\pi k}\int_{0}^{\infty} \dd t'\,\frac{e^{{\rm i}k t'}}{e^{t'} - 1}\Big] \nonumber \\
& = & \int_{0}^{\infty} |\mathcal{F}[\nu](k)|^2\,\gamma(k)\,\frac{\dd k}{k}
\eea
where:
\beq
\gamma(k) = \frac{1}{\pi k} + \frac{1}{\pi} \int_{0}^{\infty} \dd t'\,\frac{2\mathrm{sin}(kt')}{e^{t'} - 1} = \frac{1}{\pi k} + \frac{\pi k\,\mathrm{cotan}(\pi k) - 1}{\pi k} = \mathrm{cotanh}(\pi k).
\eeq
The same computation with $f(t/a_m)$ will result in replacing $\gamma(k)$ by $\gamma(a_mk)$, hence the result of Lemma~\ref{lemm1}:
\beq
\label{QA}\mathcal{Q}[\nu] = \int_{0}^{\infty} Q(k)\,|\mathcal{F}[\nu](k)|^2\,\frac{\dd k}{k},\qquad \hat{Q}(k) = (2 - r)\gamma(k) + \sum_{m = 1}^r \gamma(a_m k)
\eeq

\subsection{Proof of Corollary~\ref{cor1}}
\label{sproof1}
The cotanh function has the $k \rightarrow 0$ expansion:
\beq
\gamma(k) = \frac{1}{\pi k} + O(k)
\eeq
It follows that:
\beq
Q(k) = \frac{\chi}{\pi k} + O(k),\qquad \chi = 2 - r + \sum_{m = 1}^r \frac{1}{a_m}.
\eeq
Therefore, a necessary condition for the positivity of $Q$ is $\chi \geq 0$. We immediately observe that $Q$ is always definite positive when when $r = 1$ or $2$. When $r \geq 3$, it is only positive for a certain region of parameters $a_1,\ldots,a_m$. Its description is not easy, but since $\gamma$ is decreasing, we know that $(a_1,\ldots,a_m)$ gives a positive $Q$, then $(a_1',\ldots,\check{a}_m)$ with $\check{a}_i \leq a_i$ gives a positive $Q$. For the models related to Seifert spaces, we are only interested in $a_m$ taking integer values. We can immediately exclude the uples with $\chi < 0$. Then, the list of uples satisfying to $\chi \geq 0$ with $r \geq 3$ is finite, and positivity can be checked directly case by case:
\begin{itemize}
\item[$\bullet$] For the cases $(2,2,a_3)$, we find:
\beq
Q(k) = - \mathrm{cotanh}(\pi k) + 2\,\mathrm{cotanh}(2\pi k) + \mathrm{cotan}(\pi a_3 k) = \mathrm{tanh}(\pi k) + \mathrm{cotanh}(\pi a_3 k) > 0
\eeq
\item[$\bullet$] For $(2,3,a_3)$ with $a_3 = 3,4,5,6$, it is enough to check positivity in the case $a_3 = 6$ and use the monotonicity property stated above. We have for $(2,3,6)$:
\bea
Q(k) & = & - \mathrm{cotanh}(\pi k) + \mathrm{cotanh}(2\pi k) + \mathrm{cotanh}(3\pi k) + \mathrm{cotanh}(6\pi k) \nonumber \\
& = & 2\mathrm{tanh}^2(\pi k)\,\frac{16\,\mathrm{sinh}^4(\pi k) + 20\,\mathrm{sinh}^2(\pi k) + 5}{(4\,\mathrm{sinh}^2(\pi k) + 1)(4\,\mathrm{sinh}^2(\pi k) + 3)} > 0
\eea
\item[$\bullet$] For $(2,4,4)$, we have:
\bea
Q(k) & = & - \mathrm{cotanh}(\pi k) + \mathrm{cotanh}(2\pi k) + 2\,\mathrm{cotanh}(4\pi k) \nonumber \\
& = & \mathrm{tanh}(\pi k)\,\frac{4\,\mathrm{sinh}^2(\pi k) + 3}{2\,\mathrm{sinh}^2(\pi k) + 1} > 0
\eea
\item[$\bullet$] For $(3,3,3)$, we have:
\beq
Q(k) = -\mathrm{cotanh}(\pi k) + 3\mathrm{cotanh}(3\pi k) = \frac{4\,\mathrm{sinh}(2\pi k)}{4\,\mathrm{sinh}^2(\pi k) + 1} > 0
\eeq
\item[$\bullet$] For $(2,2,2,2)$, we find:
\beq
Q(k) = -2\,\mathrm{cotanh}(\pi k) + 4\,\mathrm{cotanh}(2\pi k) = 2\,\mathrm{tanh}(\pi k) > 0
\eeq
\end{itemize}
This concludes the proof of Corollary~\ref{cor1}.

\subsection{Proof of Theorem~\ref{1cut}}
\label{sproof2}
\noindent\textbf{Proof.} The effective potential is:
\beq
\label{effe}V^{\lambda_{{\rm eq}}}(t) = \frac{t^2}{2u} - \int \Psi(t - t')\,\dd\lambda_{{\rm eq}}(t')
\eeq
where:
\beq
\Psi(t) = (2 - r)\ln\big|\mathrm{sinh}(t/2)\big| + \sum_{m = 1}^r \ln\big|\mathrm{sinh}(t/2a_m)\big|
\eeq
We compute:
\bea
-4\Psi''(t) & = & \frac{2}{\mathrm{sinh}^2(t/2)} + \sum_{m = 1}^r \Big(\frac{1}{a_m^2\mathrm{sinh}^2(t/2a_m)} - \frac{1}{\mathrm{sinh}^2(t/2)}\Big) \nonumber \\
& = & \frac{2}{\mathrm{sinh}^2(t/2)} + \sum_{m = 1}^r \frac{\big(a_m\mathrm{sinh}(t/2a_m) + \mathrm{sinh}(t/2)\big)f_{a_m}(t)}{2a_m^2\mathrm{sinh}^2(t/2a_m)\mathrm{sinh}^2(t/2)} 
\eea
We have introduced $f_{a_m}(t) = 2\mathrm{sinh}(t/2) - 2a_m\mathrm{sinh}(t/2a_m)$. If $a_m \geq 1$, we have $f(t) > 0$ ensuing from $f(0) = 0$ and:
\beq
f'(t) = \mathrm{cosh}(t/2) - \mathrm{cosh}(t/2a_m) \geq 0
\eeq
Therefore, $\Psi$ is a concave function, hence $V^{\lambda_{\rm eq}}$ is a convex function for any fixed $u > 0$. This implies that it must achieve its minimum $C^{\lambda_{{\rm eq}}}$ in a connected set, i.e. the support must be a segment.

We claim that the density of $\lambda_{{\rm eq}}$ vanishes exactly like a squareroot at the edges. By analyticity arguments, we know that it behaves $\sim c|t - \alpha|^{(2k + 1)/2}$ at any edge $\alpha$ of the support, for some integer $k$ and irrelevant constant $c > 0$, but we claim $k = 0$. Indeed, the effective potential $V^{\lambda_{{\rm eq}}}$ behaves like $(t - \alpha)^{(2k + 3)/2}$ when $t \rightarrow \alpha$ from the exterior of the support. However, from the expression, the effective potential is $\mathcal{C}^2$ in the variable $t$ not belonging to the support, and:
\beq
\partial_{t}^2V^{\lambda_{{\rm eq}}}(t) = \frac{1}{u} - \int \Psi''(t - t')\dd\lambda_{{\rm eq}}(t')
\eeq
Since we have shown that $\Psi$ is concave:
\beq
\partial_{t}^2V^{\lambda_{{\rm eq}}}(t) \geq \frac{1}{u}
\eeq
which is only compatible with $k = 0$.

In case the equilibrium measure is unique, $\lambda_{{\rm eq}}$ is even, so its support is of the form $[-\delta,\delta]$, and after the change of variable, we find that the support of  $\check{\lambda}_{{\rm eq}}$ is $\Gamma = [1/\gamma,\gamma]$. In case the equilibrium measure is not unique, and denote $[\delta_-,\delta_+]$ the support of a given equilibrium measure $\lambda_{{\rm eq}}$. Denote $\lambda_{{\rm eq}}^{-}$ the image of $\lambda_{{\rm eq}}$ under $t \mapsto -t$. The characterization of equilbrium measures shows that any convex combination $\beta\lambda_{{\rm eq}} + (1 - \beta)\lambda_{{\rm eq}}^{-}$ is an equilibrium measure. By the previous argument its support $[\delta_{-},\delta_{+}]\cup[-\delta_{+},-\delta_{-}]$ is connected, therefore $\delta_{-} \leq 0 \leq \delta_{+}$, i.e. the support of any equilibrium measure contains $0$. 
 \hfill $\Box$

\subsection{For $\mathrm{SO}$ and $\mathrm{Sp}$ gauge groups}
\label{CvxSO}

We have to repeat the analysis of \S~\ref{S23} for the quadratic form:
\bea
Q_{{\rm BCD}}[\nu] & = & -\iint \dd\nu(t)\dd\nu(t')\,\Big[(2 - r)\ln\big|\mathrm{sinh}\big(\frac{t + t'}{2}\big)\big| + \ln\big|\mathrm{sinh}\big(\frac{t + t'}{2}\big)\big| + \nonumber \\
& & + \sum_{m = 1}^r \ln\big|\mathrm{sinh}\big(\frac{t - t'}{2a_m}\big)\big| + \ln\big|\mathrm{sinh}\big(\frac{t + t'}{2a_m}\big)\big|\Big] 
\eea
over measures $\nu$ which are the difference of two even probability measures. We compute as in \eqref{rep}:
\bea
-\iint \dd\nu(t)\dd\nu(t')\,\ln\big|\mathrm{sinh}\big(\frac{t + t'}{2}\big)\big| & = & -\iint \frac{\dd\nu(t)\dd\nu(-t') + \dd\nu(-t)\dd\nu(t')}{2}\,f(t - t') \nonumber \\
& = & \int_{0}^{\infty} \big[\mathrm{Re}\big(\mathcal{F}[\nu](k)\big)\big]^2\,\gamma(k)\,\frac{\dd k}{k}
\eea
Therefore:
\beq
\mathcal{Q}_{{\rm BCD}}[\nu] = \int_{0}^{\infty} 2 \big[\mathrm{Re}\,\mathcal{F}[\nu](k)\big]^2 \hat{Q}(k)
\eeq
Since $\nu$ is even, its Fourier transform is real and therefore, $\mathcal{Q}_{{\rm BCD}}$ is definite positive iff \eqref{QA} is strictly convex. In that case, one can prove repeating \S~\ref{sproof2} that the equilibrium measure for the BCD series is supported on $1$ segment and behaves exactly as a squareroot at the edges.

\section{Finite subgroups of $\mathrm{PSL}_{2}(\mathbb{C})$}
\label{subgop}

We are studying functional relations with the values of $\phi(x + {\rm i}0)$ and $\phi(x - {\rm i}0)$ on both sides of a cut $\sigma$ are related to values of $\phi$ at $g(x)$, where $g$ is a M\"obius transformation, and the $g$'s involved generate a finite subgroup of $\mathrm{Aut}\,\widehat{\mathbb{C}} = \mathrm{PSL}_{2}(\mathbb{C})$. The list of possibilities, and has an ADE classification: up to conjugation, $G$ must be either cyclic, dihedral, or one or the three polyhedral groups.
\begin{itemize}
\item[$\bullet$] $\mathbb{Z}_n$, cyclic group of order $n$. Generated by $r_n(z) = e^{2{\rm i}\pi/n}z$ -- $A_n$ case.
\item[$\bullet$] $\mathbb{D}_{2n}$, dihedral group of order $2n$. Generated by $r_n$ and $\iota(z) = 1/z$ -- $D_n$ case.
\item[$\bullet$] $\mathbb{A}_4$, tetrahedral group, order $12$. It is also the alternate subgroup of $\mathfrak{S}_4$. It is generated by $r_2$ and $\jmath(z) = {\rm i}\frac{z - 1}{z + 1}$ (order $3$) -- $E_6$ case.
\item[$\bullet$] $\mathbb{S}_4$, octahedral group, order $24$. It is generated by $r_4$ and $\jmath$ -- $E_7$ case.
\item[$\bullet$] $\mathbb{A}_5$, icosahedral group, order $120$. It is generated by $r_2$, $\jmath$, and $\mu(z) = \frac{2{\rm i}z + \beta^*}{-\beta z  - 2{\rm i}}$, with $\beta = (1 - \sqrt{5}) + {\rm i}(1 + \sqrt{5})$ and $\beta^*$ is its complex conjugate (order $5$) -- $E_8$ case.
\end{itemize}

\section{Case $(2,3,3)$: matrices of singularities}
\label{singes}
\label{E6sing}

We have $\mathbf{B}^{(0)}_{{\rm sp}} = \sum_{i = 1}^5 \mathbf{B}^{(0)}_{{\rm sp}}(i)\,\hat{e}_i$ with:
\bea
\mathbf{B}^{(0)}_{{\rm sp}}(1) & = & -\left(\begin{array}{cccccccc}
0 & 0 & 0 & 0 & 1 & 1 & 1 & 1 \\
0 & 1 & 0 & 1 & 1 & 0 & 1 & 0 \\
0 & 0 & 1 &  1 & 0 & 0 & 1 & 1 \\
1 & 0 & 0 & 1 & 0 & 1 & 1 & 0 \\
0 & 1 & 1 & 0 & 1 & 0 & 0 & 1 \\
1 & 1 & 0 & 0 & 1 & 1 & 0 & 0 \\
1 & 0 & 1 & 0 & 1 & 0 & 1 & 0 \\
1 & 1 & 1 & 1 & 0 & 0 & 0 & 0 
\end{array}\right),\qquad \mathbf{B}^{(0)}_{{\rm sp}}(2) = -\left(\begin{array}{cccccccc}
0 & 0 & 0 & 1 & 0 & 1 & 1 & 1 \\
0 & 0 & 1 & 0 & 1 & 0 & 1 & 1 \\
0 & 1 & 0 & 0 & 1 & 1 & 0 & 1 \\
0 & 1 & 0 & 0 & 1 & 1  & 0 & 1 \\
0 & 1 &  1 & 1 & 0 & 0 & 0 & 1 \\
1 & 0 & 0 & 0 & 1 & 1 & 1 & 0 \\
1 & 0 & 1 & 1 & 0 & 0 & 1 & 0 \\
1 & 1 & 0 & 1 & 0 & 1 & 0 & 0 \\
1 & 1 & 1 & 0 & 1 & 0 & 0 & 0
 \end{array}\right) \nonumber \\
 \mathbf{B}^{(0)}_{{\rm sp}}(3) & = & \left(\begin{array}{cccccccc}
 1 & 0 & 0 & 0 & 0 & 0 & 0 & -1 \\
 0 & 0 & 1 & 0 & 0 & -1 & 0 & 0 \\
 0 & 1 & 0 & 0 & 0 & 0 & -1 & 0 \\
 0 & 0 & 0 & 1 & -1 & 0 & 0 & 0 \\
 0 & 0 & 0 & -1 & 1 & 0 & 0 & 0 \\
 0 & 0 & 0 & -1 & 1 & 0 & 0 & 0 \\
 0 & -1 & 0 & 0 & 0 & 0 & 1 & 0 \\
 0 & 0 & -1 & 0 & 0 & 1 & 0 & 0 \\
 -1 & 0 & 0 & 0 & 0 & 0 & 0 & 1
 \end{array}\right), \nonumber \\
\mathbf{B}^{(0)}_{{\rm sp}}(4) & = & -\left(\begin{array}{cccccccc}
0 & 0 & 0 & 0 & 1 & 1 & 1 & 1 \\
0 & 0 & 1 & 1 & 0 & 0 & 1 & 1 \\
0 & 1 & 0 & 1 & 0 & 1 & 0 & 1 \\
1 & 0 & 0 & 1 & 0 & 1 & 1 & 0 \\
0 & 1 & 1 & 0 & 1 & 0 & 0 & 1 \\
1 & 0 & 1 & 0 & 1 & 0 & 1 & 0 \\
1 & 1 & 0 & 0 & 1 & 1 & 0 & 0 \\
1 & 1 & 1 & 1 & 0 & 0 & 0 & 0 
\end{array}\right),\qquad \nonumber \mathbf{B}^{(0)}_{{\rm sp}}(5) = -\left(\begin{array}{cccccccc}
0 & 0 & 0 & 1 & 0 & 1 & 1 & 1 \\
0 & 1 & 0 & 0 & 1 & 1 & 0 & 1 \\
0 & 0 & 1 & 0 & 1 & 0 & 1 & 1 \\
0 & 1 & 1 & 1 & 0 & 0 & 0 & 1 \\
1 & 0 & 0 & 0 & 1 & 1 & 1 & 0 \\
1 & 0 & 0 & 0 & 1 & 1 & 1 & 0 \\
1 & 1 & 0 & 1 & 0 & 1 & 0 & 0 \\
1 & 1 & 1 & 0 & 1 & 0 & 0 & 0
\end{array}\right) \nonumber
\eea

\section{$(2,3,4)$ case: polynomial equations}

\label{234coef}
The $32$ coefficients in the polynomial $\tilde{\mathcal{P}}(\tilde{x},y)$ (see \eqref{234Pxy}) for the spectral curve for $(2,3,4)$ can be expressed solely in terms of the four coefficients $m_k = c^{-k}\,M_k$ for $k \in \{3,4,6,8\}$.
\bea
C_1 & = & -1 - 6m_3 \nonumber \\
C_2 & = & -2(1 + 3m_3 + 6m_4 - 3m_6) \nonumber \\
C_3 & = & 0 \nonumber \\
C_4 & = & 2(-1 -12 m_3 -12m_4 + 18m_6 - 6m_8 + 18m_3^2 - 72m_3m_4 + 12m_4^2) \nonumber \\
C_5 & = & c^{-6}(1 - 6m_3 + 12m_4) \nonumber \\
C_6 & = & -6(4m_3 - 2m_4 - m_6 + 12m_3^2 + 24m_3m_4 - 18m_3m_6 - 24m_4^2 + 12m_4m_6) \nonumber \\
 C_7 & = & 12c^{-6}(m_3 + m_4 - m_6) \nonumber \\
 C_8 & = & -2c^{-12} - 6\big(5m_3 + 2m_4 - 8m_6 + 4m_8 + 96m_3m_4 - 84m_3m_6 + 24m_3m_8 - 32m_4^2 + 36m_4m_6 \nonumber \\
 & & - 6m_6^2 + 96m_3m_4^2\big) \nonumber \\
C_9 & = & c^{-12} \nonumber \\
C_{10} & = & 2c^{-6}(1 + 6m_3)(-1 + 3m_3 - 6m_4) \nonumber \\
C_{11} & = & -2c^{-12} - 6\big(1 - 10m_3 + 28m_4 - 12m_6 - 4m_8 + 4(3m_3^2 -72m_3m_4 + 93m_3m_6 - 36m_3m_8 \nonumber \\
& & + 82m_4^2 - 90m_4m_6 +36m_4m_8 - 3m_6^2) + 72(4m_3^3 - 6m_3^2m_4 - 3m_3^2m_6 + 2m_3m_4m_6 + 4m_4^3)\big) \nonumber \\
C_{12} & = &  2c^{-12} \nonumber \\
C_{13} & = & 6c^{-6}(m_3 + 6m_4 - 8m_6 + 4m_8 -12m_3^2 + 24m_3m_4 + 16m_4^2) \nonumber \\
C_{14} & = & -4c^{-12} + 3\big(1 + 4(m_3 + 6m_4 - 5m_6) + 12(2m_3^2 - 10m_3m_4 + 7m_3m_6 - 6m_3m_8 + 24m_4^2 \nonumber \\
& & - 38m_4m_6 + 8m_4m_8 + 14m_6^2 - 4m_6m_8) + 24(3m_3^2 - 18m_3^2m_4 - 6m_3^2m_8 \nonumber \\
& &  - 18m_3m_4^2 + 12m_3m_4m_6 + 16m_4^3 - 8m_4^2m_6 ) + 144m_3^2(3m_3^2 - 12m_3m_4 + 2m_4^2)\big) \nonumber \\
C_{15} & = & c^{-12}(3 + 18m_3) \nonumber \\
C_{16} & = & -3c^{-6}\big(1 + 4(5m_4 - 3m_6 + m_8) + 4(3m_3^2 + 10m_4^2) + 72m_3^3\big) \nonumber \\
C_{17} & = & -6c^{-12}(1 + 6m_3) + 3\big(1 + 4m_3 + 20m_4 - 38m_6 + 20m_8 - 4(9m_3^2 - 12m_3m_8 - 26m_4^2 \nonumber \\
& &  + 42m_4m_6 - 24m_4m_8 - 9m_6^2  + 6m_6m_8) - 24(12m_3^3 - 27m_3^2m_6 + 12m_3^2m_8 \nonumber \\
& & + 16m_3m_4^2  - 24m_3m_4m_6 + 9m_3m_6^2  - 12m_3m_4m_8 - 16m_4^3 + 10m_4^2m_6 ) \nonumber \\
& & + 144m_3(3m_3^3 - 18m_3^2m_4 - 3m_3^2m_6 + 28m_3m_4^2 - 4m_4^3) \nonumber \\
C_{18} & = & c^{-12}(1 + 18m_3 - 24m_4) \nonumber \\
C_{19} & = & c^{-6}(1 + 6m_3)\big(-1 -12m_3 - 24m_4 + 26m_6 + 72m_3(m_3 + m_4)\big) \nonumber
\eea
\bea
C_{20} & = & c^{-12}(-2 + 36m_3 + 48m_4) + 3\big(1 + 2(13m_3 + 6m_4 - 15m_6 + 6m_8) - 12(m_3^2  - 36m_3m_4 \nonumber \\
& & + 27m_3m_6 - 12m_3m_8 - 2m_4^2 + 20m_4m_6  - 4m_4m_8  - 33m_6^2 + 24m_6m_8 - 4m_8^2)\nonumber \\
& &  - 24(6m_3^3 + 6m_3^2m_4 - 96m_3^2m_6 + 36m_3^2m_8 - 60m_3m_4^2 + 114m_3m_4m_6 \nonumber \\
& & - 48m_3m_4m_8 + 15m_3m_6^2 - 20m_4^3 + 24m_4^2m_6 + 8m_4^2m_8 - 12m_4m_6^2 ) + 96(18m_3^4 \nonumber \\
& & - 36m_3^3m_4 - 27m_3^3m_6 + 54m_3^2m_4^2 + 9m_3^2m_4m_6- 24m_3m_4^3 + 2m_4^4 ) + 2592m_3^5\big) \nonumber \\
C_{21} & = & c^{-12}(2 + 30m_3 - 12m_4) \nonumber \\
C_{22} & = & 6c^{-6}(3m_3 - 4m_4 + 12m_6 - 8m_8 + 24m_3^2 + 24m_3m_4 - 72m_3m_6 + 12m_3m_8 + 16m_4^2 \nonumber \\
& & - 24m_4m_6 + 36m_3^3 - 72m_3^2m_4 - 24m_3m_4^2) \nonumber \\
C_{23} & = & -2c^{-12}(2 + 30m_3 - 12m_4) + 2\big(-39m_3 + 72m_4 - 24m_6 + 6m_8 + 6(3m_3^2 - 108m_3m_4 \nonumber \\
& & + 117m_3m_6 - 26m_3m_8 + 190m_4^2 - 240m_4m_6 + 60m_4m_8  + 90m_6^2 - 36m_6m_8 ) \nonumber \\
& & + 36(6m_3^3 - 18m_3^2m_4 + 18m_3^2m_6 - 12m_3^2m_8 - 12m_3m_4^2 + 36m_3m_4m_6 \nonumber \\
& &  - 9m_3m_6^2 - 6m_3m_6m_8 + 52m_4^3 - 36m_4^2m_6 + 6m_4m_6^2) - 216m_3(6m_3^3 \nonumber \\
& & - 3m_3^2m_6 - 28m_3m_4^2 + 30m_3m_4m_6 - 2m_4^2m_6 ) + 3888m_3^4(m_3 - 2m_4)\big) \nonumber \\
C_{24} & = & 6c^{-12}(5m_3 - m_4 + m_6) \nonumber \\
C_{25} & = & -3c^{-18} - c^{-6}\big(1 + 12(2m_3 + 5m_4 - 4m_6 + m_8) - 12(3m_3^2 -36m_3m_4 + 12m_3m_6 \nonumber \\
& & - 12m_3m_8 - 34m_4^2 + 36m_4m_6 - 3m_6^2) + 72m_3(9m_3^2 + 12m_3m_4 - 4m_4^2)\big) \nonumber \\
C_{26} & = & 12c^{-12}(-5m_3 + 6m_4 - m_6) + 6\big(14m_3 - 26m_4 + 19m_6 - 4m_8 +288m_3m_4 \nonumber \\
& & - 264m_3m_6 + 96m_3m_8 - 400m_4^2 + 540m_4m_6 - 144m_4m_8 - 96m_6^2 + 36m_6m_8 \nonumber \\
& & - 72m_3^3 + 864m_3^2m_4 - 1188m_3^2m_6+ 720m_3^2m_8 -48m_3m_4^2 \nonumber \\
& & + 144m_3m_4(9m_6 - 5m_8) - 36m_3m_6(9m_6 - 4m_8) - 576m_4^3 + 350m_4^2m_6 + 72m_4m_6^2 \nonumber \\
& & - 36m_6^3 - 1296m_3^4 + 216m_3^3(24m_4 + 3m_6 + 2m_8) - 3168m_3^2m_4^2 - 864m_3^2m_4m_6 \nonumber \\
& &  + 288m_3m_4^2(5m_4 - m_6) + 1296m_3^5 + 2592m_3^4m_4 - 864m_3^3m_4^2\big) \nonumber \\
C_{27} & = & c^{-12}\big(-1 + 6(2m_3 - 6m_4 + 3m_6) + 36m_3^2\big) \nonumber \\
C_{28} & = & -3c^{-18} + 3c^{-6}\big(1 + 4(2m_3 + 6m_4 - 3m_6) + 12(2m_3^2 + 4m_3m_4 + 3(-2m_4 + m_6)^2) \nonumber \\
& & - 144m_3^2(m_3 - 2m_4)\big) \nonumber \\
C_{29} & = & 2c^{-12}\big(1 - 6(2m_3 - 6m_4 + 3m_6) - 36m_3^2\big) - 1 + 6(14m_3 - 30m_4 + 33m_6 - 16m_8) \nonumber \\
& & + 12\big(24m_3^2 + 6m_3(19m_4 - 20m_6 + 3m_8) - 200m_4^2 + 372m_4m_6 - 144m_4m_8 \nonumber \\
& & - 231m_6^2 + 180m_6m_8 - 36m_8^2\big) + 216\big(3m_3^3 + 2m_3^2(7m_4 - 23m_6 + 9m_8) \nonumber \\
& & - 2m_3(m_4^2 8m_4m_6 + 12m_4m_8 - 17m_6^2 + 6m_6m_8)\big) - 432\big(39m_3^4 - 6m_3^3(11m_4 \nonumber \\
& & + 8m_6 - m_8) + 3m_3^2(22m_4^2 + 20m_4m_6 + m_6^2) - 12m_3m_4^2(2m_4 + m_6) + 4m_4^4\big) \nonumber \\
C_{30} & = & c^{-12}(1 + 12(m_3 + m_4 - m_6) +36m_3^2) \nonumber \\
C_{31} & = & -6c^{-18} + 2c^{-6}\big(1 - 3(m_3 - 8m_4 + 16m_6 - 8m_8) - 6(15m_3^2 - 12m_3m_4 \nonumber \\
& & - 6m_3m_6 - 6m_3m_8 + 8m_4^2 - 12m_4m_6 + 6m_6^2) + 36(9m_3^3 - 2m_3m_4^2)\big) \nonumber
\eea
\bea
C_{32} & = & 2c^{-12}\big(1 + 12(m_3 + m_4 + m_6) + 36m_3^2\big) - 1 - 12(8m_4 - 5m_6 + m_8) \nonumber \\
& & + 12\big(3m_3^2 - 3m_3(8m_4 - 5m_6 + 8m_8) - 118m_4^2 + 12m_4(9m_6 - m_8) + 3m_6(7m_6 - 4m_8)\big) \nonumber \\
& & + 72\big(12m_3^3 - 3m_3^2(18m_4 - 3m_6 + 4m_8) + m_3(-40m_4^2 + 84m_4m_6 - 48m_4m_8 \nonumber \\
& & + 6m_6m_8) - 68m_4^3 + 100m_4^2m_6 - 42m_4m_6^2 + 6m_6^3\big) - 432m_3(15m_3^3 \nonumber \\
& & + 3m_3^2(4m_4  + 9m_6 - 4m_8) + m_3(44m_4^2 + 24m_4m_6 - 9m_6^2) + 2m_4^2(-8m_4 + m_6)\big) \nonumber \\
& & + 5184m_3^3(3m_3^2 - 9m_3m_4 + 2m_4^2)
\eea

\section{$(2,3,5)$ case: additional formulas}
\label{235more}
\subsection{The basis of $\delta$'s}
\label{deltaE8}
In \S~\ref{e8e8}, we have found a $\langle$orthonormal$\rangle$ basis $\delta_{1},\ldots,\delta_{8}$ of $E \subseteq \hat{E}$. Its expression in terms of the canonical basis of $\hat{E}$ is:
\bea
\delta_1 & = & \frac{1}{60}\big(\hat{e}_0 - 6\hat{e}_1 + \hat{e}_2 - \hat{e}_3 - 9\hat{e}_4 - \hat{e}_5 - 4\hat{e}_6 - \hat{e}_7 + \hat{e}_8 - 11\hat{e}_{9} \nonumber \\
& & + \hat{e}_{10} - 6\hat{e}_{11} + \hat{e}_{12} -\hat{e}_{13} - 9\hat{e}_{14} - \hat{e}_{15} - 4\hat{e}_{16} - \hat{e}_{17} + \hat{e}_{18} - 11\hat{e}_{19} \nonumber \\
& & + \hat{e}_{20} - 6\hat{e}_{21} + \hat{e}_{22} - \hat{e}_{23} - 9\hat{e}_{24} - \hat{e}_{25} - 4\hat{e}_{26} - \hat{e}_{27} + \hat{e}_{28} - 11\hat{e}_{29}\big) \nonumber \\
\delta_2 & = & \frac{1}{60}\big(\hat{e}_0 - 6\hat{e}_1 - 9\hat{e}_2 - \hat{e}_3 + \hat{e}_4 - \hat{e}_5 - 4\hat{e}_6 - 11\hat{e}_7 + \hat{e}_8 - \hat{e}_9 \nonumber \\
& & + \hat{e}_{10} - 6\hat{e}_{11} - 9\hat{e}_{12} - \hat{e}_{13} + \hat{e}_{14} - \hat{e}_{15} - 4\hat{e}_{16} - 11\hat{e}_{17} + \hat{e}_{18} - \hat{e}_{19} \nonumber \\
& & + \hat{e}_{20} - 6\hat{e}_{21} - 9\hat{e}_{22} - \hat{e}_{23} + \hat{e}_{24} - \hat{e}_{25} - 4\hat{e}_{26} - 11\hat{e}_{27} + \hat{e}_{28} - \hat{e}_{29}\big) \nonumber \\
\delta_3 & = & \frac{1}{60}\big(3\hat{e}_0 + 3\hat{e}_2 + \hat{e}_3 - 3\hat{e}_4 + \hat{e}_5 + 8\hat{e}_6 - 5\hat{e}_7 + 3\hat{e}_8 + \hat{e}_9 \nonumber \\
& & - 3\hat{e}_{10} + 6\hat{e}_{11} + 3\hat{e}_{12} - 5\hat{e}_{13} + \hat{e}_{14} + \hat{e}_{15} + 2\hat{e}_{16} + \hat{e}_{17} + 3\hat{e}_{18} - 5\hat{e}_{19} \nonumber \\
& & + 3\hat{e}_{20} + 6\hat{e}_{21} - 3\hat{e}_{22} + \hat{e}_{23} + 3\hat{e}_{24} - 5\hat{e}_{25} + 8\hat{e}_{26} + \hat{e}_{27} - 3\hat{e}_{28} + \hat{e}_{29}\big) \nonumber \\
\delta_4 & = & -\varepsilon_{-10}(\delta_3) \nonumber \\
\delta_5 & = & \iota_2(f_2) \nonumber \\
\delta_6 & = & -\varepsilon_{10}(\delta_3) \nonumber \\
\delta_7 & = & \iota_2(f_1) \nonumber \\
\delta_8 & = & \frac{1}{60}\big(11\hat{e}_0 + 4\hat{e}_1 + 11\hat{e}_2 + 9\hat{e}_3 + 11\hat{e}_4 + 9\hat{e}_5 + 6\hat{e}_6 + 9\hat{e}_7 + 11\hat{e}_8 + 9\hat{e}_9 \nonumber \\
& & + 11\hat{e}_{10} + 4\hat{e}_{11} + 11\hat{e}_{12} + 9\hat{e}_{13} + 11\hat{e}_{14} + 9\hat{e}_{15} + 6\hat{e}_{16} + 9\hat{e}_{17} + 11\hat{e}_{18} + 9\hat{e}_{19} \nonumber \\
& & + 11\hat{e}_{20} + 4\hat{e}_{21} + 11\hat{e}_{22} + 9\hat{e}_{23} + 11\hat{e}_{24} + 9\hat{e}_{25} + 6\hat{e}_{26} + 9\hat{e}_{27} + 11\hat{e}_{28} + 9\hat{e}_{29}\big),
\eea
where $\iota_k$ is a linear operator reverting the order of the canonical basis, and $\varepsilon_{k}$ is the shift:
\beq
\forall j \in \mathbb{Z}_{30},\qquad \iota_2(\hat{e}_{j}) = \hat{e}_{2-j},\qquad \varepsilon_{k}(\hat{e}_j) = \hat{e}_{j - k}
\eeq

\subsection{Two-point function}

$\hat{w'} = \frac{1}{2274168727208063179297410229829}\,\sum_{k \in \mathbb{Z}_{30}} w'_k\,\hat{e}_k$ with the following coefficients:
\bea
w'_0 & = & 2005578886156579626234908709016 \nonumber \\
w'_1 & = & -181850671509615045845883695564 \nonumber \\
w'_2 & = &  -398832024581226871603715770986 \nonumber \\
w'_3 & = & 494023035051327637314292878686 \nonumber \\
w'_4 & = & -72481828443769717690937743838 \nonumber \\
w'_5 & = & -478705354064380740163879665333 \nonumber \\
w'_6 & = & -272221635191729696024320996996 \nonumber \\
w'_7 & = &  531983641045281729652509079403 \nonumber \\
w'_8 & = & 80216473933964133567145532937 \nonumber \\
w'_9 & = & -883637678946883204685258803332 \nonumber \\
w'_{10} & = & 41855358216453085931546308030 \nonumber \\
w'_{11} & = &  488985776484254042177463041119 \nonumber \\
w'_{12} & = & -189525571049715363422540854413 \nonumber \\
w'_{13} & = & -698861535159392744131953611950 \nonumber \\
w'_{14} & = & 416719334036478522315020112595 \nonumber \\
w'_{15} & = & -63665858634089577006092269410 \nonumber \\
w'_{16} & = & -331136157745715176730745158182 \nonumber \\
w'_{17} & = & -130237342982970710128084520404 \nonumber \\
w'_{18} & = &  265307029451540743194145338690 \nonumber \\
w'_{19} & = &  808107200558064801506713158162 \nonumber \\
w'_{20} & = & -1296344276492615387004191464157 \nonumber \\
w'_{21} & = &  -172825427679090502479437872628 \nonumber \\
w'_{22} & = &  540357624573790525657246218981 \nonumber \\
w'_{23} & = & 288569873075710680412572798533 \nonumber \\
w'_{24} & = &  -829286452152780352970281977032 \nonumber \\
w'_{25} & = & -208718755181947007992291618146 \nonumber \\
w'_{26} & = & 469048115641896378902924682251 \nonumber \\
w'_{27} & = & -353746327005159310155414152581 \nonumber \\
w'_{28} & = & -429254876353150450356202936896 \nonumber \\
w'_{29} & = & 560579424948889951524745253445 \nonumber
\eea

\newpage

\bibliographystyle{plain}

\end{document}